\documentclass[twoside,onecolumn,11pt,a4paper]{book}
\usepackage{anuthesis}
\usepackage{fancyhdr}
\usepackage{graphicx}
\usepackage{epstopdf}
\usepackage{setspace}
\usepackage{color}

\begin{document}
\pagenumbering{roman}

\begin{titlepage}
\title{\textbf{The impact of magnetic geometry on wave modes in cylindrical plasmas}\\[2cm]}
\author{\textbf{Lei Chang}\\[6cm]
\textbf{A thesis submitted for the degree of}\\
\textbf{Doctor of Philosophy} \\
\textbf{of The Australian National University}\\[1cm]}
\date{\textbf{October 2013}}
\maketitle
\end{titlepage}

\sloppy
\chapter*{Declaration}
\addcontentsline{toc}{chapter}{Declaration}

This thesis is an account of research undertaken between November 2009 and October 2013 at The Research School of Physics and Engineering, College of Physical and Mathematical Sciences, The Australian National University, Canberra, Australia. Except where acknowledged in the customary manner, the material presented in this thesis is, to the best of my knowledge, original and has not been submitted in whole or part for a degree in any university.

This thesis is supervised by a supervision panel including A/Prof Matthew J. Hole (principal supervisor), Prof Boris N. Breizman (advisor, The University of Texas at Austin), A/Prof Boyd D. Blackwell (co-supervisor for Australian Institute of Nuclear Science and Engineering), Dr Cormac S. Corr (advisor), Emeritus Prof Robert L. Dewar (advisor), and Emeritus Prof Rod W. Boswell (advisor).

\vspace{20mm}
\hspace{80mm}\rule{40mm}{.15mm}\par
\hspace{80mm} Lei Chang\par
\hspace{80mm} October 2013

\sloppy
\chapter*{Dedication}
\addcontentsline{toc}{chapter}{Dedication}
To my son, Tianrui Chang. 

\chapter*{Acknowledgements}
\addcontentsline{toc}{chapter}{Acknowledgements}

Foremost I would like to thank Matthew for his huge amount of time and effort dedicated to my project. I almost knew nothing about fusion/helicon plasma when started my PhD, and he taught me from the very beginning with great patience. The way he explains physics by analogies is intuitional and humorous, making the study with him enjoyable. The great funding support he provided for me to attend conferences, summer and winter schools, and to visit overseas institutes, is appreciated from my heart. Life is like running on an up-and-down hill road and so is PhD, filled with frustration and excitation. I am very grateful to his many times consoling encouragement when I felt frustrated with my PhD. By the way, running with him around the Lake Burley Griffin and Black Mountain during lunch time is one of the most memorable and enjoyable experiences I had in Canberra. It is usually hard to do research and administration at the same time, but Matthew manages to do it and performs very well. Being the leader of the Plasma Theory and Modelling (PTM) group, he makes every member efficient in doing research and also his own research productive. I am grateful to his unconscious influence regarding this fascinating management ability. 

Deep appreciation also goes to Boris whose sharp mind in physics and mathematics makes doing research with him very enjoyable and productive. I really appreciate the academic attitude he taught me: research for solving problems, not papers, which will effect my whole academic career in the future. His inputs regarding the gap eigenmode study of radially localised helicon waves and shear Alfv\'{e}n waves enhance the quality of this thesis remarkably. I feel very much indebted to the numerous discussions he had with me through the Skype and email. His considerate arrangements for my visit at the University of Texas at Austin are also deeply appreciated. 

I am also deeply grateful to Bob, who provided me (together with Matthew) the opportunity to study my PhD in PTM. Bob made considerate arrangements for my first arrival, and offered many enlightening discussions later on during my study. His answers are always short and sometimes even ``too fast to catch" but all key to solving problems. I feel lucky to benefit from his knowledgeable and brilliant mind. 

The project could not go well without the experimental support from Boyd, Cormac and Juan. Their data makes my modelling work trustworthy and more meaningful. Special thanks go to Boyd and Cormac for their kind support for my top-up scholarship application from the Australian Institute of Nuclear Science and Engineering (AINSE), and many inspiring discussions regarding the modelling work on the MAGnetised Plasma Interaction Experiment (MAGPIE). I also appreciate the encouragement that Cormac gave to me when I was unconfident about my PhD and felt blue with my life. 

I am also ever grateful to other colleagues who provided direct support for my PhD. Guangye kindly provided the ElectroMagnetic Solver (EMS) code and many instructions for its usage, with which Alex helped me getting started. Inspiring and brilliant suggestions always came from Michael, talking with whom saved me much time and made me confident and sunshine, and Graham who is so excellent in numerical computation that my project benefits remarkably, especially the TwO-fluid Electromagnetic Flowing pLasma (TOEFL) model computation. Greg made a number of constructive suggestions regarding solving ordinary and partial differential equations from purely mathematical perspective, enlightening me impressively. Rod, Christine and Trevor offered many insightful suggestions about the wave modelling study in helicon plasmas. Jason, Mat and Ashley kindly offered much help in using Mathematica. Zhisong helped me running the EMS code with automatic file saving during frequency scans. Also, many thanks go to John, Uyen, Julia, Maxine, Karen and Heeok for their great administrative support, and Julie and James for their IT support.

Furthermore, I would like to thank my dear teachers in China. I have been very grateful to the recommendation letters provided by Lie, Baoliang and Yonggui, which played an important role in helping me getting the ANU offer. Lie continuously encouraged me through my whole PhD study, and has been kindly seeking for job opportunities for me regarding my research. 

Last but not least, I would like to say ``thank you very much" to my family. It has been $22$ years now since I started my primary school in $1991$, and $18$ years now since I had the last Mid-Autumn Day with my parents and little brother. I feel extremely indebted to their eternal love and firm support these years, and guilty for my long-time absence. Surely the one I am most grateful to is my wife, Huijie, for her great support, as always sincere understanding, eternal love, and our little son, Tianrui. ``I am only here today because of you. You are the reason I am. You are all my reasons." I also deeply appreciate my parents-in-law for taking care of Tianrui when I was away for my PhD. 

This thesis received funding support from the Chinese Scholarship Council through scholarship $2009611029$, the AINSE through Postgraduate Research Award, the Australian Institute of Physics through Student Conference Support, the ANU Vice-Chancellor's Travel Grant, and Plasma Research Laboratory. 
\chapter*{Publications}
\addcontentsline{toc}{chapter}{Publications}

This thesis has resulted in three publications in peer reviewed journals and a manuscript in preparation, as listed below. Some of the results presented in the following chapters have been adapted from the material in these publications and the manuscript.

\begin{flushleft}
\textbf{L. Chang}, M. J. Hole, and C. S. Corr

\emph{A flowing plasma model to describe drift waves in a cylindrical helicon discharge}

\textbf{Physics of Plasmas} 18, 042106 (2011)
\end{flushleft}

\begin{flushleft}
\textbf{L. Chang}, M. J. Hole, J. F. Caneses. G. Chen, B. D. Blackwell, and C. S. Corr

\emph{Wave modeling in a cylindrical non-uniform helicon discharge}

\textbf{Physics of Plasmas} 19, 083511 (2012)
\end{flushleft}

\begin{flushleft}
\textbf{L. Chang}, B. N. Breizman and M. J. Hole

\emph{Gap eigenmode of radially localised helicon waves in a periodic structure}

\textbf{Plasma Physics and Controlled Fusion} 55, 025003 (2013)
\end{flushleft}

\begin{flushleft}
\textbf{L. Chang}, B. N. Breizman and M. J. Hole

\emph{Gap eigenmode of shear Alfv\'{e}n waves in a periodic structure}

\textbf{(In preparation)}
\end{flushleft}
\chapter*{Abstract}  % the * means don't put a number in the title
\addcontentsline{toc}{chapter}{Abstract} % but add it to the table of contents
Both space and laboratory plasmas can be associated with static magnetic field, and the field geometry varies from uniform to non-uniform. This thesis investigates the impact of magnetic geometry on wave modes in cylindrical plasmas. The cylindrical configuration is chosen so as to explore this impact in a tractable but experimentally realisable configuration. Three magnetic geometries are considered: uniform, focused and rippled. 

For a uniform magnetic field, wave oscillations in a plasma cylinder with axial flow and azimuthal rotation are modelled through a two-fluid flowing plasma model. The model provides a qualitatively consistent description of the plasma configuration on a Radio Frequency (RF) generated linear magnetised plasma (WOMBAT, Waves On Magnetised Beams And Turbulence [Boswell and Porteous, Appl. Phys. Lett. $50$, $1130$ ($1987$)]), and yields agreement between measured and predicted dependences of the wave oscillation frequency with axial field strength. The radial profile of the density perturbation predicted by this model is consistent with the data. Parameter scans show that the dispersion curve is sensitive to the axial field strength and the electron temperature, and the dependence of the oscillation frequency with electron temperature matches the experiment. These results consolidate earlier claims that the density and floating potential oscillations are a resistive drift mode, driven by the density gradient. This, to our knowledge, is the first detailed physics modelling of plasma flows in the diffusion region away from the RF source. 

For a focused magnetic field, wave propagations in a pinched plasma (MAGPIE, MAGnetised Plasma Interaction Experiment [Blackwell et al., Plasma Sources Sci. Technol. $21$, $055033$ ($2012$)]) are modelled through an ElectroMagnetic Solver (EMS) based on Maxwell's equations and a cold plasma dielectric tensor.[Chen et. al., Phys. Plasmas $13$, $123507$ ($2006$)] The solver produces axial and radial profiles of wave magnitude and phase that are consistent with measurements, for an enhancement factor of $9.5$ to the electron-ion Coulomb collision frequency and a $12\%$ reduction in the antenna radius. It is found that helicon waves have weaker attenuation away from the antenna in a focused field compared to a uniform field. This may be consistent with observations of increased ionisation efficiency and plasma production in a non-uniform field. The relationship between plasma density, static magnetic field strength and axial wavelength agrees well with a simple theory developed previously. Moreover, the wave amplitude is lowered and the power deposited into the core plasma decreases as the enhancement factor to the electron-ion Coulomb collision frequency increases, possibly due to the stronger edge heating for higher collision frequencies. 

For a rippled magnetic field, the spectra of radially localised helicon (RLH) waves [Breizman and Arefiev, Phys. Rev. Lett. $84$, $3863$ ($2000$)] and shear Alfv\'{e}n waves (SAW) in a cold plasma cylinder are investigated. A gap-mode analysis of the RLH waves is first derived and then generalised to ion cyclotron range of frequencies for SAW. The EMS is employed to model the spectral gap and gap eigenmode. For both the RLH waves and SAW, it is demonstrated that the computed gap frequency and gap width agree well with the theoretical analysis, and a discrete eigenmode is formed inside the gap by introducing a defect to the system's periodicity. The axial wavelength of the gap eigenmode is close to twice the system's periodicity, which is consistent with Bragg's law, and the decay length agrees well with the analytical estimate. Experimental realisation of a gap eigenmode on a linear plasma device such as the LArge Plasma Device (LAPD) [Gekelman et al., Rev. Sci. Instrum. 62, 2875 (1991)] may be possible by introducing a symmetry-breaking defect to the system's periodicity. Such basic science studies could provide the possibility to accelerate the science of gap mode formation and mode drive in toroidal fusion plasmas, where gap modes are introduced by symmetry-breaking due to toroidicity, plasma ellipticity and higher order shaping  effects.

These studies suggest suppressing drift waves in a uniformly magnetised plasma by increasing the field strength, enhancing the efficiency of helicon wave production of plasma by using a focused magnetic field, and forming a gap eigenmode on a linear plasma device by introducing a local defect to the system's periodicity, which is useful for understanding the gap-mode formation and interaction with energetic particles in fusion plasmas.

\tableofcontents
\listoffigures
\listoftables
\pagenumbering{arabic}
\setcounter{page}{1}

\pagenumbering{arabic} % switch to Arabic numerals for page numbers
\setcounter{page}{1}  % set page number to 1
\chapter{Introduction}\label{chp1}

``A plasma is a quasineutral gas of charged and neutral particles which exhibits collective behaviour'' \cite{Chen:1984aa} It is often referred to as the ``fourth" state of matter, following the well-known solid, liquid and gaseous states, because of the special collective behaviour involved. It was first identified by Sir William Crookes in $1879$ in a Crookes tube, who named it ``radiant matter".\cite{Crookes:1879aa} The term ``plasma" was coined by Irving Langmuir in $1927$ considering that the electrified fluid carries high velocity electrons, ions and impurities in a similar way of the blood plasma which carries red and white corpuscles and germs.\cite{Langmuir:1928aa, Smith:1971aa} The Greek word ``$\pi\lambda\acute{\alpha}\sigma\mu\alpha$" (``plasma") means ``moldable substance". Since the mercury arc that Irving Langmuir used to study oscillations in ionised gases diffuses through the whole glass chamber, and molds itself, this may be another reason for Irving Langmuir to call the ionised gas ``plasma".\cite{Goldston:1995aa}

Although the plasma rarely exists in our natural life due to its high temperature, more than $90\%$ of the matter of the universe is believed to be composed of plasmas, e. g. stellar interiors and atmospheres, gaseous nebulae and much of the interstellar medium, excluding the more speculative nature of dark matter.\cite{Chen:1984aa, Goedbloed:2004aa} The field of plasma physics can be dated back to $1920$s when Langmuir, Tonks and their collaborators worked on gas discharges. It was significantly advanced later by the nuclear fusion program and industrial applications of plasma processing.\cite{Boswell:1997aa, Chen:1997aa} Plasma physics is the underlying science of space and solar physics, astrophysics, and finds applications in diverse areas of MagnetoHydroDynamics (MHD) energy conversion and ion propulsion, solid state plasmas and gas lasers.\cite{Chen:1984aa}

Both space and laboratory plasmas can be associated with static magnetic field, and the field geometry varies from uniform to non-uniform. This thesis investigates the impact of magnetic geometry on wave modes in cylindrical plasmas. The cylindrical configuration is chosen so as to explore this impact in a tractable but experimentally realisable configuration. Three magnetic geometries are considered: uniform, focused and rippled. 

This chapter first introduces the plasma source employed in the thesis, and then illustrates four magnetic geometries. Specifically, the dispersion relation of helicon waves, which drive the helicon discharge and will be considered in detail in the following chapters, is derived from a cold-plasma dielectric tensor perspective. Then the concept of helicon discharge is given, together with its typical applications. We select low temperature helicon plasma sources for comparison to experiments, both because of availability, and diagnostic opportunities. Finally, examples of helicon discharge with uniform (WOMBAT, Waves On Magnetised Beams And Turbulence\cite{Boswell:1987aa}) and focused magnetic fields (MAGPIE, MAGnetised Plasma Interaction Experiment\cite{Blackwell:2012aa}), large linear plasma device with multiple magnetic mirrors (LAPD, LArge Plasma Device\cite{Gekelman:1991aa}), and magnetic confinement fusion with toroidal magnetic field (tokamak) are given. Gap eigenmodes that will be formed in a linear plasma with rippled magnetic field, by introducing a local defect to the system's periodicity, are applicable to understanding the gap-mode formation and its interaction with energetic particles in tokamaks.\cite{Heidbrink:2008aa} 

\section{Plasma source}\label{wav1}
Although a cold plasma model, which assumes zero-temperature frictionless fluids of ions and electrons, ignores physics such as pressure and flow, it is tractable in its mathematical analysis and provides a remarkably accurate description of the common small-amplitude perturbations that are possible for a hot plasma.\cite{Stix:1992aa} Particularly, it is applicable to plasmas and physics for which the thermal effects and particle motions are ignorable, e. g low-temperature plasma and long-time scale problems. For the helicon discharge studied in this thesis, the temperature is low, and so we utilise the cold plasma model to describe helicon waves and shear Alfv\'{e}n waves. The approach that is followed is essentially that of Stix, and Gurnett and Bhattacharjee.\cite{Stix:1992aa, Gurnett:2005aa}.

\subsection{Helicon wave}
\subsubsection{Conductivity tensor and dielectric tensor}
We start from the first of Maxwell's equations
\begin{equation}\label{eq1_1}
\nabla\times\mathbf{B}=\mu_0 \mathbf{j}+\mu_0\epsilon_0\frac{\partial\mathbf{E}}{\partial t}=\mu_0\frac{\partial \mathbf{D}}{\partial t},
\end{equation}
where the plasma current $\mathbf{j}$ and the vacuum displacement $\epsilon_0\partial\mathbf{E}/\partial t$ have been combined into the electric displacement $\mathbf{D}$. Here, $\mathbf{E}$ and $\mathbf{B}$ are the electric and magnetic fields, respectively, $\mu_0$ is the permeability of free space, $\epsilon_0$ is the permittivity of free space, and $t$ is time. The conductivity tensor $\mathbf{\tilde{\sigma}}$ and the dielectric tensor $\mathbf{\tilde{\epsilon}}$ are defined as: 
\begin{equation}\label{eq1_2}
\mathbf{j}=\mathbf{\tilde{\sigma}}\cdot\mathbf{E},~\mathbf{D}=\epsilon_0\mathbf{\tilde{\epsilon}}\cdot\mathbf{E}, 
\end{equation}
respectively. We substitute Eq.~(\ref{eq1_2}) into Eq.~(\ref{eq1_1}) and consider perturbations of the general form $\exp[i(\mathbf{k}\cdot\mathbf{r}-\omega t)]$, with $\mathbf{k}$ the wave vector, $\mathbf{r}$ the space vector, and $\omega$ the wave frequency. The relation between $\mathbf{\tilde{\sigma}}$ and $\mathbf{\tilde{\epsilon}}$ is thus
\begin{equation}\label{eq1_3}
\mathbf{\tilde{\epsilon}}=\mathbf{\tilde{1}}-\frac{\mathbf{\tilde{\sigma}}}{i\omega\epsilon_0},
\end{equation}
with $\mathbf{\tilde{1}}$ the unit tensor and $i$ the unit imaginary number. The plasma is assumed to be uniform, unbounded and immersed in a uniform static magnetic field $\mathbf{B_0}$. For a cold plasma, the equilibrium velocities of particles are zero, and the equilibrium electric field vanishes. Therefore, a set of linear equations that govern the plasma system is formed: 
\begin{equation}\label{eq1_4}
\mathbf{E}=\mathbf{E_1},~\mathbf{B}=\mathbf{B_0}+\mathbf{B_1},~\mathbf{v_s}=\mathbf{v_{s1}},~n_s=n_{s0}+n_{s1},
\end{equation}
where $\mathbf{v_s}$ is the particle velocity and $n_s$ is the plasma density. The subscript $s$ denotes the plasma species, namely electrons and ions, $0$ denotes the equilibrium values, and $1$ denotes the small-amplitude perturbations. The linearised current density and momentum equation are thus:
\begin{equation}\label{eq1_5}
\mathbf{j_1}=\sum_s q_s n_{s0} \mathbf{v_{s1}},
\end{equation}
\begin{equation}\label{eq1_6}
m_s\frac{\partial\mathbf{v_{s1}}}{\partial t}=q_s\mathbf{E_1}+q_s\mathbf{v_{s1}}\times\mathbf{B_0},
\end{equation}
with $q_s$ and $m_s$ the particle charge and mass respectively. Substituting the solved $\mathbf{v_{s1}}$ from Eq.~(\ref{eq1_6}) into Eq.~(\ref{eq1_5}) and recalling $\mathbf{j_1}=\mathbf{\tilde{\sigma}}\cdot\mathbf{E_1}$, we get the conductivity tensor
\begin{equation}\label{eq1_7}
\mathbf{\tilde{\sigma}}=\sum_s\frac{n_{s0}q_s^2}{m_s}\left[
\begin{array}{ccc}
\frac{-i\omega}{\omega_{cs}^2-\omega^2}&\frac{\omega_{cs}}{\omega_{cs}^2-\omega^2}&0\\
\frac{-\omega_{cs}}{\omega_{cs}^2-\omega^2}&\frac{-i\omega}{\omega_{cs}^2-\omega^2}&0\\
0&0&\frac{i}{\omega}
\end{array}
\right],
\end{equation}
where $\omega_{cs}=q_s |\mathbf{B_0}|/m_s$ is the cyclotron frequency. According to Eq.~(\ref{eq1_3}), the dielectric tensor is thereby\cite{Stix:1992aa}
\begin{equation}\label{eq1_8}
\mathbf{\tilde{\epsilon}}=\left[
\begin{array}{ccc}
S&-iD&0\\
iD&S&0\\
0&0&P
\end{array}
\right].
\end{equation}
Here, the quantities $S$ (for sum), $D$ (for difference), and $P$ (for plasma) are defined as:
\begin{equation}\label{eq1_9}
S=(R+L)/2,~D=(R-L)/2,\\
\end{equation}
\begin{equation}\label{eq1_10}
R=1-\sum_s\frac{\omega_{ps}^2}{\omega(\omega+\omega_{cs})},~L=1-\sum_s\frac{\omega_{ps}^2}{\omega(\omega-\omega_{cs})},~P=1-\sum_s\frac{\omega_{ps}^2}{\omega^2},
\end{equation}
and $\omega_{ps}=\sqrt{q_s^2 n_{s0}/({\epsilon_0 m_s})}$ is the plasma frequency. It will be shown later (through Eq.~(\ref{eq1_17}) and Eq.~(\ref{eq1_18})) that the parameters $R$ and $L$ stand for right-hand circularly polarised (RHCP) and left-hand circularly polarised (LHCP) modes, respectively. Equations~(\ref{eq1_8})-~(\ref{eq1_10}) represent the well-known cold-plasma dielectric tensor which has been used widely for plasma wave analyses, and will be employed below to derive the dispersion relation of helicon waves. 

\subsubsection{Whistlers: unbounded helicon waves}
\begin{figure}[ht]
\begin{center}
\hspace{-0.2cm}\includegraphics[width=0.7\textwidth,angle=0]{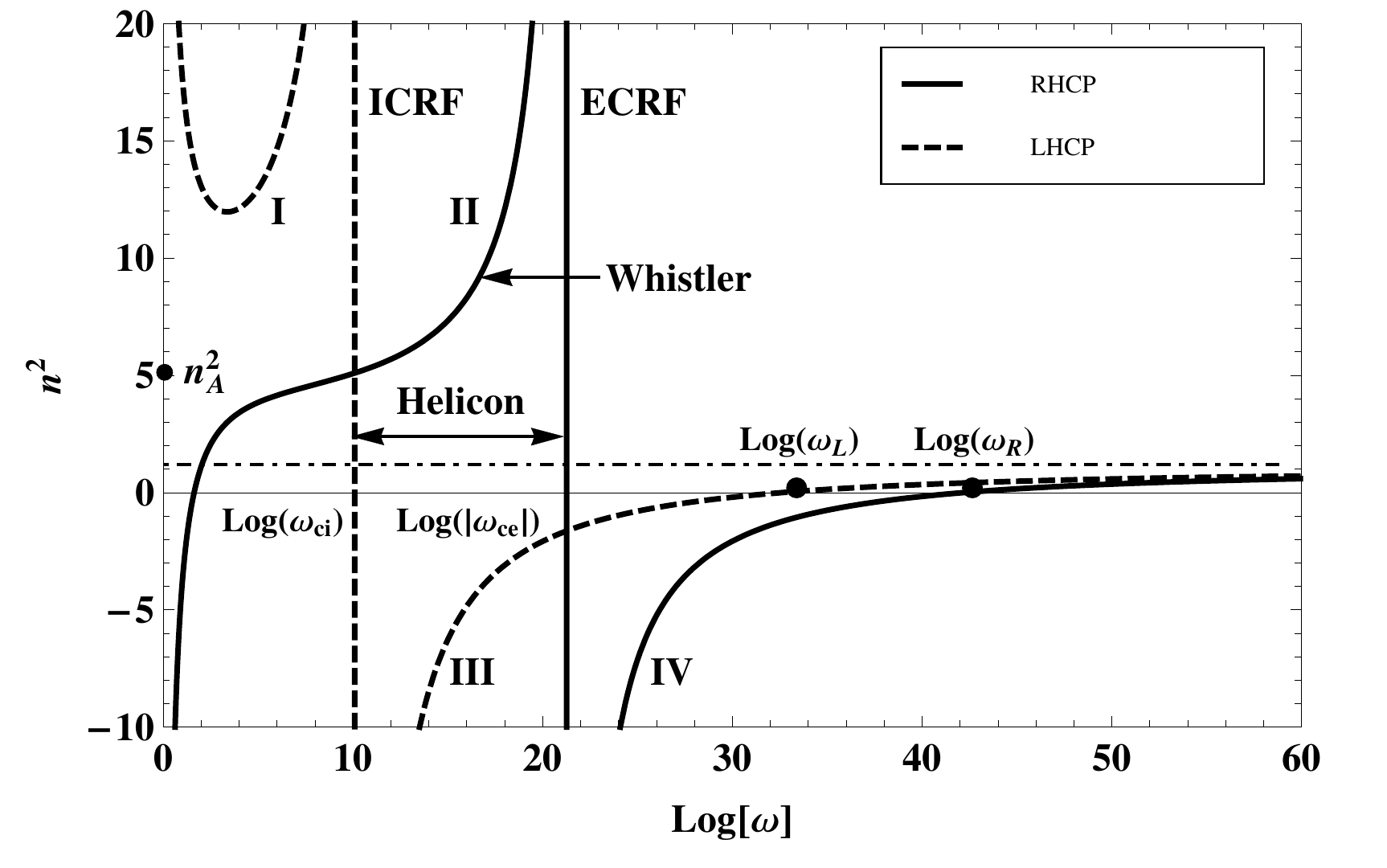}\\
\end{center}
\caption{Dispersion curves of waves propagating parallel to the static magnetic field in a cold, unbounded, and uniform plasma: I, shear (or slow) Alfv\'{e}n waves; II, compressional (or fast) Alfv\'{e}n waves which is also called whistler mode; III and IV are free space modes for $\omega>\omega_{L}$ (cut-off frequency for the LHCP mode) and $\omega>\omega_{R}$ (cut-off frequency for the RHCP mode), respectively.}
\label{fg1_1}
\end{figure}

We first study the normal modes of this unbounded plasma system. Combining Eq.~(\ref{eq1_1}) with the Faraday's law
\begin{equation}\label{eq1_11}
\nabla\times\mathbf{E_1}=-\frac{\partial\mathbf{B_1}}{\partial t},
\end{equation}
we get the homogeneous-plasma wave equation
\begin{equation}\label{eq1_12}
\mathbf{k}\times(\mathbf{k}\times\mathbf{E_1})+\frac{\omega^2}{c^2}\mathbf{\tilde{\epsilon}}\cdot\mathbf{E_1}=0,
\end{equation}
which can be written in a more convenient form
\begin{equation}\label{eq1_13}
\mathbf{n}\times(\mathbf{n}\times\mathbf{E_1})+\mathbf{\tilde{\epsilon}}\cdot\mathbf{E_1}=0
\end{equation}
by introducing the dimensionless index of refraction vector $\mathbf{n}=\mathbf{k}c/\omega$. The magnitude of $n=|\mathbf{n}|=c/(\omega/k)$ is the ratio of the speed of light $c=1/\sqrt{\mu_0\epsilon_0}$ to the wave phase velocity $v_{\mathrm{ph}}=\omega/k$, with $k=|\mathbf{k}|$ the wave number. Without losing the generality, we choose a 3D Cartesian coordinate system $(\hat{x},~\hat{y},~\hat{z})$ and assume $\mathbf{n}$ to be in the $\hat{x}-\hat{z}$ plane and $\mathbf{B_0}$ along the $\hat{z}$ axis. The angle between $\mathbf{n}$ and $\mathbf{B_0}$ is $\vartheta$. Equation~(\ref{eq1_13}) can be written in a matrix form\cite{Stix:1992aa}
\begin{equation}\label{eq1_14}
\left[
\begin{array}{ccc}
S-n^2 \cos^2{\vartheta}&-iD&n^2\sin{\vartheta}\cos{\vartheta}\\
iD&S-n^2&0\\
n^2\sin{\vartheta}\cos{\vartheta}&0&P-n^2\sin^2{\vartheta}
\end{array}
\right]\left[
\begin{array}{c}
E_{1\hat{x}}\\
E_{1\hat{y}}\\
E_{1\hat{z}}
\end{array}
\right]=0,
\end{equation}
which has nontrivial solutions of $\mathbf{E_1}$ only if the determinant of the matrix equals zero. The zero-value determinant results in
\begin{equation}\label{eq1_15}
\tan^2{\vartheta}=\frac{-P(n^2-R)(n^2-L)}{(S n^2-R L)(n^2-P)},
\end{equation}
describing the normal modes (no source term in Eq.~(\ref{eq1_13})) of the system. For modes propagating parallel to the static magnetic field, namely $\vartheta=0$, the electric field eigenvectors associated with each of the three roots of Eq.~(\ref{eq1_15}) are:
\begin{equation}\label{eq1_16}
P=0:~\mathbf{E_1}=(0,~0,~E_{1\hat{z}}),
\end{equation}
\begin{equation}\label{eq1_17}
n^2=R:~\mathbf{E_1}=(E_{1\hat{x}},~i E_{1\hat{x}},~0),
\end{equation}
\begin{equation}\label{eq1_18}
n^2=L:~\mathbf{E_1}=(E_{1\hat{x}},~-i E_{1\hat{x}},~0).
\end{equation}
Here, the $\hat{y}$ component of the electric field has been expressed in form of the $\hat{x}$ component. Equation~(\ref{eq1_16}) represents the electron plasma oscillations which occur at the plasma frequency according to Eq.~(\ref{eq1_10}), i. e. $\omega^2=\omega_{pi}^2+\omega_{pe}^2\approx\omega_{pe}^2$ ($m_e\ll m_i$), and do not propagate. This is the same as that in an unmagnetised plasma because the presence of the static magnetic field does not affect particle motions (driven by $E_{1\hat{z}}$) in the parallel direction. This mode is called longitudinal mode due to the parallel wave vector to the electric field. Equations~(\ref{eq1_17}) and (\ref{eq1_18}) represent transverse modes because of $\mathbf{k}\perp\mathbf{E_1}$, and from the phases of $E_{1\hat{x}}$ and $E_{1\hat{y}}$ we can see that they are RHCP and LHCP modes, respectively. The dispersion relations of these two modes are plotted in Fig.~\ref{fg1_1} for a singly ionised argon plasma, as an example, with density $n_{s0}=10^{18}~\mathrm{m^{-3}}$ and field strength $|\mathbf{B_0}|=0.01$~T. 
While Fig.~\ref{fg1_1} is the straight visualisation of Eqs.~(\ref{eq1_17})-(\ref{eq1_18}), as done by Ginzburg,\cite{Ginzburg:1970aa} the singularity at $\omega=0$ is ``unreal" and can be removed by introducing the quasineutral condition of $q_e n_{e0}+q_i n_{i0}=0$. In the fact, for this low frequency range ($\omega\ll\omega_{ci}$) of waves, careful analysis shows that\cite{Gurnett:2005aa}
\begin{equation}\label{eq1_19}
n^2\approx 1+\frac{\omega_{pi}^2}{\omega_{ci}^2}=1+\frac{c^2}{v_A^2}=n_A^2.
\end{equation}
Here, $v_A=B_0/\sqrt{\mu_0 n_{i0} m_i}$ and $n_A$ are the phase velocity and refraction of index of Alfv\'{e}n waves, as labeled in Fig.~\ref{fg1_1}. Branch I is called shear (or slow) Alfv\'{e}n waves (SAW), which will be used to form a gap eigenmode in Chapter~\ref{chp5}. Due to the same direction of the ion cyclotron motion and wave polarisation, this branch sees the ion cyclotron resonance when $\omega\rightarrow\omega_{ci}$, forming the so-called ion cyclotron range of frequency (ICRF) waves. The ICRF waves have an important role in heating ions in fusion plasmas.\cite{Swanson:1985aa} Similarly, the branch II, named compressional (or fast) Alfv\'{e}n waves, sees the electron cyclotron resonance when $\omega\rightarrow|\omega_{ce}|$, because of the same direction of the electron cyclotron motion and wave polarisation. The electron cyclotron range of frequency (ECRF) waves have been used widely in studying the energy transport, driving the current, and diagnosing the radial profile of electron temperature. Waves polarised in the opposite direction of particle cyclotron motions do not resonate with these particles, so that the LHCP branch does not see the electron cyclotron motion and the RHCP branch does not see the ion cyclotron motion. Branch II is also called whistlers, which for $\omega_{ci}\ll\omega\ll |\omega_{ce}|$ are viewed as unbounded helicon waves.\cite{Boswell:1997aa} The dispersion relation of whistlers for $\omega_{ci}\ll\omega\ll |\omega_{ce}|$ can be obtained by neglecting the ion terms and assuming $|\omega_{ce}|\ll\omega_{pe}$ in Eq.~(\ref{eq1_17})
\begin{equation}\label{eq1_20}
n^2=\frac{\omega_{pe}^2}{\omega(|\omega_{ce}|-\omega)}\approx \frac{\omega_{pe}^2}{\omega |\omega_{ce}|},
\end{equation}
which further gives the corresponding group velocity
\begin{equation}\label{eq1_21}
v_g=2c\frac{\omega^{1/2}(|\omega_{ce}|-\omega)^{3/2}}{|\omega_{ce}|\omega_{pe}}\approx 2c\frac{\omega^{1/2}|\omega_{ce}|^{1/2}}{\omega_{pe}}.
\end{equation}
Equation~(\ref{eq1_21}) shows that the group velocity of whistlers increases with the wave frequency. This implies that a received frequency of a finite duration burst will decrease in tone, leading to the name ``whistler". It was first observed around the second half of the World War I,\cite{Boswell:1997aa} reported later by Barkhausen in $1919$,\cite{Barkhausen:1919aa, Barkhausen:1930aa} and explained by Storey in $1953$.\cite{Storey:1953aa} It can be used to measure the electron density due to $n^2 \propto n_e$, and believed to be the only diagnostic for the electron density at high altitudes above the Earth's ionosphere, before direct in situ measurements were available.\cite{Gurnett:2005aa} Interestingly, there exists another type of whistlers which, however, have a rising tone with the increasing time. They occur near the ion cyclotron frequency on the branch I, and are thereby called ion cyclotron whistlers.\cite{Gurnett:1965aa} The ion cyclotron whistlers cannot be observed from the ground but satellites outside of the ionosphere.\cite{Gurnett:2005aa}

\subsubsection{Helicon waves in a cylindrical uniform plasma}\label{uni1}
When the plasma is bounded, e. g. in a cylindrical geometry, the wave field of whistlers needs to meet the either conducting or insulating boundary conditions, and their pure electromagnetic features cannot be kept. Then they become helicon waves, named after the associated helical force lines which rotate and carry electrons.\cite{Boswell:1997aa} The word ``helicon" was first suggested by Aigrain in $1960$ to describe an electromagnetic wave which propagates in solid metals at low temperatures and with frequencies $\omega_{ci}\ll\omega\ll |\omega_{ce}|$.\cite{Aigrain:1960aa} The dispersion relation of helicon waves in a cylindrical plasma with uniform density was first derived by KMT,\cite{Klozenberg:1965aa} and more recently by Chen and Arnush.\cite{Chen:1991aa, Chen:1997ab, Arnush:1997aa} We shall follow the steps of the latter.

For a cylindrical coordinate system ($r$, $\theta$, $z$), the exponential factor changes to $\exp[i(m\theta+k_z z-\omega t)]$ with $k_z$ the axial wave number and $m$ the azimuthal mode number. We introduce a total current $\mathbf{j_{t}}=\partial\mathbf{D}/\partial t$ and recall $\mathbf{D}=\epsilon_0\mathbf{\tilde{\epsilon}}\cdot\mathbf{E}$ to get
\begin{equation}\label{eq1_22}
-i\omega\epsilon_0 \mathbf{E_1}=\mathbf{\tilde{\epsilon}}^{-1}\cdot\mathbf{j_{t1}}=
\left[
\begin{array}{ccc}
\frac{S}{R L}&\frac{i D}{R L}&0\\
-\frac{i D}{R L}&\frac{S}{R L}&0\\
0&0&\frac{1}{P}
\end{array}
\right]
\left[
\begin{array}{c}
j_{t1r}\\
j_{t1\theta}\\
j_{t1z}
\end{array}
\right],
\end{equation}
which can be equivalently written as
\begin{equation}\label{eq1_23}
-i\omega\epsilon_0 \mathbf{E_1}=\alpha_c \mathbf{j_{t1}}+i\alpha_h(\mathbf{\hat{z}}\times\mathbf{j_{t1}})+\alpha_d[\mathbf{\hat{z}}\times(\mathbf{\hat{z}}\times\mathbf{j_{t1}})].
\end{equation}
Here, the relation $S^2-D^2=R L$ (according to Eq.~(\ref{eq1_9})) has been used, $\alpha_c=1/P$ is the conduction or polarisation current, $\alpha_h=-D/(R L)$ is the Hall current or $\mathbf{E}\times\mathbf{B}$ drift, $\alpha_d=1/P-S/(R L)$ is the displacement current, and $\mathbf{\hat{z}}$ is the unit vector along $z$.\cite{Chen:1997ab} Assuming that the plasma is sufficiently dense ($|\omega_{ce}|\ll\omega_{pe}$) to ensure the ignorable displacement current compared to the plasma current and the wave frequency is in range of $\sqrt{\omega_{ci}|\omega_{ce}|}\ll\omega\ll |\omega_{ce}|$ to ignore the ion current,\cite{Breizman:2000aa} we neglect the ion terms in Eqs.~(\ref{eq1_10}) and get
\begin{equation}\label{eq1_24}
R=-\frac{\omega_{pe}^2}{\omega(\omega+\omega_{ce})},~L=-\frac{\omega_{pe}^2}{\omega(\omega-\omega_{ce})},~P=-\frac{\omega_{pe}^2}{\omega^2}.
\end{equation}
Equation~(\ref{eq1_24}) results in $\alpha_c=-\omega^2/\omega_{pe}^2$, $\alpha_h=-\omega\omega_{ce}/\omega_{pe}^2$, $\alpha_d=0$, and simplifies Eq.~(\ref{eq1_23}) to
\begin{equation}\label{eq1_25}
\mathbf{E_1}=\frac{\omega_{ce}}{\epsilon_0\omega_{pe}^2}(i\delta_{rt}\mathbf{j_{t1}}+\mathbf{\hat{z}}\times\mathbf{j_{t1}})
\end{equation}
with $\delta_{rt}=-\omega/\omega_{ce}$. Combining Eq.~(\ref{eq1_25}) with Eq.~(\ref{eq1_1}) (neglect the vacuum displacement $\epsilon_0\partial\mathbf{E_1}/\partial t$) and Eq.~(\ref{eq1_11}), we have
\begin{equation}\label{eq1_26}
\delta_{rt}\nabla\times\nabla\times\mathbf{B_1}-k_z\nabla\times\mathbf{B_1}+k_w^2\mathbf{B_1}=0
\end{equation}
with $k_w^2=-\mu_0 \omega n_{e0} q_e/B_0=\delta_{rt} k_s^2$. Here, $k_w$ is the wave number of helicon waves in free space (namely whistlers according to Eq.~(\ref{eq1_20})), and $k_s=\omega_{pe}/c$ is the skin number standing for the decay constant of electromagnetic waves penetrating into the plasma. Equation~(\ref{eq1_26}) can be factored into\cite{Klozenberg:1965aa}
\begin{equation}\label{eq1_27}
(\beta_1-\nabla\times)(\beta_2-\nabla\times)\mathbf{B_1}=0,  
\end{equation}
with $\beta_1$ and $\beta_2$ the separation constants representing the roots of 
\begin{equation}\label{eq1_28}
\delta_{rt}\beta^2-k_z\beta+k_w^2=0. 
\end{equation}

\begin{figure}[h]
\begin{center}$
\begin{array}{ccccc}
\varrho_{0}=-\infty&\varrho_{0}=-1&\varrho_{0}=0&\varrho_{0}=1&\varrho_{0}=\infty\\
\hspace{-0.1cm}\includegraphics[width=0.18\textwidth,angle=0]{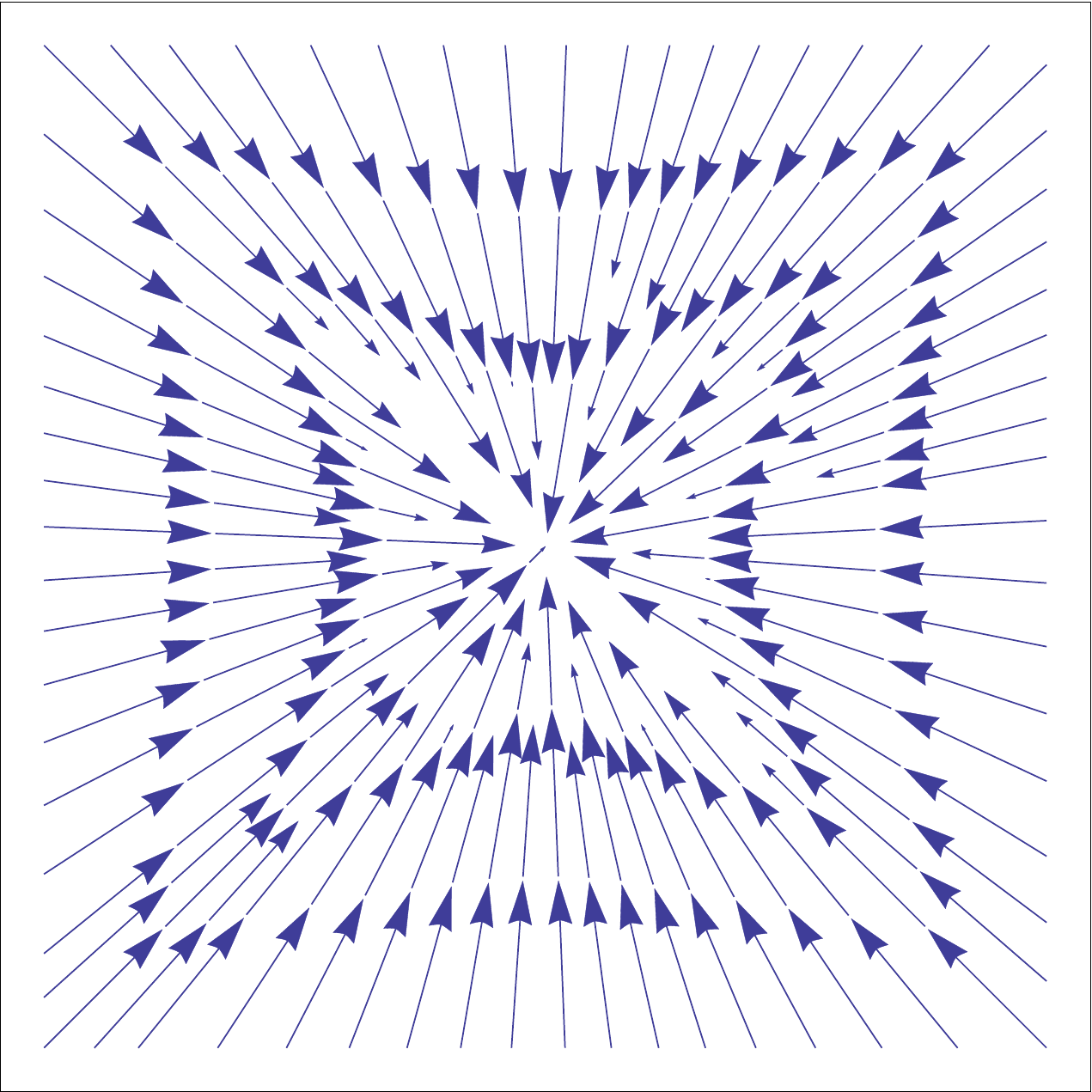}&\includegraphics[width=0.18\textwidth,angle=0]{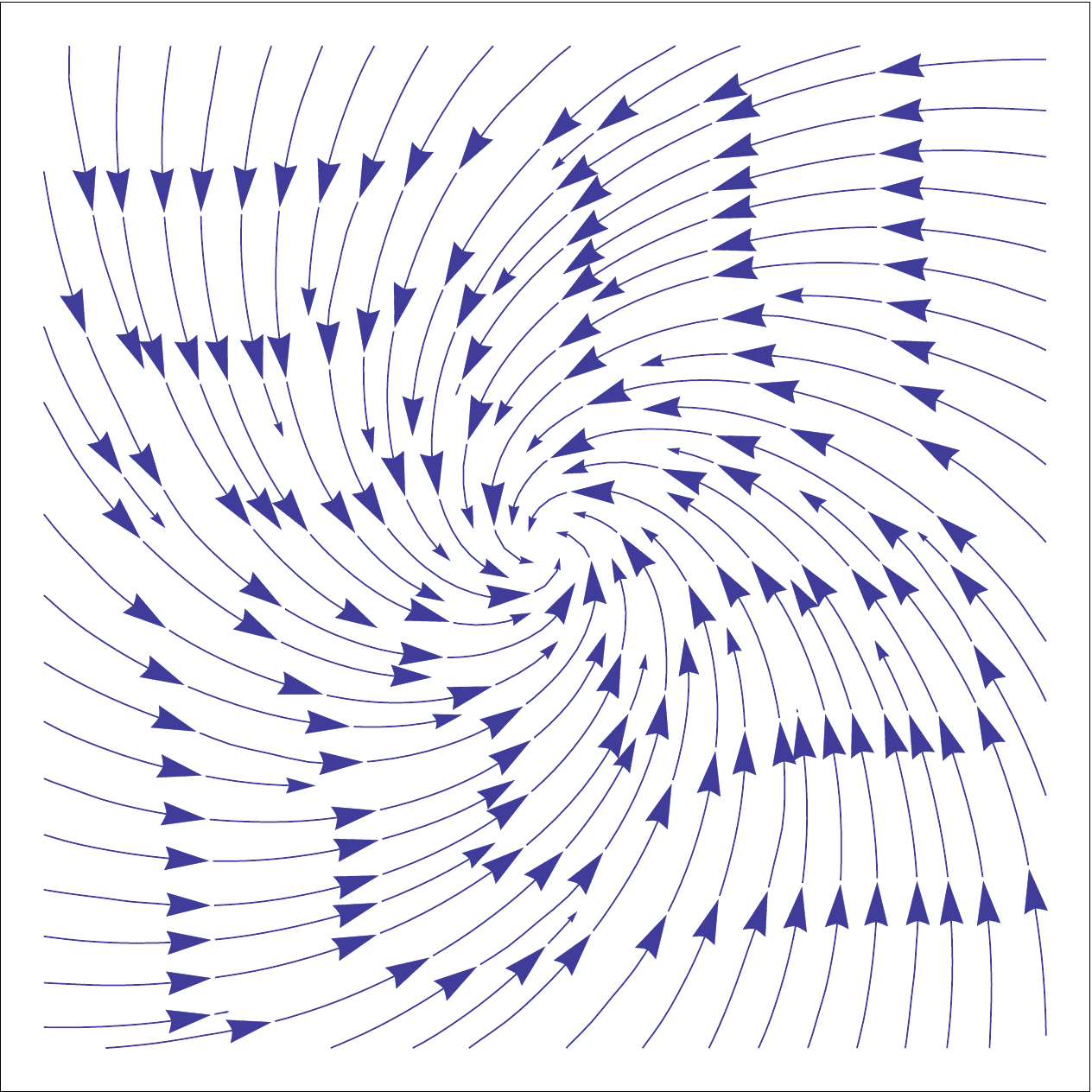}&\includegraphics[width=0.18\textwidth,angle=0]{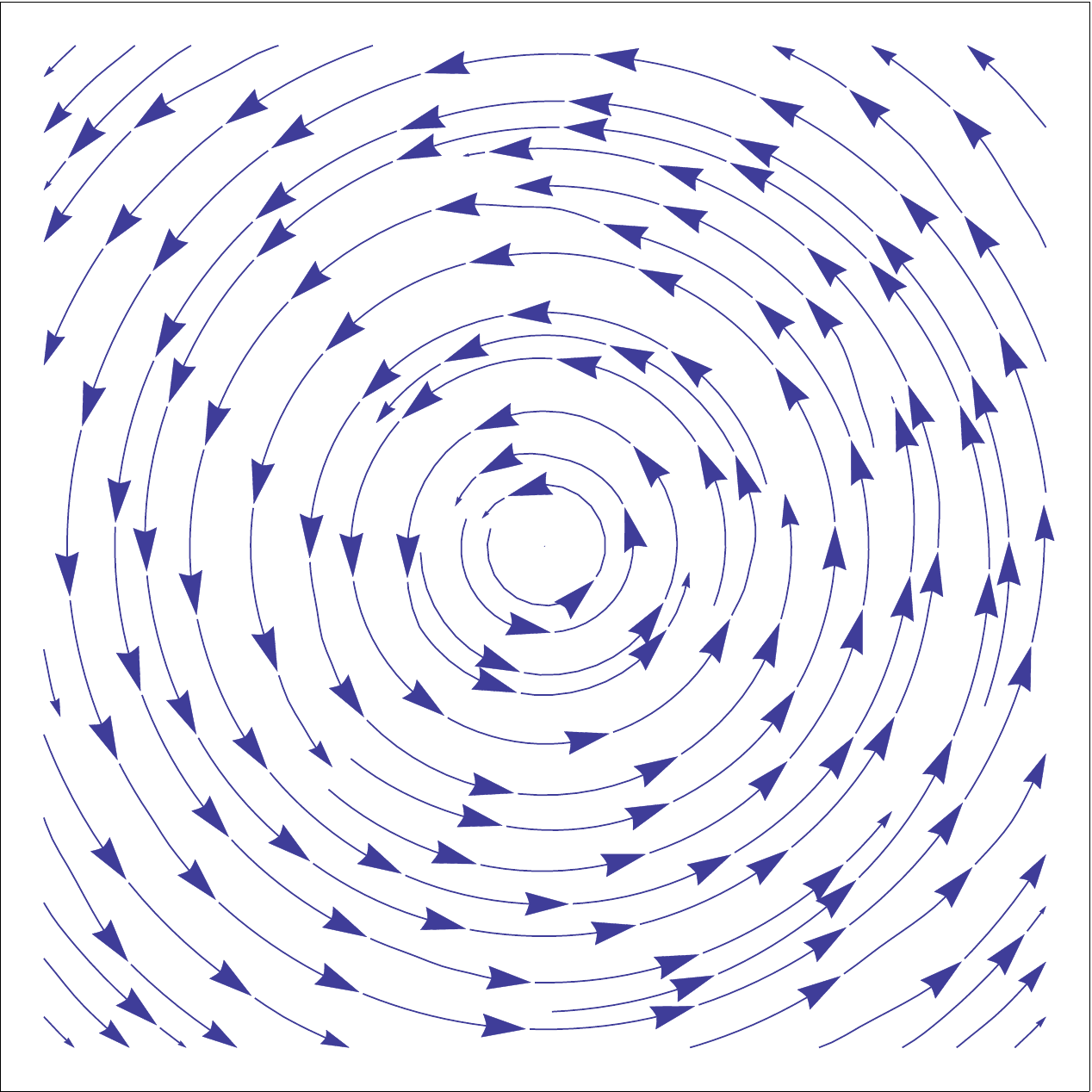}&\includegraphics[width=0.18\textwidth,angle=0]{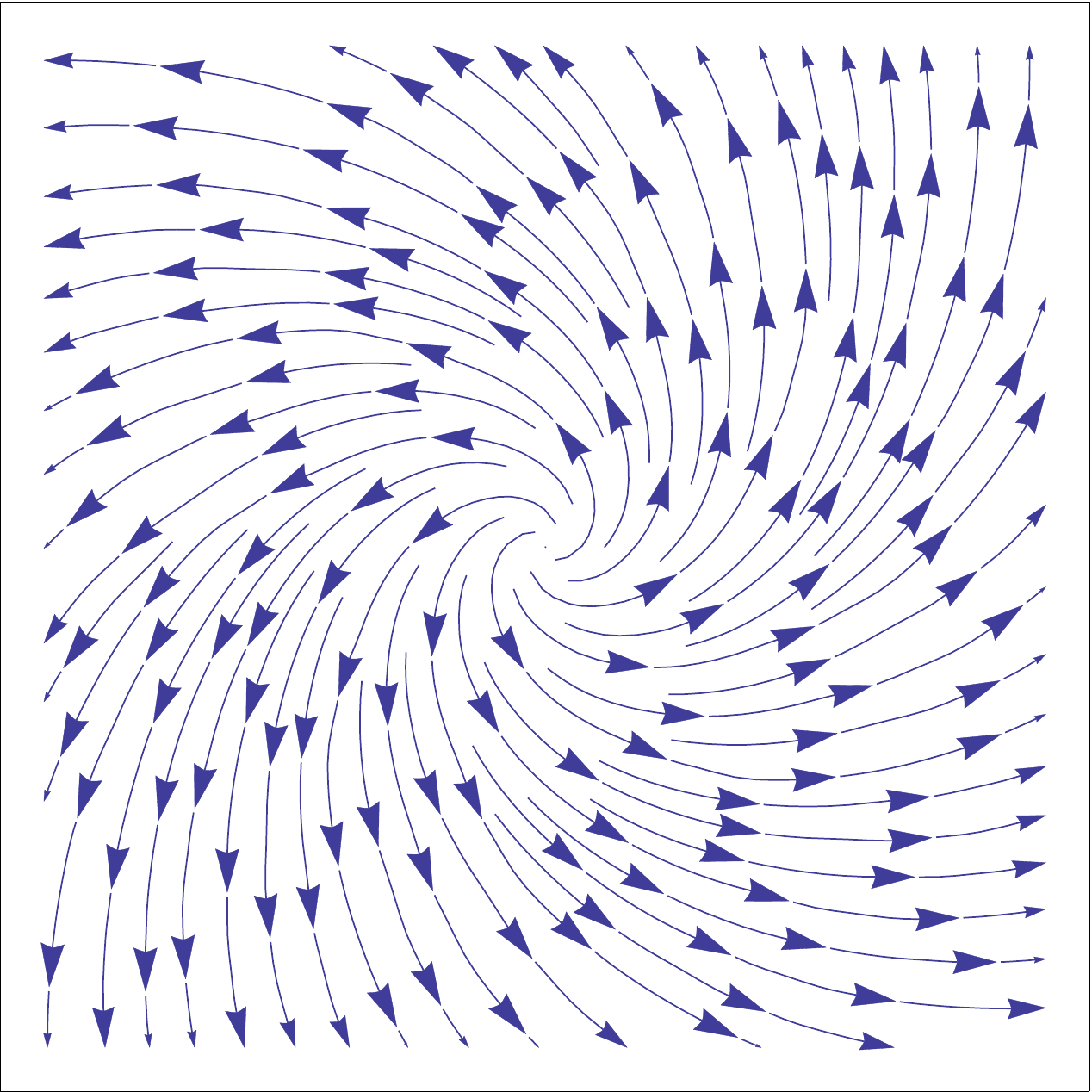}&\includegraphics[width=0.18\textwidth,angle=0]{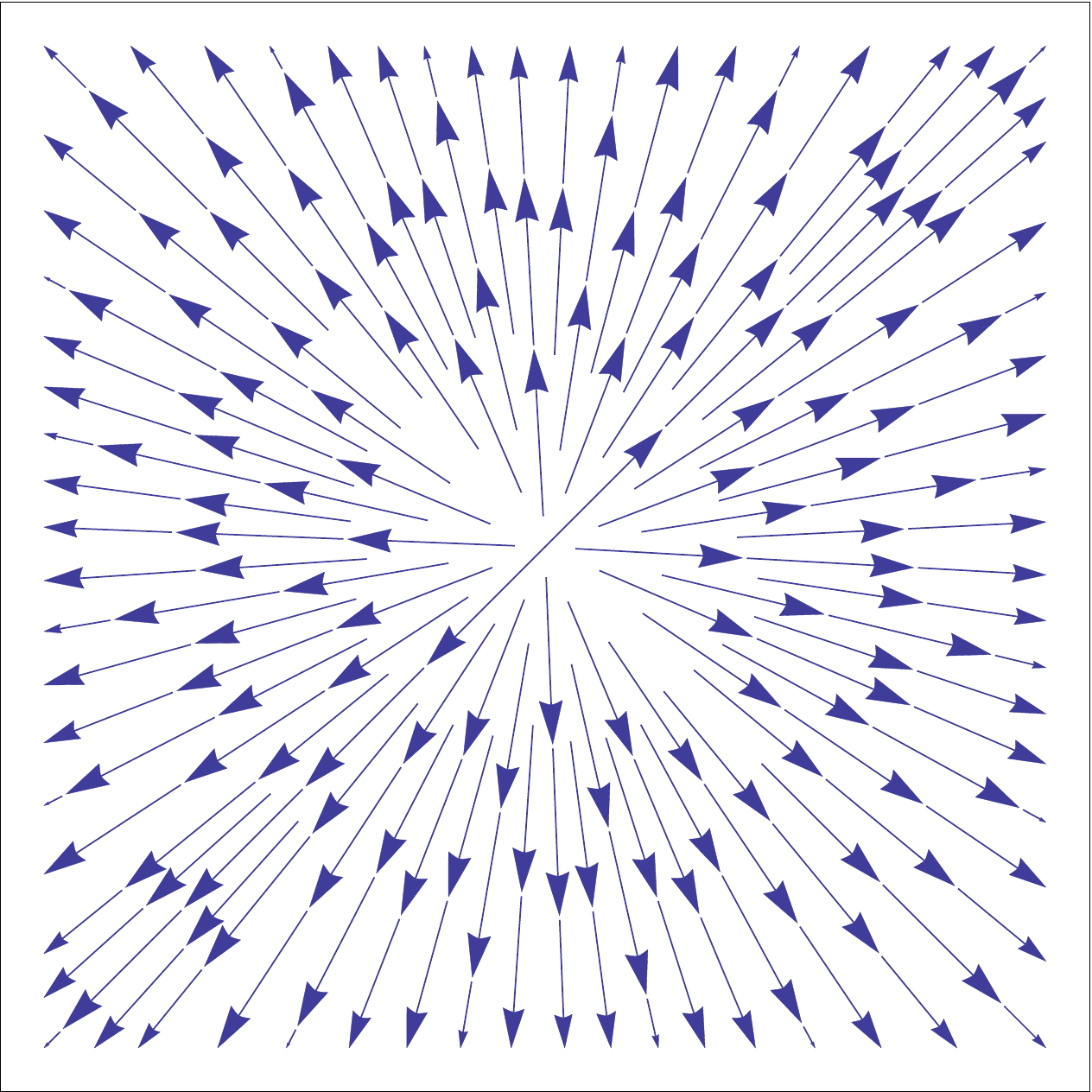}\\
\hspace{-0.1cm}\includegraphics[width=0.18\textwidth,angle=0]{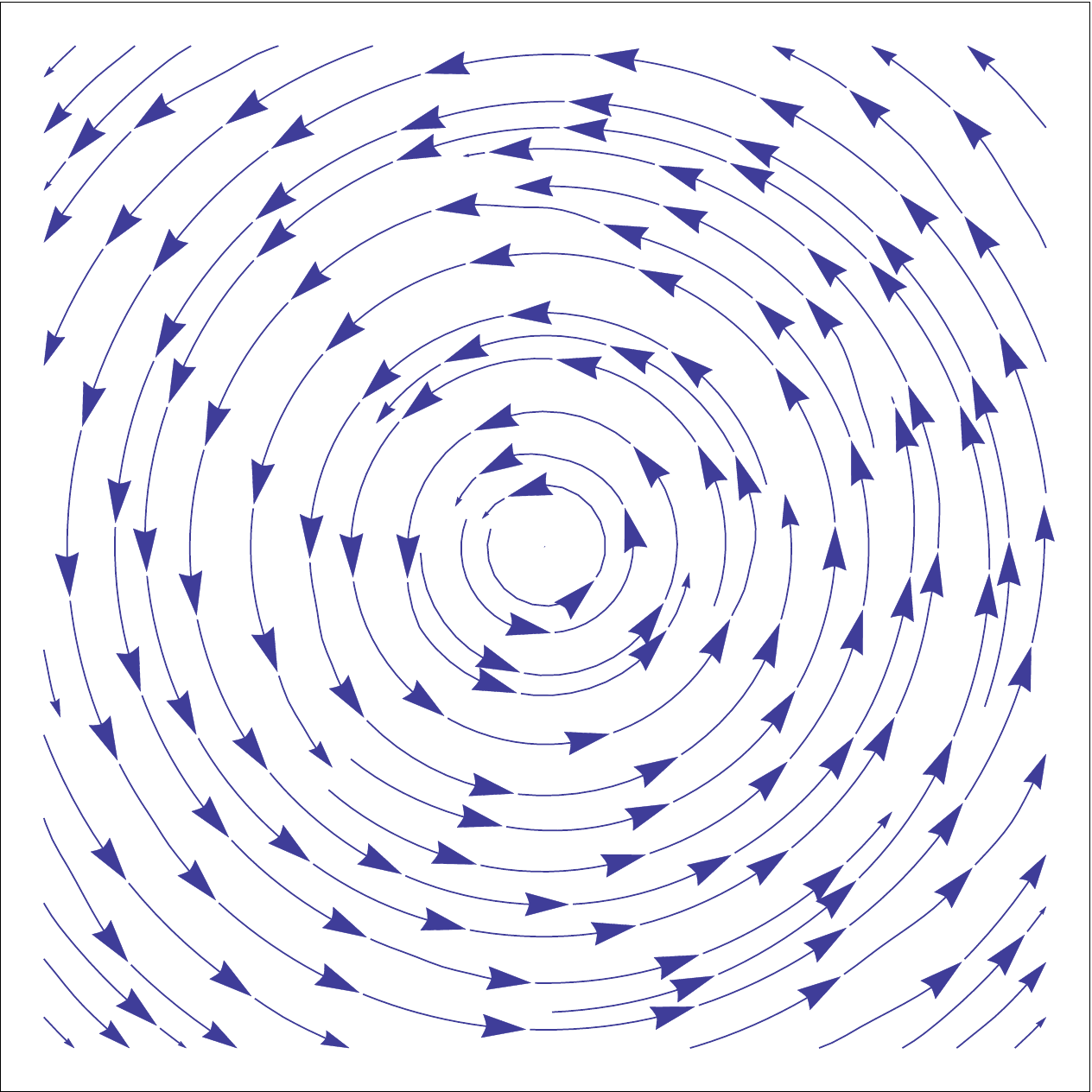}&\includegraphics[width=0.18\textwidth,angle=0]{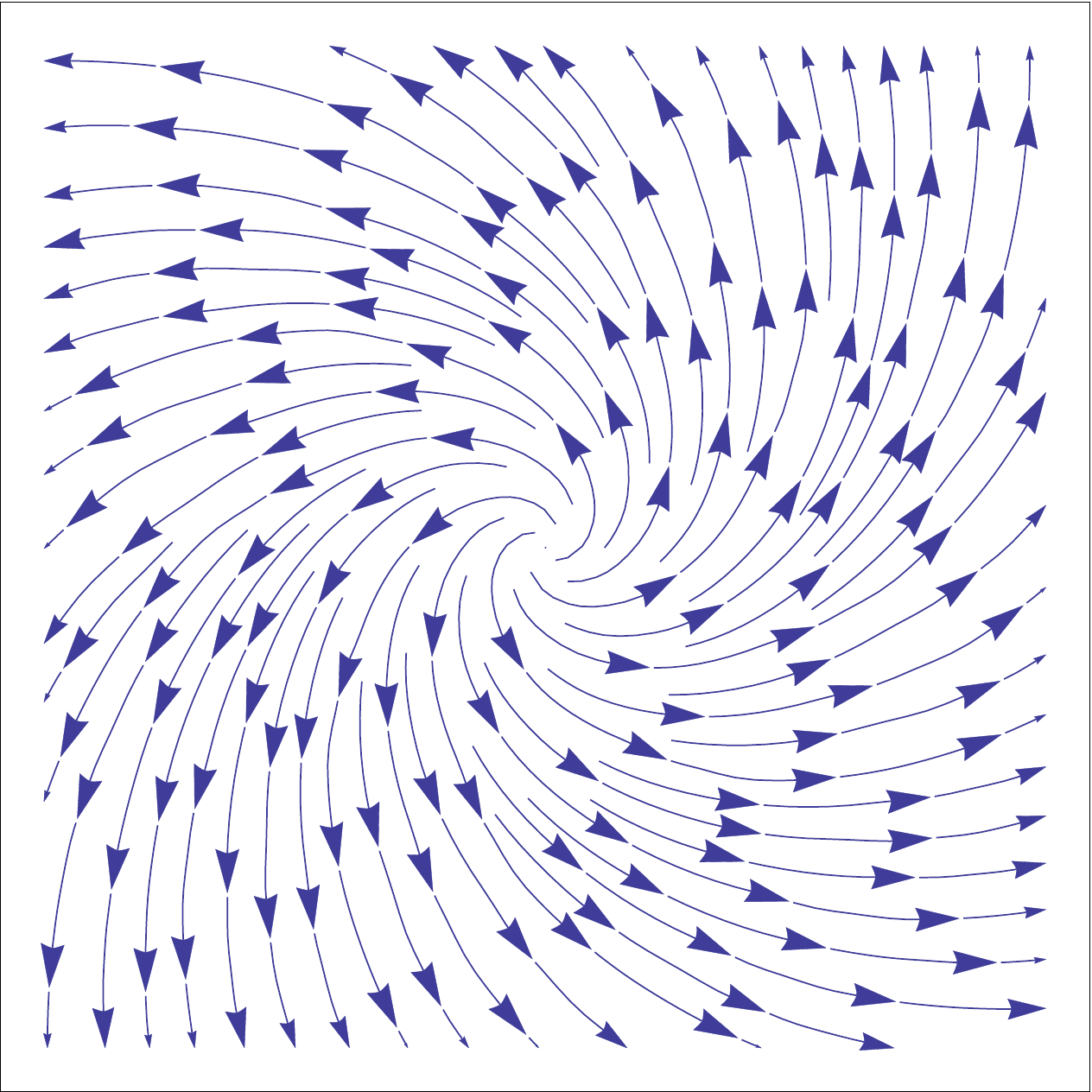}&\includegraphics[width=0.18\textwidth,angle=0]{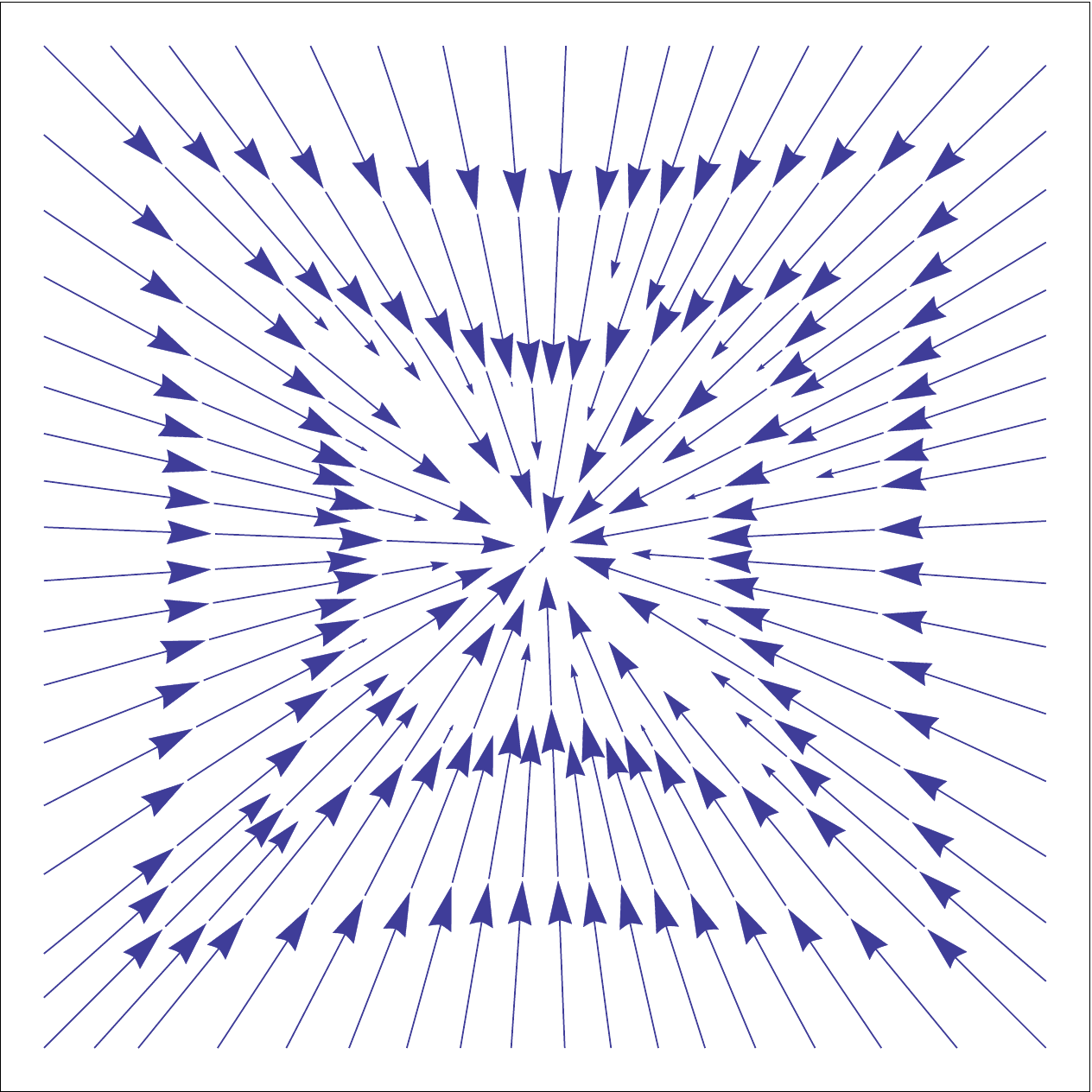}&\includegraphics[width=0.18\textwidth,angle=0]{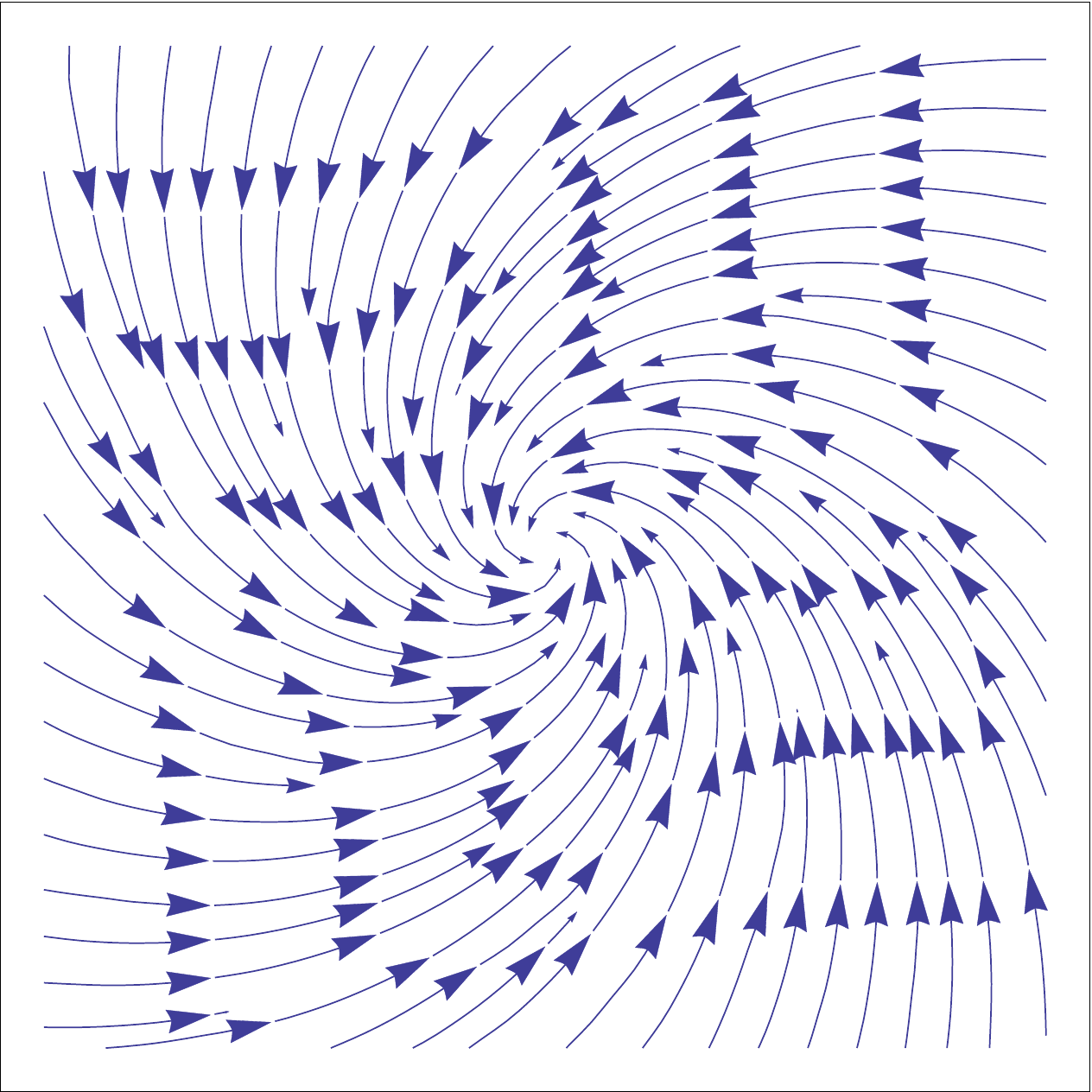}&\includegraphics[width=0.18\textwidth,angle=0]{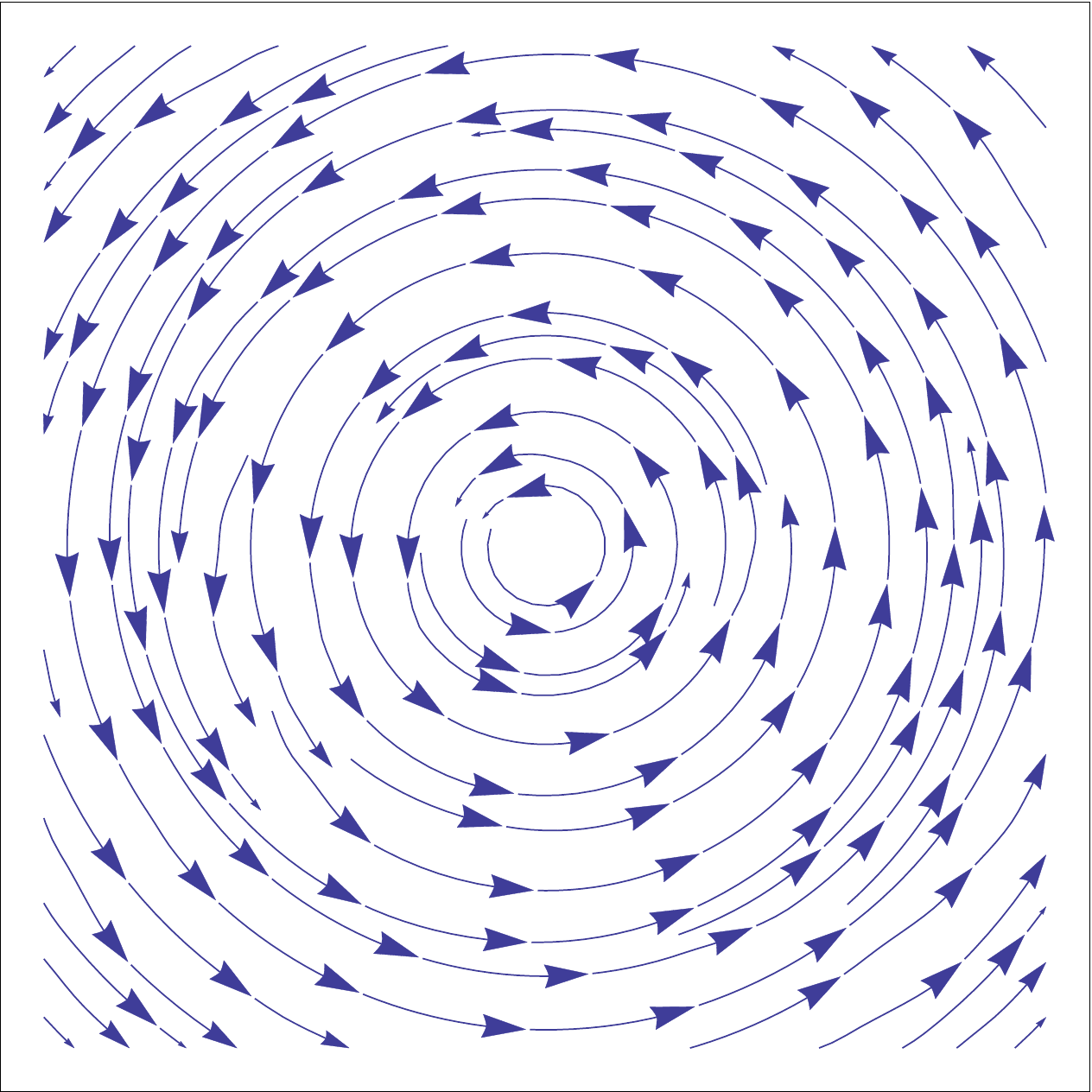}
\end{array}$
\end{center}
\caption{Electromagnetic field line patterns for the $m=0$ mode as a function of $\varrho_{0}$ where $\varrho_{0}=B_{1r}/B_{1\theta}$: upper row shows the magnetic field line pattern and lower row shows the electric field line pattern. The axes represent the cross section edges of the plasma cylinder. The patterns for $\varrho_{0}=0$ represent a helicon mode.}
\label{fg1_2}
\end{figure}

\begin{figure}[h]
\begin{center}$
\begin{array}{ccccc}
\varrho_{1}=1/3&\varrho_{1}=9/11&\varrho_{1}=1&\varrho_{1}=11/9&\varrho_{1}=3\\
\hspace{-0.1cm}\includegraphics[width=0.18\textwidth,angle=0]{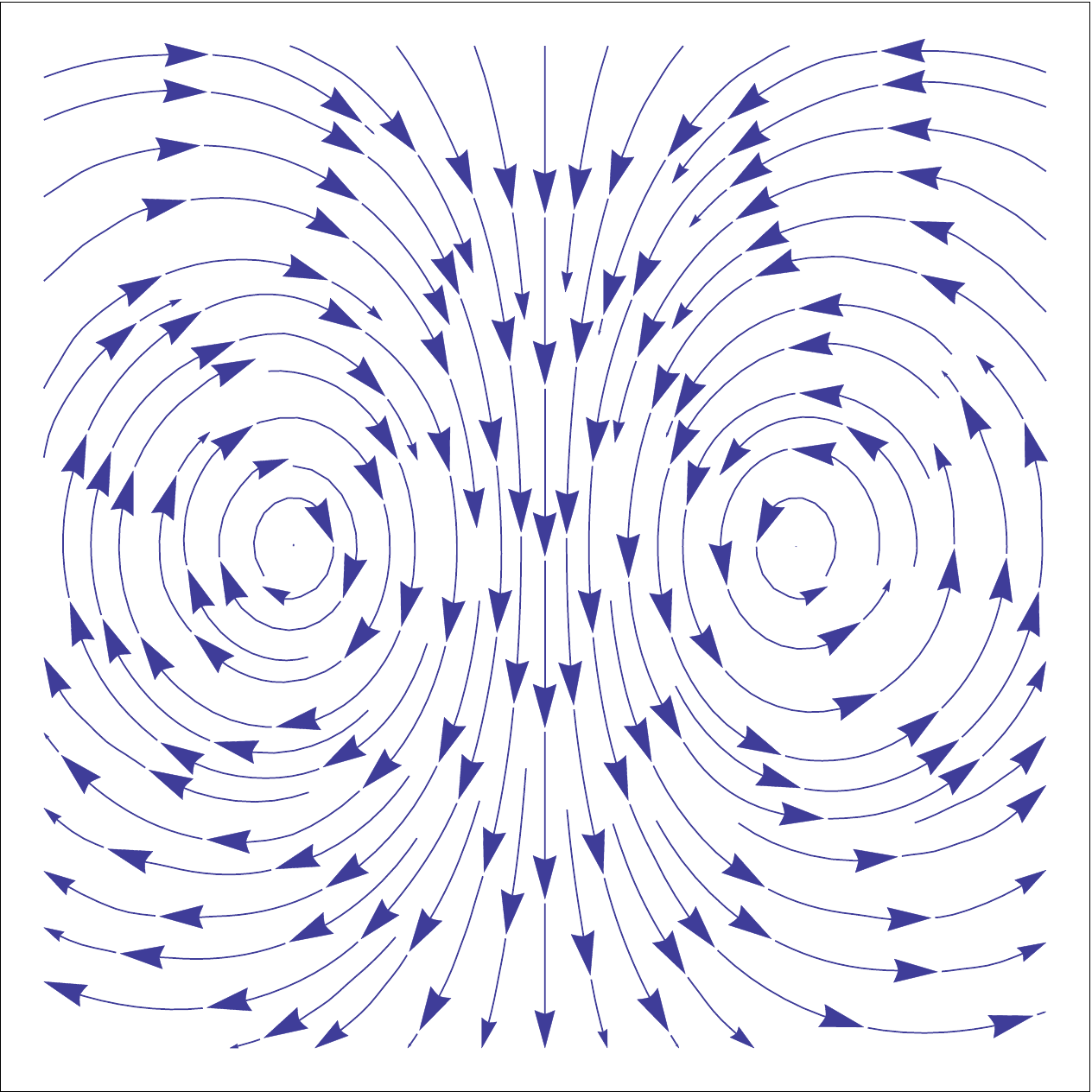}&\includegraphics[width=0.18\textwidth,angle=0]{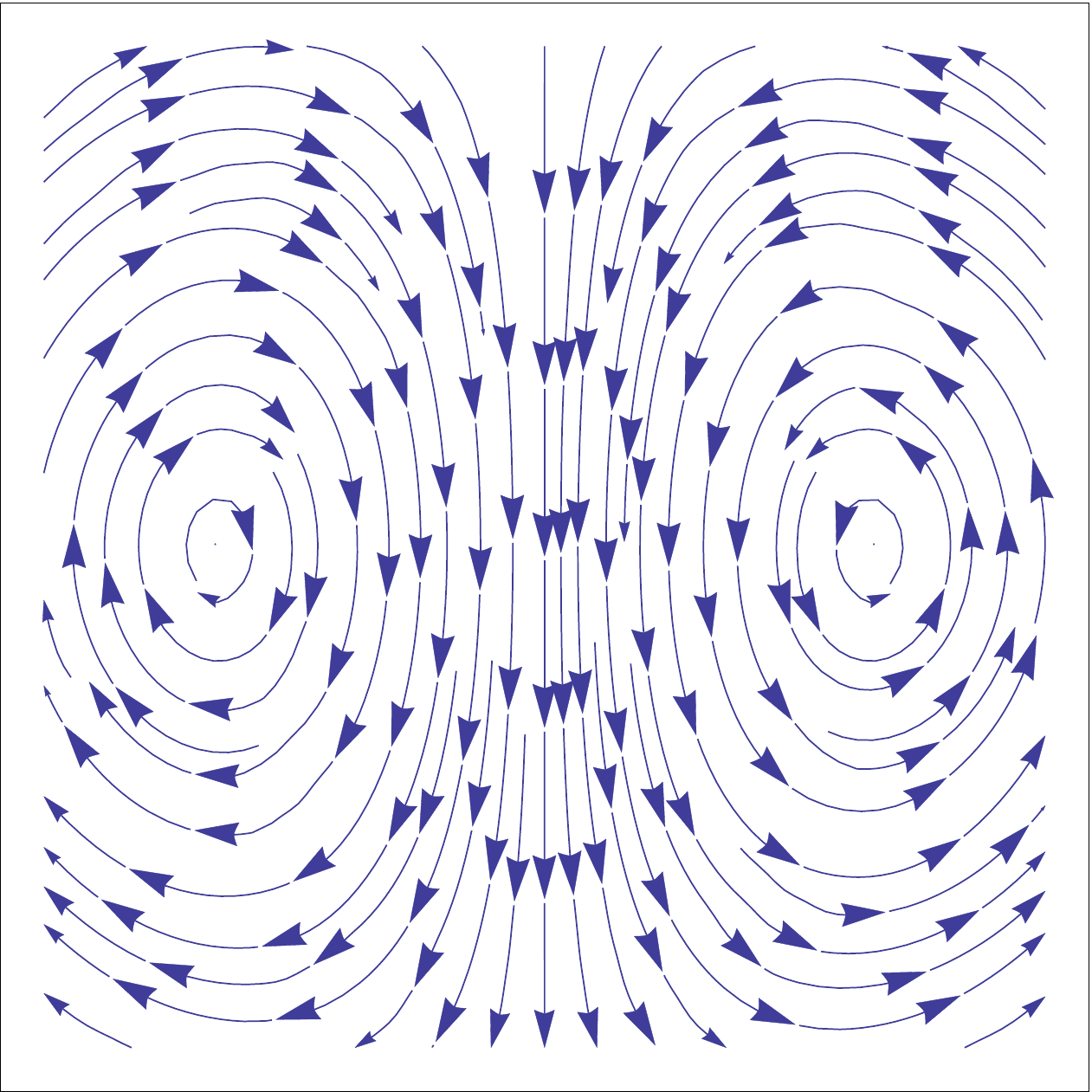}&\includegraphics[width=0.18\textwidth,angle=0]{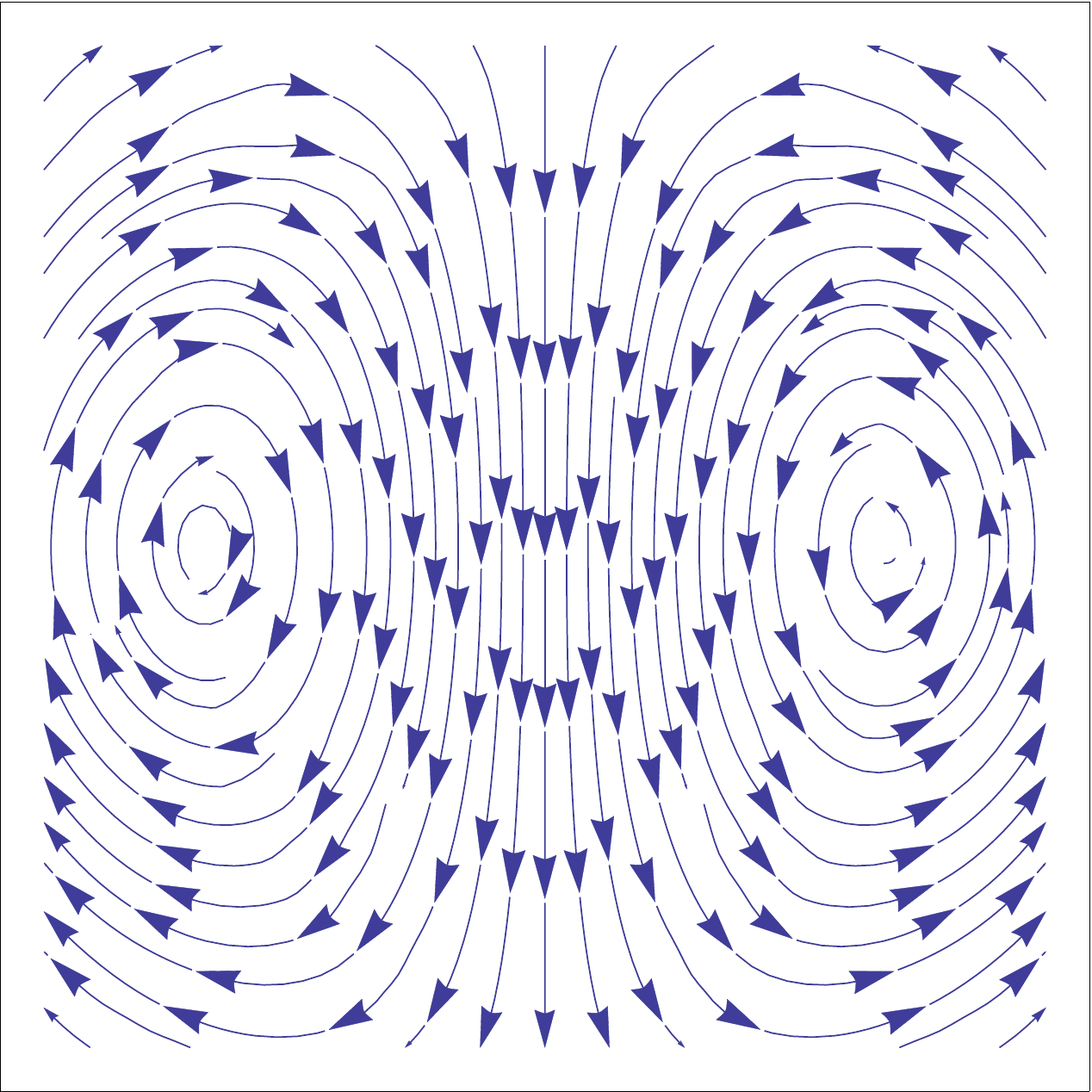}&\includegraphics[width=0.18\textwidth,angle=0]{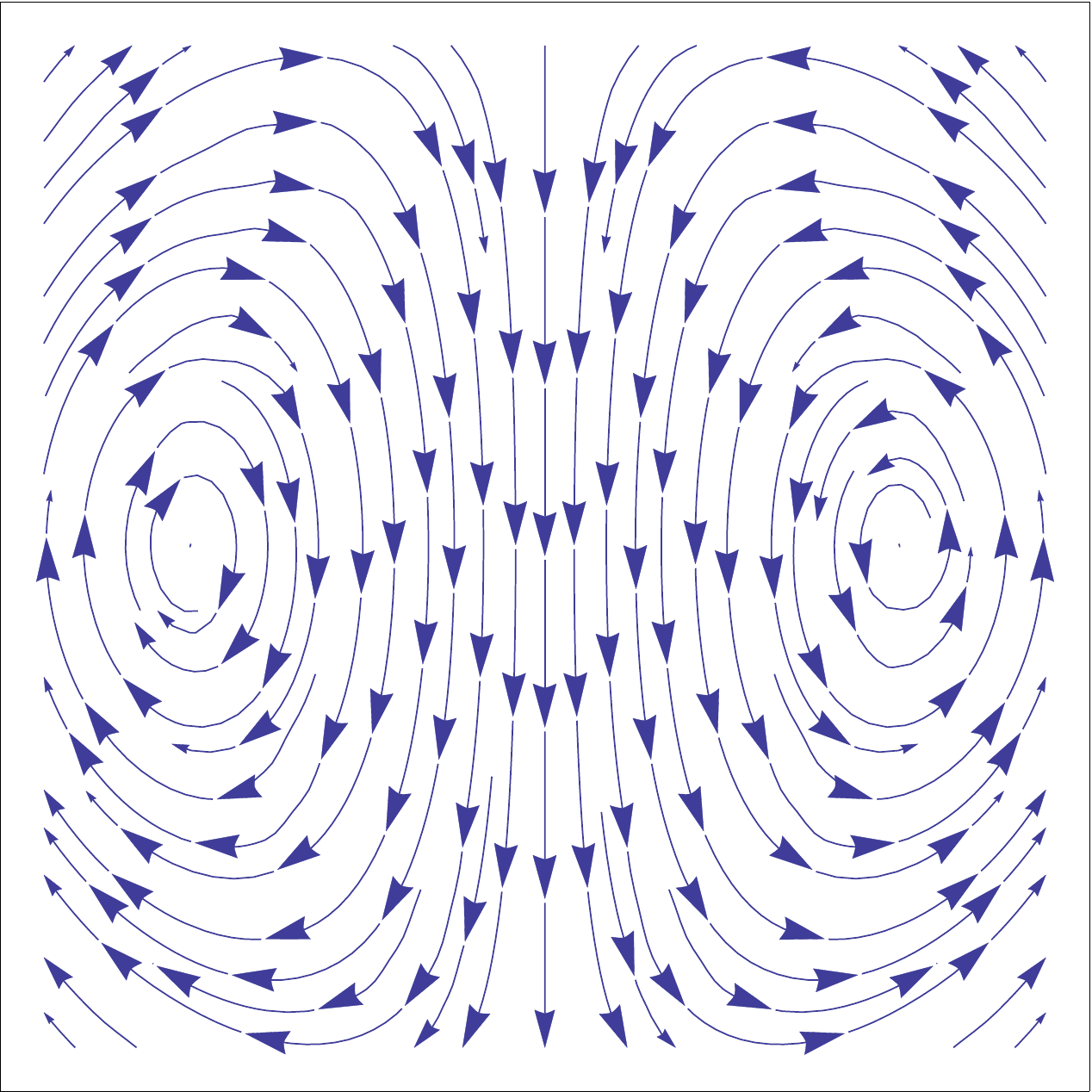}&\includegraphics[width=0.18\textwidth,angle=0]{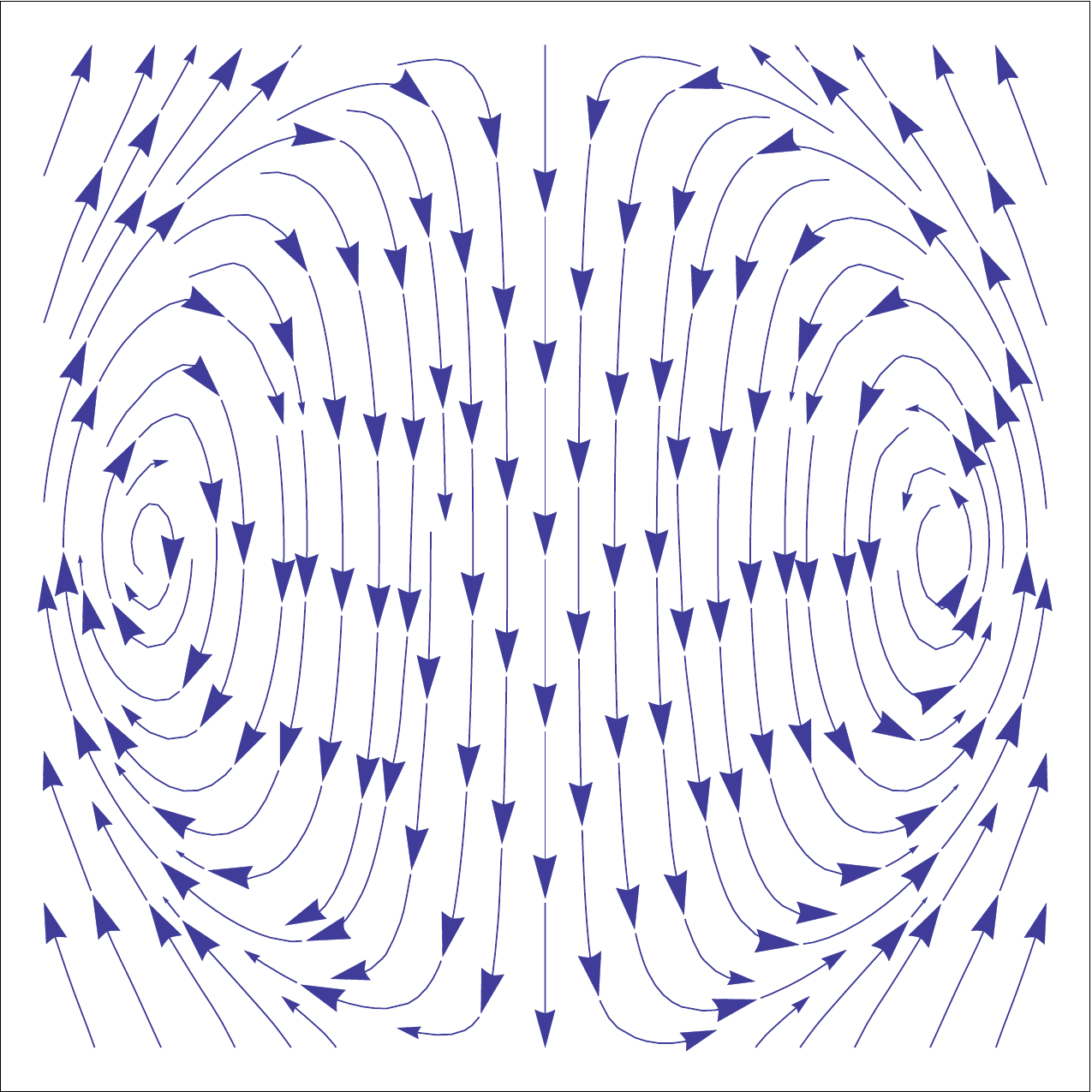}\\
\hspace{-0.1cm}\includegraphics[width=0.18\textwidth,angle=0]{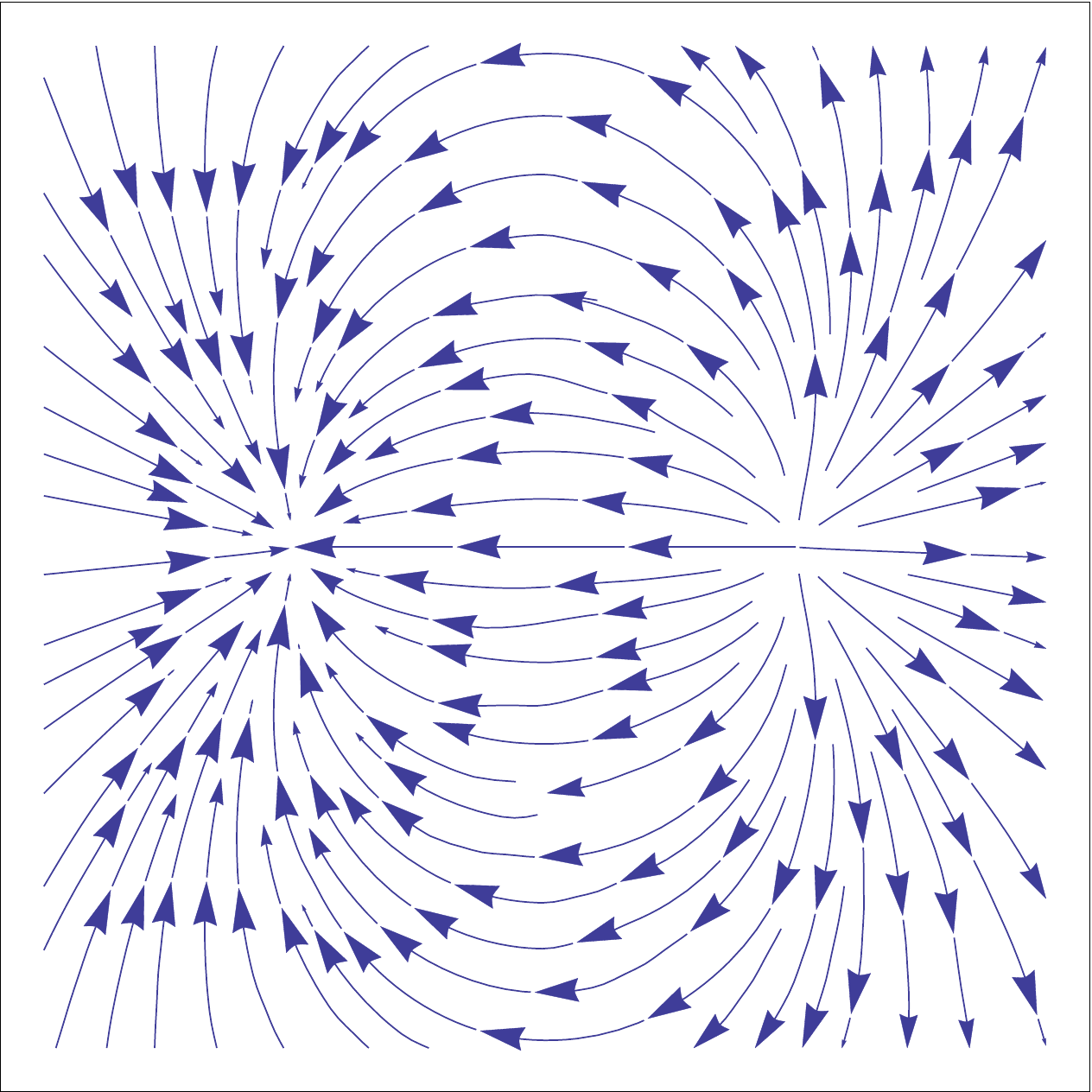}&\includegraphics[width=0.18\textwidth,angle=0]{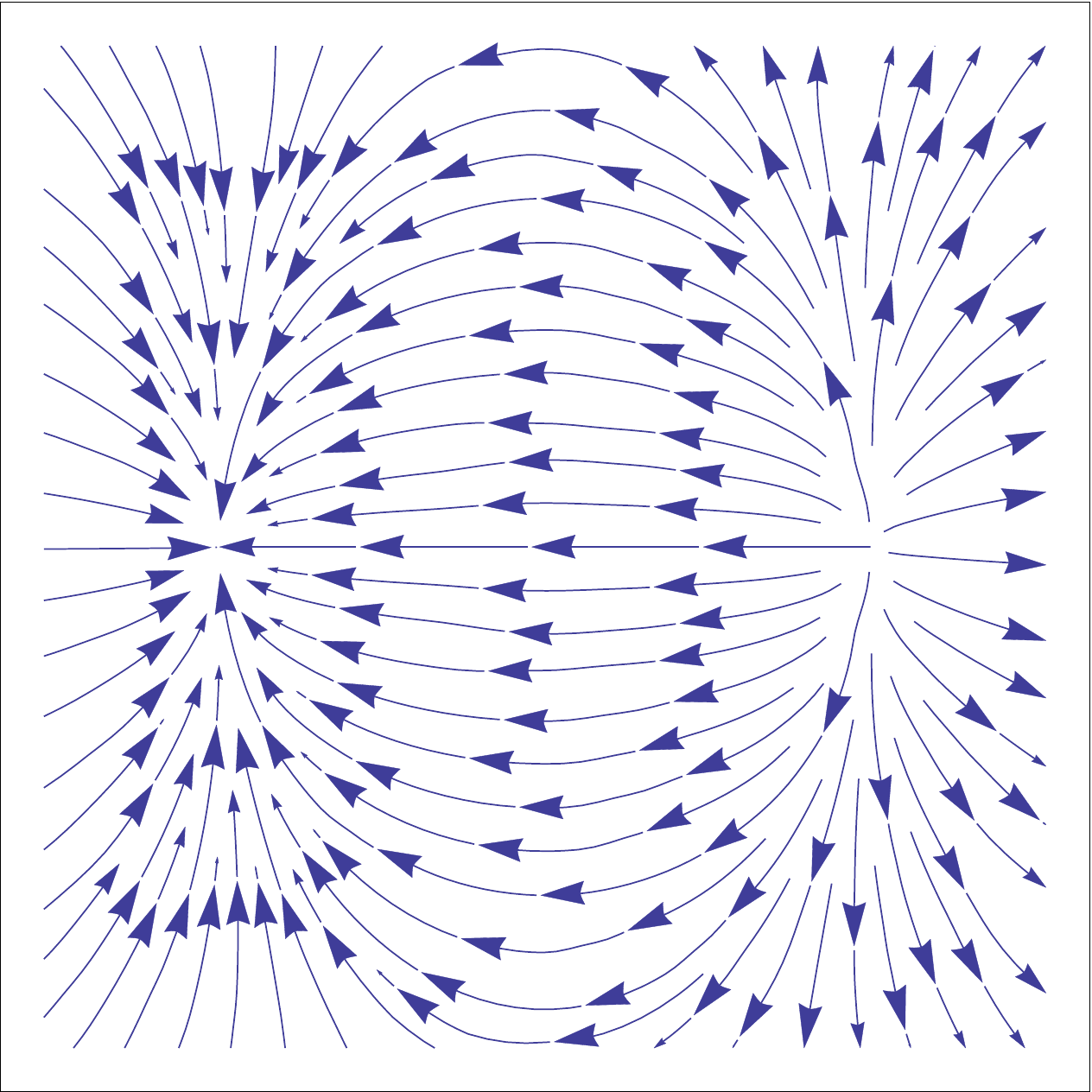}&\includegraphics[width=0.18\textwidth,angle=0]{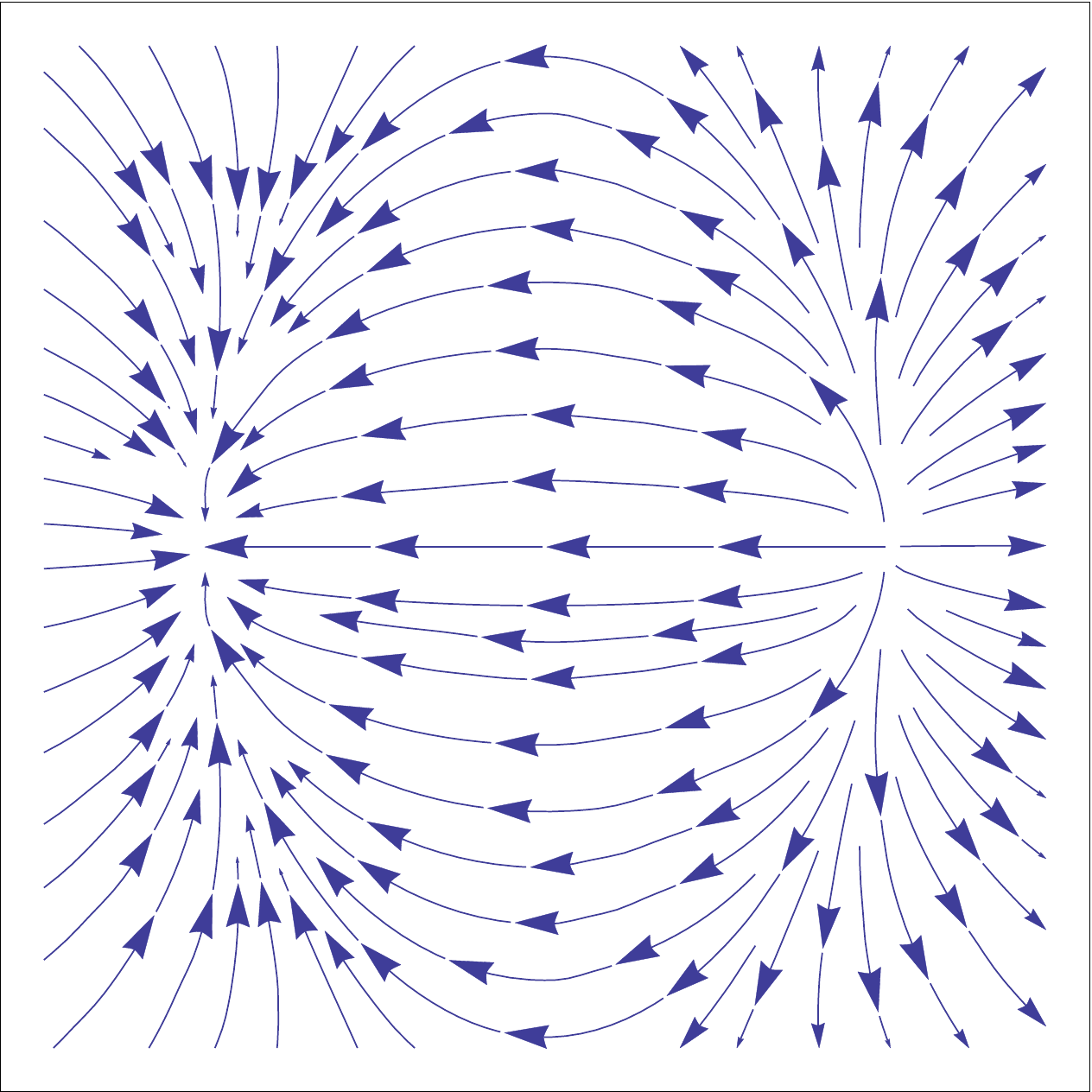}&\includegraphics[width=0.18\textwidth,angle=0]{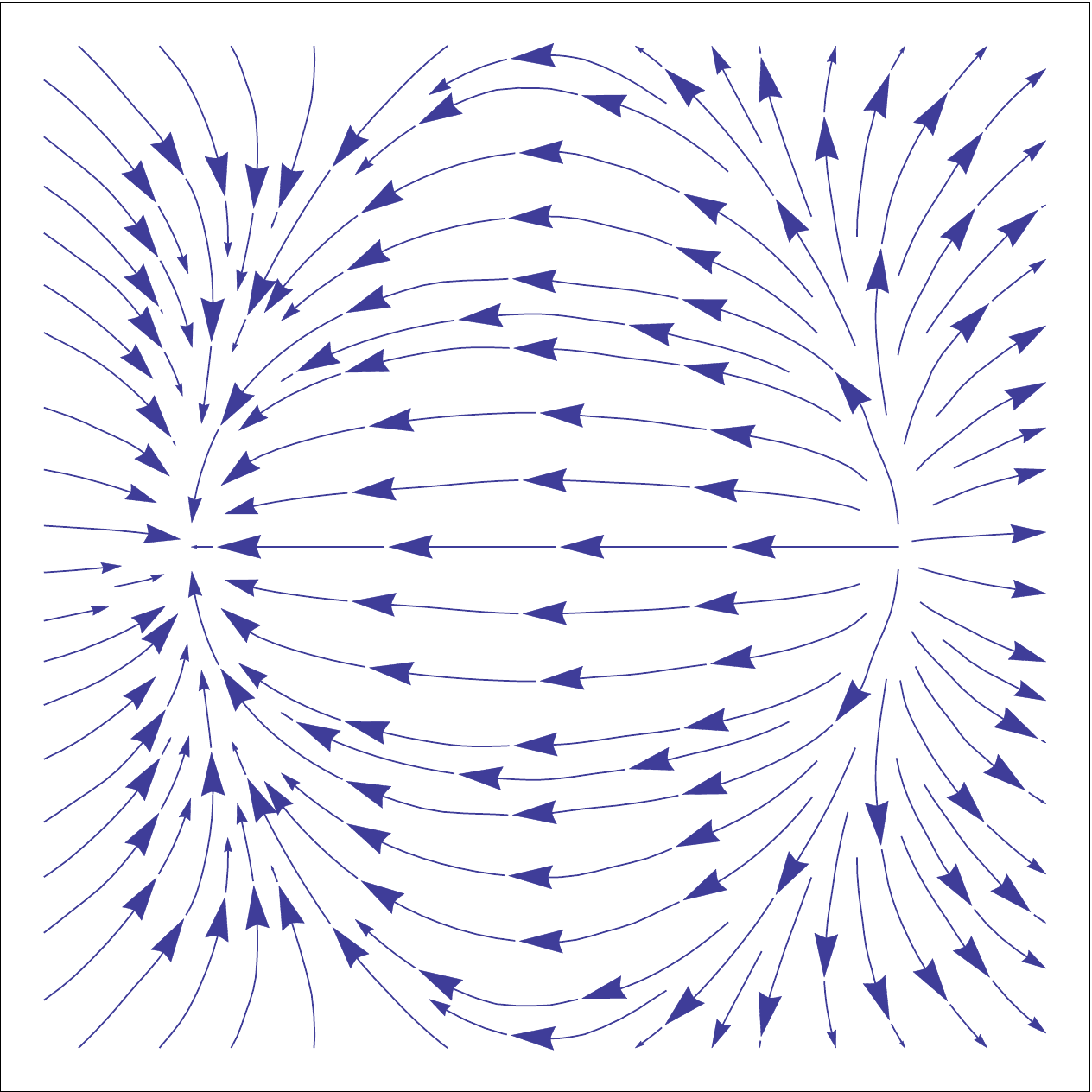}&\includegraphics[width=0.18\textwidth,angle=0]{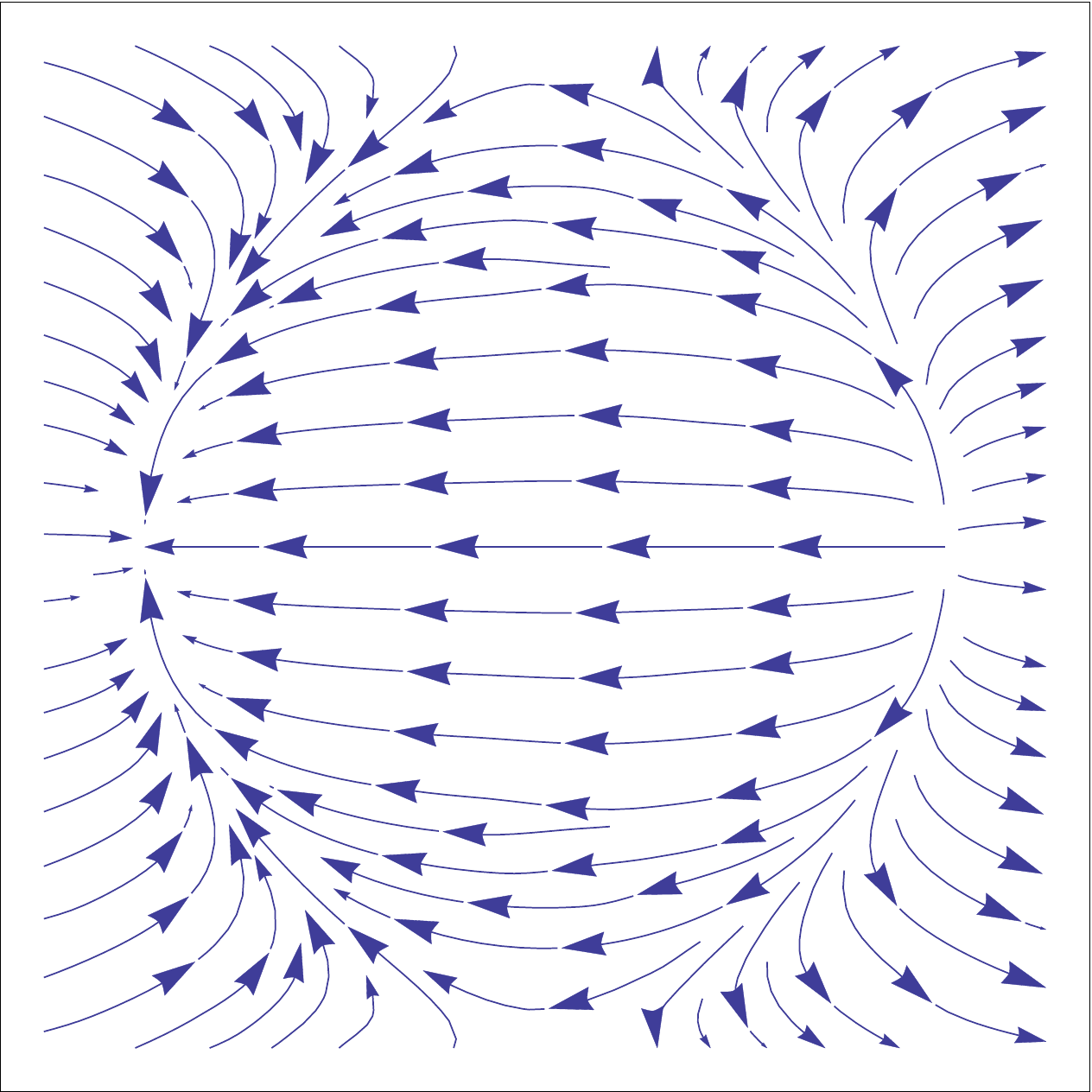}
\end{array}$
\end{center}
\caption{Electromagnetic field line patterns for the $m=1$ mode as a function of $\varrho_{1}$ where $\varrho_{1}=(1+k_z/\beta)/(1-k_z/\beta)$: upper row shows the magnetic field line pattern and lower row shows the electric field line pattern. The axes represent the cross section edges of the plasma cylinder. The patterns for $\varrho_{1}=1$ represent a helicon mode.}
\label{fg1_3}
\end{figure}

\begin{figure}[ht]
\begin{center}$
\begin{array}{ll}
(a)~~~~~~~~~~~~~~~~~m=0&(b)~~~~~~~~~~~~~~~~~m=1\\
\includegraphics[width=0.5\textwidth,angle=0]{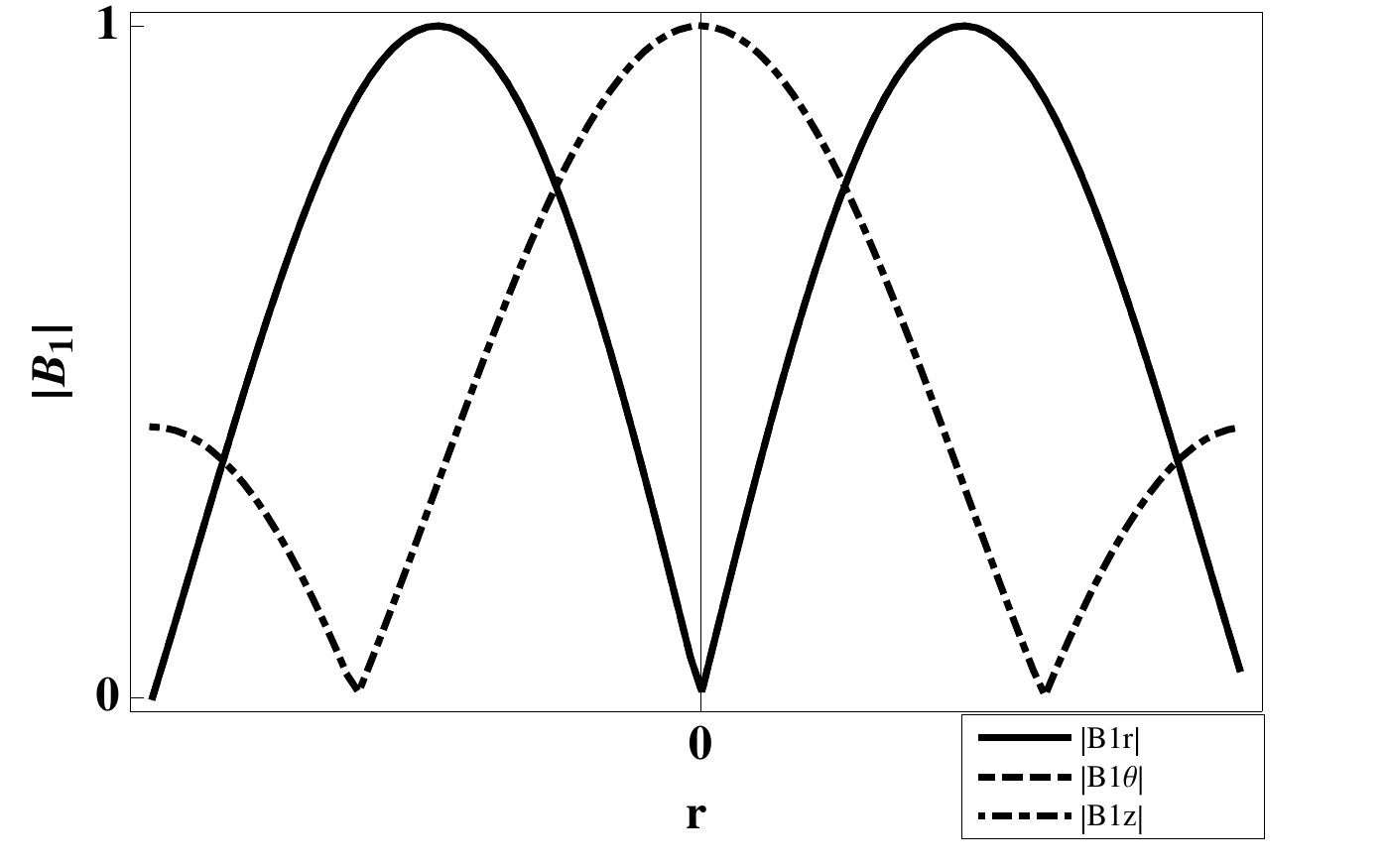}&\includegraphics[width=0.5\textwidth,angle=0]{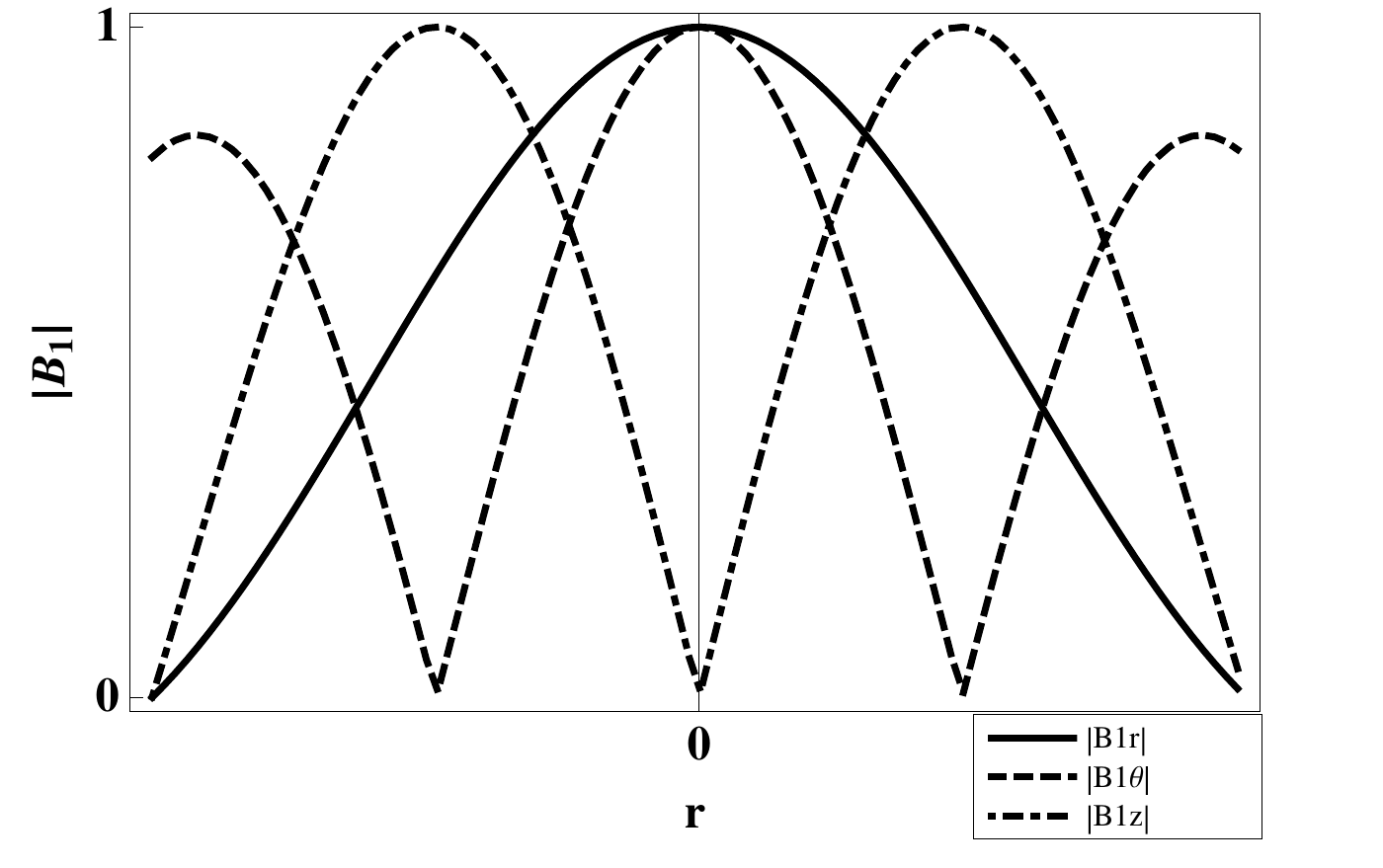}
\end{array}$
\end{center}
\caption{Radial structures of the $m=0$ mode (a) and $m=1$ mode (b).}
\label{fg1_4}
\end{figure}

In the limit of $m_e=0$, which yields $\delta_{rt}=0$, there is only one root of Eq.~(\ref{eq1_28}),
\begin{equation}\label{eq1_29}
\beta=\frac{k_w^2}{k_z}=-\frac{\omega}{k_z}\frac{\mu_0 n_{e0} q_e}{B_0}.
\end{equation}
This is the usual dispersion relation of helicon waves in a bounded uniform plasma. The root $\beta$, as will be seen in Eq.~(\ref{eq1_36}), is determined by the boundary condition and the mode structure of $\mathbf{B_1}$. Taking the curl of Eq.~(\ref{eq1_26}) and employing $\nabla\cdot\mathbf{B_1}=0$, we get
\begin{equation}\label{eq1_30}
\nabla^2\mathbf{B_1}+\beta^2\mathbf{B_1}=0.
\end{equation}
The axial component of Eq.~(\ref{eq1_30}) is
\begin{equation}\label{eq1_31}
\frac{\partial^2 B_{1z}}{\partial r^2}+\frac{1}{r}\frac{\partial B_{1z}}{\partial r}+\left(\beta^2-k_z^2-\frac{m^2}{r^2}\right)B_{1z}=0, 
\end{equation}
which has the solution
\begin{equation}\label{eq1_32}
B_{1z}=A J_m(T r)
\end{equation}
due to the finite $B_{1z}$ at the origin. Here, $A$ is the amplitude constant, $J_m$ is the Bessel function of the first kind of order $m$, and $T^2=\beta^2-k_z^2$. The other two components of $\mathbf{B_1}$ can be obtained from Eq.~(\ref{eq1_26}) ($\delta_{rt}=0$): 
\begin{equation}\label{eq1_33}
B_{1r}=\frac{i A}{2T}[(\beta+k_z)J_{m-1}(T r)+(\beta-k_z)J_{m+1}(T r)],
\end{equation}
\begin{equation}\label{eq1_34}
B_{1\theta}=-\frac{A}{2T}[(\beta+k_z)J_{m-1}(T r)-(\beta-k_z)J_{m+1}(T r)].
\end{equation}
Equations~(\ref{eq1_32})-(\ref{eq1_34}) describe the radial and azimuthal components of the wave magnetic field, from which we can also compute the electric wave field through Eq.~(\ref{eq1_11}):
\begin{equation}\label{eq1_35}
E_{1r}=\frac{\omega}{k_z}B_{1\theta},~E_{1\theta}=-\frac{\omega}{k_z}B_{1r},~E_{1z}=0. 
\end{equation}
Here, the axial component of $\mathbf{E}$ vanishes because of $m_e=0$ assumed in Eq.~(\ref{eq1_25}), namely $E_{1z}=-(i\omega m_e/q_e^2 n_{e0}) j_{t1z}=0$. A conducting boundary requires $E_{1\theta}(a)=E_{1z}(a)=0$ which thereby gives $B_{1r}(a)=0$, while an insulating boundary requires $j_{1r}(a)=0$. Here, $a$ is the radius of the boundary wall. Neglect the displacement current in Eq.~(\ref{eq1_1}) ($\omega/c^2\ll 1$) and combine it with Eq.~(\ref{eq1_26}) ($\delta_{rt}=0$), we get $\mathbf{j_1}=(\beta/\mu_0)\mathbf{B_1}$ which implies a vanishing radial component of $\mathbf{B}$, i. e. $B_{1r}(a)=(\mu_0/\beta) j_{1r}(a)=0$. Hence, there exists a universal boundary condition for both conducting and insulating boundaries for $m_e=0$, namely, 
\begin{equation}\label{eq1_36}
(\beta+k_z)J_{m-1}(T a)+(\beta-k_z)J_{m+1}(T a)=0. 
\end{equation}
The root $\beta$ can then be solved from Eq.~(\ref{eq1_36}) for a given mode structure. For the lowest two azimuthal modes, i. e. $m=0$ and $m=1$ which are usually observed in helicon discharges, Eq.~(\ref{eq1_36}) gives $J_1(T a)=0$ and $J_1(T a)=(T a k_z/2\beta)[J_2(T a)-J_0(T a)]$, respectively, which also vanishes for $a k_z\ll 1$, a long, thin tube. The lowest root of $J_1(T a)=0$ is $T a=3.83$ which yields $\beta\approx3.83/a$ for $k_z\ll 1/a$. To illustrate the helicon wave, we shall plot the electromagnetic field line patterns of the $m=0$ and $m=1$ modes. Putting back the exponential factor, we can write the real parts of the wave field of these two modes as:
\begin{equation}\label{eq1_37}
\hspace{-5.25cm}
\begin{array}{l}
(m=0)\\
B_{1r}(r, z, t)=\frac{A}{T}k_z J_1(Tr)\sin(k_z z-\omega t),\\
B_{1\theta}(r, z, t)=\frac{A}{T}\beta J_1(Tr)\cos(k_z z-\omega t),\\
B_{1z}(r, z, t)=A J_0(Tr)\cos(k_z z-\omega t),\\
E_{1r}(r, z, t)=\frac{A}{T}\frac{\omega}{k_z}\beta J_1(Tr)\cos(k_z z-\omega t),\\
E_{1\theta}(r, z, t)=-\frac{A}{T}\omega J_1(Tr)\sin(k_z z-\omega t),\\
E_{1z}(r, z, t)=0,
\end{array}
\end{equation}
\begin{equation}\label{eq1_38}
\begin{array}{l}
(m=1)\\
B_{1r}=-\frac{A}{2T}[(\beta+k_z)J_0(Tr)+(\beta-k_z)J_2(Tr)]\sin(\theta+k_z z-\omega t),\\
B_{1\theta}=-\frac{A}{2T}[(\beta+k_z)J_0(Tr)-(\beta-k_z)J_2(Tr)]\cos(\theta+k_z z-\omega t),\\
B_{1z}=A J_1(Tr)\cos(\theta+k_z z-\omega t),\\
E_{1r}=-\frac{A}{2T}\frac{\omega}{k_z}[(\beta+k_z)J_0(Tr)-(\beta-k_z)J_2(Tr)]\cos(\theta+k_z z-\omega t),\\
E_{1\theta}=\frac{A}{2T}\frac{\omega}{k_z}[(\beta+k_z)J_0(Tr)+(\beta-k_z)J_2(Tr)]\sin(\theta+k_z z-\omega t),\\
E_{1z}=0.
\end{array}
\end{equation}
The cross sections of these two modes can be visualised easily by comparing the radial and azimuthal components in Eq.~(\ref{eq1_37}) and Eq.~(\ref{eq1_38}). For $m=0$, we introduce $\varrho_{0}=B_{1r}/B_{1\theta}$, and plot the electromagnetic field line pattern as a function of $\varrho_{0}$. As shown in Fig.~\ref{fg1_2}, the magnetic field line pattern changes from purely radial to spiral and to purely circular as $|\varrho_{0}|$ is decreased. 
The electric field line pattern shows a reverse trend, consistent with Faraday's law. Due to the usually true $|k_z/\beta|\ll 1$ in helicon discharges (inclusion of cylindrical boundary decreases $k_z$ over the case of infinite plane wave propagation),\cite{Blevin:1966aa, Chen:1991aa} which implies $|\varrho_{0}|\ll 1$ and the dominance of $E_{1r}$ over $E_{1\theta}$, the best way to excite this mode is through coupling the $E_{1r}$ component. Here, $E_{1r}$ and $E_{1\theta}$ stand for the electrostatic and electromagnetic components, respectively. For $m=1$, we introduce $\varrho_{1}=(1+k_z/\beta)/(1-k_z/\beta)$ and set $k_z z-\omega t=0$. Figure~\ref{fg1_3} shows the corresponding magnetic and electric field line patterns which stay almost the same with $\varrho_{1}$ increased from $1/3$ to $9/11$, $1$, $11/9$ and $3$. 
The corresponding values of $k_z/\beta$ are: $-1/2$, $-1/10$, $0$, $1/10$ and $1/2$, respectively. As time advances or $z$ increases, the field line pattern rotates about its centre point but the shape does not change. To show the variation of wave field strength in radius, we plot the wave magnetic field over radius for the $m=0$ and $m=1$ modes. Here, all components are normalised to their own maximum values to show a clear radial structure. Figure~\ref{fg1_4}(a) shows the result of the $m=0$ mode, which implies that the profiles of $B_{1r}$ and $B_{1\theta}$ are overlapped and peak off axis, while the profile of $B_{1z}$ peaks on axis. 
By contrast, for the $m=1$ mode as shown in Fig.~\ref{fg1_4}(b), the profiles of $B_{1r}$ and $B_{1\theta}$ peak on axis, while the profile of $B_{1z}$ peaks off axis and vanishes at the origin. The difference between Fig.~\ref{fg1_4}(a) and Fig.~\ref{fg1_4}(b) helps identifying $m=0$ and $m=1$ modes in experiments. 

Although the approximation of $m_e=0$ works for most scenarios of helicon discharges, the inertia of electrons is found to strongly affect the high order radial modes.\cite{Boswell:1972aa} Actually, it has been verified experimentally that the inertia of electrons brings in measurable effect on the dispersion relation of waves whose frequencies are at least as low as $0.1|\omega_{ce}|$.\cite{Blevin:1968ab} For $m_e\neq 0$, Eq.~(\ref{eq1_28}) has two roots according to the quadratic formula: 
\begin{equation}\label{eq1_39}
\beta_{1}=\frac{k_z+\sqrt{k_z^2-4\delta_{rt} k_w^2}}{2\delta_{rt}},~\beta_{2}=\frac{k_z-\sqrt{k_z^2-4\delta_{rt} k_w^2}}{2\delta_{rt}}.
\end{equation}
They can be simplified for $\delta_{rt} k_w^2\ll k_z^2$:
\begin{equation}\label{eq1_40}
\beta_1\approx \frac{k_w^2}{k_z}=-\frac{\omega}{k_z}\frac{\mu_0 n_{e0} q_e}{B_0},~\beta_2\approx \frac{k_z}{\delta_{rt}}=-\frac{k_z}{\omega}\omega_{ce}, 
\end{equation}
with $\beta_1$ standing for the helicon mode (same as $\beta$, see Eq.~(\ref{eq1_29})) and $\beta_2$ evidently for an electron cyclotron mode (or TG mode, shortened for the initials of Trivelpiece and Gould who first discussed it with cylindrical boundaries\cite{Trivelpiece:1959aa}). The solution of Eq.~(\ref{eq1_27}) is eventually the combination of the following two Helmholtz equations:
\begin{equation}\label{eq1_41}
\nabla^2\mathbf{B_{1a}}+\beta_1^2\mathbf{B_{1a}}=0,~\nabla^2\mathbf{B_{1b}}+\beta_2^2\mathbf{B_{1b}}=0,
\end{equation}
where $\mathbf{B_1}=\mathbf{B_{1a}+\mathbf{B_{1b}}}$. A complete boundary condition for Eq.~(\ref{eq1_41}) was given by Chen and Arnush\cite{Chen:1997ab} for a conducting cylinder
\begin{equation}\label{eq1_42}
\begin{array}{l}
\beta_1 T_1 J_m(T_1 a)\left[(\beta_2+k_z)J_{m-1}(T_2 a)+(\beta_2-k_z)J_{m+1}(T_2 a)\right]=\\
\beta_2 T_2 J_m(T_2 a)\left[(\beta_1+k_z)J_{m-1}(T_1 a)+(\beta_1-k_z)J_{m+1}(T_1 a)\right]
\end{array}
\end{equation}
with $T_1^2=\beta_1^2-k_z^2$ and $T_2^2=\beta_2^2-k_z^2$. From Eq.~(\ref{eq1_42}), we can solve the two roots of $\beta_1$ and $\beta_2$ for a given mode structure, and then obtain the corresponding dispersion relation. 

\subsubsection{Helicon waves in a cylindrical non-uniform plasma}\label{rlh1}
In this section, we consider a plasma cylinder with radial density gradient, which is more realistic for helicon discharges than a uniform density profile that has been used so far.\cite{Chen:1997aa} The effect of the radial density gradient on helicon discharges was first recognised experimentally by Lehane and Thonemann\cite{Lehane:1965aa} and theoretically by Blevin and Christiansen\cite{Blevin:1966aa, Blevin:1968aa}, and examined in more detail by Chen and his colleagues.\cite{Chen:1994aa, Sudit:1994aa} Although Blevin and Christiansen achieved an analytical dispersion relation for a specific radially non-uniform density profile, most previous studies rely on numerical computations for arbitrary radial non-uniformity in plasma density. In $2000$, Breizman and Arefiev presented a theoretical analysis to explore the effect of the radial density gradient on helicon waves, and discovered a new surface-type helicon mode. Though their analysis is based on a step-like radial profile of plasma density, the analytical dispersion relation they obtained is actually valid for general radially non-uniform density profiles, as long as the plasma column is long, thin ($k_z\ll 1/a$) and sufficiently dense ($|\omega_{ce}|\ll\omega_{pe}$). The surface-type helicon mode is called so because it is caused by a localised ``surface" current, formed by the radial density jump. This radially localised helicon (RLH) mode may shed light on the high efficiency of helicon wave production of plasmas. Since the RLH mode will be employed in Chapter~\ref{chp4} to form a gap eigenmode in a plasma cylinder with periodic magnetic field, a short overview is given below.\cite{Breizman:2000aa} 

For $\sqrt{\omega_{ci}|\omega_{ce}|}\ll\omega\ll |\omega_{ce}|\ll\omega_{pe}$, we rewrite Eqs.~(\ref{eq1_9})-(\ref{eq1_10}) as:
\begin{equation}\label{eq1_43}
S=\frac{\omega_{pe}^2}{\omega_{ce}^2},~D=-\frac{\omega_{pe}^2}{\omega\omega_{ce}},~P=-\frac{\omega_{pe}^2}{\omega^2},
\end{equation}
and combine Eq.~(\ref{eq1_1}) with Eq.~(\ref{eq1_11}) to get the linear wave equation
\begin{equation}\label{eq1_44}
\nabla\times\nabla\times\mathbf{E_1}=\frac{\omega^2}{c^2}\mathbf{\tilde{\epsilon}}\cdot\mathbf{E_1}. 
\end{equation}
The three components of Eq.~(\ref{eq1_44}) in cylindrical coordinates are:
\begin{equation}\label{eq1_45}
\hspace{-0.5cm}\left(\frac{m^2}{r^2}+k_z^2\right)E_{1r}+\frac{i m}{r}\left(\frac{\partial E_{1\theta}}{\partial r}+\frac{E_{1\theta}}{r}\right)+i k_z\frac{\partial E_{1z}}{\partial r}=\frac{\omega^2}{c^2}\left(S E_{1r}-i D E_{1\theta}\right),
\end{equation}
\begin{equation}\label{eq1_46}
\hspace{-0.5cm}i m\frac{\partial}{\partial r}\left(\frac{E_{1r}}{r}\right)-\frac{\partial}{\partial r}\left(\frac{\partial E_{1\theta}}{\partial r}+\frac{E_{1\theta}}{r}\right)+k_z^2 E_{1\theta}- k_z\frac{m}{r}E_{1z}=\frac{\omega^2}{c^2}\left(S E_{1\theta}+i D E_{1r}\right),
\end{equation}
\begin{equation}\label{eq1_47}
\hspace{-0.5cm}\frac{i k_z}{r}\frac{\partial}{\partial r}\left(r E_{1r}\right)-k_z\frac{m}{r}E_{1\theta}-\frac{1}{r}\frac{\partial}{\partial r}\left(r\frac{\partial E_{1z}}{\partial r}\right)+\frac{m^2}{r^2}E_{1z}=\frac{\omega^2}{c^2}P E_{1z}. 
\end{equation}
For a small longitudinal wave number ($k_z\ll \mathrm{min}(\omega_{pe}/c;1/a)$) and a sufficiently dense plasma ($\omega\ll |\omega_{ce}|\ll\omega_{pe}$), both Eq.~(\ref{eq1_45}) and Eq.~(\ref{eq1_46}) read
\begin{equation}\label{eq1_48}
\frac{i m}{r}E_{1r}\approx\frac{\partial E_{1\theta}}{\partial r}+\frac{1}{r}E_{1\theta}. 
\end{equation}
As will be shown, the radial nonuniformity of plasma density has a surprisingly strong effect on the structure of helicon modes with $m\neq 0$, and this effect is most pronounced in the limit of small $k_z$ considered here. For $m\neq 0$, expanding the quasineutrality ($\nabla\cdot\mathbf{D_1}=0$) gives
\begin{equation}\label{eq1_49}
\frac{1}{r}\frac{\partial}{\partial r}\left[r\left(S E_{1r}-i D E_{1\theta}\right)\right]+\frac{i m}{r}\left(S E_{1\theta}+i D E_{1r}\right)+i k_z P E_{1z}=0.
\end{equation}
Equations~(\ref{eq1_47})-(\ref{eq1_49}) are combined to form a closed set which can be reduced to:
\begin{equation}\label{eq1_50}
\frac{1}{r}\frac{\partial}{\partial r}\left(r\frac{\partial E}{\partial r}\right)-\frac{m^2}{r^2}E=-\frac{\omega^2}{c^2}P E_{1z},
\end{equation}
\begin{equation}\label{eq1_51}
\frac{1}{r}\frac{\partial}{\partial r}\left[S r\frac{\partial}{\partial r}\left(E_{1z}-E\right)\right]-\frac{m}{r}\left(\frac{m}{r}S-\frac{\partial D}{\partial r}\right)\left(E_{1z}-E\right)-k_z^2 P E_{1z}=0.
\end{equation}
Here, a new unknown function has been introduced, $E=E_{1z}-(k_z r/m)E_{1\theta}$, which measures the radial component of the perturbed magnetic field. Equations~(\ref{eq1_50})-(\ref{eq1_51}) indicate two modes: a helicon mode with wave equation
\begin{equation}\label{eq1_52}
\frac{1}{r}\frac{\partial}{\partial r}\left(r\frac{\partial E}{\partial r}\right)-\frac{m^2}{r^2}E=\frac{m}{k_z^2 r}\frac{\omega^2}{c^2}\frac{E\partial D/\partial r}{1-(m\partial D/\partial r)/(k_z^2 r P)},
\end{equation}
and a TG mode with wave equation
\begin{equation}\label{eq1_53}
\frac{1}{r}\frac{\partial}{\partial r}\left(S r \frac{\partial E_{1z}}{\partial r}\right)-\frac{m}{r}\left(\frac{m}{r}S-\frac{\partial D}{\partial r}\right)E_{1z}-k_z^2 P E_{1z}=0.
\end{equation}
For the helicon mode, Eq.~(\ref{eq1_52}) can be solved for $\omega$ based on a step-like radial profile of plasma density
\begin{equation}\label{eq1_54a}
n_s=\left\{
\begin{array}{l}
n_{-},~r<r_0\\
n_{+},~r>r_0
\end{array}
\right.
\end{equation}
and the corresponding radial profile of electric field
\begin{equation}\label{eq1_54b}
E=E_0\left\{
\begin{array}{l}
(r/r_0)^{|m|},~r<r_0\\
(r/r_0)^{-|m|},~r>r_0
\end{array}
\right.
\end{equation}
where $r_0$ is the radius of the density discontinuity and $E_0$ is a constant electric field. Under the limit of $\omega_{pe}/c\gg 1/a$, an analytical dispersion relation can be obtained
\begin{equation}\label{eq1_55}
\omega=2\frac{m}{|m|}\frac{\omega_{ce}k_z^2 c^2}{\omega_{pe}^2(n_{+})-\omega_{pe}^2(n_{-})}.
\end{equation}
The mode involves a perturbed ``surface" current which is localised near the peak of the eigenfunction $E(r)$ and distinguishes it from the modes studied previously.\cite{Klozenberg:1965aa, Sudan:1967aa} This RLH mode can couple strongly to the antenna current during helicon discharges. 

\begin{figure}[ht]
\begin{center}
\hspace{-0.2cm}\includegraphics[width=0.7\textwidth,angle=0]{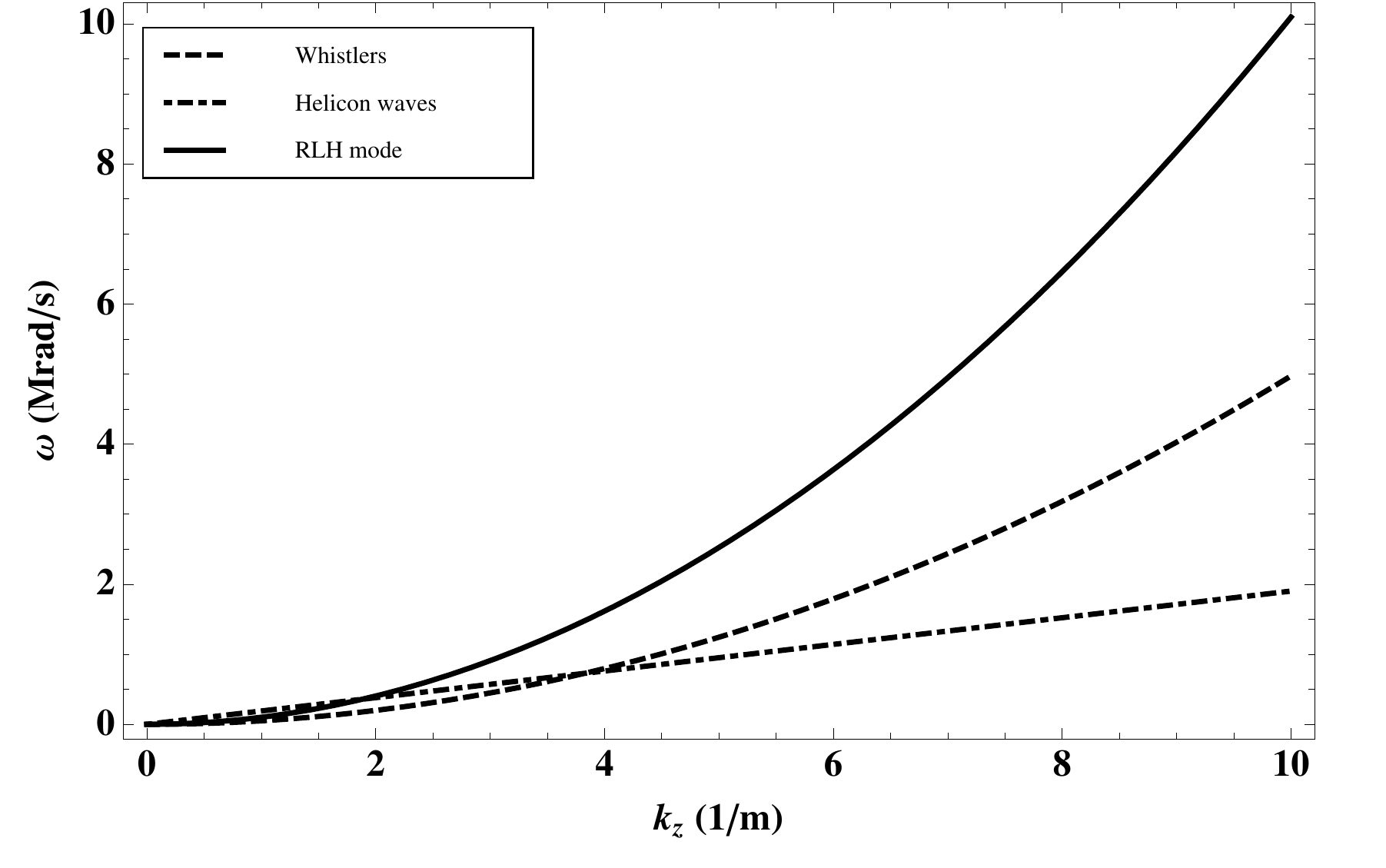}
\end{center}
\caption{Dispersion relations of whistlers (uniform plasma in unbounded slab), helicon waves (uniform plasma in cylinder) and RLH mode (non-uniform plasma in cylinder) for a singly ionised argon plasma with peak density $n_{s0}=10^{18}$ and field strength $|B_0|=0.01$~T.}
\label{fg1_5}
\end{figure}

For comparison, the dispersion relations of whistlers (Eq.~(\ref{eq1_20})), helicon waves (Eq.~(\ref{eq1_29})) and the RLH mode (Eq.~(\ref{eq1_55})) are plotted together in Fig.~\ref{fg1_5}. 
The plasma parameters employed here are the same to those used for Fig.~\ref{fg1_1}. For helicon waves, the root $\beta$ in Eq.~(\ref{eq1_29}) has been set to $3.83/a$, taking the lowest root of $J_1(T a)=0$. For the RLH mode, a numerical form factor $\Gamma$ has been introduced to Eq.~(\ref{eq1_55}), which then has a new form $\omega=\Gamma\omega_{ce}c^2 k_z^2/\omega_{pe}^2$, for a Gaussian radial profile of plasma density. Fitting this new form with the computed dispersion relation results in $\Gamma=-2.03$.\cite{Chang:2013aa} Figure~\ref{fg1_5} indicates that the dispersion curves of these three modes approach to each other when $k_z\rightarrow 0$. 

\subsection{Helicon discharge}\label{dsc1}
\begin{figure}[ht]
\begin{center}
\includegraphics[width=1\textwidth,angle=0]{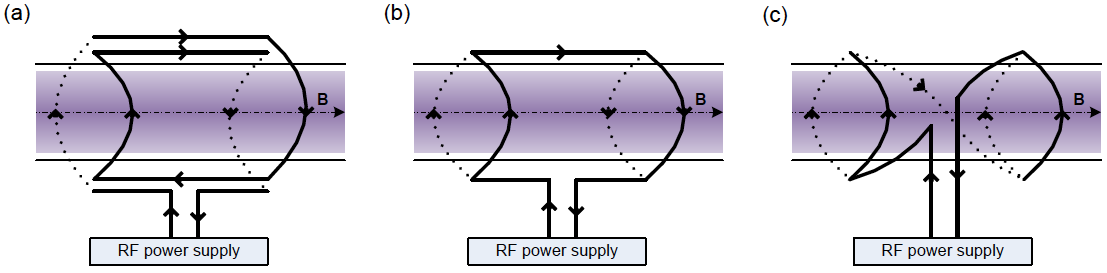}
\end{center}
\caption{RF antennas for driving helicon discharges: (a) double-saddle antenna used to produce the first helicon plasma,\cite{Boswell:1970aa} (b) plane-polarised Nagoya Type III antenna,\cite{Watari:1978aa, Okamura:1986aa}, and (c) twisted Nagoya Type III antenna (half-turn helical).\cite{Shoji:1986aa, Shoji:1987aa, Shoji:1988aa}}
\label{fg1_6}
\end{figure}

A helicon discharge usually refers to a plasma discharge driven by helicon waves which, as shown in Sec.~\ref{wav1}, are RHCP electromagnetic waves propagating in a bounded magnetised plasma with frequencies between the ion and electron cyclotron frequencies.\cite{Boswell:1997aa, Carter:2002aa} Although helicon waves in a cylinder can be either RHCP or LHCP, the former is often dominant.\cite{Boswell:1970aa, Chen:1992aa, Chen:1996ab, Chen:1997aa} A radio frequency (RF) antenna is usually employed to excite helicon waves and drive the helicon discharge.\cite{Lieberman:2005aa} Figure~\ref{fg1_6} shows the schematic of the first RF antenna (double-saddle) used to successfully drive a helicon discharge,\cite{Boswell:1970aa} together with those of other two widely used RF antennas: plane-polarised Nagoya Type III\cite{Watari:1978aa, Okamura:1986aa} and twisted Nagoya Type III (half-turn helical).\cite{Shoji:1986aa, Shoji:1987aa, Shoji:1988aa}
The RF antenna is connected to a RF power supply and wrapped around an insulating chamber (usually a glass tube). The time-varying current on the antenna wire induces a time-varying magnetic field, which then induces an electrostatic field in the opposite direction of the current through Faraday's law. The induced electrostatic field is nulled by the build up of space charge. Legs of the antenna opposite to the machine axis have opposite induced space charge, giving rise to a transverse electric field.\cite{Chen:1996ab} This transverse field couples the transverse $m=1$ mode structure, as shown in Fig.~\ref{fg1_3} ($\varrho_1=1$). The spatially oscillating electric field accelerates free electrons in the neutral gas and ionises it through collisional (high pressure) and collisionless (low pressure) heating mechanisms.\cite{Chen:1997aa, Lafleur:2011aa} For an inductively coupled discharge which does not have a static magnetic field, the formed surface plasma shields the wave field from the antenna and allows the propagation of waves only within a short distance (known as skin depth) into the column.\cite{Chen:1984aa, Lieberman:2005aa} If, however, an external magnetic field is applied axially through the plasma column, the dielectric scalar of the plasma becomes a tensor (see Eqs.~(\ref{eq1_8})-(\ref{eq1_10})), and waves of a certain frequency range will be allowed to propagate inside the plasma and not confined within the skin depth.\cite{Stix:1992aa, Boswell:1997aa, Swanson:2003aa} This frequency range of waves are typically helicon waves. Because they can propagate deep inside the plasma, which increases the efficiency of the wave energy deposition significantly, helicon waves generate plasmas with densities typically an order higher than the inductively coupled discharge when operated at similar pressures and input RF powers.\cite{Scime:2008aa} 

An enhancement in collision frequency is usually required to fit experimental data to theoretical calculations for helicon discharges,\cite{Boswell:1970aa, Lee:2011aa, Chang:2012aa} and the high efficiency in helicon wave production of plasmas has not yet been completely understood. The pursuit of this understanding has stimulated a large amount of research and promoted the discovery of various applications using helicon plasma sources.\cite{Chen:1997aa, Lafleur:2011aa} These include, to name a few, processing semiconductor circuits,\cite{Chen:1991aa, Chen:1997aa} plasma propulsion for space travel,\cite{Arefiev:2004ab, Ziemba:2005aa, Charles:2009aa, Batishchev:2009aa} gas laser media and plasma lenses for high energy particle beam,\cite{Chen:1991aa} magnetised plasma opening switches,\cite{Breizman:2000aa} laser plasma sources,\cite{Zhu:1989aa} and possibly electrodeless beam source and laser accelerator.\cite{Chen:1996ab} The helicon plasma can be also used as a research tool for measuring Hall coefficients in semiconductors,\cite{Boswell:1970ab} studying Landau and Cherenkov damping,\cite{Boswell:1970ab} measuring Doppler-shifted cyclotron damping,\cite{Christopoulos:1974aa} studying magnetic fusion,\cite{Loewenhardt:1991aa} driving RF current,\cite{petrzilka:1994aa}, studying intense-beam plasma interactions,\cite{Breizman:2000aa} understanding ionospheric phenomena,\cite{Boswell:1970ab, Breizman:2000aa} and studying Alfv\'{e}n wave propagations.\cite{Hanna:2001aa}

Helicon wave production of plasma was first achieved by Boswell in $1970$,\cite{Boswell:1970aa} following ten years history of helicon propagation study first in solid state plasmas\cite{Bowers:1961aa, Rose:1962aa, Lehane:1965aa} and then in gaseous plasmas.\cite{Harding:1965aa} The basic theory of helicon waves was also studied extensively in the $1960$s.\cite{Woods:1962ab, Woods:1964aa, Klozenberg:1965aa, Davies:1969ab, Davies:1970aa} The early history of helicon studies prior to $1997$ is reviewed by Boswell and Chen.\cite{Boswell:1997aa, Chen:1997aa} Two questions remain unresolved: the reason for the high efficiency of helicon discharges and the dominance of the RHCP mode over the LHCP mode.\cite{Chen:1997aa} This thesis first models the wave instability in helicon plasmas with consideration of plasma flows, which have been observed recently in multiple helicon devices.\cite{Blackwell:2012aa, Scime:2007aa} Second, the thesis studies the propagation of helicon waves inside a pinched plasma and supports the hypothesis that multiple radial modes could be excited simultaneously,\cite{Chen:1996ab, Mori:2004aa, Chen:1996aa, Light:1995aa} which may contribute to the understanding of the ``strangely" high efficiency. Due to the robust field line patterns of $m=1$ mode, which does not vary with axial position as discussed in Sec.~\ref{uni1}, helicon waves (particularly the RLH mode) will be utilised to form a gap eigenmode in the downstream region of a helicon discharge. 

\section{Magnetic geometry}
The magnetic geometry of plasma confinement varies from uniform to focused, diverged, rippled, toroidal, dipolar, irregular et al., depending on the specific applications. Here, we consider four geometries: uniform as typically used for the helicon discharge, focused to get a pinched plasma for material processing, rippled to form a spectral gap and a gap eigenmode through breaking the periodicity locally, and toroidal mainly for magnetic confinement fusion where the gap eigenmode found in the rippled field has applications. 

\subsection{Uniform: WOMBAT}\label{wmb}
\begin{figure}[ht]
\begin{center}$
\begin{array}{c}
\hspace{-0.1cm}\includegraphics[width=0.7\textwidth,height=0.2\textwidth,angle=0]{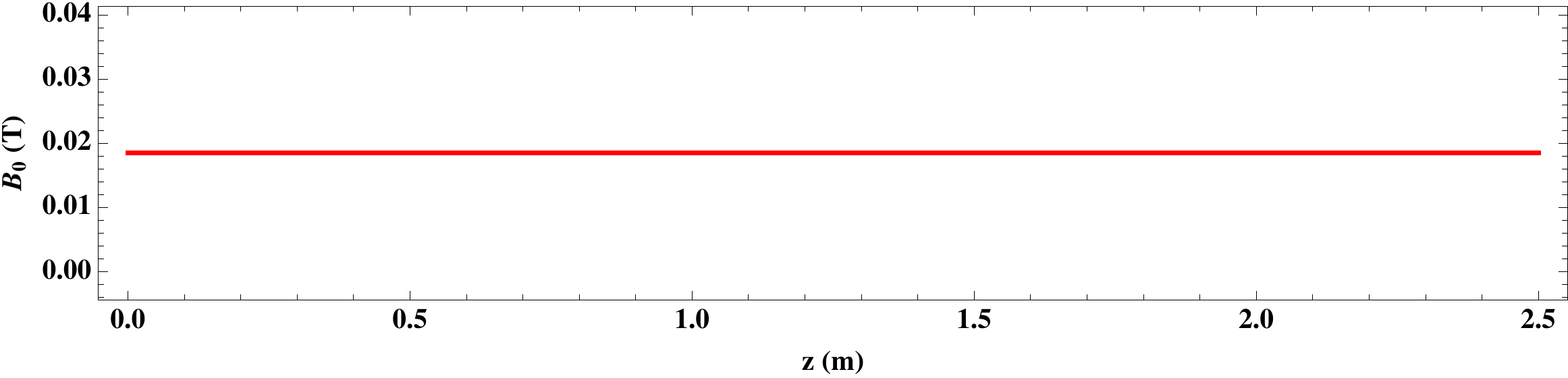}\\
\includegraphics[width=0.8\textwidth,angle=0]{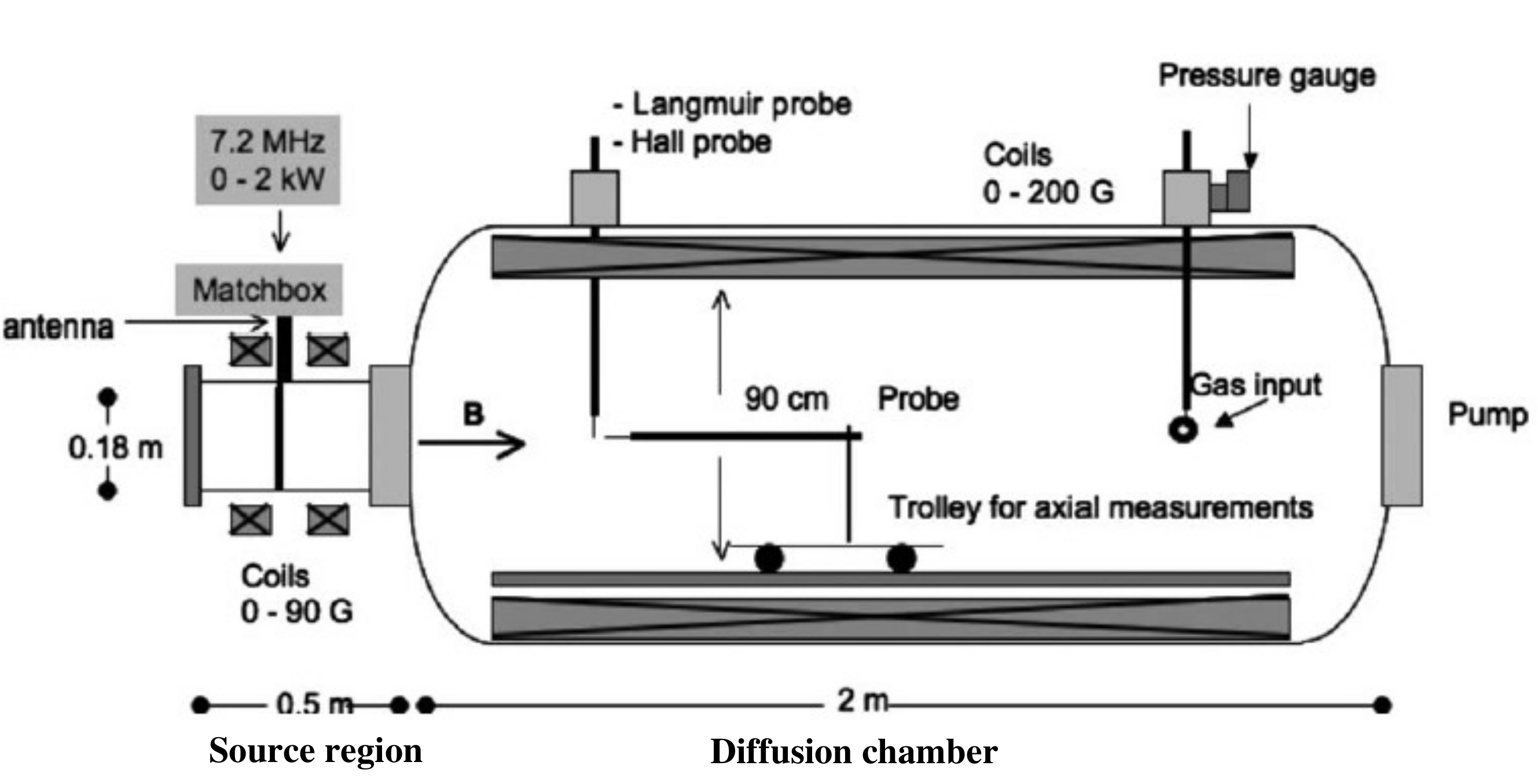}
\end{array}$
\end{center}
\caption{A schematic of the WOMBAT (Waves On Magnetised Beams And Turbulence) plasma experiment and the employed magnetic field.\cite{Corr:2009aa}}
\label{fg1_7}
\end{figure}

For a uniform magnetic geometry, we consider the WOMBAT,\cite{Boswell:1987aa} which was designed to provide a plasma environment to study wave physics including chaos, turbulence, wave saturation and the related driving mechanisms. This device will be employed in Chapter~\ref{chp2} for investigating drift waves in a flowing plasma cylinder.\cite{Chang:2011aa} Figure~\ref{fg1_7} shows a schematic and the employed magnetic field. 
It comprises a glass source tube and a large stainless steel diffusion chamber, which are attached to each other on axis. The source tube is $50$~cm long and $18$~cm in diameter, while the diffusion chamber is $200$~cm long and $90$~cm in inner diameter. A large solenoid is employed inside the diffusion chamber to provide a steady magnetic field, which can be up to $0.02$~T and highly uniform along the axis. To generate the plasma, a single loop antenna ($20$~cm in diameter, $1$~cm wide and $0.3$~cm thick) surrounding the source tube coupled the RF power to the argon gas. The $7.2$~MHz electric current passing through the loop creates a time varying magnetic field, which in turn induces azimuthal current in the argon gas and leads to break down and formation of the plasma.\cite{Boswell:1970ab} The base pressure of the diffusion chamber was maintained at $4 \times 10^{-6}$~Torr by a turbomolecular pump and a rotary pump. In this work, the large solenoid was used to produce a high density blue core in the diffusion chamber. The dispersion relation of resistive drift waves, different from that of helicon waves shown in Eq.~(\ref{eq1_20}), Eq.~(\ref{eq1_29}) and Eq.~(\ref{eq1_55}), is given in Eq.~(\ref{eq2_12}) coming from a two-fluid model.\cite{Hole:2002aa, Hole:2001aa}

The PCEN (Plasma CENtrifuge\cite{Dallaqua:1998aa}), for which a two-fluid electrostatic flowing plasma model (used in Chapter~\ref{chp2}) was developed,\cite{Hole:2002aa} is a typical vacuum arc centrifuge designed for separating metal isotopes. It also employs a uniform magnetic field but with faster normalised rotation frequency, higher temperature and higher axial velocity than those of the WOMBAT. Moreover, the plasma source is not a helicon discharge but a DC discharge, namely the plasma is generated by a discharge between cathode plate and anode mesh. A schematic of the PCEN can be found in \cite{Hole:2002aa}. 

\subsection{Focused: MAGPIE}\label{mgp}
\begin{figure*}[h]
\begin{center}$
\begin{array}{c}
\hspace{3cm}\includegraphics[width=0.7\textwidth,height=0.2\textwidth,angle=0]{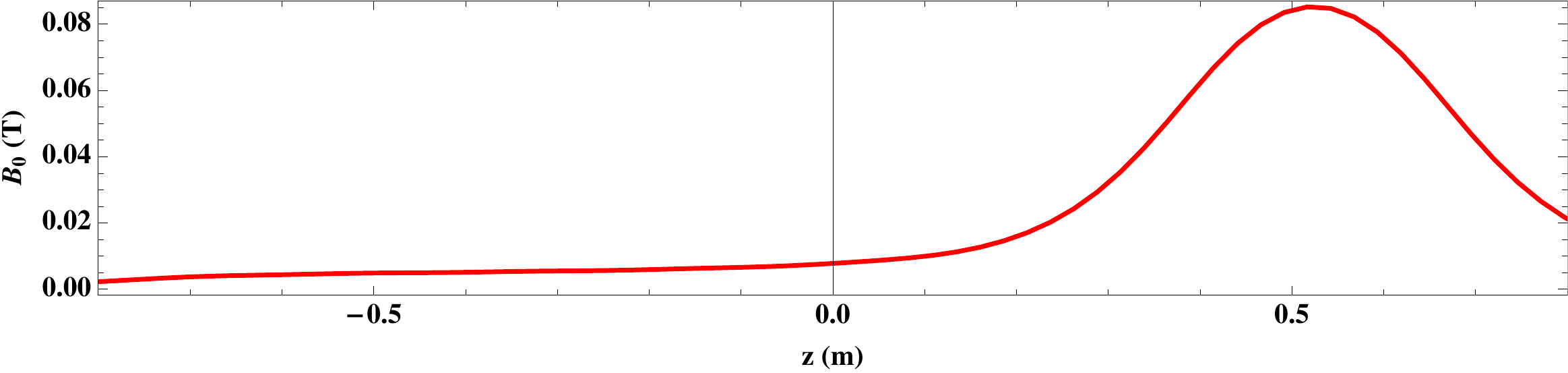}\\
\includegraphics[width=1\textwidth,angle=0]{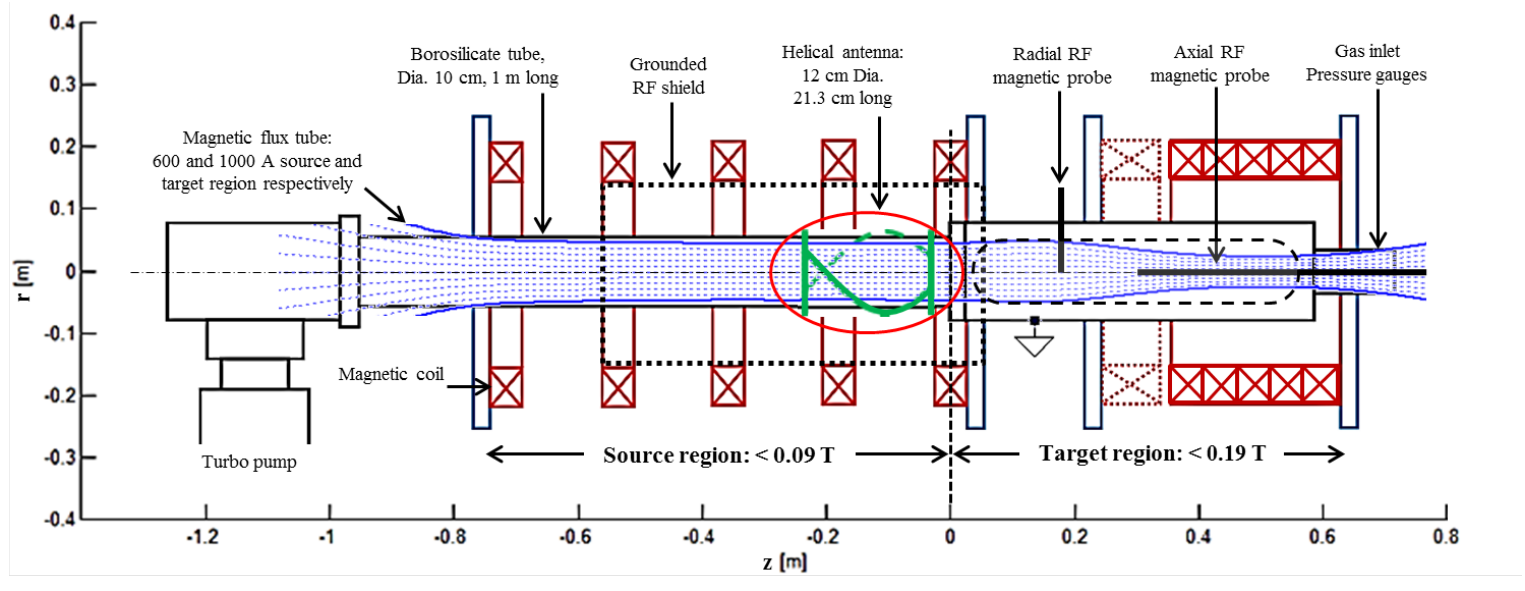}
\end{array}$
\end{center}
\caption{A schematic of the MAGPIE (MAGnetised Plasma Interection Experiment), together with the employed magnetic field.\cite{Blackwell:2012aa} A circle denotes the position of the helicon antenna which is left hand half-turn helical. The dot-dashed line is the machine and coordinate system axis, defining $r=0$~m. The coordinate system is right-handed with $\theta=0$ chosen to be the zenith angle. The dotted coil pair is an optional coil to control the magnetic field gradient, which is not used in the present work. }
\label{fg1_8}
\end{figure*}

To illustrate a focused magnetic geometry, the MAGPIE is employed. It is a linear plasma-material interaction machine which was recently built in the Plasma Research Laboratory at the Australian National University, and designed for studying basic plasma phenomena, testing materials in near-fusion plasma conditions, and developing potential diagnostics applicable for the edge regions of a fusion reactor.\cite{Blackwell:2012aa} Similar to other helicon devices,\cite{Carter:2002aa} MAGPIE mainly consists of a dielectric glass tube surrounded by an antenna, a vacuum pumping system, and a gas feeding system, together with a power supply system connected to the antenna, and various diagnostics. Figure~\ref{fg1_8} shows a schematic and the employed magnetic field, and introduces a cylindrical ($r$, $\theta$, $z$) coordinate system. The plasma is formed in the region under the antenna ($-0.243<z<-0.03$~m) and the near field to the antenna.\cite{Chen:1996aa} Following convention, however, we define the whole glass tube ($-1<z<0$~m) as the source region and the compressed field region ($0<z<0.7$~m) as the target region (or equivalently ``diffusion region" in some references). In MAGPIE, the $z<-0.243$~m region is named ``upstream" and $z>-0.03$~m ``downstream". 
A glass tube of length $1$~m and radius $0.05$~m is used to contain source plasmas in MAGPIE. A left hand half-turn helical antenna, $0.213$~m in length and $0.06$~m in radius, is wrapped around the tube and connected to a tuning box which can be adjusted between $7$ and $28$~MHz, a directional coupler, a $5$~kW RF amplifier,  and a $150$~W pre-amplifying unit. For the study presented in Chapter~\ref{chp3}, a RF power of $2.1$~kW, a frequency of $13.56$~MHz, a pulse width of $1.5$~ms and a duty circle of $1.5$~\% is used. The antenna current is measured by a Rogowski-coil-type current monitor. For these experiments an antenna current of magnitude $I_a=38.8$~A was measured. A grounded stainless steel cylindrical mesh surrounding the whole source region is employed to protect users. The source region is connected on-axis to the aluminium target chamber which is $0.7$~m in length and $0.08$~m in radius. Gases are fed through the downstream end of the target chamber, and drawn to the upstream end of the source tube by a $170$~L/s turbo pump. Gas pressures are measured in the target chamber by a hot cathode Bayard-Alpert Ionization gauge ($<0.01$~Pa), a Baratron pressure gauge ($0.01$--$10$~Pa) and a Convectron ($0.1$~Pa--$101.33$~kPa). In this experiment, argon gas is used with a filling pressure of $P_B=0.41$~Pa. The two regions, source and target, are surrounded by a set of water cooled solenoids, with internal radius of $0.15$~m. These source and target sets of solenoids are powered by two independent $1000$~A, $20$~V DC power supplies, providing flexibility in the axial configuration of the static magnetic field, e.~g. maximum of $0.09$~T and $0.19$~T in the source and target regions, respectively. The non-uniform field configuration is expected to provide a flexible degree of radial confinement, better plasma transport from the source tube to the target chamber, and possible increased plasma density according to previous studies.\cite{Blackwell:2012aa, Boswell:1997aa, Chen:1992aa, Chen:1997aa, Gilland:1998aa, Guo:1999aa} Wave propagations in the pinched plasma of the MAGPIE will be studied in Chapter~\ref{chp3}.\cite{Chang:2012aa} 

\subsection{Rippled: LAPD}\label{lpd1}
\begin{figure}[h]
\begin{center}$
\begin{array}{c}
\hspace{1.4cm}\includegraphics[width=0.7\textwidth,height=0.2\textwidth,angle=0]{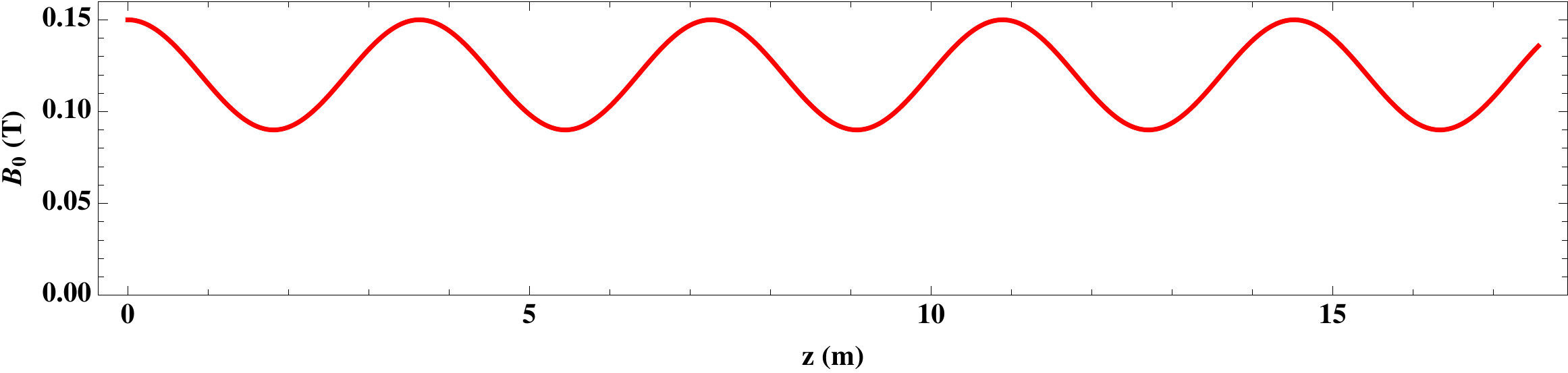}\\
\includegraphics[width=1\textwidth,angle=0]{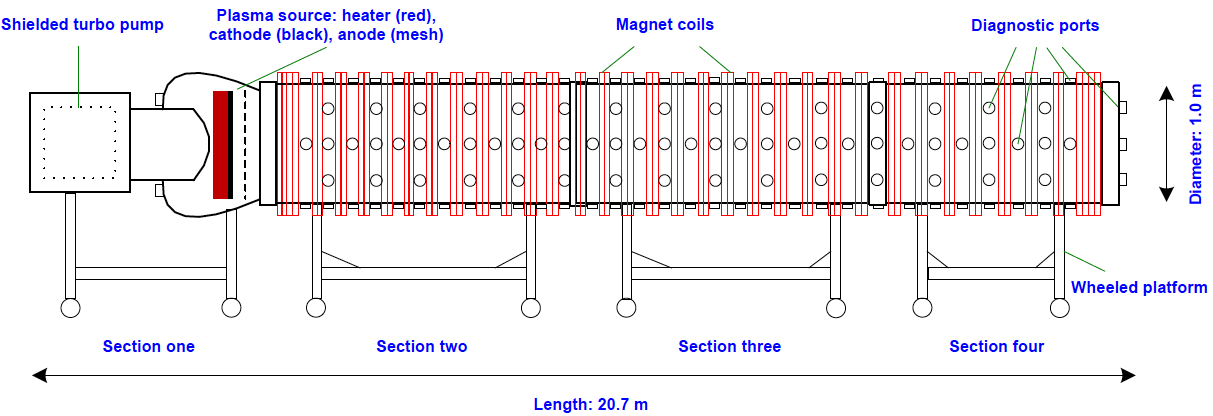}
\end{array}$
\end{center}
\caption{A schematic of the LAPD (LArge Plasma Device) and the employed magnetic field.\cite{Gekelman:1991aa, Zhang:2008aa}}
\label{fg1_9}
\end{figure}

We use the LAPD to illustrate a rippled magnetic geometry. The LAPD is a large, linear plasma research device designed to study space plasma processes,\cite{Gekelman:1991aa} and more recently fusion research. It is ideal for studying the fundamental properties of plasmas on a large scale length, e. g. propagation of low frequency whistlers and SAW.\cite{Gekelman:1991aa, Zhang:2008aa} As shown in Fig.~\ref{fg1_9}, the LAPD consists of four stainless-steel chamber sections, each of which is mounted on a wheeled platform to allow separate maintenance.  
The diameter and length of the machine are $1$~m and $20.7$~m, respectively.\cite{Zhang:2008aa} The first bell-shaped section mounts a heater, a cathode and an anode. The heater is made of hollow alumina ceramic tubes which are threaded with tungsten filaments. This special design is expected to provide a uniform heating for the cathode with temperature slightly above the cathode emission temperature. When applied a negative voltage, the heated cathode emits electrons through both the thermionic electron emission and the DC discharge (towards the anode). The emitted electrons then go through the grid anode and ionise the neutral gas filled in the system, forming a plasma column along the system's axis. The plasma column is $18$~m long and with densities up to $2\times10^{19}~\mathrm{m}^{-3}$.\cite{Gekelman:1991aa} Radial expansions of the plasma are confined by an externally applied magnetic field, which is generated by a set of $68$ magnet coils surrounding the chamber, as shown in Fig.~\ref{fg1_9}. The field strength can be up to $0.3$~T, and varied with axial position by varying the current on each of the magnet coils, making it capable to configure a periodic array of magnetic mirrors.\cite{Zhang:2008aa} Therefore, the LAPD is a promising device for identifying a spectral gap of plasma waves, and also an appropriate candidate for observing a gap eigenmode.\cite{Chang:2013aa} Discussions about identifying the gap eigenmodes of RLH waves and SAW will be given in Chapter~\ref{chp4} and Chapter~\ref{chp5}, respectively. Moreover, there are $450$ radial ports on the vacuum chamber thus providing excellent access for diagnostics.\cite{Gekelman:1991aa} 

\subsection{Toroidal: tokamak}\label{tmc1}
\begin{figure}[ht]
\begin{center}$
\begin{array}{ll}
(a)&(b)\\
\includegraphics[width=0.48\textwidth,angle=0]{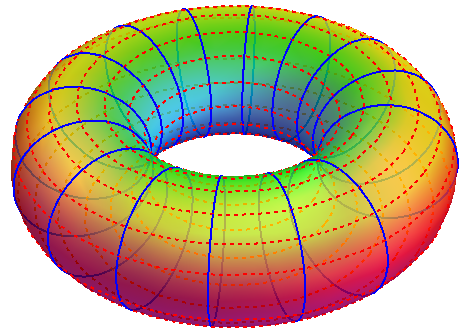}&\includegraphics[width=0.48\textwidth,angle=0]{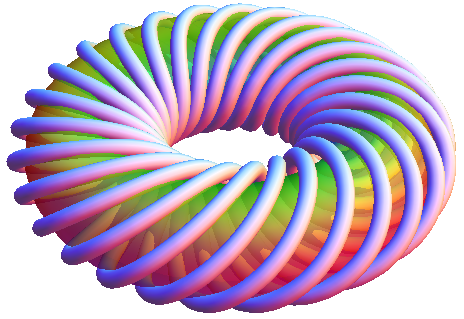}
\end{array}$
\end{center}
\caption{A schematic of the tokamak: poloidal (solid) and toroidal (dashed) magnetic field lines (a) form a helical magnetic field nest confining the plasma torus (b).}
\label{fg1_10}
\end{figure}

Finally, a toroidal magnetic geometry is illustrated in term of a tokamak. Controlled thermonuclear fusion can be an environmentally attractive and sustainable energy source, providing a large scale base-load power.\cite{Freidberg:2007aa} However, realising the controlled thermonuclear fusion faces many challenges, which in science is the combined requirement of confining a sufficient quantity of plasma for a sufficiently long time at a sufficiently high temperature to make net fusion power. The Lawson criterion formulates this requirement specifically as $\hat{n}\tau_{\mathrm{E}} \hat{T}>5\times 10^{21}~\mathrm{m}^{-3}$~s keV, with $\hat{n}$ and $\hat{T}$ the peak ion density and temperature in the plasma and $\tau_{\mathrm{E}}$ the energy confinement time.\cite{Lawson:1957aa, Wesson:2011aa} We focus on the magnetic confinement fusion (MCF): the most achievable scheme. To avoid the loss of charged particles parallel to the magnetic field line, the magnetic field is usually shaped into a torus. A tokamak is an axisymmetric torus with a poloidal magnetic field that is produced by a toroidal electric current flowing inside the plasma, and a very strong longitudinal field parallel to the current.\cite{Artsimovich:1972aa} The word ``tokamak" is a transliteration from an acronym of Russian words: \textbf{to}roidalnaya \textbf{ka}mera and \textbf{ma}gnitnaya \textbf{k}atushka, meaning ``toroidal chamber" and ``magnetic coils".\cite{Wesson:2011aa} Figure~\ref{fg1_10} shows a schematic of the tokamak: two components of the magnetic field (poloidal and toroidal) forming a helical magnetic field nest confining the plasma torus. 
The world's largest tokamak is being under construction at the Cadarache facility in the south of France: International Thermonuclear Experimental Reactor (ITER), which aims to demonstrate the scientific and technological feasibility of producing commercial energy from the controlled thermonuclear fusion. It is designed to produce $500$~MW of fusion power from $50$~MW of input power, with the energy gain of $Q=10$.\cite{iter:2007aa} 

Given that fusion plasmas are energetically rich, complex exothermic physical systems, charged particles can still be lost even if the toroidal magnetic configuration is perfectly designed. Populations of charged particles can be accelerated well above thermal speeds through processes such as resonant wave absorption, charge exchange with injected high energy neutral beams, magnetic reconnection, and the fusion process itself. The importance of these energetic populations is that they can drive electromagnetic waves which, in turn, can eject the same driving particles from confinement.\cite{Fasoli:2007aa} The most easily excited modes usually reside in their spectral gaps, which are formed by the various periodicities in the toroidally confined plasma.\cite{Heidbrink:2008aa} These gap eigenmodes are introduced by symmetry-breaking due to toroidicity, plasma ellipticity and higher order shaping  effects. Therefore, it is of practical interest to study the formation of gap eigenmodes and their interactions with energetic particles. We shall investigate two types of gap eigenmodes: RLH waves in the whistler frequency range, which can be driven unstable by energetic electrons,\cite{Chang:2013aa} and SAW with frequency below the ion cyclotron frequency, which can be excited by energetic ions in tokamaks. 

\section{Aims of the thesis}
The aim of thesis is to study the impact of magnetic geometry on spontaneous and driven wave modes in cylindrical helicon-source plasmas. Three questions will be addressed: 

(1) what are the plasma-driven wave modes in a rotating, flowing plasma with uniform magnetic field? 

(2) how do wave propagation characteristics of antenna-driven modes change with ramping magnetic field, and can this non-uniformity enhance the plasma density? 

(3) can gap eigenmodes be formed in a plasma cylinder with rippled magnetic field, by breaking the field periodicity locally?

\chapter{Drift waves in a uniformly magnetised plasma with flows}\label{chp2}

A two-fluid model developed originally to describe wave oscillations in the vacuum arc centrifuge, a cylindrical, rapidly rotating, low temperature and confined plasma column, is applied to interpret plasma oscillations in a RF generated linear magnetised plasma (WOMBAT) with similar density and field strength. Compared to typical centrifuge plasmas, WOMBAT plasmas have slower normalised rotation frequency, lower temperature and lower axial velocity. Despite these differences, the two-fluid model provides a consistent description of the WOMBAT plasma configuration and yields qualitative agreement between measured and predicted wave oscillation frequencies with axial field strength. In addition, the radial profile of the density perturbation predicted by this model is consistent with the data. Parameter scans show that the dispersion curve is sensitive to the axial field strength and the electron temperature, and the dependence of oscillation frequency with electron temperature matches the experiment. These results consolidate earlier claims that the density and floating potential oscillations are a resistive drift mode, driven by the density gradient. To our knowledge, this is the first detailed physics model of flowing plasmas in the diffusion region away from the RF source. Possible extensions to the model, including temperature non-uniformity and magnetic field oscillations, are also discussed.

\section{Introduction}\label{int2}
Oscillations have been observed in the electric probe signals of many laboratory plasmas, including theta pinches,\cite{Rostoker:1961aa, Freidberg:1978aa} magnetic mirrors,\cite{Perkins:1963aa, Kuo:1964aa, Hooper:1983aa} Q machines,\cite{Motley:1963aa} vacuum arc centrifuge \cite{Krishnan:1981aa} and helicon plasmas.\cite{Boswell:1983aa, Boswell:1984aa, Degeling:1999aa, Greiner:1999aa, Sun:2005aa} Theta pinches and magnetic mirrors behave vastly different from the latter because of their high beta and significant magnetic curvature, which result in highly different plasma and field geometry. In contrast, Q machine, plasma centrifuge and helicon plasmas have low beta and negligible magnetic curvature, and have similar plasma properties. We exploit this similarity by deploying a two-fluid collisional model, initially developed to describe vacuum arc centrifuge plasmas, to describe the configuration and oscillations in a helicon plasma. Our goal is to better describe a class of electrostatic oscillations observed, but not yet conclusively identified, in the diffusion region of helicon plasmas, away from the RF source. 

The helicon plasma we study is provided by the WOMBAT device, whose schematic has been given in Fig.~\ref{fg1_7}. The WOMBAT has been used to study helicon waves,\cite{Ellingboe:1996aa, Degeling:1999aa, Corr:2007aa} which have attracted great interest in the past decades due to their ability to produce high plasma densities. \cite{Boswell:1984ab} Such plasmas also provide a rich environment for the experimental study of drift waves and their impact on anomalous transport,\cite{Schroder:2004aa} which is important to understand in fusion plasmas. Although there have been many publications on helicon wave physics, less attention has been devoted to the physics of low frequency oscillations in the diffusion region away from the source.\cite{Schroder:2005aa} Light \emph{et al.} \cite{Light:2001aa, Light:2002aa} observed a low frequency electrostatic instability only above a critical magnetic field and identified it as a mixture of drift waves and a Kelvin-Helmholtz instability. Degeling \emph{et al.} \cite{Degeling:1999aa, Degeling:1999ab} observed relaxation oscillations in the kilohertz range that were associated with the various types of mode coupling in helicon discharges. More recently, a cylindrical linear model which treats global eigenmodes \cite{Ellis:1980aa} has been employed to study drift waves in the magnetically confined plasma device VINETA,\cite{Schroder:2005aa, Schroder:2004aa, Grulke:2007aa} however, the model does not involve the $\mathbf{E} \times \mathbf{B}$ rotation and neglects ion parallel motion which may influence the characteristics of low frequency oscillations.\cite{Horton:1984aa} Sun \emph{et al.} \cite{Sun:2005aa} observed low frequency ($13$ kHz) oscillations near the end of helicon source region, and identified them as resistive-drift Alfv\'{e}n waves. Sutherland \emph{et al.} \cite{Sutherland:2005aa} observed low frequency ion cyclotron waves that were highly localised along the axial center of the WOMBAT plasma device. Their analysis of spectral measurements suggested the possible existence of a four wave interaction, where energy is down-converted to the ion cyclotron mode from the helicon pump.

Low frequency oscillations in helicon plasmas can be broadly classified into two types: the Kelvin-Helmholtz instability and drift waves. The Kelvin-Helmholtz instability is driven by a velocity shear in a mass flow or a velocity difference across the interface between two fluids. \cite{Miloshevsky:2010aa} This can occur in helicon plasmas, because the streaming velocity of the ion fluid is much slower than that of electrons. The drift wave is a universal instability driven by a plasma pressure gradient perpendicular to the ambient field.\cite{Chen:1984aa} It can arise in fully ionised, magnetically confined and low-beta plasmas,\cite{Hendel:1968aa} and has been observed in both linear and toroidal field geometries.\cite{Okabayashi:1977aa, Pecseli:1983aa, Liewer:1985aa, Klinger:1992aa, Poli:2006aa} In WOMBAT plasmas, the velocity shear is small and the density and temperature gradients produce large pressure gradients, suggestive of large drive of drift waves. We thus restrict attention to this class of modes. 

In this chapter, a two-fluid model, which was developed originally for the vacuum arc centrifuge by Hole \emph{et al.} \cite{Hole:2002aa, Hole:2001ab} to explain oscillations observed in the density and electric potential, is applied to study low frequency oscillations observed in the WOMBAT. We show that the equilibrium and perturbed density profiles and the space potential profile from the model and data are consistent. The model predicts unstable modes with a similar global mode radial structure, and a frequency comparable to observed signals. The chapter is organised as follows: Sec.~\ref{exp2} gives a brief description of the diagnostics on WOMBAT, together with typically measured parameters, Sec.~\ref{mdl2} introduces the two-fluid model and Sec.~\ref{phy2} discusses the wave physics revealed by this model and the data. Finally, Sec.~\ref{cnl2} discusses the possible extensions to this model and presents concluding remarks. 

\section{Diagnostics and measured parameters}\label{exp2}
To measure spatial and temporal profiles in the plasma, an uncompensated, translating Langmuir probe was inserted radially into the diffusion chamber, 50 cm from the source and diffusion chamber interface. The central wire of the probe was fed through an alumina support which was in turn shielded by a 6 mm diameter grounded steel tubing covering the whole extent of the probe length up to the probe tips. The removable probe tip was made of a $0.2$~mm diameter and 8 mm long nickel wire. Radial translation of the probe was set using a computer controlled stepper motor arrangement that allowed the probe tip position to be selected with an accuracy of a few micrometers. To determine the plasma $I(V)$ characteristics, the bias voltage on the Langmuir probe was swept between $-50$~V and $+30$~V using a \emph{Labview} program. The plasma density ($n_i$), electron temperature ($T_e$), plasma potential ($V_p$) and floating potential ($V_f$) were determined from the $I(V)$ characteristics of the cylindrical probe. For the present high-density plasma discharge, the effect that RF fluctuations have on the floating potential, and hence the $I(V)$ characteristics, was found to be negligible. The reproducibility for $T_e$ was within $0.5$~eV, and the signal-to-noise ratio was improved by averaging many $I(V)$ curves. The effect that fluctuations due to waves will have on time-averaged $I(V)$ characteristics will be far more dominant than the RF fluctuations. These wave fluctuations would lead to error magnification in the second derivative of the $I(V)$ curve and hence the electron energy distribution function. For this reason, the plasma parameters were obtained directly from the $I(V)$ curve and not by the Druyvesteyn method.\cite{Biloiu:2010aa}

Typical Langmuir probe measurements of WOMBAT argon plasmas are shown in Table~\ref{tb2_1}.
\begin{table}[ht]
\caption{Typical parameters of WOMBAT and PCEN (Plasma CENtrifuge).}
\begin{center}
\begin{tabular}{lll}
parameter                                                       & WOMBAT                             & PCEN \cite{Dallaqua:1998aa,Hole:2002aa} \\
\hline
$n_i$ (on axis)                                                 & $1.4\times 10^{19}\ \mathrm{m^{-3}}$   & $5.2\times10^{19}\ \mathrm{m^{-3}}$\\
$T_e$                                                           & $1.5$ eV                           & $2.9$ eV\\
$T_i$                                                           & $0.1$ eV                           & $2.9$ eV\\
$m_i$                                                           & $40$ amu (Ar)                           & $24.31$ amu (Mg)\\
$B_z$                                                           & $0.0185$ T                         & $0.05$ T\\
$Z$                                                             & $1.0$                              & $1.5$\\
$V_{z0}$                                                        & $200\ \mathrm{m~s^{-1}}$               & $10^4\ \mathrm{m~s^{-1}}$\\
$\omega_0$                                                      & $1.42\ \mathrm{krad~s^{-1}}$           & $184\ \mathrm{krad~s^{-1}}$\\
$\omega_{ci}= \frac{B_z e Z}{m_i}$                              & $44.5\ \mathrm{krad~s^{-1}}$           & $295\ \mathrm{krad~s^{-1}}$\\
$\Omega_{i0}=\frac{\omega_0}{\omega_{ci}}$                      & $0.032$                            & $0.59$\\
$\Psi=(\frac{T_i}{T_e}+Z)\frac{k_B T_e}{m_i \omega_{ci}^2 R^2}$ & $0.54$                           & $1.6$ \\
$\delta_{rs}=\frac{e Z n_{i0}}{B_z} \frac{\eta_L}{\gamma_E}$         & $0.0235$                           & $0.03$\\
$R$(characteristic radius)                                      & $6$ cm                             & $1.43$ cm
\end{tabular}
\end{center}
\label{tb2_1}
\end{table}
Other parameters, obtained for similar helicon plasmas,\cite{Kline:2003aa, Scime:2007aa} and measured by laser induced fluorescence, include: the ion temperature $T_i$, the bulk rotation frequency $\omega_0$, and axial streaming velocity $V_{z0}$. The parameter $m_i$ is the argon ion mass and the charge number $Z$ is taken to match similar argon plasma conditions. \cite{Plihon:2005aa} The remaining parameters are introduced in section~\ref{mdl2}. The antenna driving frequency is $7.2$~MHz, which is much higher than the bulk rotation frequency and wave frequency. Finally, plasma parameters for the PCEN device are taken from published work.\cite{Dallaqua:1998aa, Hole:2002aa} It should be noted that the operating gas in the PCEN work is magnesium while here it is argon.

\section{Plasma model}\label{mdl2}

\subsection{Model assumptions}\label{asm2}

The two-fluid model developed by Hole \emph{et al.} \cite{Hole:2002aa, Hole:2001aa} is based on the following plasma assumptions:

(1) Ions of different charge can be treated as a single species with average charge $Z$.

(2) The plasma is quasi-neutral, giving that $n_e=Z n_i$.

(3) The steady-state plasma is azimuthally symmetric and has no axial structure.

(4) The effects induced by plasma fluctuations on the externally applied field can be neglected.

(5) Both finite Larmor radius (FLR) and viscosity effects are negligible.

(6) For the range of frequencies considered here, the electron inertia can be neglected. 

(7) The ion and electron temperatures, $T_i$ and $T_e$, are uniform across the plasma column.

(8) The steady-state ion density distribution has a form of $n_{0}=n_i(0)e^{-(r/R)^2}$, which is a Gaussian profile. Here, $n_i(0)$ is the on-axis ion density, and $R$ is the characteristic radius at which the density is $1/e$ of its on-axis value.

(9) The steady-state velocities of ions and electrons can be written as $\mathbf{v_i}=(0, \omega_i r, v_{iz})$ and $\mathbf{v_e}=(0, \omega_e(r) r, v_{ez}(r))$, respectively, where $\omega_i$ is the ion rigid rotor rotation frequency, $v_{iz}$ is the ion uniform axial streaming velocity, $\omega_e(r)$ is the electron rotation frequency, and $v_{ez}(r)$ is the electron streaming velocity. While treated in other work \cite{Hole:2001aa}, radial diffusion of both ions and electrons due to electron-ion collision is negligible.

Here, length and time are normalised to $R$ and $1/\omega_{ci}$ respectively. A normalised cylindrical coordinate system is then developed, with $(\bar{r},~\theta,~\bar{z})=(r/R,~\theta,~z/R)$ and $\tau=\omega_{ci}t$, where $\bar{r}$ and $\bar{z}$ are the normalised radial and axial positions, respectively. 

\subsection{Two-fluid equations}\label{eqt2}
The model comprises the motion and continuity equations of ion and electron fluids, written respectively as
\begin{equation}\label{eq2_1}
\frac{\partial \mathbf{u_i}}{\partial \tau}+(\mathbf{u_i} \cdot \mathbf{\nabla}) \mathbf{u_i} = -\psi(Z \mathbf{\nabla} \chi+\lambda_T \mathbf{\nabla} l_i)+\mathbf{u_i} \times \mathbf{\hat{\bar{z}}}+\delta_{rs} n_{cl} \mathbf{\tilde{\xi}} \cdot (\mathbf{u_e}-\mathbf{u_i}),
\end{equation}
\begin{equation}\label{eq2_2}
\psi Z(-\mathbf{\nabla} l_i+\mathbf{\nabla} \chi)-\mathbf{u_e} \times \mathbf{\hat{\bar{z}}}+\delta_{rs} n_{cl} \mathbf{\tilde{\xi}} \cdot (\mathbf{u_i}-\mathbf{u_e})=0,
\end{equation}
\begin{equation}\label{eq2_3}
-\frac{\partial l_i}{\partial \tau}=\mathbf{\nabla} \cdot \mathbf{u_i}+\mathbf{u_i} \cdot \mathbf{\nabla} l_i,
\end{equation}
\begin{equation}\label{eq2_4}
-\frac{\partial l_i}{\partial \tau}=\mathbf{\nabla} \cdot \mathbf{u_e}+\mathbf{u_e} \cdot \mathbf{\nabla} l_i,
\end{equation}
with terms defined as follows: 
\[\mathbf{u_i}=\frac{\mathbf{v_i}}{\omega_{ci} R}=(\bar{r} \varphi_{\bar{r}i}, \bar{r} \Omega_i, u_{i\bar{z}}), \mathbf{u_e}=\frac{\mathbf{v_e}}{\omega_{ci}R}=(\bar{r} \varphi_{\bar{r}e}, \bar{r} \Omega_e, u_{e\bar{z}}),\]
\[\lambda_T=\frac{T_i}{T_e}, \psi=\frac{k_B T_e}{m_i \omega_{ci}^2 R^2}, \chi=\frac{e \phi}{k_B T_e}, \mathbf{\tilde{\xi}}=\mathrm{diag}(\xi_\bot, \xi_\bot, 1),\]
\[l_i=\mathrm{ln} \frac{n_i}{n_i(0)}, n_{cl}=\frac{\nu_{ei}}{\nu_{ei}(0)}, \delta_{rs}=\frac{e Z n_{i0}}{B_z} \frac{\eta_L}{\gamma_E}.\]
Here, subscript $i$ and $e$ refer to ion and electron parameters respectively, $\varphi_{\bar{r}}$ is the normalised radial velocity divided by $\bar{r}$, $\Omega$ is the normalised rotation frequency, $u_{\bar{z}}$ indicates the normalised axial velocity, $\lambda_T$ is the ratio between ion and electron temperatures, $\psi$ is a convenient constant which for $\lambda_T=1$ becomes the square of the normalised ion thermal velocity, $\chi$ is a normalised electric potential $\phi$, $l_i$ is the logarithm of the ratio of the ion density $n_i$ to its on-axis value $n_i(0)$, and $n_{cl}$ is the ratio of the electron-ion collision frequency $\nu_{ei}$ to its on-axis value $\nu_{ei}(0)$. Also, $\delta_{rs}$ is the normalised resistivity parallel to the magnetic field, where $\eta_L$ is the electrical resistivity of a Lorentz gas and $\gamma_E$ is the ratio of the conductivity of a charge state $Z$ to that in a Lorentz gas.\cite{Spitzer:1962aa}

\subsection{Steady-state solution}\label{std2}
In cylindrical geometry with a purely axial constant field ${\mathbf B}=(0, 0, B_z)$, the steady-state solution of this model is given by
\begin{equation}\label{eq2_5}
\chi_0(\bar{r})=\chi_c+\left[\frac{\Omega_{i0}}{2\psi Z}(1+\Omega_{i0})+\frac{\lambda_T}{Z}\right]\bar{r}^2, 
\end{equation}
with
\begin{equation}\label{eq2_6}
\Omega_{e0}=\Omega_{i0}(1+\Omega_{i0})+2 \psi (\lambda_T+Z),
\end{equation}
where $\chi_c$ is an arbitrary reference potential. The axial current in this model is unconstrained, and can arbitrarily be set to zero ($u_{i\bar{z}0}=u_{e\bar{z}0}$), consistent with WOMBAT boundary conditions. 

\begin{figure}[ht]
\begin{center}$
\begin{array}{c}
\begin{array}{cc}
\includegraphics[width=0.5\textwidth,height=0.38\textwidth,angle=0,scale=0.98]{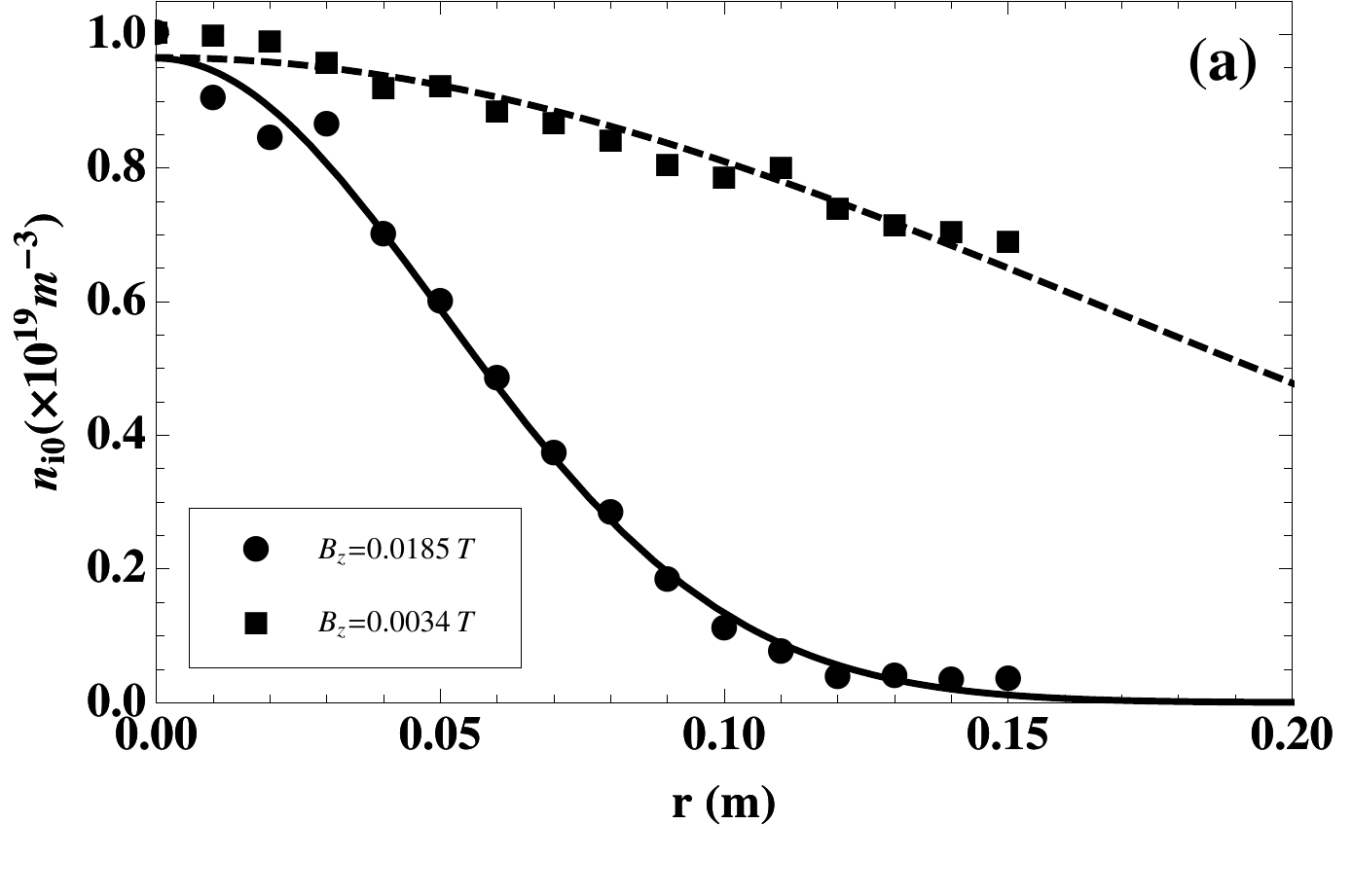}&\hspace{-0.4cm}\includegraphics[width=0.5\textwidth,height=0.38\textwidth,angle=0,scale=1.01]{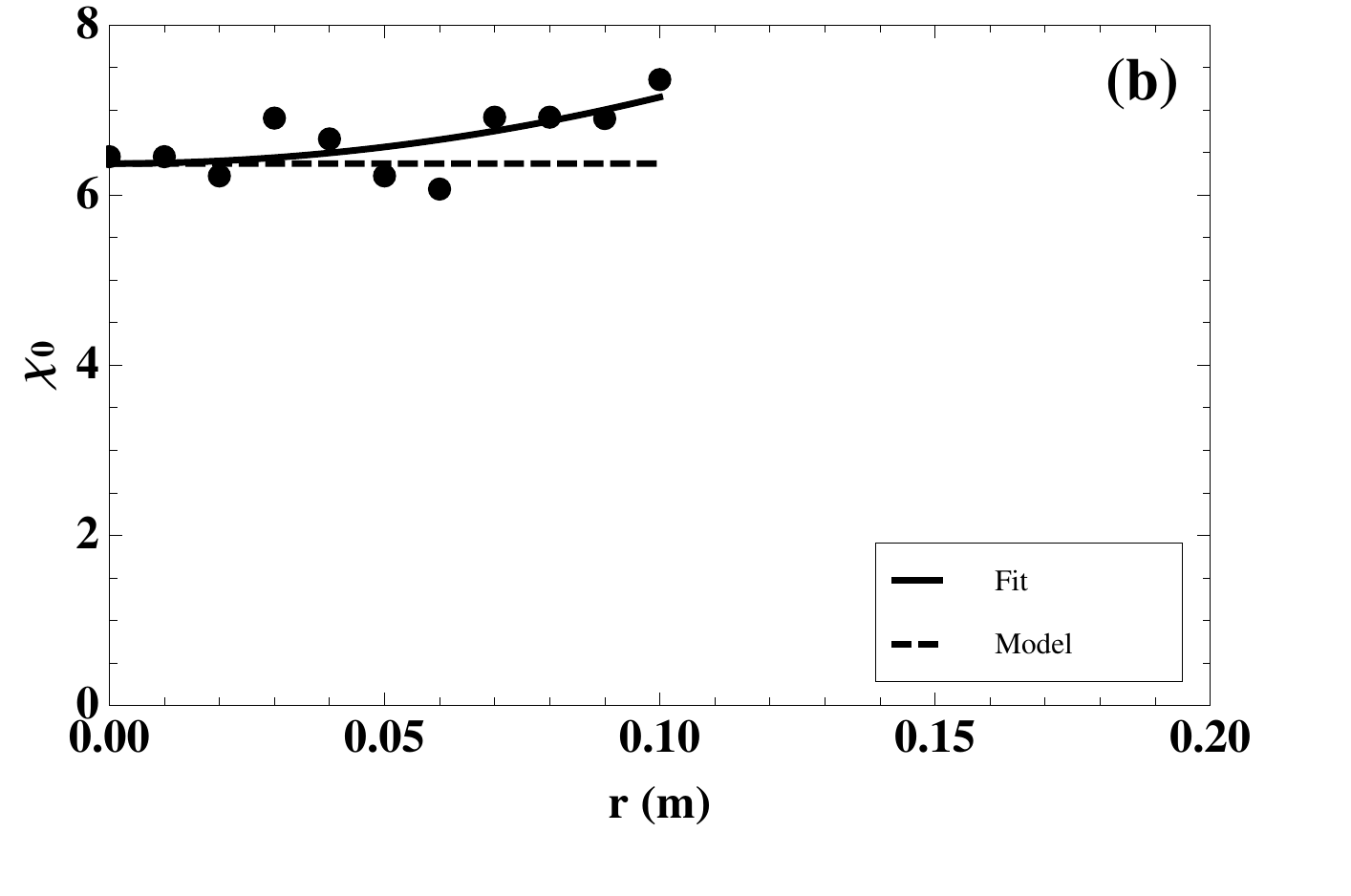}\\
\end{array}
\\
\begin{array}{c}
\includegraphics[width=0.5\textwidth,height=0.38\textwidth,angle=0,scale=1.025]{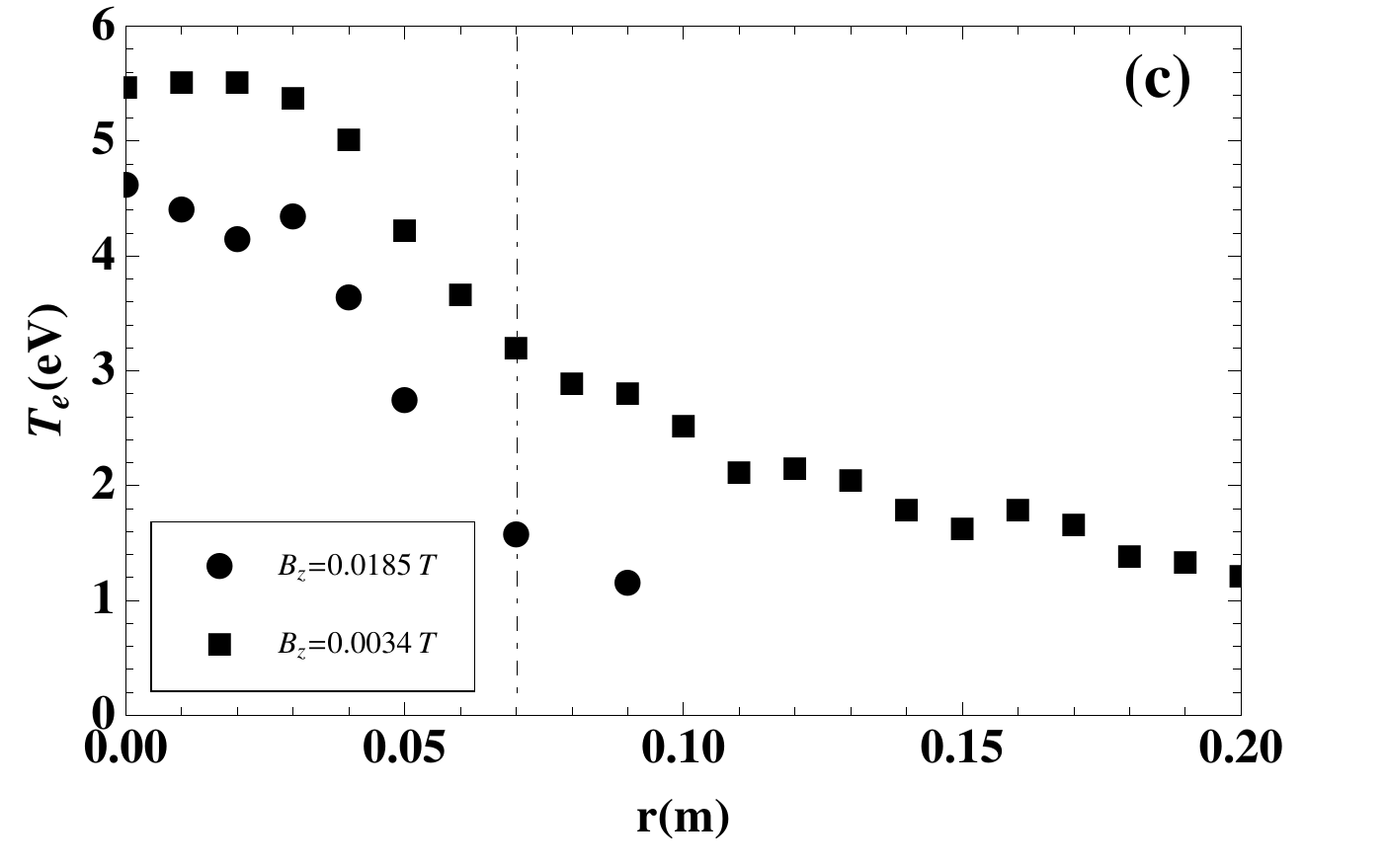}
\end{array}
\end{array}$
\end{center}
\caption{Plasma configuration of WOMBAT: (a) Gaussian fit of the equilibrium density profile, for $B_z=0.0185$~T and $B_z=0.0034$~T respectively;\cite{Corr:2007aa} (b) the radial variation of the normalised space potential (the measurable electric potential is about $\phi_0=8.6\times 10^{-5}T_e\chi_0$~V) from the fit procedure (solid line), the model (dashed) and the data (dots),\cite{Corr:2009aa} which was taken under the conditions of $B_z=0.0068$~T, $0.6$~mTorr and $1708$~W in the experiment; (c) the radial profile of the electron temperature at two field strengths: $B_z=0.0185$~T and $B_z=0.0034$~T. \cite{Corr:2007aa} The dashed vertical line in (c) corresponds to the radial location of fluctuation measurements. }
\label{fg2_1}
\end{figure}

We have compared the steady-state solution to experimental data. Measurements of the fluctuation frequency from probe measurements suggest rigid rotation, which is consistent with the spectroscopic data obtained by Scime \emph{et al}..\cite{Scime:2007aa} Figure~\ref{fg2_1}(a) shows the equilibrium density profile of the WOMBAT plasma for two different axial field strengths, $B_z=0.0185$~T and $B_z=0.0034$~T, respectively.\cite{Corr:2007aa} 
Overlaid are best fits to the profile using a Gaussian profile ($n_{i0}=n_i(0)e^{-(r/R)^2}$): these show reasonable agreement over the range of the data. The stronger field produces better confinement with a smaller characteristic radius $R$. Figure~\ref{fg2_1}(b) shows the radial variation of the space potential in WOMBAT for a slightly different field strength, $B_z=0.0068$~T.  \cite{Corr:2009aa} Overlaid is a best fit to the data using the parabolic potential profile in Eq.~(\ref{eq2_5}), with the arbitrary reference potential ($\chi_c$) and gradient $G=\frac{\Omega_{i0}}{2\psi Z}(1+\Omega_{i0})+\frac{\lambda_T}{Z}$ free parameters. Although the fit, for which $G=0.5$, is reasonable, the scatter in the data is large. For the same $\chi_c$, we have over plotted the model potential profile with $G=0.04$. Within the bulk of the plasma, out to the characteristic radius $r=0.08$~m, model and observed potential profiles broadly agree. Figure~\ref{fg2_1}(c) shows the radial profile of $T_e$ in the WOMBAT plasma.\cite{Corr:2007aa} The constant electron temperature of the model is an approximation to the radially varying experimental profile. For the model, we have chosen the $T_e$ value at the position at which the frequency was determined, $r=7$~cm. 

\subsection{Normal mode analysis}\label{ana2}
To compute the normal modes of the system, Hole \emph{et al.}\cite{Hole:2002aa} apply a linear perturbation treatment with plasma parameters $\zeta$ taking the form
\begin{equation}\label{eq2_7}
\zeta (\tau, \bar{r}, \theta, \bar{z})=\zeta_0(\bar{r})+\varepsilon \zeta_1(\bar{r}) e^{i(m \theta+k_{\bar{z}} \bar{z}-\omega \tau)}.
\end{equation}
Here, $\varepsilon$ is the perturbation parameter, $m$ the azimuthal mode number, $k_{\bar{z}}$ the axial wave number and $\omega$ the angular frequency. We consider a complex value of $\omega$ so as to study both the mode frequency and growth rate, i. e. $\omega=\omega^r+i\omega^i$. To first order in $\varepsilon$ the system of Eq.~(\ref{eq2_1})-(\ref{eq2_4}) reduces to
\begin{equation}\label{eq2_8}
\left(
\begin{array}{clcr}
\psi [l_{i1}'(y)-X_1'(y)]\\
y \varphi_{\bar{r}i1}'(y)\\
y \varphi_{\bar{r}e1}'(y)-i m \Psi X_1'(y)\\
\Psi \delta_{rs} \xi_\bot e^{-y} X_1'(y)\\
0
\end{array}
\right)
=
\tilde{\mathbf{A}}
\left(
\begin{array}{clcr}
l_{i1}(y)\\
X_1(y)\\
\varphi_{\bar{r}i1}(y)\\
\varphi_{\bar{r}e1}(y)\\
u_{e\bar{z}}(y)
\end{array}
\right),
\end{equation}
where $\tilde{\mathbf{A}}$ is the matrix
\small
\begin{equation}\label{eq2_9}
\begin{array}{lll}
\tilde{\mathbf{A}} & = &
\left(
\begin{array}{ccccc}
\frac{m \Psi C}{2 \varpi y}                                                 & 0                   & \frac{i \varpi}{2}-\frac{i C^2}{2 \varpi} & \frac{i C}{2 \varpi} & 0\\
\frac{i}{2}\left[\varpi-\frac{\Psi}{\varpi}(\frac{m^2}{y}+k_{\bar{z}}^2)\right] & 0                   & -1+y-\frac{m C}{2 \varpi}                       & \frac{m}{2 \varpi}   & 0\\
\frac{i}{2}(\varpi-m \Omega_{i0}^2-2m \Psi)                                & 0                   & 0                                                     & -1+y                       & -\frac{i k_{\bar{z}}}{2}\\
0                                                                                & \frac{i m \Psi}{2y} & 0                                                     & -\frac{1}{2}               & 0\\
0                                                                                & -i k_{\bar{z}} \Psi & 0                                                     & 0                          & 0\\
\end{array}
\right)\\
\\
&+&\delta_{rs}
\left(
\begin{array}{ccccc}
0                                                                                & 0 & -\frac{\xi_\bot e^{-y}}{2}                   & \frac{\xi_\bot e^{-y}}{2}                   & 0\\
0                                                                                & 0 & 0                                            & 0                                           & 0\\
0                                                                                & 0 & -\frac{i m \xi_\bot e^{-y}}{2}               & \frac{i m \xi_\bot e^{-y}}{2}               & 0\\
\xi_\bot e^{-y}(-\frac{m \Psi}{2 \varpi y}+\frac{\Omega_{i0}^2}{2}+\Psi)   & 0 & \frac{i C \xi_\bot e^{-y}}{2 \varpi}   & -\frac{i \xi_\bot e^{-y}}{2 \varpi}   & 0\\
\frac{e^{-y}k_{\bar{z}} \Psi}{\varpi}                                      & 0 & 0                                            & 0                                           & -e^{-y}
\end{array}
\right).
\end{array}
\end{equation}
\normalsize
Here, $\varpi=\omega-m \Omega_{i0}-k_{\bar{z}} u_{\bar{z} 0}$ is the frequency in the frame of the ion fluid, $C=1+2\Omega_{i0}$, $\Psi=(\lambda_T+Z)\psi$, $y=\bar{r}^2$ and we have introduced a new dependent variable $X_1(y)$, where
\begin{equation}\label{eq2_10}
X_1(y)=\frac{Z}{Z+\lambda_T}[l_{i1}(y)-\chi_1(y)]. 
\end{equation}
For large axial wavelength modes of the resistive plasma column, for which $k_{\bar{z}}^2\leq\delta_{rs}$, this model can be reduced to a second order differential equation
\begin{equation}\label{eq2_11}
\left(\frac{\varpi^2-C^2}{\varpi\Psi}\right)L(N_c)[g_1(y)]=0,
\end{equation}
where
\[L(N_c)=y \frac{\partial^2}{\partial y^2}+(1-y)\frac{\partial}{\partial y}+\left(\frac{N_c}{2}-\frac{m^2}{4y}\right),\]
\[N_c=\frac{(\varpi^2-C^2)[m+\frac{i}{2}f(y)]}{\varpi-m \Omega_{i0}^2+i \Psi f(y)}+\frac{m C}{\varpi},\]
and $f(y)=F^2 e^{y}$ with the normalised axial wave number $F=k_{\bar{z}}/\sqrt{\delta_{rs}}$. For odd $m$ modes, the boundary conditions for Eq.~(\ref{eq2_11}) are $g_1(0)=0$ and $g_1(Y)=0$ with the infinite radius $Y$ representing the edge of plasma column. For even $m$, these conditions become $g_1'(0)=0$ and $g_1(Y)=0$. We seek unstable solutions to Eq.~(\ref{eq2_11}), for which $\varpi^i>0$. The solutions $\varpi=\pm C$ are stable and hence discarded.  

\section{Wave physics}\label{phy2}
The experimental results indicate that a $m=1$ drift wave mode dominates the density fluctuations, and that the mode appears to propagate purely azimuthally in direction of the electron diamagnetic drift.\cite{Shinohara:2001ab} There is no clear radial component of the mode propagation. The observed phase velocity of the mode is approximately $200$~m/s in the region of maximum density perturbation. 

\subsection{Dispersion curve}\label{dsp2}
\begin{figure}[ht]
\begin{center}$
\begin{array}{cc}
\hspace{-0.3cm}\includegraphics[width=0.5\textwidth,height=0.38\textwidth,angle=0]{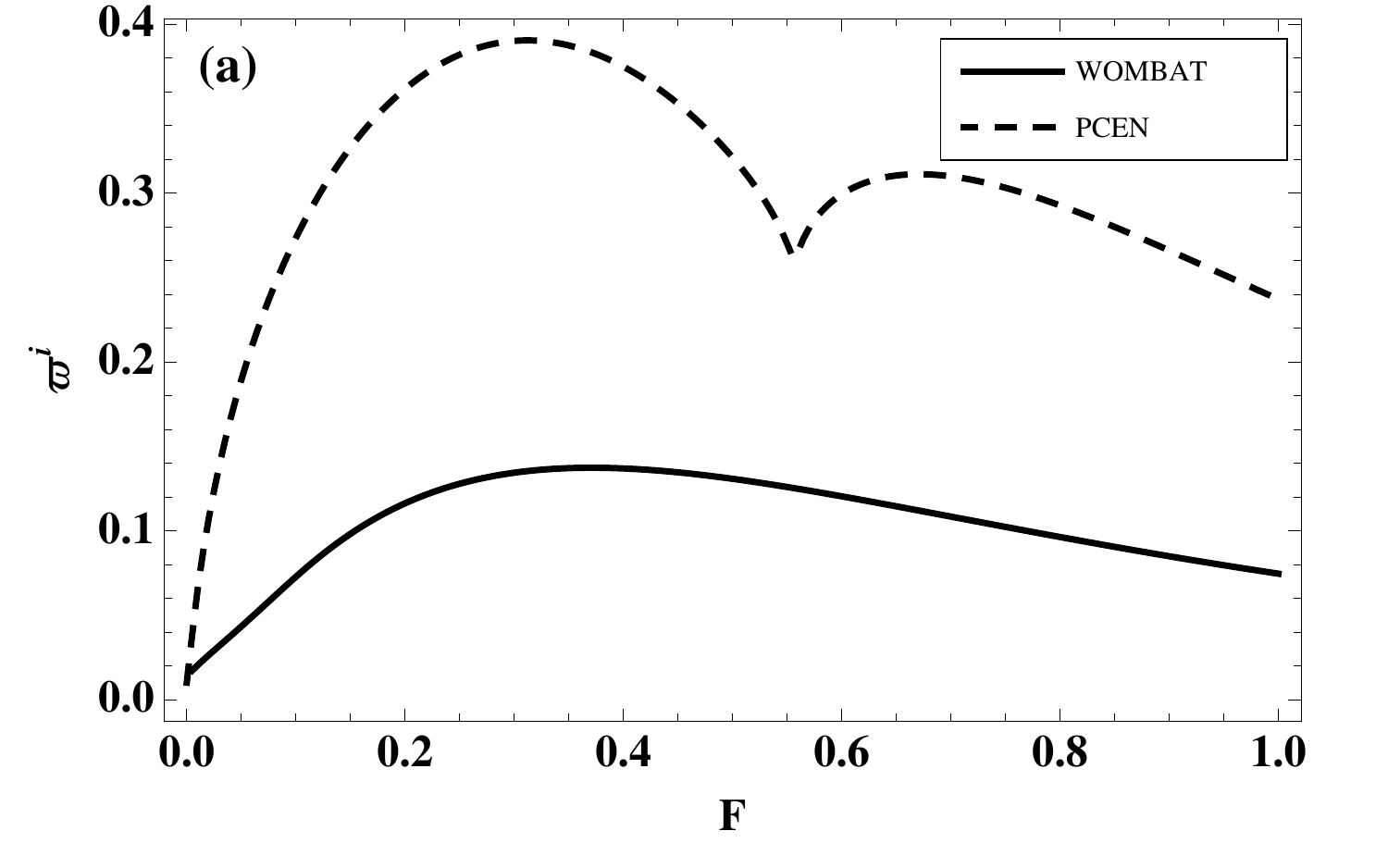}&\hspace{-0.5cm}\includegraphics[width=0.495\textwidth,height=0.375\textwidth,angle=0]{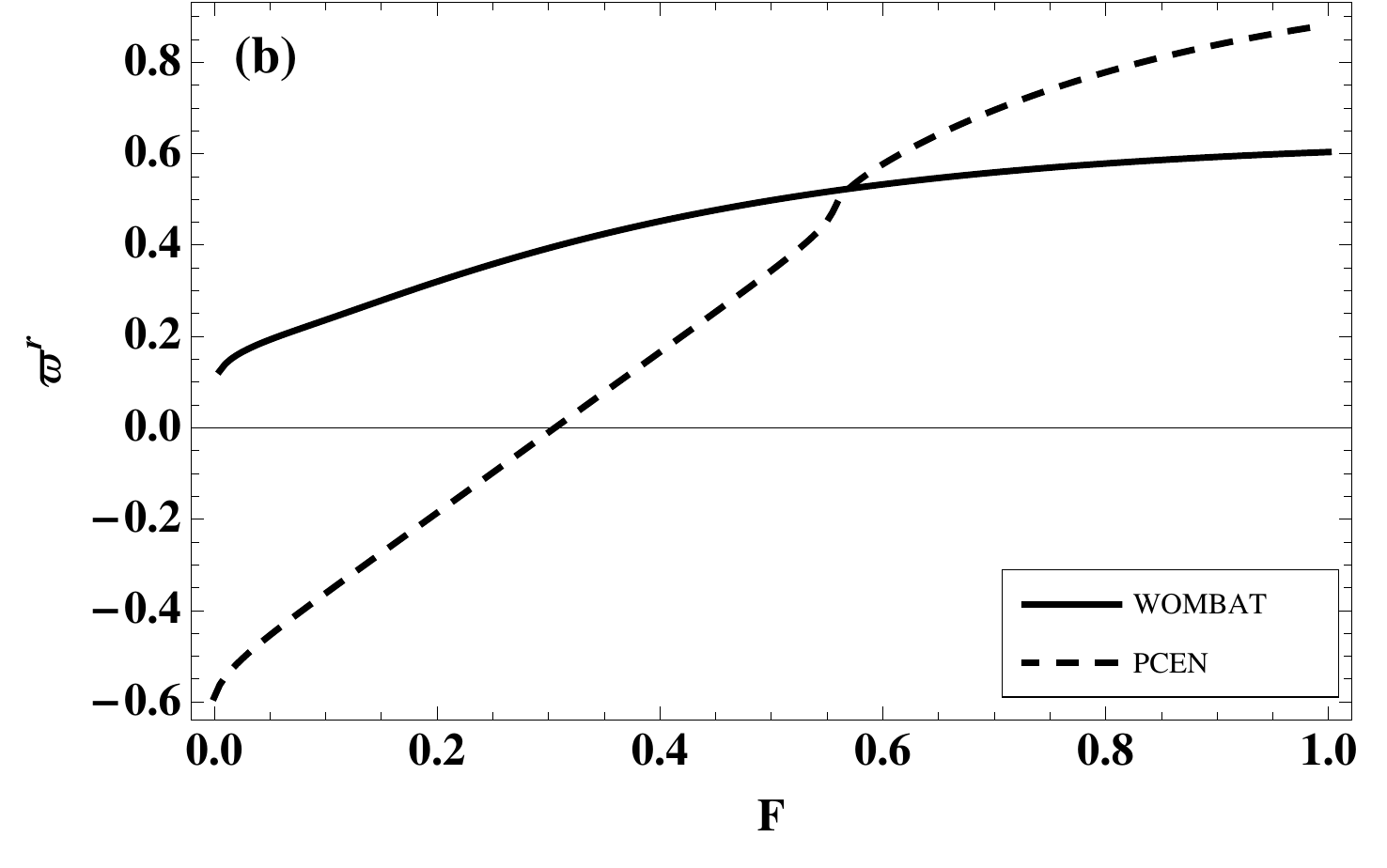}
\end{array}$
\end{center}\caption{Dispersion curves of PCEN (dashed line) and WOMBAT (solid line), generated by the model based on the conditions shown in Table~\ref{tb2_1}: (a) normalised growth rate $\varpi^i$ (the measurable growth rate is about $2.95\times 10^5 \varpi^i~\mathrm{rad}~\mathrm{s}^{-1}$ for PCEN and $4.45\times 10^4 \varpi^i~\mathrm{rad}~\mathrm{s}^{-1}$ for WOMBAT); (b) normalised frequency $\varpi^r$ (the measurable frequency is about $2.95\times 10^5(\varpi^r+0.59 m+3.4\times10^{-2} k_{z})~\mathrm{rad}~\mathrm{s}^{-1}$ for PCEN and $4.45\times 10^4(\varpi^r+0.032 m+4.5\times 10^{-3} k_{z})~\mathrm{rad}~\mathrm{s}^{-1}$ for WOMBAT) vs normalised axial wavenumber $F=k_{\bar{z}}/\sqrt{\delta_{rs}}$ (the measurable axial wavenumber is about $k_z=12.1F~\mathrm{m}^{-1}$ for PCEN and $k_z=2.6F~\mathrm{m}^{-1}$ for WOMBAT).}
\label{fg2_2}
\end{figure}

As in Hole \emph{et al.},\cite{Hole:2002aa} we have solved Eq.~(\ref{eq2_11}) by a shooting method but using our new code written in Mathematica. For $m=1$ the boundary conditions are $g_1(0)=g_1(Y)=0$. As the differential equation is homogeneous, the gradient at the edge $g_1'(Y)$ is arbitrary: we have chosen $g_1'(Y)=1$. To solve for given $F$, we choose a trial $\varpi$ and march the solution from the edge to the core. The complex frequency $\varpi$ is then adjusted until the on-axis boundary condition is satisfied. We commence the procedure at $F=0$, for which an analytical solution for $\varpi$ is available. This is found from the solution to 
\begin{equation}\label{eq2_12}
N_c=\frac{m(\varpi^2-C^2)}{\varpi-m \Omega_{i0}^2}+\frac{m C}{\varpi}=2n+|m|,
\end{equation}
where $n$ is the number of radial nodes in the plasma. 
Figure~\ref{fg2_2} shows the dispersion curve for the plasma conditions of the PCEN in Table~\ref{tb2_1}.   
We have compared the dispersion curve to Hole \emph{et al.}, \cite{Hole:2002aa} and found it to be identical, thus validating our numerics. For PCEN plasmas, the peak of normalised growth rate $\varpi^i=0.39$ lies at $F=0.3$, with normalised frequency $\varpi^r=0.007$: the wave is thus near stationary in the frame of the ion fluid. The mode crossings located at $F=0.55$ are associated with the centrifugal instability. Also shown is the dispersion curve for WOMBAT plasma conditions of Table~\ref{tb2_1}. For WOMBAT plasmas, the peak growth rate of $\varpi^i=0.14$ occurs at $F=0.37$, for which $\varpi^r=0.44$. There is no mode crossing in the dispersion curve for WOMBAT plasmas. Also, $\varpi^r>0$, and so in the laboratory frame $\omega^r=\varpi^r+m \Omega_{i0}+k_{\bar{z}} u_{\bar{z} 0}$, approximately $\omega^r \approx \varpi^r+\Omega_{i0}$, such that the frequency of unstable waves is always larger than the sum of the plasma rotation frequency and axial velocity, and the wave propagates in the $+~\theta$ direction (electron diamagnetic drift direction). We attribute the difference between dispersion curves to the very low ion temperature $T_i$ and slow normalised rotation frequency $\Omega_{i0}$ in WOMBAT plasmas, compared with PCEN plasmas. Indeed, this can be seen by varying the parameters from PCEN to WOMBAT conditions. If $\Psi$ is held fixed and $\Omega_{i0}$ lowered, the growth rate drops linearly with $\Omega_{i0}$, and the dispersion curve shifts to lower $k_{\bar{z}}$. Conversely, if $\Omega_{i0}$ is held fixed and $\Psi$ lowered, the dispersion curve expands to higher $k_{\bar{z}}$, and the growth rate drops. Thus, the decrease in growth rate is due to both the decrease in rotation and temperature, while the expansion of the dispersion curve to higher $k_{\bar{z}}$ is due to the lower temperature. The corresponding increase in $\varpi^r$ at low $k_{\bar{z}}$ occurs because as rotation drops, the Doppler shifted frequency, $\varpi^r=\omega^r-m \Omega_{i0}-k_{\bar{z}} u_{\bar{z} 0}$, increases.

\subsection{Dispersion curve sensitivity with plasma parameters}\label{sns2}
\begin{figure}[ht]
\begin{center}$
\begin{array}{cc}
\hspace{-1 cm} \vspace{-0.1 cm}\includegraphics[width=0.5\textwidth,height=0.38\textwidth]{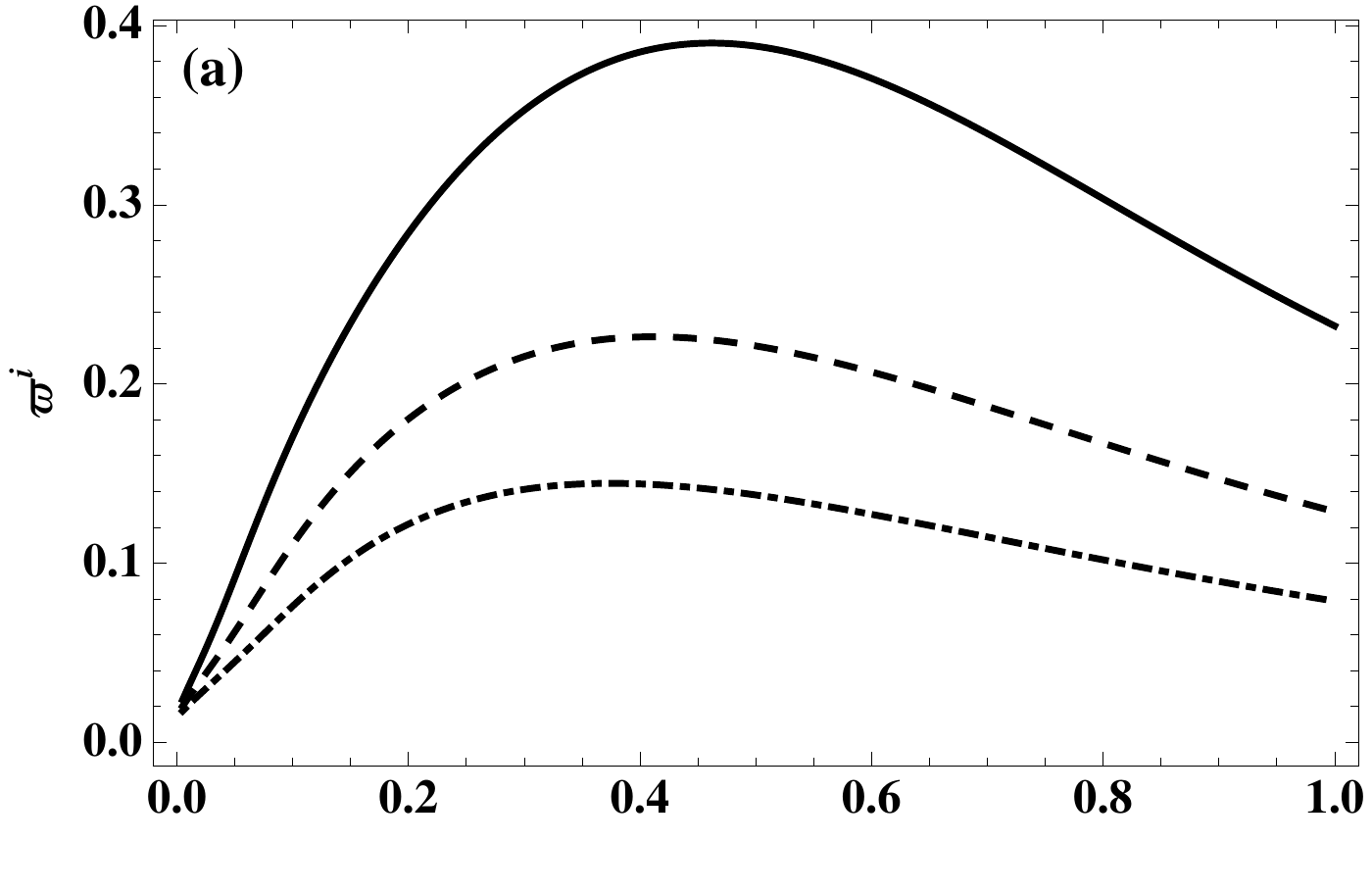}& \vspace{-0.15cm}\hspace{-0.9cm}\includegraphics[width=0.5\textwidth,height=0.375\textwidth]{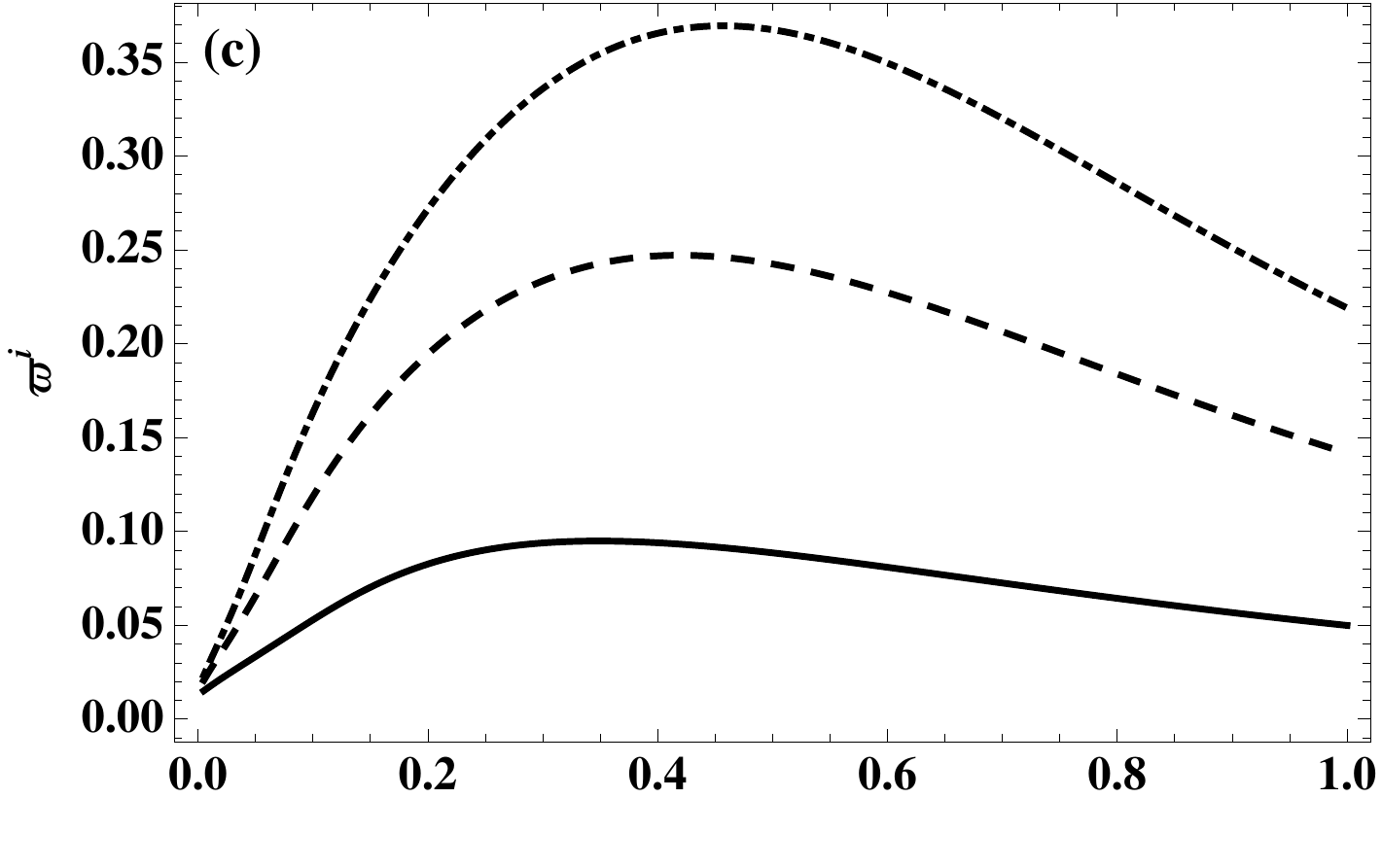}\\
\hspace{-0.6 cm} \includegraphics[width=0.537\textwidth,height=0.38\textwidth]{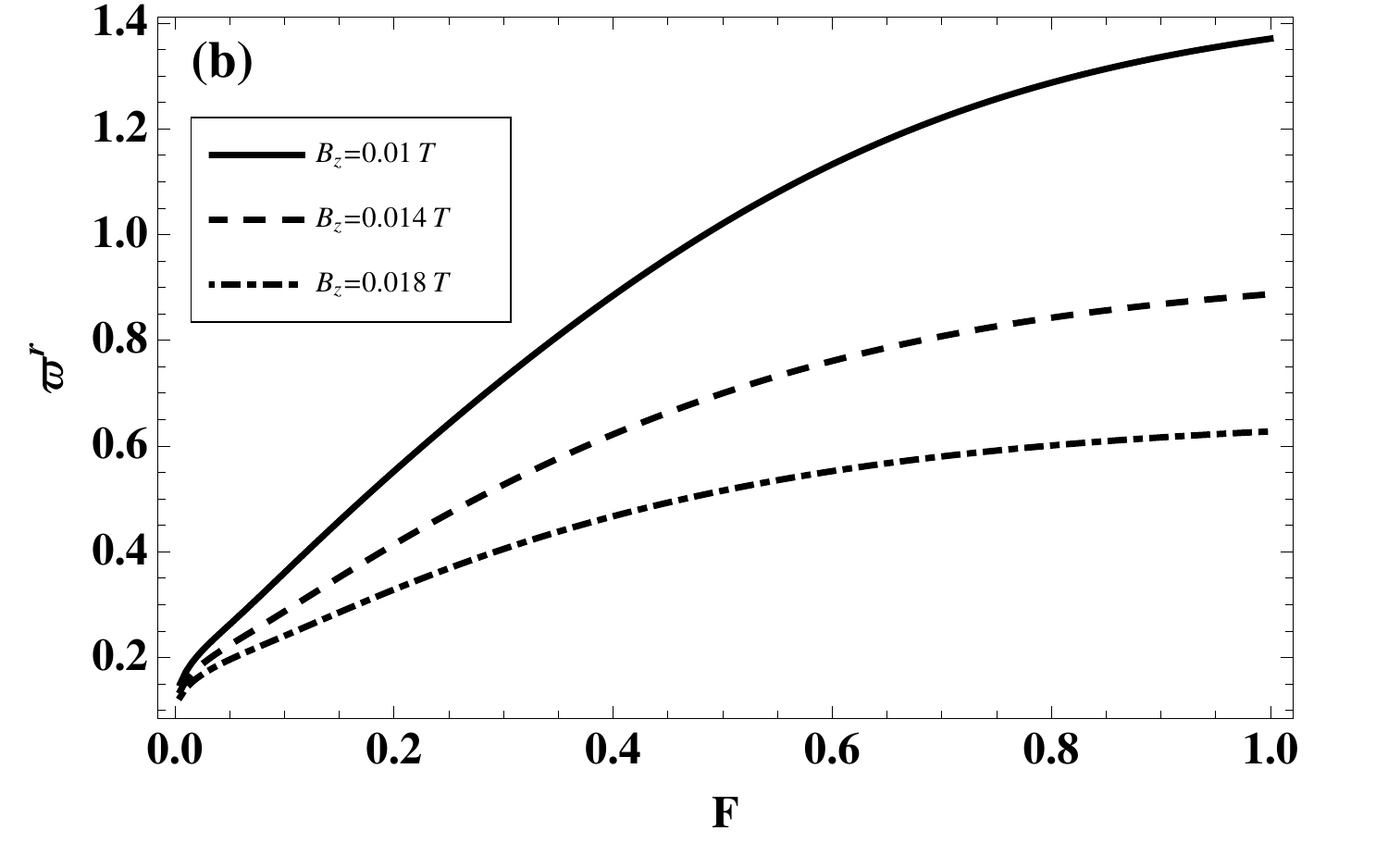} & \hspace{-0.5cm}\includegraphics[width=0.52\textwidth,height=0.375\textwidth]{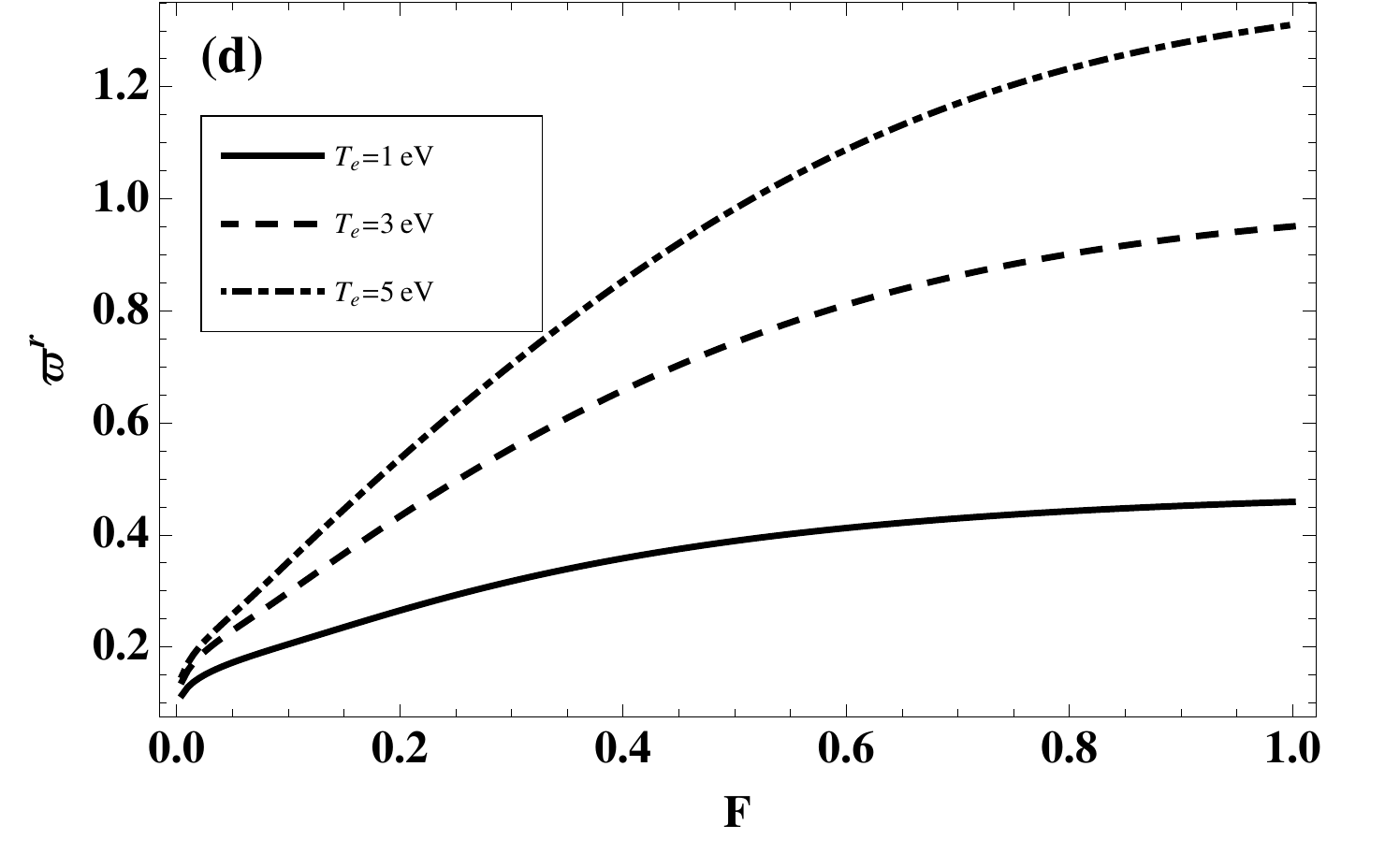}
\end{array}$
\end{center}
\caption{Results of dispersion curve sensitivity with $B_z$ and $T_e$: (a), (b) scanning results of $B_z$; (c), (d) scanning results of $T_e$. Here, the measurable growth rate is about $4.45\times 10^4 \varpi^i~\mathrm{rad}~\mathrm{s}^{-1}$and the measurable frequency is about $4.45\times 10^4(\varpi^r+0.032 m+4.5\times 10^{-3} k_{z})~\mathrm{rad}~\mathrm{s}^{-1}$, with the measurable axial wavenumber about $k_z=2.6F~\mathrm{m}^{-1}$.}
\label{fg2_3}
\end{figure}

We have examined the sensitivity of the WOMBAT plasma dispersion curve with $\Omega_{i0}$, $B_z$ and $T_e$. For $\Omega_{i0}<0.05$ the dispersion curve changes by less than $1\%$. Figure~\ref{fg2_3} shows the change in dispersion curve for different $B_z$ and $T_e$.  
Figure~\ref{fg2_3}(a), (b) shows that $\varpi^i$ and $\varpi^r$ decrease with increasing $B_z$: the physical growth rate $\omega^i=\varpi^i \omega_{ci}$ also obeys this trend while $\omega^r=\varpi^r \omega_{ci}$ exhibits the same trend in the range $F>0.5$. Also, the peak growth rate shifts to lower $k_{\bar{z}}$, and so for constant $R$ the axial wavelength of the most unstable mode is increased. In contrast, Fig.~\ref{fg2_3}(c), (d) shows that $\varpi^i$ and $\varpi^r$ increase with increasing $T_e$, and the peak growth rate shifts to larger $k_{\bar{z}}$. The dependence of growth rate with pressure gradient, and frequency with inverse field strength is consistent with a resistive drift wave.\cite{Chen:1984aa}

\subsection{Wave oscillation frequency with $B_z$ and $T_e$}\label{frq2}
In this section we compute the variation in wave frequency at the maximum growth rate with field strength, for two choices of electron temperature: a constant $T_e=2.13$~eV, independent of $B_z$; and a $B_z$ dependent $T_e$. As $T_e$ data is only available at two field strengths (Fig.~\ref{fg2_1}(c)):  $B_z=0.0185$~T, $T_e=3.2$~eV; and $B_z=0.0034$~T, $T_e=1.5$~eV, we have used linear interpolation to compute $T_e$ at other field strengths. Fluctuation data is available at five field strengths: $B_z=0.0129$~T, $B_z=0.0146$~T, $B_z=0.0162$~T, $B_z=0.0179$~T and $B_z=0.0195$~T. We have solved the dispersion curve for each case, and selected $\varpi^r$ corresponding to the maximum $\varpi^i$ to calculate the frequency in the laboratory frame. The maximum error in the electron temperature measurements is $\pm 0.5$~eV which was determined from the reproducibility of the experimental measurements. This is also the minimum temperature that we could measure.

\begin{figure}[h]
\begin{center}$
\begin{array}{c}
\includegraphics[width=0.8\textwidth,angle=0, scale=1.05]{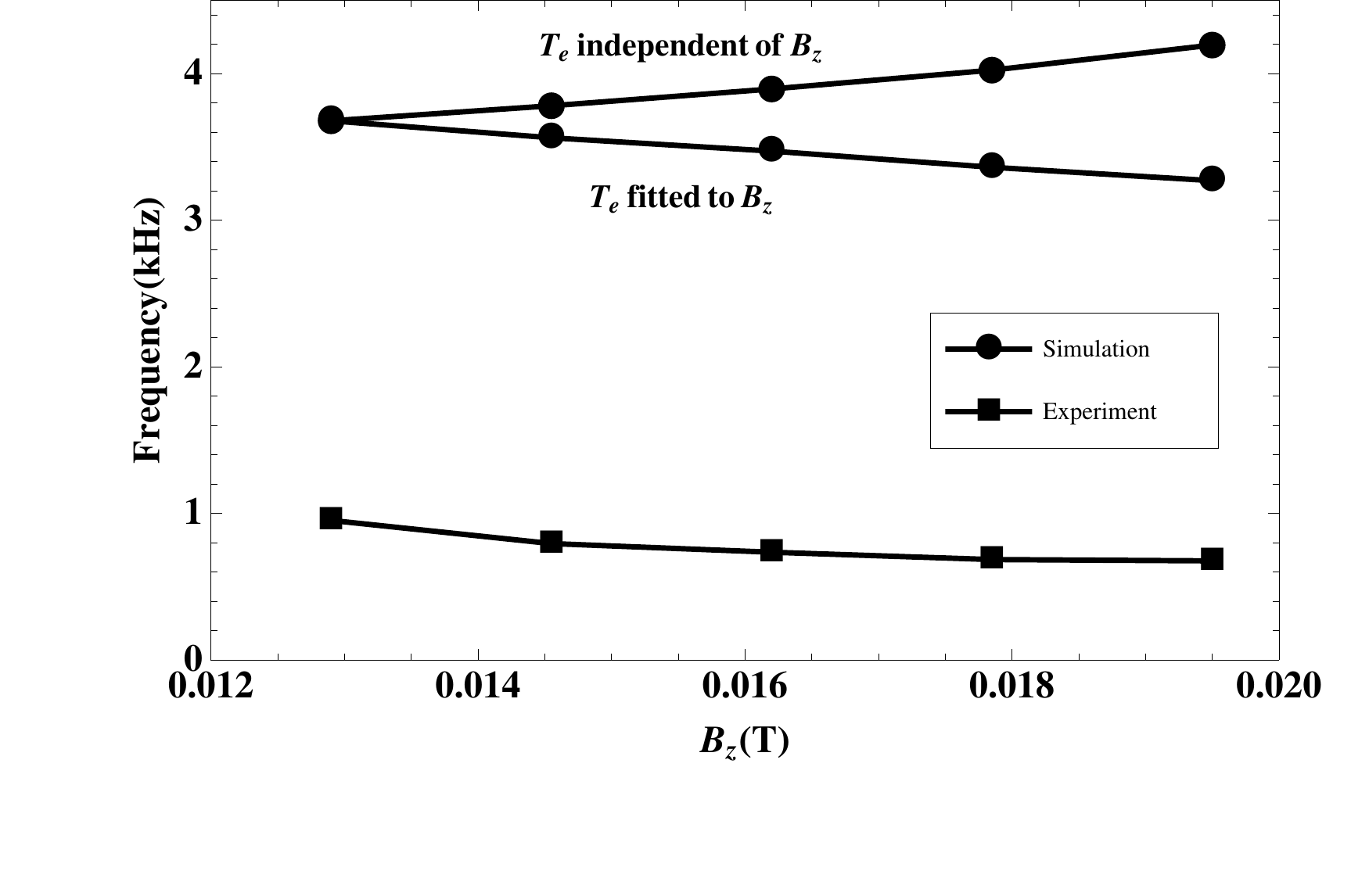}
\end{array}$
\end{center}
\caption{Wave oscillation frequency with $B_z$ and $T_e$}
\label{fg2_4}
\end{figure}

Figure~\ref{fg2_4} shows the measured and predicted oscillation frequencies. 
For both curves, the predicted and measured frequencies differ by a factor of $3.5$. There are various possible reasons for this gap. The most likely is that the model assumes flat $T_e$ and $T_i$ profiles, whereas WOMBAT plasmas exhibit a non-uniform $T_e$ profile. It may also be the case that the $T_i$ profile is non-uniform. Second, there may be slip between the rotation of WOMBAT plasmas and the oscillation frequency. Third, we have neglected the fluctuations in the magnetic field, which may affect the dispersion curve. Finally, the calculation of $T_e$ for each $B_z$ is a linear interpretation between the two known data, which may bring errors and imprecisions into the prediction. Figure~\ref{fg2_4} also reveals that two trends predicted by using a constant $T_e=2.13$~eV and a $B_z$ dependent $T_e$ are divergent. Predictions for the $B_z$ dependent $T_e$ profile exhibit the same trend as the data, suggesting the model may have correctly captured the $T_e$ dependence with $B_z$. Finally, to ensure that this model is in principle capable to reproduce the data, we adjusted $T_e$ to fit the data points. We found that a value of $T_e=0.5$~eV at low field dropping to $T_e=0.3$~eV at high field was able to reproduce the observed frequency.

\subsection{Perturbed density profile}\label{dst2}
\begin{figure}[ht]
\begin{center}$
\begin{array}{c}
\includegraphics[width=0.7\textwidth,angle=0]{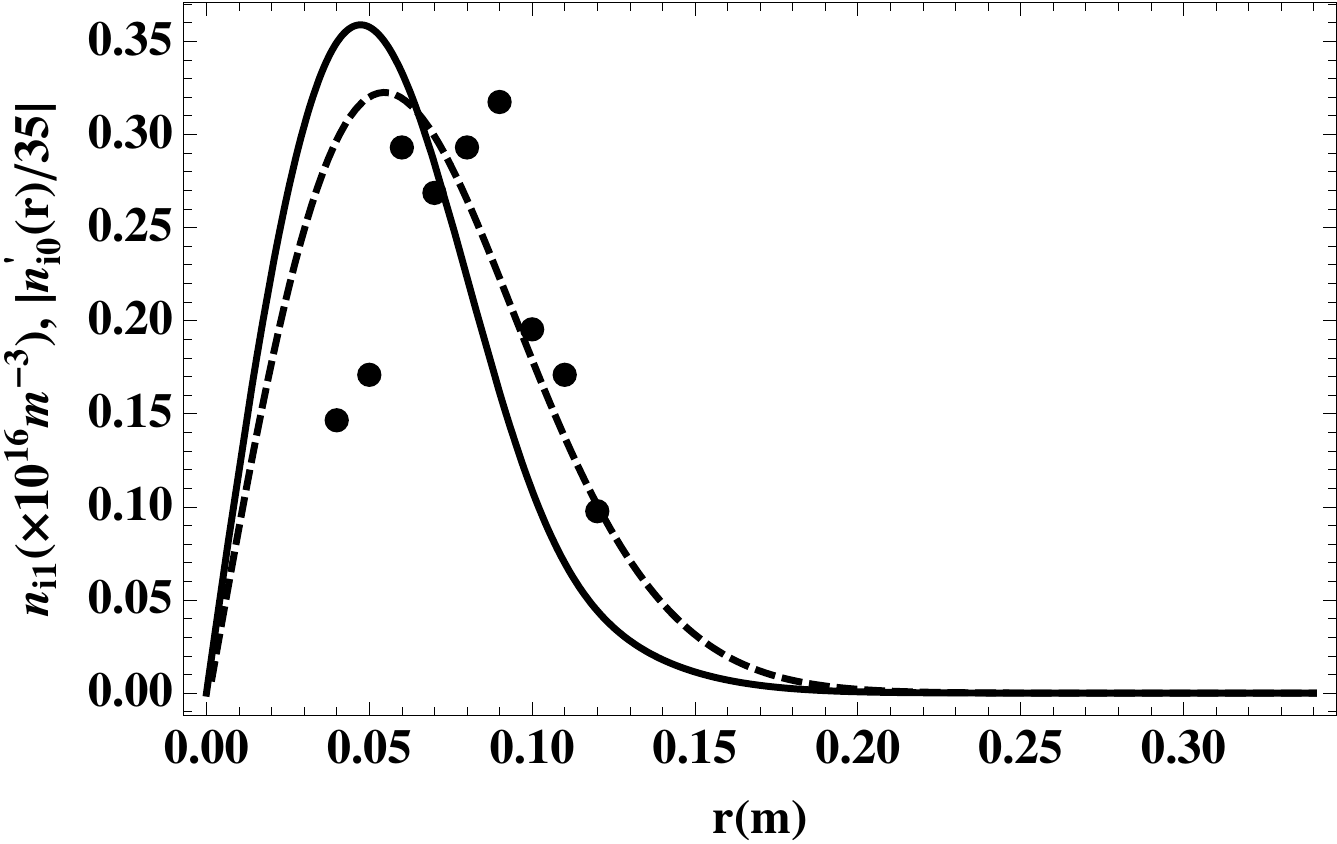}
\end{array}$
\end{center}
\caption{Comparison between the measured (dots) and predicted (solid line) radial profiles of the perturbed density $n_{i1}$, together with the equilibrium density gradient $|n'_{i0}(r)|$ (dashed line).}
\label{fg2_5}
\end{figure}

Figure~\ref{fg2_5} shows the measured and predicted radial profiles of the perturbed density $n_{i1}$, together with the equilibrium density gradient $|n'_{i0}(r)|$. 
The data used here were taken under the conditions of $B_z=0.01$~T, $R=7.6$~cm and $T_e=2.46$~eV. The linear perturbation treatment gives no solution for the absolute magnitude of $n_{i1}$, and so we have fitted for the amplitude. Inspection of Fig.~\ref{fg2_5} suggests $n_{i1}$ has a single peak, and therefore consistent with $n=0$ of the perturbed mode. Also, the peak of the eigenfunction occurs in the region where the equilibrium density gradient is large, suggesting that the instability has resistive drift-type characteristics.\cite{Hendel:1968aa} The exact radial position of the peak in $n_{i1}$ however does not agree. This could be due to uncertainty in the core position of the plasma, or the assumption of constant $T_e$ in the model.

\begin{figure}[ht]
\begin{center}$
\begin{array}{c}
\includegraphics[width=0.7\textwidth,angle=0]{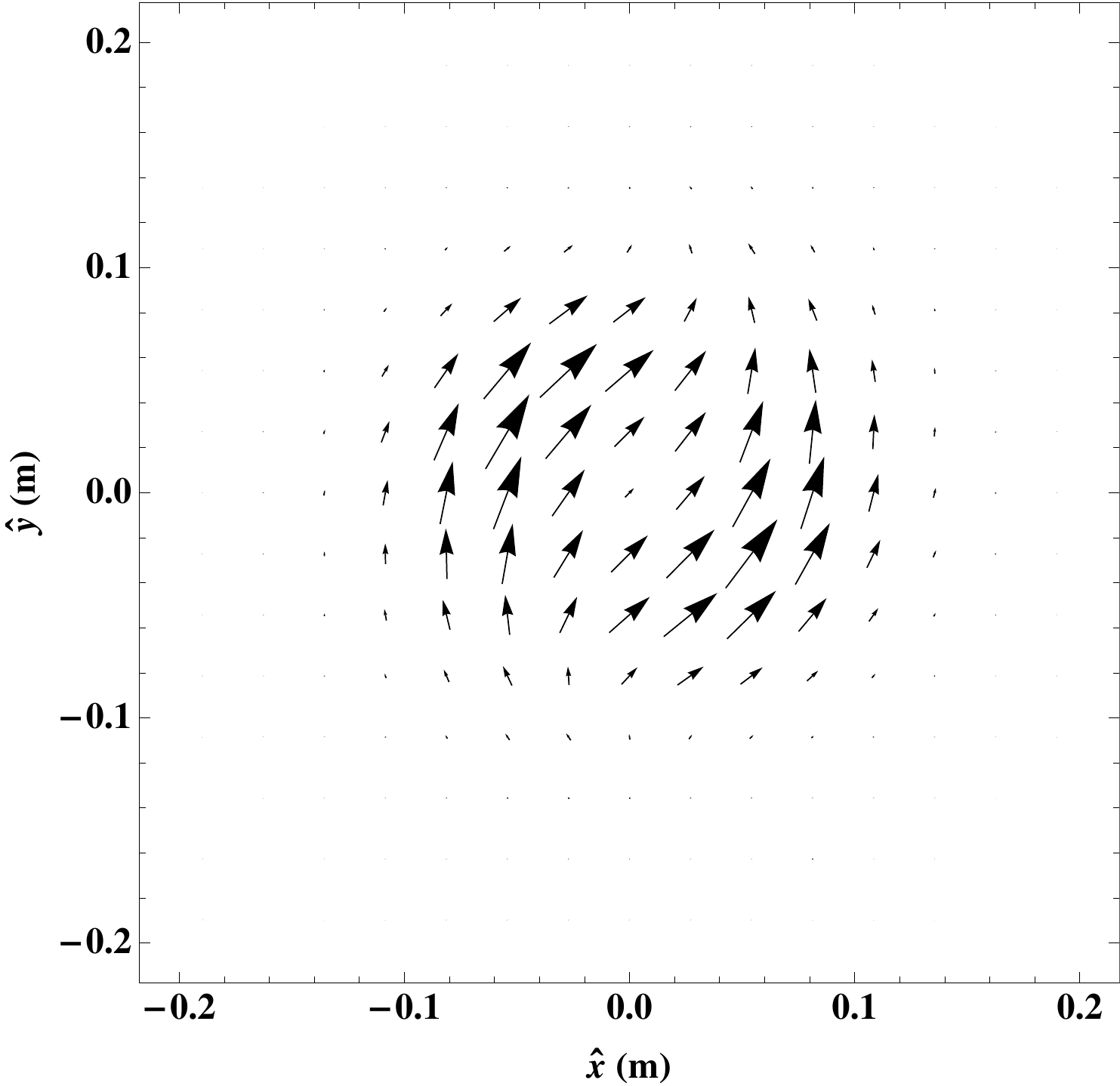}
\end{array}$
\end{center}
\caption{Vector field plot of the real part of the perturbed mass flow for mode $(m=1, n=0)$ in a cross-section of the WOMBAT plasma.}
\label{fg2_6}
\end{figure}

Finally, to complement our visualisation of the mode structure, we have computed the vector field of the linear perturbed mass flow by $m_i (n_{i1} \mathbf{u_{i0}}+n_{i0} \mathbf{u_{i1}})$. The perturbed velocity components ${\mathbf u_{i1}}=(\bar{r} \varphi_{\bar{r}i1}, \bar{r} \Omega_{i1}, u_{i\bar{z}1})$, ${\mathbf u_{e1}}=(\bar{r} \varphi_{\bar{r}e1}, \bar{r} \Omega_{e1}, u_{e\bar{z}1})$ and perturbed density $n_{i1}$ can be computed from the solution of $g_1(y)$ and following equations:
\begin{equation}\label{eq2_13}
l_{i1}(y)=\frac{-g_1(y)}{1+\frac{i}{\Psi}\left[\frac{m \Omega_{i0}^2+2m \Psi-\varpi}{f(y)-2 i m}\right]},
\end{equation}
\begin{equation}\label{eq2_14}
\chi_1(y)=-\frac{\lambda_T}{Z}l_{i1}(y)-(1+\frac{\lambda_T}{Z})g_1(y),
\end{equation}
\begin{equation}\label{eq2_15}
\begin{array}{rcl}
\vspace{0.3cm}\varphi_{\bar{r}i1}(y) & = & \frac{2\varpi}{i(\varpi^2-C^2)-\varpi \delta_{rs} \xi_\bot e^{-y}}\{\Psi[l_{i1}'(y)-X_1'(y)]\\
&&-\frac{m \Psi C}{2\varpi y}l_{i1}(y)-(\frac{i C}{2\varpi}+\frac{\delta_{rs} \xi_\bot e^{-y}}{2})\varphi_{\bar{r}e1}(y)\},
\end{array}
\end{equation}
\begin{equation}\label{eq2_16}
\begin{array}{rcl}
\vspace{0.3cm}\varphi_{\bar{r}e1}(y) & = &\frac{2\varpi}{\varpi+i \delta_{rs} \xi_\bot e^{-y}}[\frac{i m \Psi}{2y}X_1(y)-\Psi \delta_{rs} \xi_\bot e^{-y} X_1'(y)\\
\vspace{0.3cm}&&+\delta_{rs} \xi_\bot e^{-y}(-\frac{m \Psi}{2 \varpi y}+\frac{\Omega_{i0}^2}{2}+\Psi)l_{i1}(y)\\
&&+\frac{i C \delta_{rs} \xi_\bot e^{-y}}{2\varpi}\varphi_{\bar{r}i1}(y)],
\end{array}
\end{equation}
\begin{equation}\label{eq2_17}
\Omega_{i1}(y)=\frac{1}{\varpi}\left\{m \Psi \frac{l_{i1}(y)}{y}+i[\varphi_{\bar{r}e1}(y)-C\varphi_{\bar{r}i1}(y)]\right\}, 
\end{equation}
\begin{equation}\label{eq2_18}
u_{i\bar{z}1}(y)=\frac{\sqrt{\delta_{rs}} F \Psi}{\varpi}l_{i1}(y).
\end{equation}
Figure~\ref{fg2_6} shows the flow vector field at time $t=0$. 
As time advances, the mass flow vector field rotates in the clockwise direction, which is the direction of electron diamagnetic drift with $B_z$ into the page. Here, the coordinate $\hat{x}$ and $\hat{y}$ label the cross-section of the plasma, with $\hat{x}=r\cos\theta$ and $\hat{y}=r\sin\theta$. The $\theta$ vector points clockwise with $\hat{z}$ into the page. The mass flow is zero on axis ($\hat{x}=0,~\hat{y}=0$), consistent with the boundary conditions for density. The symmetric mass flow represents a flow around a stagnant core, with the largest flow in regions of peak density gradient. While we have only studied linear oscillatory response in this work, a similar nonlinear flow pattern may contribute to rotation damping. 

\section{Conclusions}\label{cnl2}
In this chapter, we employed a two-fluid model, which was developed originally for describing the wave oscillations observed in PCEN, \cite{Hole:2002aa, Hole:2001aa, Hole:2001ab} to study the low frequency oscillations in WOMBAT. To ensure that this model is consistent with WOMBAT, the measured and predicted plasma configurations were compared, including the equilibrium density profile and the space potential profile. These show that although the density and space potential profiles agree, the temperature profile is not flat, as assumed by the model. Next, dispersion curves were generated for WOMBAT plasmas. Compared to the more rapidly rotating centrifuge plasma, the drift wave instability, unstable at larger wavelengths, has a normalised frequency much larger than the normalised frequency for the centrifuge. The difference between dispersion curves is principally due to the low rotation speed of WOMBAT plasmas compared to those of the centrifuge. 

Our study of the wave oscillation frequency with $B_z$ and $T_e$ reveals that the measured fluctuation frequency is a factor of $3.5$ lower than the predicted frequency, and the predicted trend of oscillation frequency with $B_z$ for inferred $T_e$ matches the data. The discrepancy between the measured and predicted fluctuation frequencies may be attributable to the limitations of assumed temperature uniformity across plasma column. Weaker model limitations may be our neglect of effects induced by plasma fluctuations on the externally applied field. These effects are included in a TwO-fluid Electromagnetic Flowing pLasma (TOEFL) model developed in Appendix~\ref{app}, for which, however, the computation has not yet been finished. Data limitations include the simultaneous measurements of plasma rotation profile and probe frequency, and the measured dependency of $T_e$ with $B_z$. 

Finally, we find that the measured and predicted perturbed density profiles have a single peak in the radial direction, indicating the perturbed mode has $n=0$, and the peak position is in the region of maximum equilibrium density gradient. This qualitative agreement consolidates earlier claims that the mode is a drift mode, driven by the density gradient of the plasma. To our knowledge, this is the first detailed physics model of flowing plasmas in the diffusion region away from the RF source. 

\chapter{Wave propagations in a pinched plasma}\label{chp3}

A RF field solver\cite{Chen:2006aa} based on Maxwell's equations and a cold plasma dielectric tensor is employed to describe wave phenomena observed in a cylindrical non-uniform helicon discharge. The experiment is carried out on a recently built linear plasma-material interaction machine: the MAGPIE (see Fig.~\ref{fg1_8}),\cite{Blackwell:2012aa} in which both plasma density and static magnetic field are functions of axial position. The field strength increases by a factor of $15$ from source to target plate, and the plasma density and electron temperature are radially non-uniform. With an enhancement factor of $9.5$ to the electron-ion Coulomb collision frequency, a $12\%$ reduction in the antenna radius, and the same other conditions as employed in the experiment, the solver produces axial and radial profiles of wave amplitude and phase that are consistent with measurements. A numerical study on the effects of axial gradient in plasma density and static magnetic field on wave propagations is performed, revealing that the helicon wave has weaker attenuation away from the antenna in a focused field compared to a uniform field. This may be consistent with observations of increased ionization efficiency and plasma production in a non-uniform field. We find that the relationship between plasma density, static magnetic field strength and axial wavelength agrees well with a simple theory developed previously. A numerical scan of the enhancement factor to the electron-ion Coulomb collision frequency from $1$ to $15$ shows that the wave amplitude is lowered and the power deposited into the core plasma decreases as the enhancement factor increases, possibly due to the stronger edge heating for higher collision frequencies.

\section{Introduction}\label{int3}
To date, most helicon studies have treated devices with uniform static magnetic fields, however, many applications require operation with axial magnetic field variations.\cite{Mori:2004aa} A few researchers have investigated helicon plasma sources with non-uniform magnetic fields, and have found that the plasma density increased when a cusp or non-uniform magnetic field was placed in the vicinity of the helicon antenna.\cite{Boswell:1997aa, Chen:1992aa, Chen:1997aa, Gilland:1998aa} However, detailed examination of the reasons for this enhanced plasma density has not yet been conducted, although fast electrons and improved confinement are mentioned as possible contributors. Guo et al.\cite{Guo:1999aa} furthered this study by looking at the effects of non-uniform magnetic field on source operations, and found that a strong axial gradient in density associated with non-uniform field configuration can contribute to the absorption of wave fields and a high ionization efficiency. Takechi et al.\cite{Takechi:1999aa} also suggested that there may be a close relationship between plasma density profile and RF wave propagation and absorption regions, finding the density uniformity in the radial direction improved markedly with the cusp field. Therefore, studying the effects of various static magnetic field configurations on helicon wave propagation is of significant importance to producing desired plasma profiles and understanding the role of magnetic field in helicon plasma generations. 

This chapter is dedicated to modelling the wave field observed in MAGPIE, and investigating helicon wave propagation in the non-uniform magnetised plasma of this machine, in which both the static magnetic field and its associated plasma density are functions of axial position. The plasma density and electron temperature are also dependent on radius. Independent measurements of electron temperature $T_e$ in MAGPIE at lower field and power conditions show that $T_e$ does not change substantially along $z$. We have thus assumed $T_e$ is independent of $z$ in the present study. The static magnetic field is almost independent of $r$ (Eq.~(\ref{eq3_5})). A RF field solver,\cite{Chen:2006aa} based on Maxwell's equations and a cold plasma dielectric tensor, is employed in this study. The motivations of our work are to explain the wave field measurements in MAGPIE, and to study the effects of magnetic field configuration on helicon wave propagation. The rest of the chapter is organised as follows: Sec.~\ref{exp3} describes the diagnostic tools, together with the measured static magnetic field, plasma density and temperature profiles; Sec.~\ref{sml3} provides an overview of the employed theoretical model and the numerical code, together with comparisons between computed and measured wave fields; Sec.~\ref{num3} is dedicated to a numerical study of the effects of plasma density and static magnetic field profiles on the wave propagation characteristics; and Sec.~\ref{cll3} studies the physics meaning of the enhancement factor to the electron-ion Coulomb collision frequency, and the effects of the direction of static magnetic field on wave propagations. Finally, Sec.~\ref{cnl3} presents concluding remarks and future work for continuing research. 

\section{Experiment}\label{exp3}
\subsection{Plasma profile diagnostics}\label{prf3}
A passively compensated Langmuir probe was employed in our experiment to measure the plasma density and electron temperature, calculated from the $I(V)$ curve obtained by an Impedans Data Acquisition system.\cite{Impedans:2012aa} The probe comprises a platinum wire of diameter $0.1$~mm, and a surrounding alumina insulator. The length of the insulator is $6$~mm shorter than that of the platinum wire so that the exposed platinum wire forms the probe tip. Electron currents were drawn to clean the probe during regular intervals of argon discharges. The probe is located at $z=0.17$~m as shown in Fig.~\ref{fg1_8}.
\begin{figure}[h]
\begin{center}$
\begin{array}{l}
(a)\\
\includegraphics[width=0.7\textwidth]{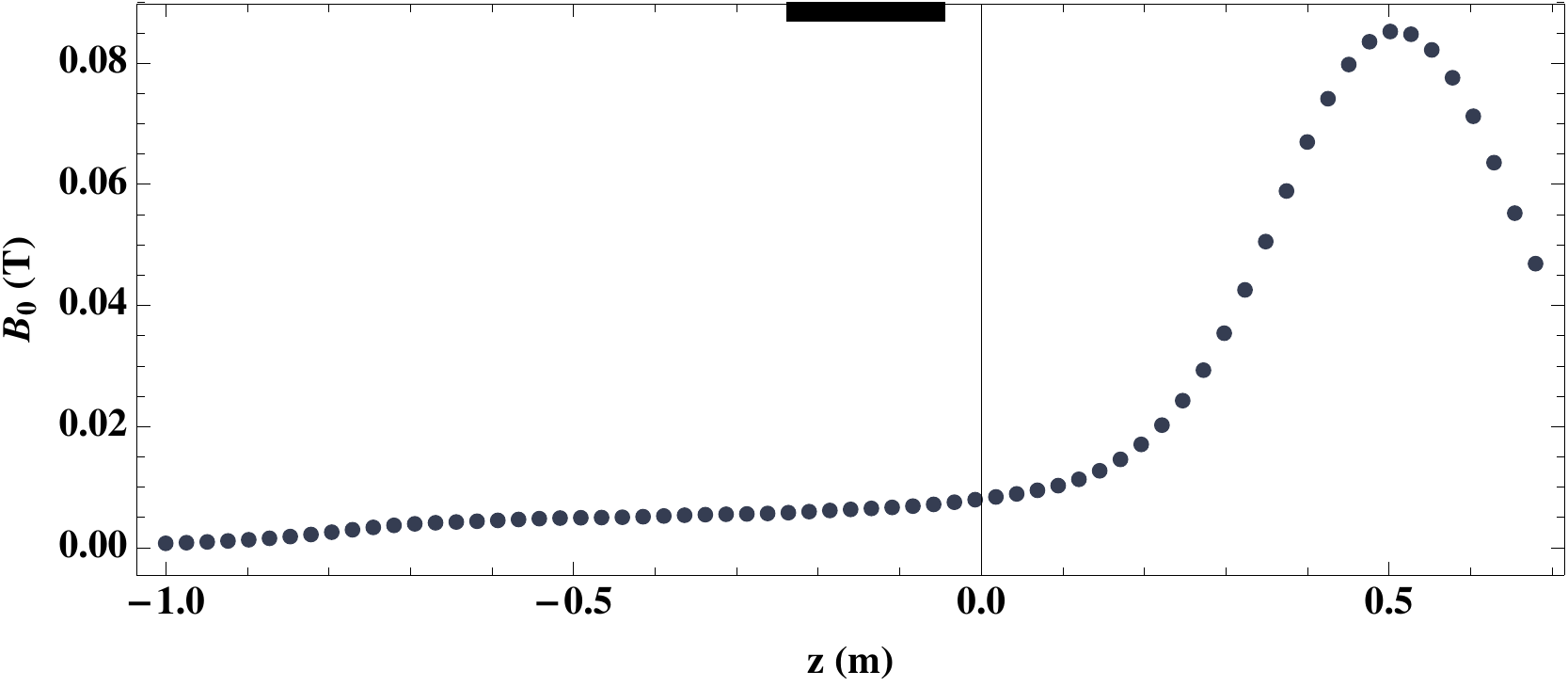}\\
(b)\\
\hspace{1.3cm}\includegraphics[width=0.6\textwidth]{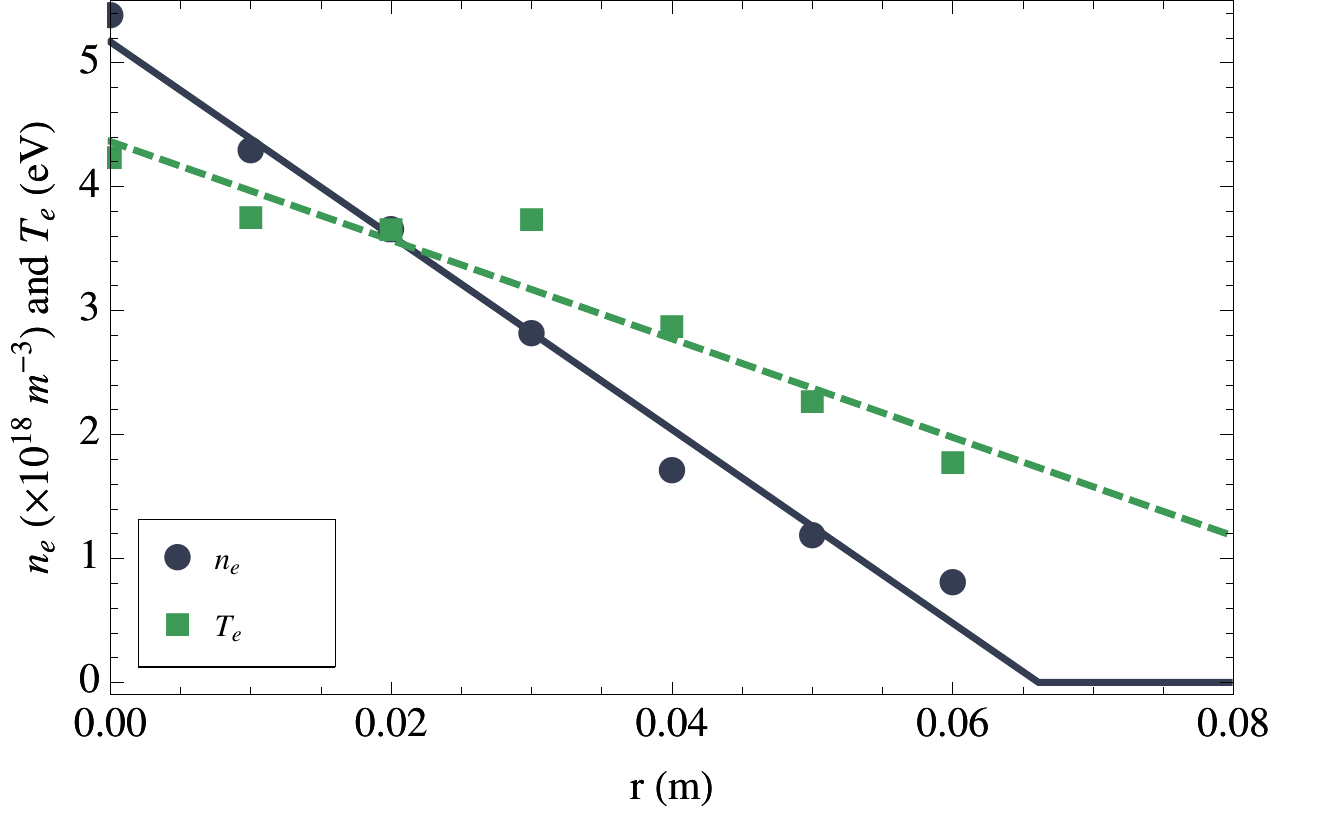}
\end{array}$
\end{center}
\caption{Typical measured profiles: (a) axial profile of static magnetic field on axis, (b) radial profiles of plasma density (dots) and electron temperature (squares) at $z=0.17$~m, together with their fitted lines, solid and dashed, respectively. The solid bar in (a) denotes the antenna location. }
\label{fg3_1}
\end{figure}

Typical measured axial profile of field strength and radial profiles of plasma density and electron temperature in MAGPIE are shown in Fig.~\ref{fg3_1}. As shown in Fig.~\ref{fg3_1}(a), the increase in field strength, $B_0(z)$, from antenna end ($z=-0.243$~m) to field peak ($z=0.51$~m) is a factor of $15$. The axial profile of plasma density, $n_e(z)$, is assumed to be proportional to $B_0(z)$, consistent with generally accepted knowledge that the density follows the magnetic field linearly.\cite{Lieberman:2005aa, Boswell:2012aa} Figure~\ref{fg3_1}(b) shows the radial profiles of plasma density, $n_e (r)$, and electron temperature, $T_e(r)$, measured at $z=0.17$~m and fitted with straight lines. During the density fitting procedure, in order to avoid negative fitted values, the density was set to zero in the region of $0.066\leq r \leq 0.08$~m. We assume the total density profile is separable, such that $n_e(r, z)=n_e(r)\times n_e(z)$. The fitted lines in $n_e(r)$ and $T_e(r)$, and the measured $B_0(z)$ data will be used in Sec.~\ref{sml3} to constrain wave field simulations. 

\subsection{Wave field diagnostics}\label{dig3}
Helicon wave fields were measured by a two-axis ``B dot" or Mirnov probe. Details about the probe can be found in Blackwell et al..\cite{Blackwell:2012aa} To measure the axial profiles of $B_r$ and $B_z$, the probe was inserted on axis from the end of the target chamber. The probe is long enough to measure $B_r$ and $B_z$ in the range $-0.25<z<0.7$~m. Two perpendicular magnetic field components ($B_r$ and $B_z$ in this case) can be sampled simultaneously. To measure the radial profiles of the three magnetic wave components, $B_r$, $B_\theta$ and $B_z$, the probe was inserted radially at $z=0.17$~m, and rotated about its axis to measure $B_\theta$ and $B_z$. The B-dot probe couples inductively to the magnetic components of the helicon wave and electrostatically to the RF time varying plasma potential. To limit our measurements to the inductively coupled response, a current balun was employed to screen the electrostatic response. Further information about the procedure to eliminate the electrostatic response of the probe can be found in Franck et al..\cite{Franck:2002aa} Both axial and radial profiles of wave phase were measured through a phase-comparison method, similar to Light et al..\cite{Light:1995aa} To measure the variation in wave phase with axial position, the signal from an on-axis axially inserted probe was compared to the phase of the antenna current. A similar procedure was conducted to measure the variation in wave phase with radial position at $z=0.17$~m. It should be noted that all probe diagnostics are intrusive, and can affect the plasma parameters, and hence the wave fields. 

\section{Simulation}\label{sml3}
A RF field solver (or ElectroMagnetic Solver, EMS\cite{Chen:2006aa}) based on Maxwell's equations and a cold-plasma dielectric tensor is employed in this study to interpret the RF waves measured in MAGPIE. This solver has been used successfully in explaining wave phenomena in two other machines: a helicon discharge machine at The University of Texas at Austin \cite{Lee:2011aa} and the LAPD at the University of California at Los Angles.\cite{Zhang:2008aa} Since these two machines have similar plasma density, temperature and particle species to those on the MAGPIE, the EMS should be applicable to describing the wave observations in MAGPIE. Details of the solver can be found in Chen et al.,\cite{Chen:2006aa} while a brief overview is given below. 

\subsection{EMS}\label{ems3}
The Maxwell's equations that this solver employs to determine the RF wave field in a helicon discharge are eventually Eq.~(\ref{eq1_1}) and Eq.~(\ref{eq1_11}) but including an external antenna current $\mathbf{j_a}$:
\begin{equation}\label{eq3_1}
\bigtriangledown\times\mathbf{E}=i \omega\mathbf{B}, 
\end{equation}
\begin{equation}\label{eq3_2}
\frac{1}{\mu_0}\bigtriangledown\times\mathbf{B}=-i \omega \mathbf{D}+\mathbf{j_a}.
\end{equation}
Here, we have dropped the subscript ``$1$" for simplicity. Because the wave field is driven by the antenna, the wave frequency is the same to the antenna driving frequency $\omega$. The quantities $\mathbf{D}$ and $\mathbf{E}$ are linked via the cold-plasma dielectric tensor $\mathbf{\tilde{\epsilon}}$ (Eq.~(\ref{eq1_8})) that represents vacuum, glass and plasma. In the vacuum and glass regions, the dielectric tensor is $\mathbf{\tilde{\epsilon}}\equiv\epsilon_\ast \mathbf{\tilde{\delta}}$, where $\mathbf{\tilde{\delta}}$ is the Kronecker symbol and $\epsilon_\ast$ is a scalar. The term $\epsilon_\ast$ equals to $1$ and $\epsilon_g$ for vacuum and glass regions, respectively, where $\epsilon_g$ is the dielectric constant of glass. In the plasma region, the relation between $\mathbf{D}$ and $\mathbf{E}$ has the form\cite{Ginzburg:1970aa}
\begin{equation}\label{eq3_3}
\mathbf{D}=\varepsilon_0[S\mathbf{E}-i D(\mathbf{E}\times\mathbf{b})+(P-S)(\mathbf{E}\cdot\mathbf{b})\mathbf{b}],
\end{equation}
where $\mathbf{b}\equiv\mathbf{B_0}/B_0$ is the unit vector along the static magnetic field. To take into account the collisions of particles in the discharge, we introduce a phenomenological collision term $\nu_{s}$ for each species into $\mathbf{\tilde{\epsilon}}$, and rewrite Eq.~(\ref{eq1_9}) and Eq.~(\ref{eq1_10}) to:\cite{Ginzburg:1970aa}
\begin{equation}\label{eq3_4}
\begin{array}{l}
\vspace{0.3cm}S=1-\sum\limits_{s}\frac{\omega^2_{p s}(\omega+i\nu_s)}{\omega[(\omega+i\nu_s)^2-\omega^2_{cs}]},\\
\vspace{0.3cm}D=\sum\limits_{s}\frac{\omega^2_{ps}\omega_{c s}}{\omega[(\omega+i\nu_s)^2-\omega^2_{cs}]},\\
P=1-\sum\limits_{s}\frac{\omega^2_{ps}}{\omega(\omega+i\nu_s)}.
\end{array}
\end{equation}
For plasma parameters typical of MAGPIE, the electron-neutral collision frequency is found to be an order of magnitude smaller than the electron-ion collision frequency, and two orders smaller than the electron-ion collision frequency required to match the experimental results (Sec.~\ref{cmp3}). Thus, electron-neutral collisions are neglected. Because $\nu_{ee}$ and $\nu_{ii}$ do not contribute to the momentum exchange between electron and ion fluids,\cite{Chen:1984aa} collision frequencies for electrons and ions species are $\nu_{e}=\nu_{ei}=2.91\times 10^{-12}n_e T^{-3/2}_e \mathrm{ln}\Lambda$ and $\nu_{i}=\nu_{ie}=m_e m_i^{-1} \nu_{ei}$, respectively, from which we can see $\nu_{ie}\ll\nu_{ei}$. Here, $T_e$ and $n_e$ are given in eV and $\mathrm{m^{-3}}$, respectively, and the Coulomb logarithm is calculated to be ln$\Lambda=12$. Singly ionised argon ions are assumed in this study, so that $q_i=-q_e=|e|$.

The static magnetic field is assumed to be axisymmetric with $B_{0r}~\ll~B_{0z}$ and $B_{0\theta}=0$. It is therefore appropriate to use a near axis expansion for the field, namely $B_{0z}$ is only dependent on $z$ and 
\begin{equation}\label{eq3_5}
B_{0r}(r,~z)=-\frac{1}{2}r\frac{\partial B_{0z} (z)}{\partial z}. 
\end{equation}
The antenna, as described in Sec.~\ref{mgp}, is a left hand half-turn helical antenna. We assume that the antenna current is divergence free, to eliminate the capacitive coupling. Fourier components of the antenna current density are given by:
\begin{equation}\label{eq3_6}
\begin{array}{l}
\vspace{0.5cm}j_{ar}=0, \\
\vspace{0.3cm}j_{a\theta}=I_a\frac{e^{i m \pi}-1}{2}\delta(r-R_a)\left\{\frac{i}{m \pi}[\delta(z-z_a)+\delta(z-z_a-L_a)]\right.\\
\vspace{0.5cm}\hspace{0.9cm}\left.+\frac{H(z-z_a)H(z_a+L_a-z)}{L_a}e^{-i m \pi[1+(z-z_a)/L_a]}\right\}, \\
\vspace{0.3cm}j_{az}=I_a\frac{e^{-i m \pi[1+(z-z_a)/L_a]}}{\pi R_a}\frac{1-e^{i m \pi}}{2}\delta(r-R_a)\\
\hspace{0.9cm}\times H(z-z_a)H(z_a+L_a-z), \\
\end{array}
\end{equation}
where $L_a$ is the antenna length, $R_a$ the antenna radius, $z_a$ the distance between the antenna and the endplate in the source region, and $H$ the Heaviside step function. Note that the antenna geometry selects only odd harmonic mode number $m$, as indicated by Chen et al..\cite{Chen:2006aa} 

\begin{figure}[h]
\begin{center}$
\begin{array}{c}
\includegraphics[width=0.9\textwidth,angle=0]{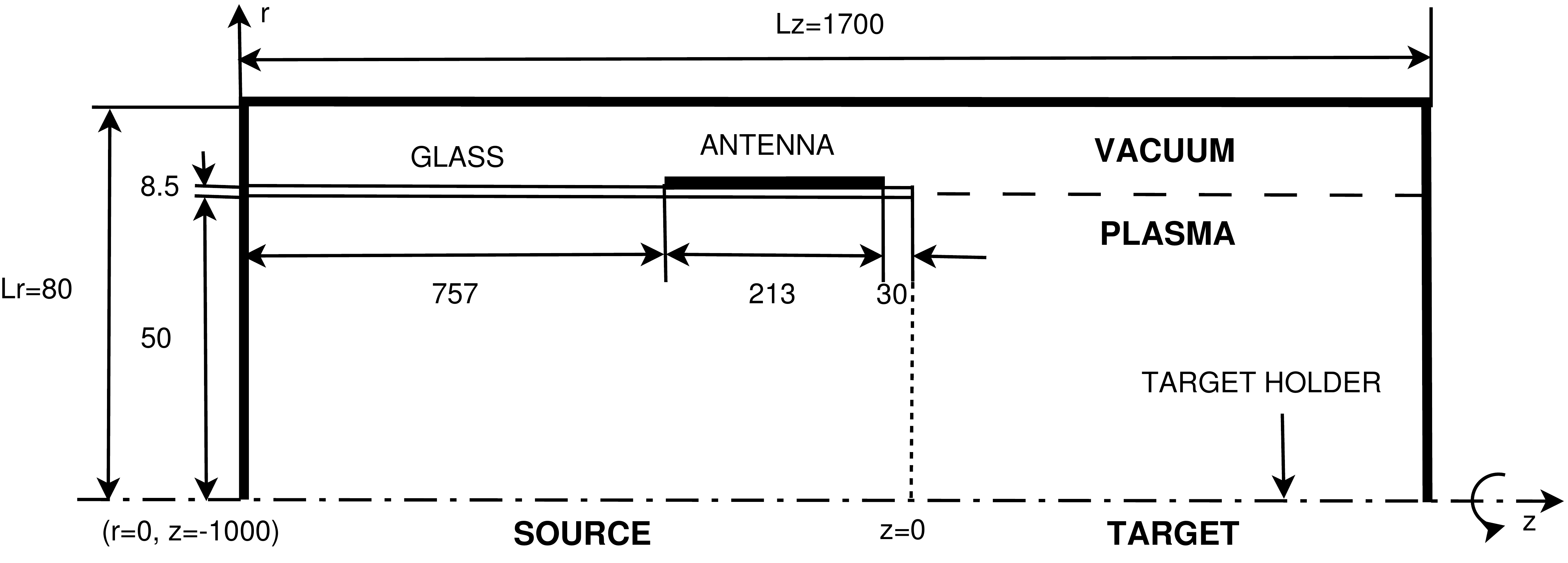}
\end{array}$
\end{center}
\caption{Computational domain employed to simulate the experimental setup shown in Fig.~\ref{fg1_8}. Here all dimensions are given in millimetres. }
\label{fg3_2}
\end{figure}

Equations~(\ref{eq3_1}) and (\ref{eq3_2}) are Fourier transformed with respect to the azimuthal angle and then solved (for an azimuthal mode number $m$) by a finite difference scheme on a 2D domain ($r$, $z$). The computational domain is shown in Fig.~\ref{fg3_2}.
The enclosing chamber is assumed to be ideally conducting so that the tangential components of $\mathbf{E}$ vanish at the chamber walls, i. e. 
\begin{equation}\label{eq3_7}
\begin{array}{l}
\vspace{0.3cm}E_\theta(L_r, z)=E_z(L_r, z)=0, \\
\vspace{0.3cm}E_r(r, 0)=E_\theta(r, 0)=0, \\
E_r(r, L_z)=E_\theta(r, L_z)=0,  
\end{array}
\end{equation}
with $L_r$ and $L_z$ the radius and the length of the chamber, respectively. Moreover, all field components must be regular on axis, thus, $B_\theta |_{r=0}=0$ and $(rE_\theta)|_{r=0}=0$ for $m=~0$; $E_z|_{r=0}=0$ and $(rE_\theta)|_{r=0}=0$ for $m\neq 0$.\cite{Zhang:2008aa} In the present work, we choose the fundamental odd mode number $m=1$, which is preferentially excited in the helicon discharge launched by a left hand half-turn helical antenna.\cite{Chen:1996aa, Light:1995aa, Light:1995ab}. In the experiment, there is a radial air gap ( $0.055<r<0.0585$~m) between the antenna and the glass tube, which is taken as glass region in the computational domain. We found that simulated results are insensitive to the dielectric constant in the glass region $0.05<r<0.055$~m by varying the constant from $1$ to $10$. As no change was detected in the wave field, we therefore expanded the glass area radially to fill this air gap. The thickness of the antenna is approximately $0.002-0.003$~m. 

\subsection{Computed and measured wave fields}\label{cmp3}
\begin{figure*}[h]
\begin{center}$
\begin{array}{l}
(a)\\
\hspace{0.35 cm}\includegraphics[width=0.8\textwidth,angle=0]{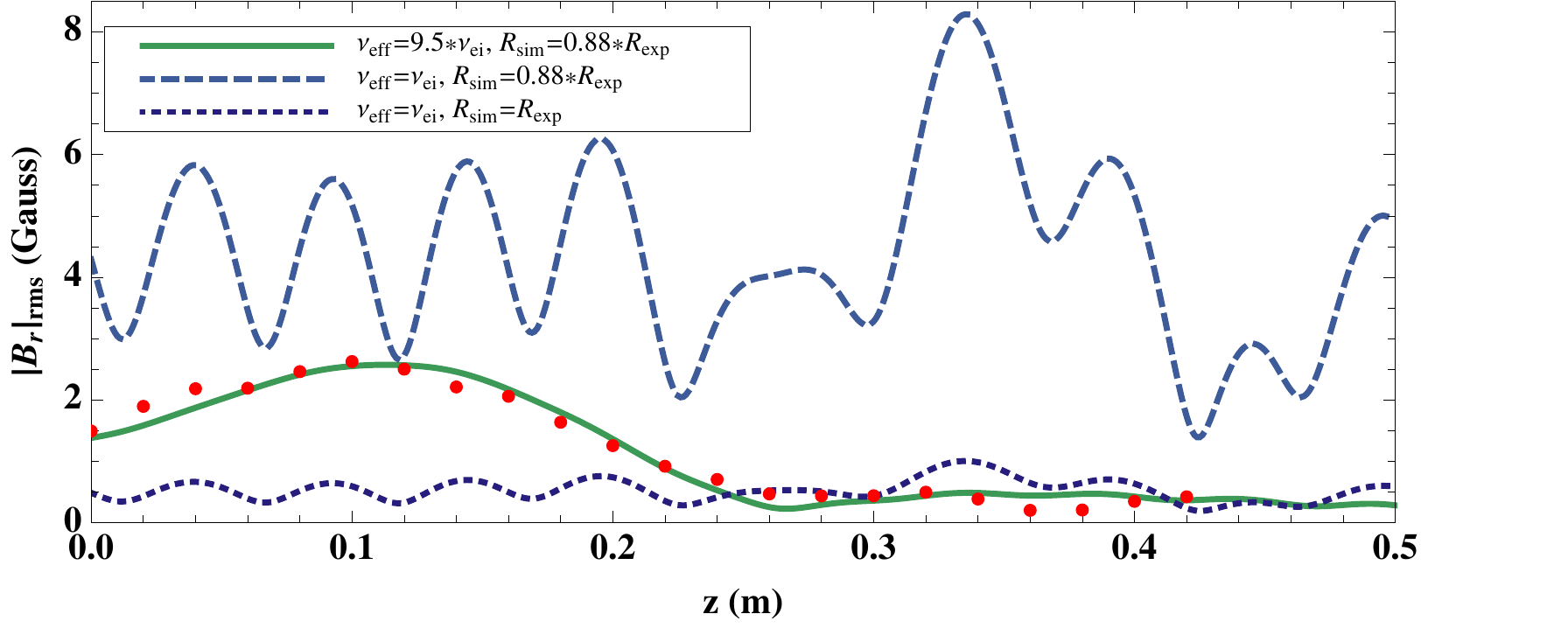}\\
(b)\\
\includegraphics[width=0.737\textwidth,angle=0]{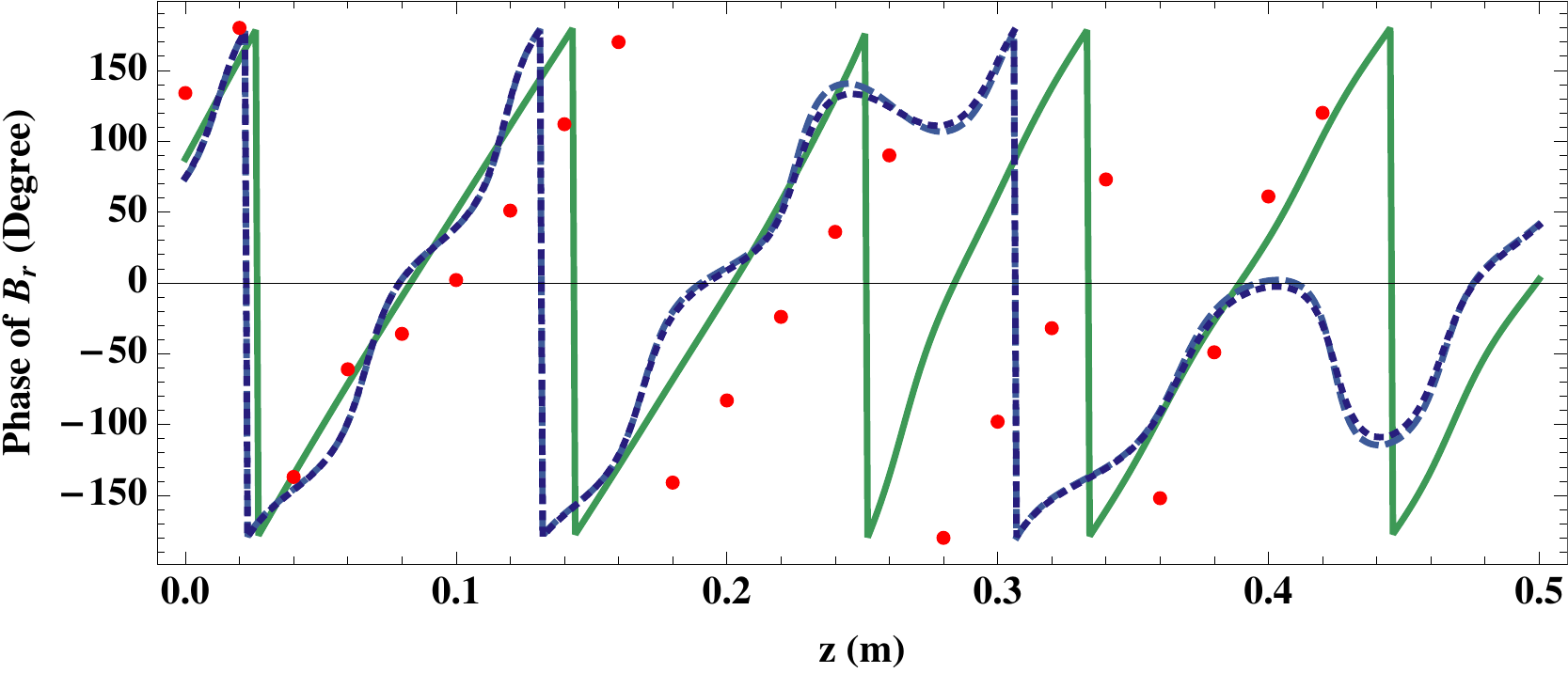}
\end{array}$
\end{center}
\caption{Variations of magnetic wave field in axial direction (on-axis): (a) $|B_r|_{\mathrm{rms}}$, (b) phase of $B_r$. Computed results (lines: dotted for $\nu_{\mathrm{eff}}=\nu_{ei}$ and $R_{\mathrm{sim}}=R_{\mathrm{exp}}$, dashed for $\nu_{\mathrm{eff}}=\nu_{ei}$ and $R_{\mathrm{sim}}=0.88R_{\mathrm{exp}}$, and solid for $\nu_{\mathrm{eff}}=9.5\nu_{ei}$ and $R_{\mathrm{sim}}=0.88R_{\mathrm{exp}}$) are compared with experimental data (dots). }
\label{fg3_3}
\end{figure*}

Based on the measured field strength configuration and plasma profiles shown in Fig.~\ref{fg3_1}, simulations are performed. Figure~\ref{fg3_3} shows the axial profiles of the computed $B_r$ amplitude and phase on axis, and their comparisons with experimental data. With the collisionality set to $\nu_{\mathrm{eff}}=\nu_{ei}$, where $\nu_{\mathrm{eff}}$ is the effective collision frequency, and the antenna radius set to match the experiment, the predicted wave field is $\sim 30\%$ of the measured value, and the profile a poor match to the experiment. It is possible to obtain better agreement by varying the collisionality, which strongly affects the profile but leaves the magnitude largely unchanged, and the antenna radius, which strongly affects the field amplitude and leaves the radial and axial profiles of $\mathbf{B}$ unchanged. A qualitative match between measurement and simulation of the axial variation of $B_r$ is found using an enhancement in collisionality of $\nu_{\mathrm{eff}}=\varrho_\mathrm{en} (\nu_{ei}+\nu_{ie})\approx \varrho_\mathrm{en} \nu_{ei}$ with $\varrho_\mathrm{en}=9.5$, and an adjustment in antenna dimension of $R_{\mathrm{sim}}=\iota_\mathrm{ad} R_{\mathrm{exp}}$ with $\iota_\mathrm{ad}=0.88$. Since the collision frequency is proportional to $T_e^{-3/2}$, if the electron temperature was wrong, it has to be $4.5$ times smaller in order to get the agreement. Calculation of the axial gradient of the computed phase variation shows a travelling wave, with a good agreement with data.

\begin{figure*}[h]
\begin{center}$
\begin{array}{ll}
(a)&(b)\\
\includegraphics[width=0.45\textwidth,angle=0]{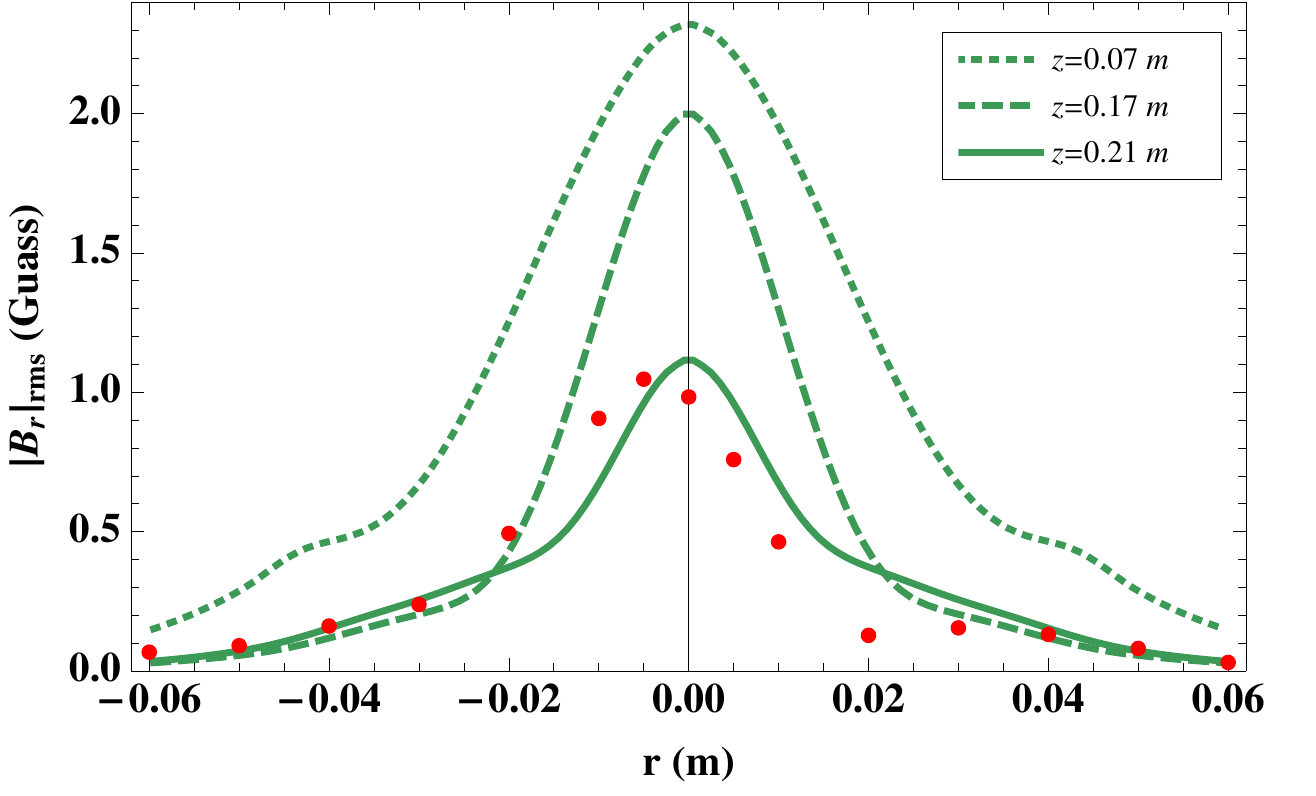} &\includegraphics[width=0.455\textwidth,angle=0]{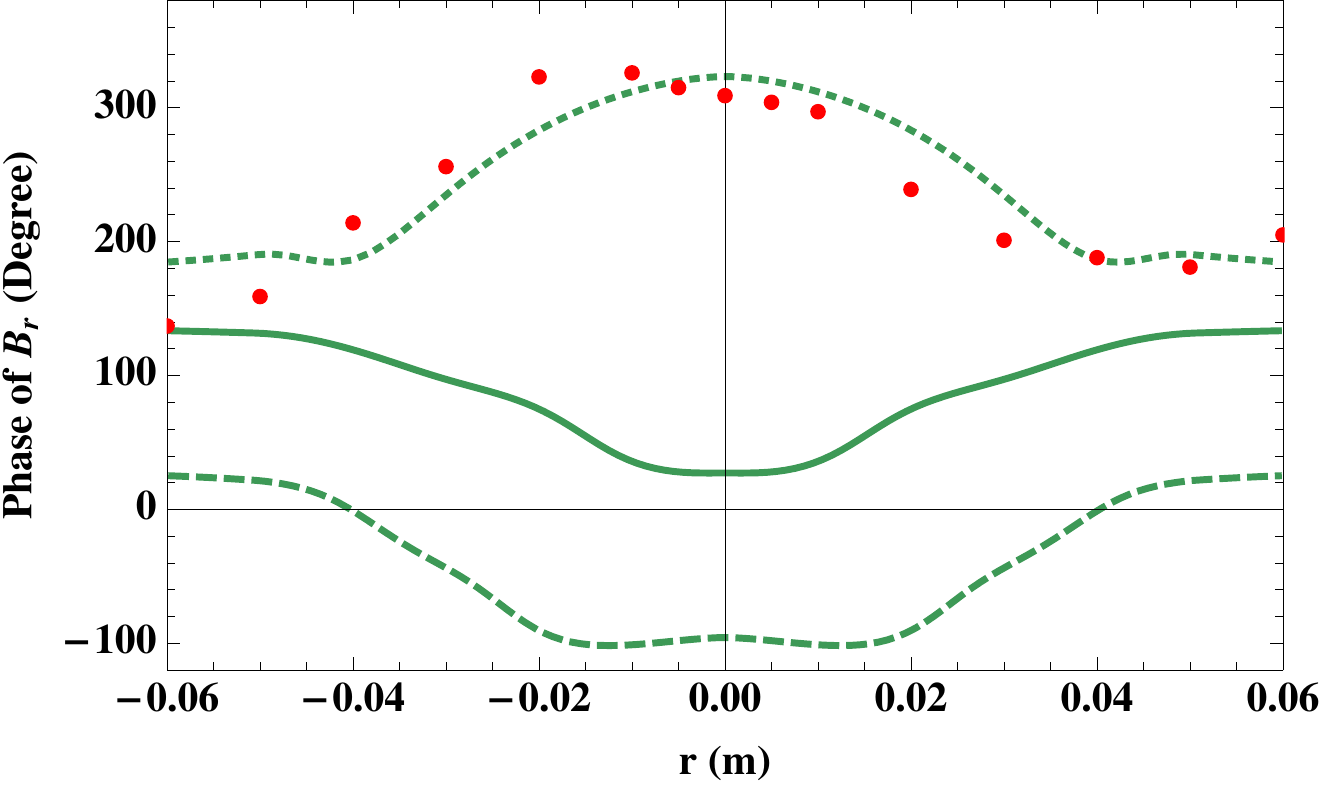}\\
(c)&(d)\\
\includegraphics[width=0.455\textwidth,angle=0]{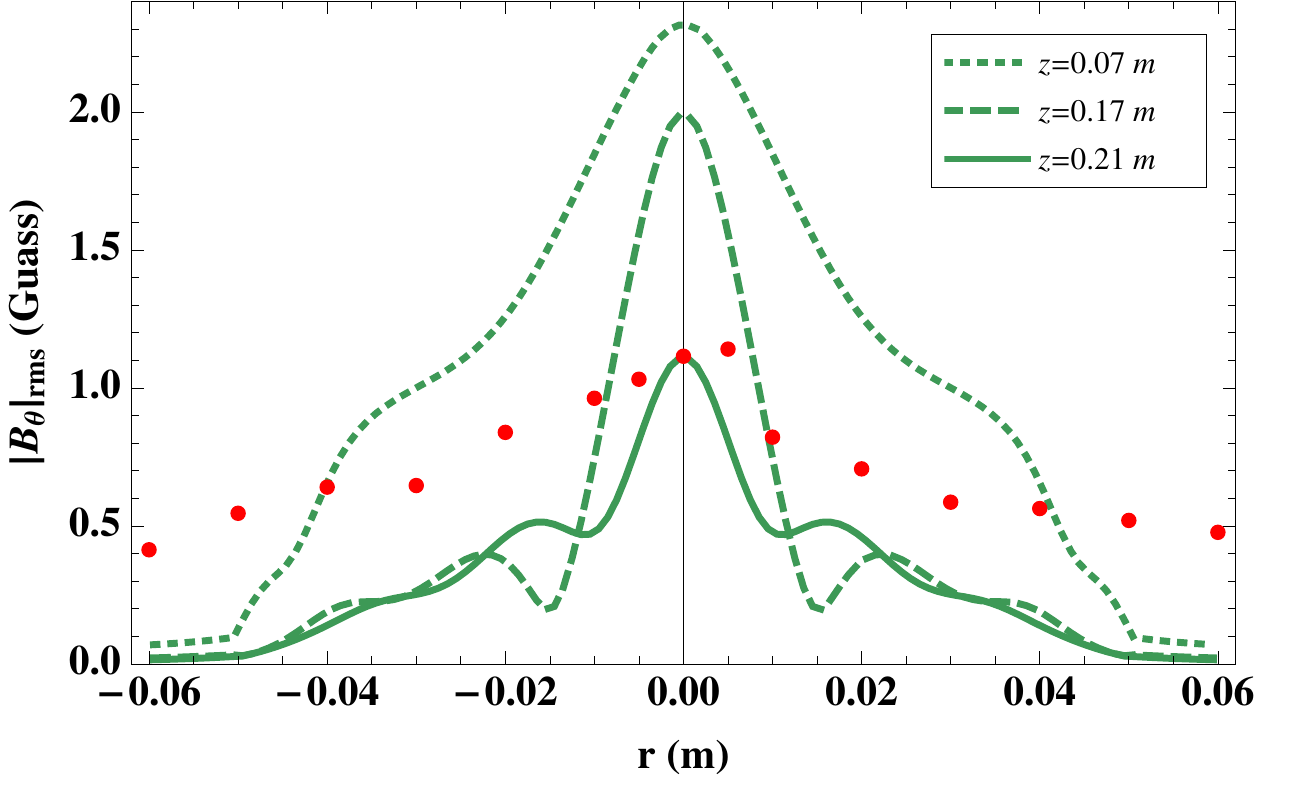} &\includegraphics[width=0.45\textwidth,angle=0]{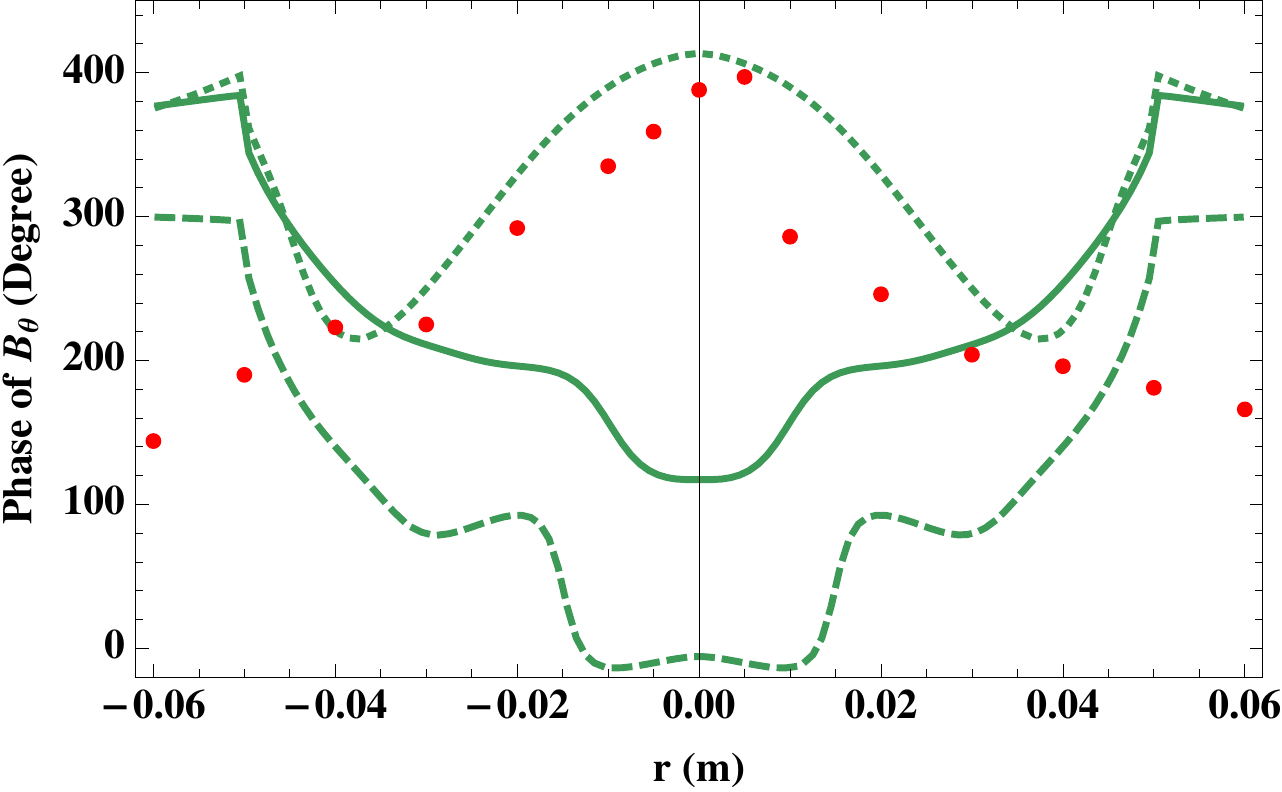}\\
(e)&(f)\\
\includegraphics[width=0.45\textwidth,angle=0]{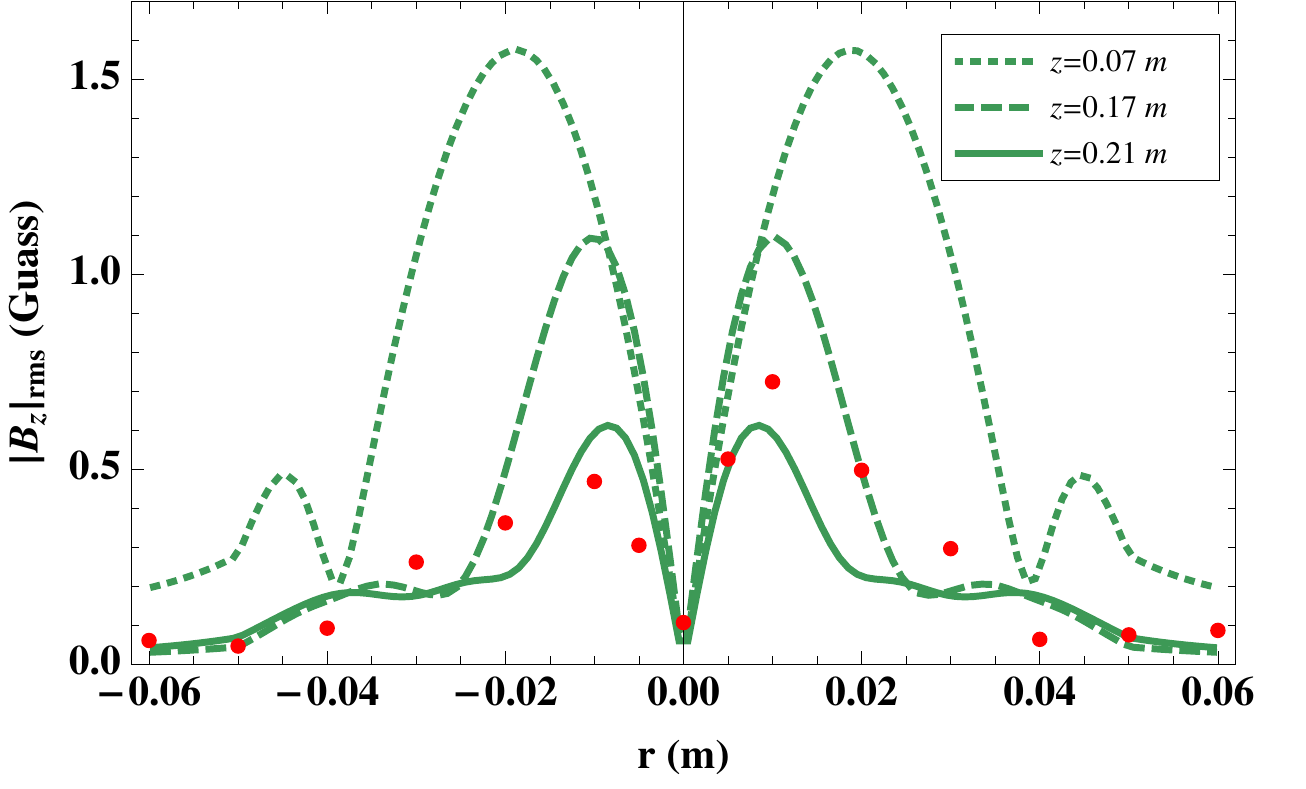} &\includegraphics[width=0.455\textwidth,angle=0]{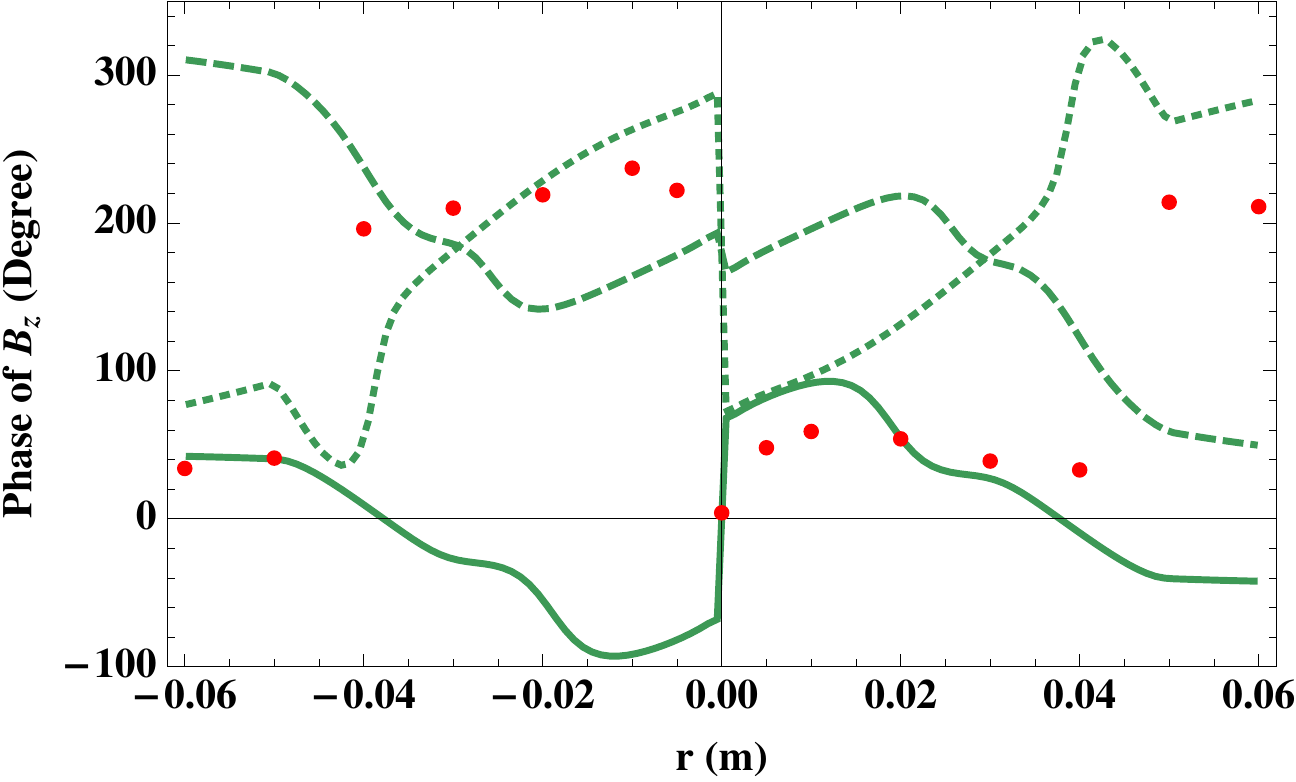}
\end{array}$
\end{center}
\caption{Variations of magnetic wave field in radial direction: (a), (c) and (e) are $|B_r|_{\mathrm{rms}}$, $|B_\theta|_{\mathrm{rms}}$ and $|B_z|_{\mathrm{rms}}$, respectively; (b), (d) and (f) are the corresponding phase variations. Dots are experimental data while lines (dotted: $z=0.07$~m, dashed: $z=0.17$~m, solid: $z=0.21$~m) are simulated results. }
\label{fg3_4}
\end{figure*}

The local minimum observed both experimentally and numerically in the axial profile of $|B_r|_{\mathrm{rms}}$ around $z=0.27$~m has been also observed in many other devices,\cite{Guo:1999aa, Degeling:2004aa, Light:1995aa, Mori:2004aa} for both uniform and non-uniform field cases. For the uniform field case, it has been suggested that the spatial modulation of the helicon wave amplitude is not caused by reflections from the end boundaries, but by a simultaneous excitation of two radial modes.\cite{Chen:1996ab, Chen:1996aa, Light:1995aa, Mori:2004aa} Similarly, the minimum observed in MAGPIE cannot be explained by standing waves, because the amplitude becomes much smaller at bigger $z$ (suggesting strong damping), and the phase advances with increasing $z$ (denoting a travelling wave). Further, radial profiles of the wave field in Fig.~\ref{fg3_4} feature a possible superposition of the first and second radial modes of the $m=1$ azimuthal mode. Therefore, we speculate that the minimum observed here may be also due to the simultaneous excitation of two fundamental radial modes. We will show later that radial gradient in plasma density is essential for the excitation of this local minimum under the present experimental conditions.  

Figure~\ref{fg3_4} shows the radial profiles of computed wave fields for $\nu_{\mathrm{eff}}=9.5 \nu_{ei}$ and $R_{\mathrm{sim}}=0.88R_{\mathrm{exp}}$ at three axial positions in the target region, together with the experimental data measured at $z=0.17$~m. The predicted wave field amplitude profile at $z=0.17$~m is consistent with the data, but the magnitude is nearly double the measured value, and the phase profile is a poor match. We have also computed the wave fields at axial locations with best agreement to the amplitude ($z=0.21$~m) and phase ($z=0.07$~m). We justify this freedom of choice by the experimental uncertainty in axial density profile, and the numerical sensitivity identified in the radial profile of the wave field with axial position. Inspection of Fig.~\ref{fg3_4} reveals that it is possible to find a reasonable agreement to the wave amplitude and phase profile, albeit independently. As expected, all calculations show the wave mode structure of $m=1$ through $|B_z(r=0)|_{\mathrm{rms}}\approx 0$, consistent with the antenna parity. The near-null minima in both $|B_\theta(r)|_{\mathrm{rms}}$ and $|B_z(r)|_{\mathrm{rms}}$ suggest a likely simultaneous excitation of the first and second radial modes. This may account for the minimum in Fig.~\ref{fg3_3}(a), and is consistent with conclusions from others.\cite{Chen:1996ab, Mori:2004aa, Chen:1996aa, Light:1995aa, Light:1995ab} Mori et al.\cite{Mori:2004aa} suggested that the superposition feature is associated with a wave focusing effect caused by the non-uniform magnetic field in the target region. 

\section{Numerical profile scans}\label{num3}
It has previously been shown that the plasma density can be further increased by introducing a cusp or non-uniform static magnetic field in the vicinity of the helicon antenna.\cite{Boswell:1997aa, Chen:1992aa, Chen:1997aa, Gilland:1998aa} To shed light on the increased plasma production, we perform a detailed numerical study on the effects of radial and axial plasma density gradients and axial magnetic field gradient on wave propagation characteristics. The enhancement of $\varrho_\mathrm{en}=9.5$ to $\nu_{ei}$ and the adjustment of $\iota_\mathrm{ad}=0.88$ to $R_{\mathrm{exp}}$ are still employed in this section because they provide a good agreement with the measured wave field. 

\subsection{Axial profile of plasma density}\label{den3}
\begin{figure*}[h]
\begin{center}$
\begin{array}{l}
(a)\\
\hspace{0.4cm}\includegraphics[width=0.865\textwidth]{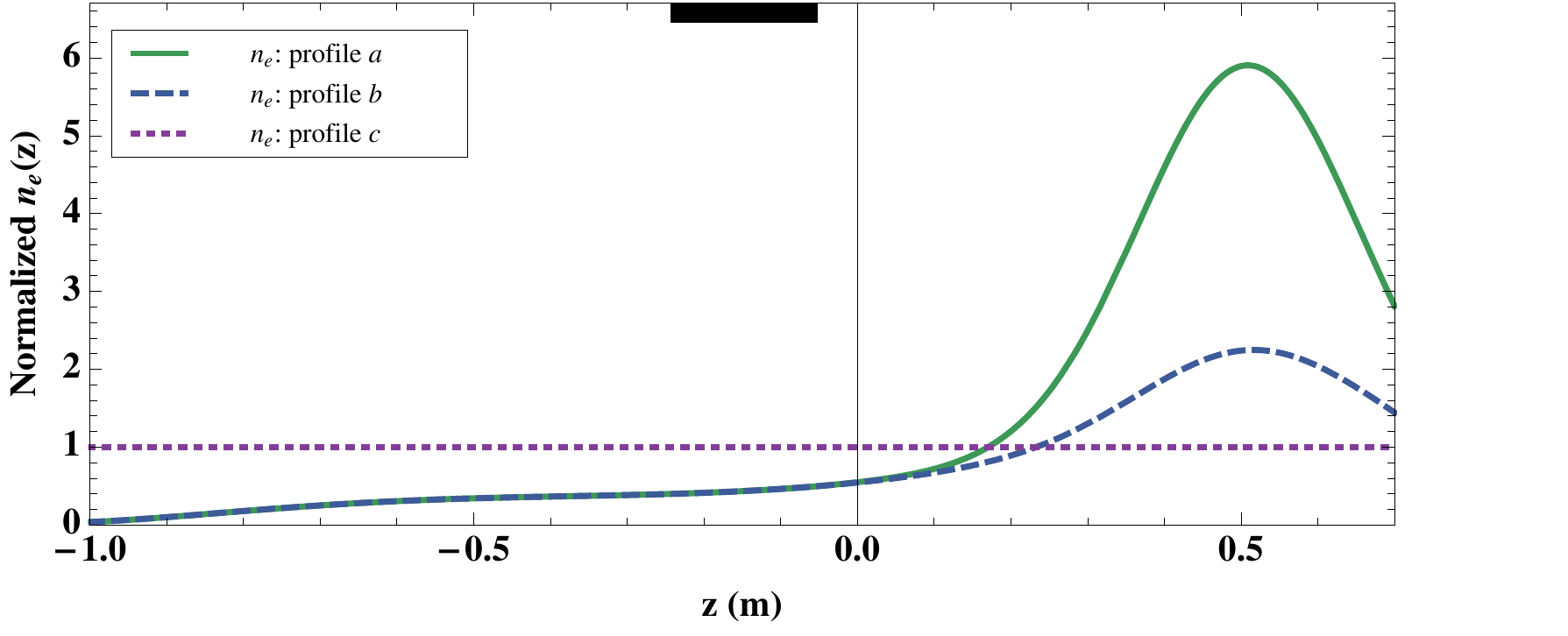}\\
(b)\\
\includegraphics[width=0.855\textwidth]{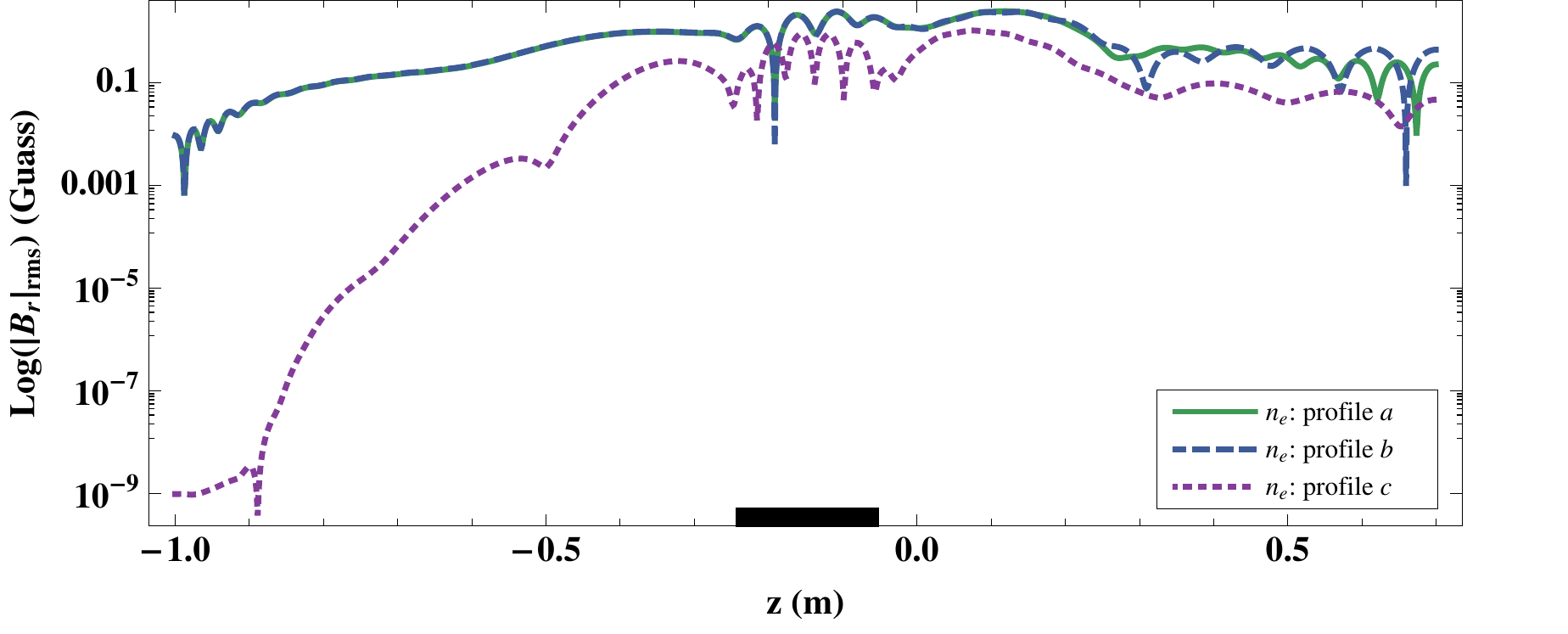}\\
(c)\\
\includegraphics[width=0.8\textwidth]{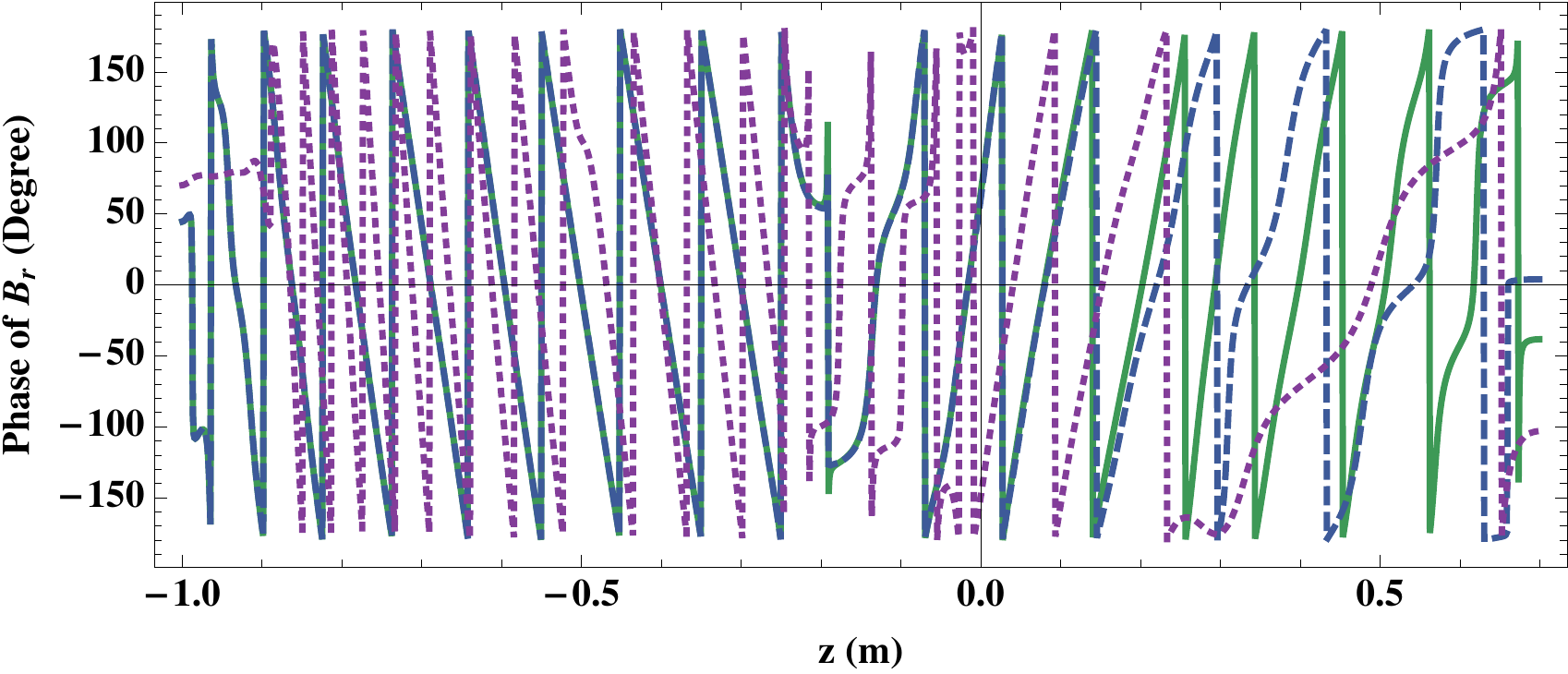}
\end{array}$
\end{center}
\caption{Normalised axial profiles of plasma density and the corresponding variations of magnetic wave field in axial direction: (a), normalised axial profiles of plasma density, (b) log scale of $|B_r|_{\mathrm{rms}}$, and (c) phase of $B_r$. They are all measured on axis. In (a), the solid line is associated with density profile linear with $B_0(z)$, the dashed line is the same to the solid one except in region of $0<z<0.7$~m where the density maximum is adjusted to match preliminary experimental observations, and the dotted line represents a $z$-independent density profile.}
\label{fg3_5}
\end{figure*}

We first study the effect of varying the axial gradient of plasma density, which has been assumed to be linear with the static magnetic field so far, on wave propagations by comparing the wave fields from three different on-axis density profiles shown in Fig.~\ref{fg3_5}(a). 
Other conditions are kept the same as previous sections. The computed wave fields in axial direction (on-axis) are shown in Fig.~\ref{fg3_5}(b) and Fig.~\ref{fg3_5}(c). A log scale in the amplitude has been employed to see the wave propagation features clearly. We can see from the phase variations (Fig.~\ref{fg3_5}(c)) that as density is decreased in the target region the wavelength increases, which is consistent with a simple theory developed previously (see Eq.~(\ref{eq1_29})), \cite{Chen:1996ab} 
\begin{equation}\label{eq3_8}
\frac{3.83}{R_p}=\frac{\omega}{k}\frac{n_e e \mu_0}{B_0}.
\end{equation}
Thus, if $\omega$, $R_p$ (plasma radius) and $B_0$ are all fixed, $k$ is proportional to $n_e$, which means that the wavelength becomes larger at lower density. Here, the value of $3.83$ is the first non-zero Bessel root of $J_1(r)=0$, representing the first radial mode, which is assumed to be dominant in our case. Moreover, for all density profiles shown in Fig.~\ref{fg3_5}(a), the wavelength is bigger in region of $0<z<0.6$~m than that in other regions, indicating a locally increased phase velocity. Further inspection of Fig.~\ref{fg3_5}(b) and Fig.~\ref{fg3_5}(c) shows that density increasing in proportion to the static magnetic field has little effect on RF absorption, while the density level near the antenna affects the wave amplitude significantly at all axial locations.

\subsection{Axial profile of static magnetic field}\label{mgf3}
\begin{figure*}[h]
\begin{center}$
\begin{array}{l}
(a)\\
\hspace{0.2cm}\includegraphics[width=0.85\textwidth,angle=0]{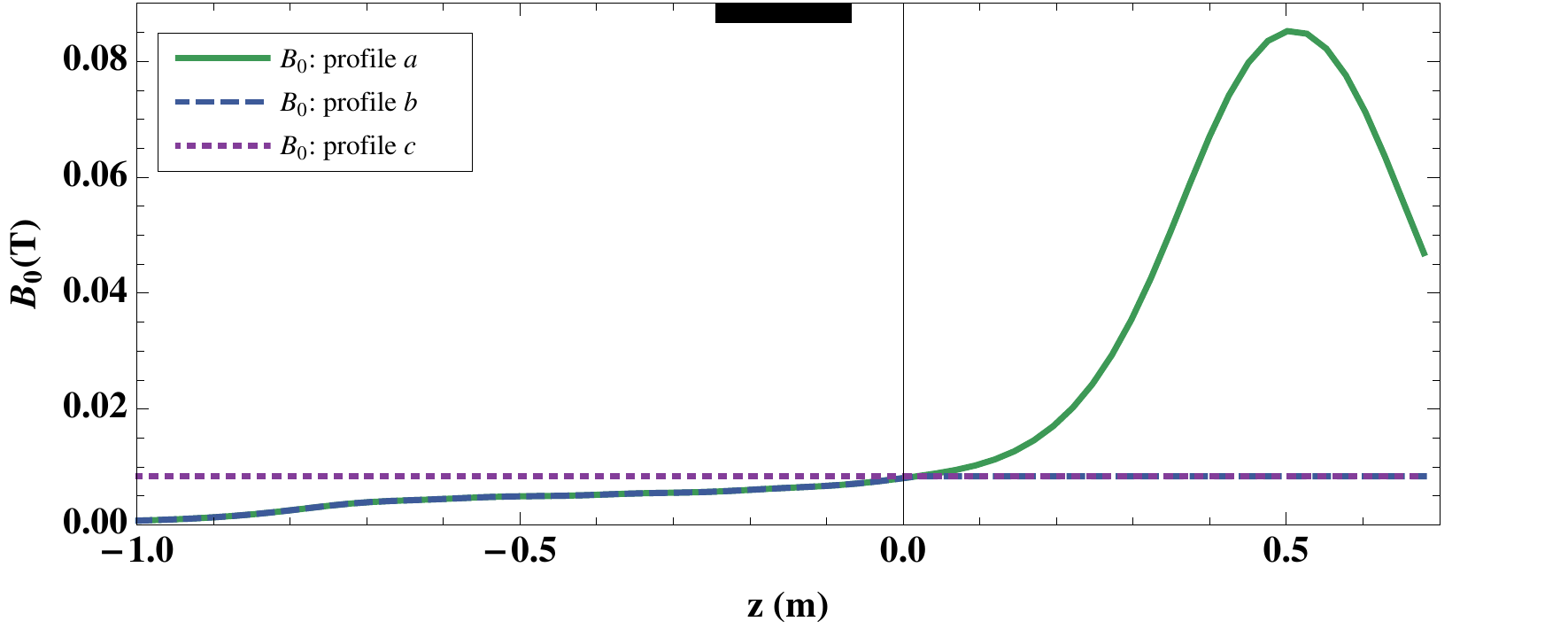}\\
(b)\\
\includegraphics[width=0.86\textwidth,angle=0]{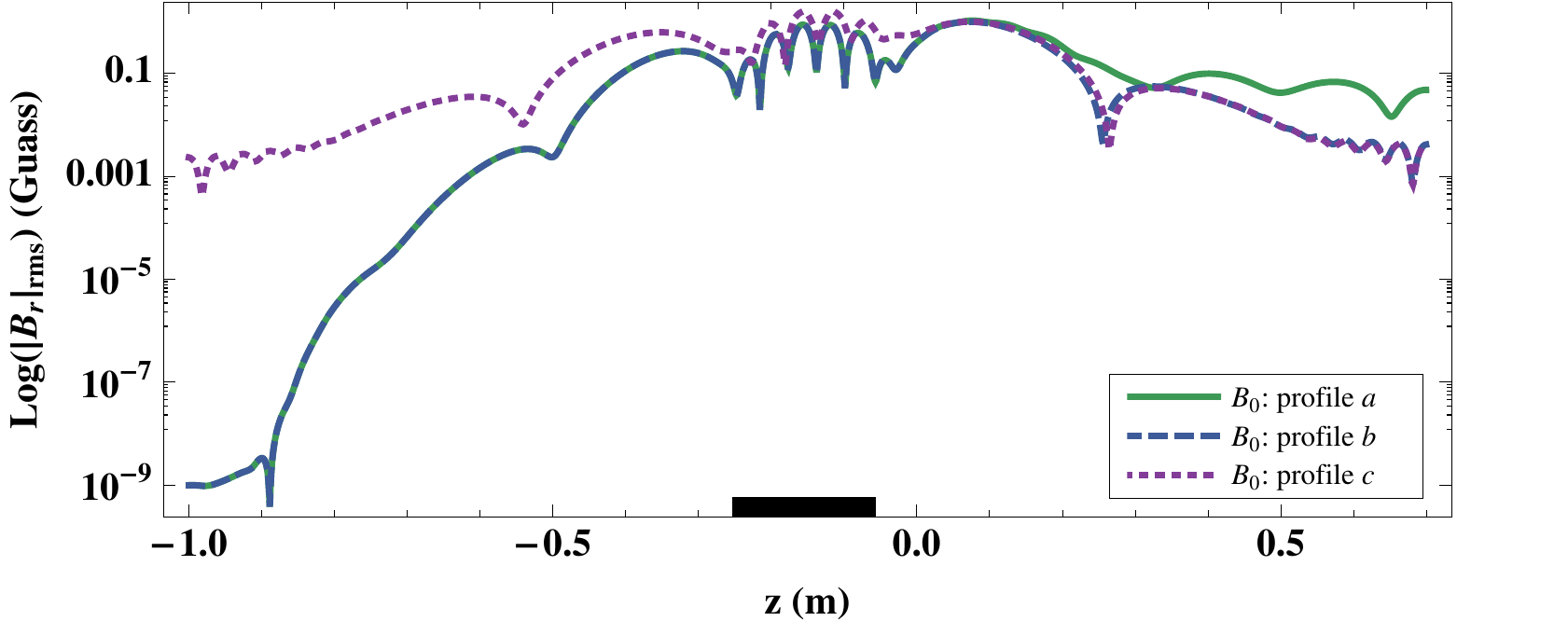}\\
(c)\\
\hspace{0.08cm}\includegraphics[width=0.795\textwidth,angle=0]{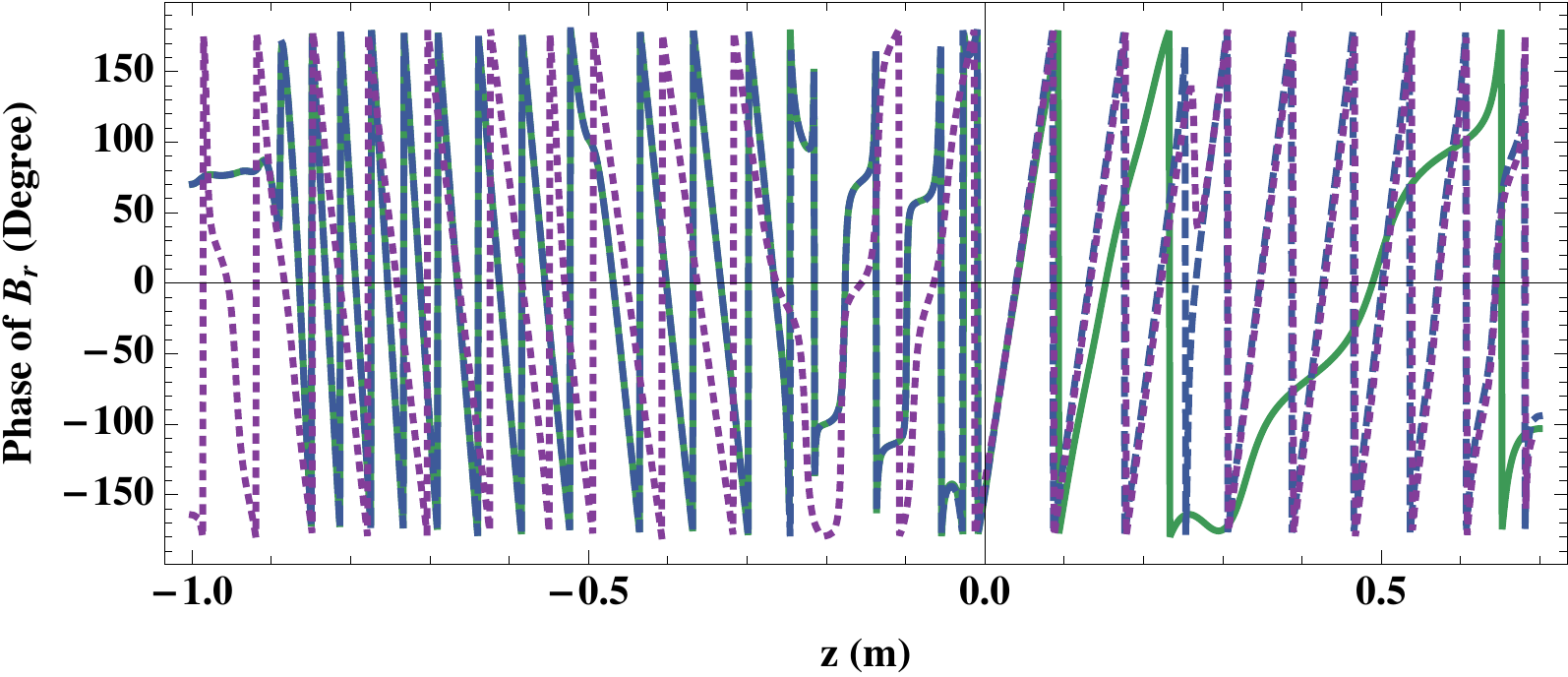}
\end{array}$
\end{center}
\caption{Axial profiles of static magnetic field and the corresponding variations of magnetic wave field in axial direction: (a) axial profiles of static magnetic field, (b) log scale of $|B_r|_{\mathrm{rms}}$ (on-axis), and (c) phase of $B_r$ (on-axis). In (a), the solid line shows original experimental data (Fig.~\ref{fg3_1}(a)), while the dashed line shows the same except being flattened in region of $z>0$~m and the dotted line is flattened everywhere.}
\label{fg3_6}
\end{figure*}

Second, following Sec.~\ref{den3}, we keep the axially uniform density profile and study the effects of axial gradient in static magnetic field, which is radially near uniform according to Eq.~(\ref{eq3_5}). Three employed field profiles are shown in Fig.~\ref{fg3_6}(a), 
which enable us to study the effects of field gradient in target and source regions seperately on wave propagations. Comparison between solid and dashed lines in Fig.~\ref{fg3_6}(b) and Fig.~\ref{fg3_6}(c) shows that the axial gradient in magnetic field in the target region may increase the propagation distance of helicon waves, consistent with Mori et al.'s conclusion that a focused non-uniform magnetic field provides easier access for helicon wave propagations than a uniform field,\cite{Mori:2004aa} or perhaps the increased field strength in that region increases the collisional damping length of helicon waves. The simple theory in Eq.~(\ref{eq3_8}) is satisfied again here: with decreased field strength in the target region, the wavelength becomes shorter. Although the difference between dashed and dotted field profiles is small as shown in Fig.~\ref{fg3_6}(a), the computed field amplitudes are significantly different. Figure~\ref{fg3_6}(b) and Fig.~\ref{fg3_6}(c) show that with a uniform field profile, the wave amplitude is much bigger than that with non-uniform field profile for $z<0$~m. Furthermore, waves keep their travelling features till the left endplate for uniform $B_0$, whereas for non-uniform $B_0(z)$ the wavelength becomes smaller when approaching left, and the waves are not travelling at all when $B_0(z)$ is low enough ($z<-0.9$~m). 

\subsection{Radial profile of plasma density}\label{rdn3}
\begin{figure*}[h]
\begin{center}$
\begin{array}{l}
(a)\\
\hspace{1.3cm}\includegraphics[width=0.7\textwidth,angle=0]{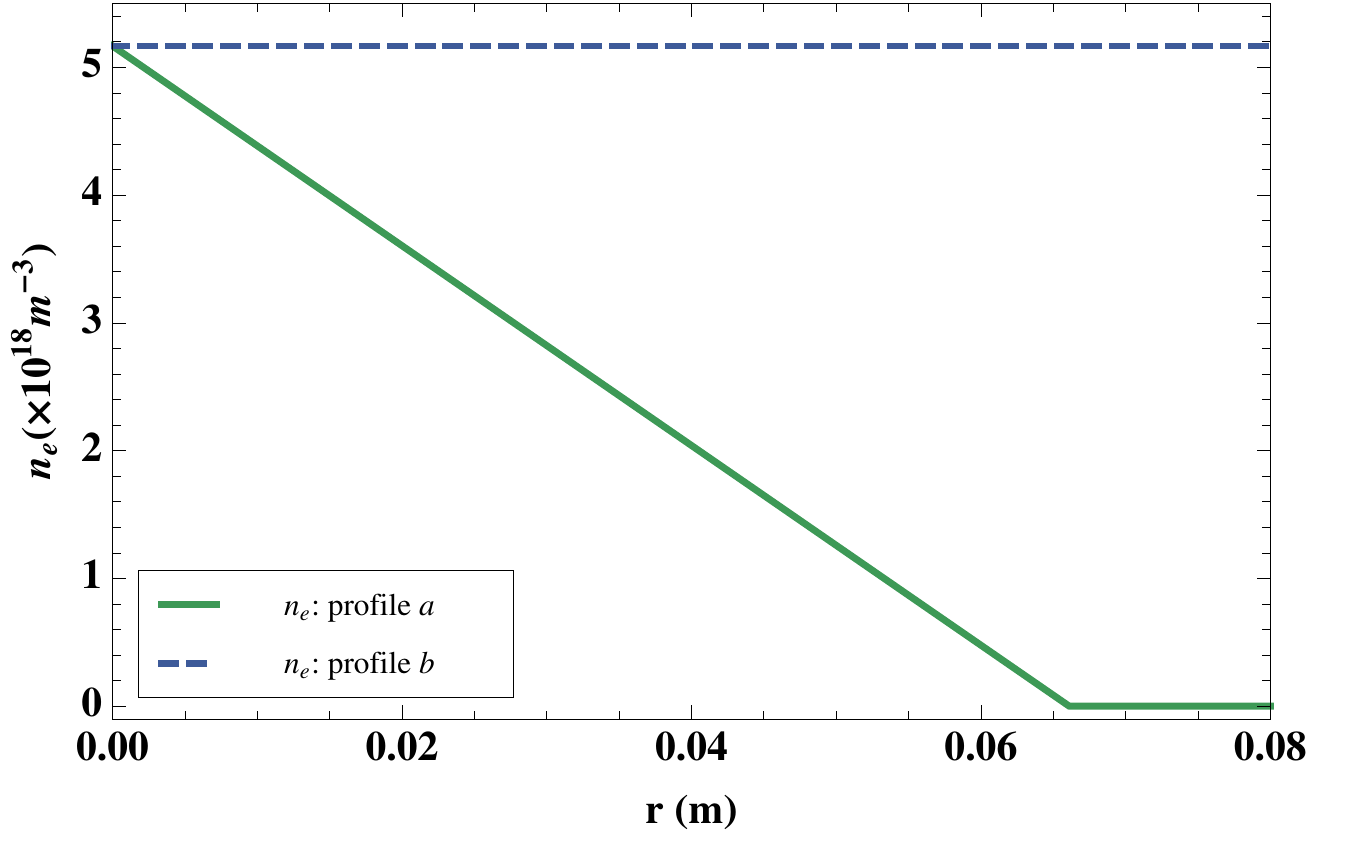}\\
(b)\\
\hspace{0.08cm}\includegraphics[width=0.85\textwidth,angle=0]{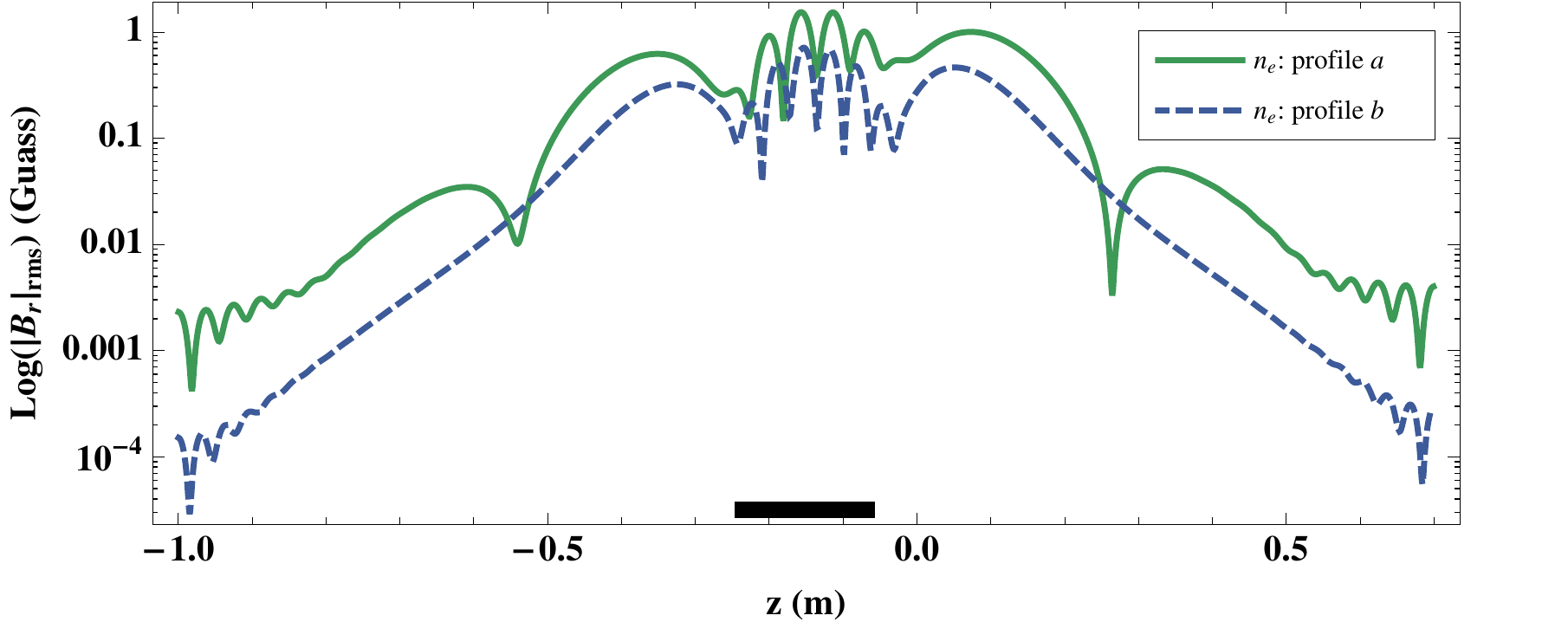}\\
(c)\\
\includegraphics[width=0.8\textwidth,angle=0]{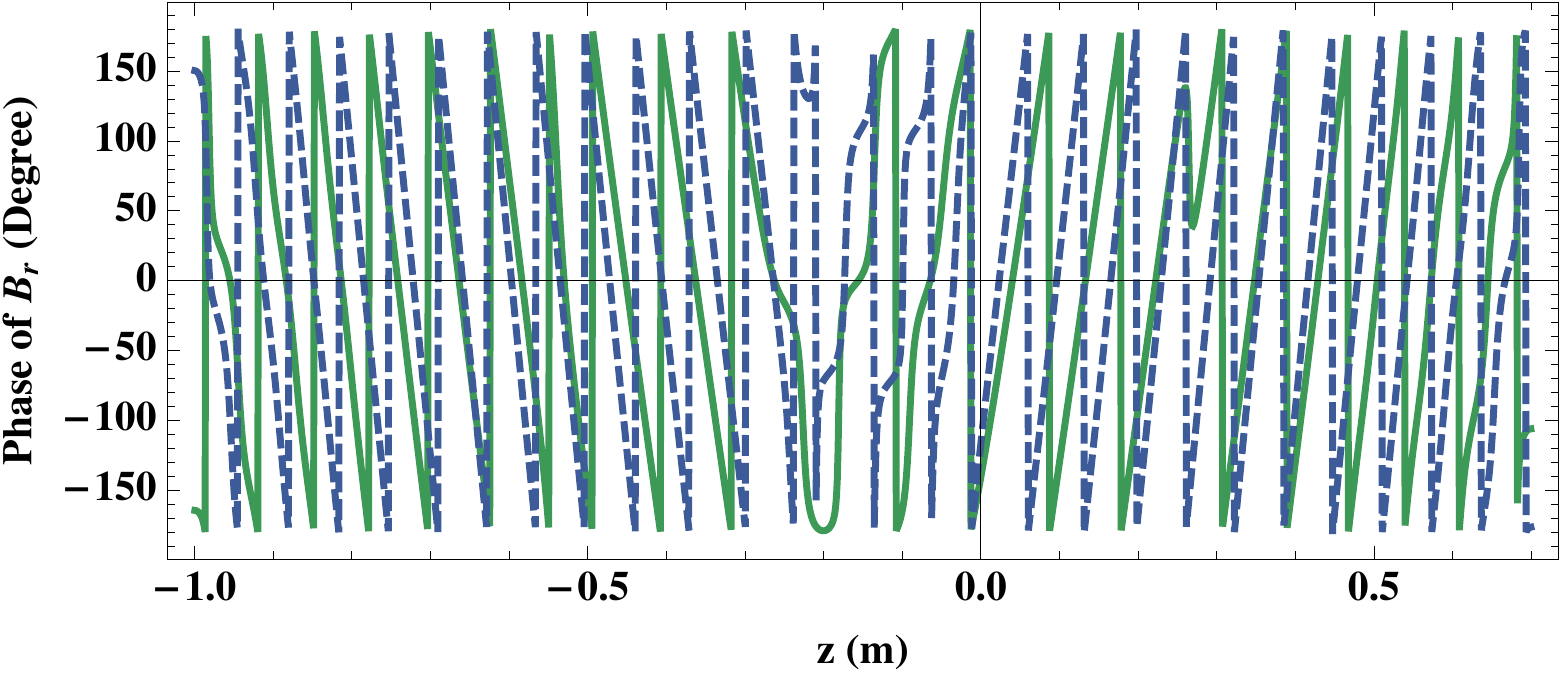}
\end{array}$
\end{center}
\caption{Radial profiles of plasma density and the corresponding variations of magnetic wave field in axial direction: (a) radial profiles of plasma density, (b) log scale of $|B_r|_{\mathrm{rms}}$, (c) phase of $B_r$. All parameters are measured on axis.  }
\label{fg3_7}
\end{figure*}

Now, we keep the plasma density and static magnetic field both uniform in the axial direction, and study the effects of radial gradient in plasma density. The two density profiles employed are shown in Fig.~\ref{fg3_7}(a), 
with and without radial gradient, and their corresponding results are shown in Fig.~\ref{fg3_7}(b) and Fig.~\ref{fg3_7}(c). We can see first that the local minimum in wave amplitude profiles, e. g. at $z=-0.52$~m and $z=0.27$~m, disappear when the radial density profile is flat, suggesting that the radial gradient in plasma density is essential to have a local minimum under the present conditions. Second, the wave amplitude is much bigger in both target and source regions for plasma density with radial gradient, suggesting that a radial gradient in density may be useful to maximise the plasma production. 

\section{Collisionality and field direction}\label{cll3}
\subsection{Enhancement of electron-ion collision frequency}\label{frq3}
\begin{figure*}[h]
\begin{center}$
\begin{array}{l}
\includegraphics[width=0.9\textwidth,angle=0]{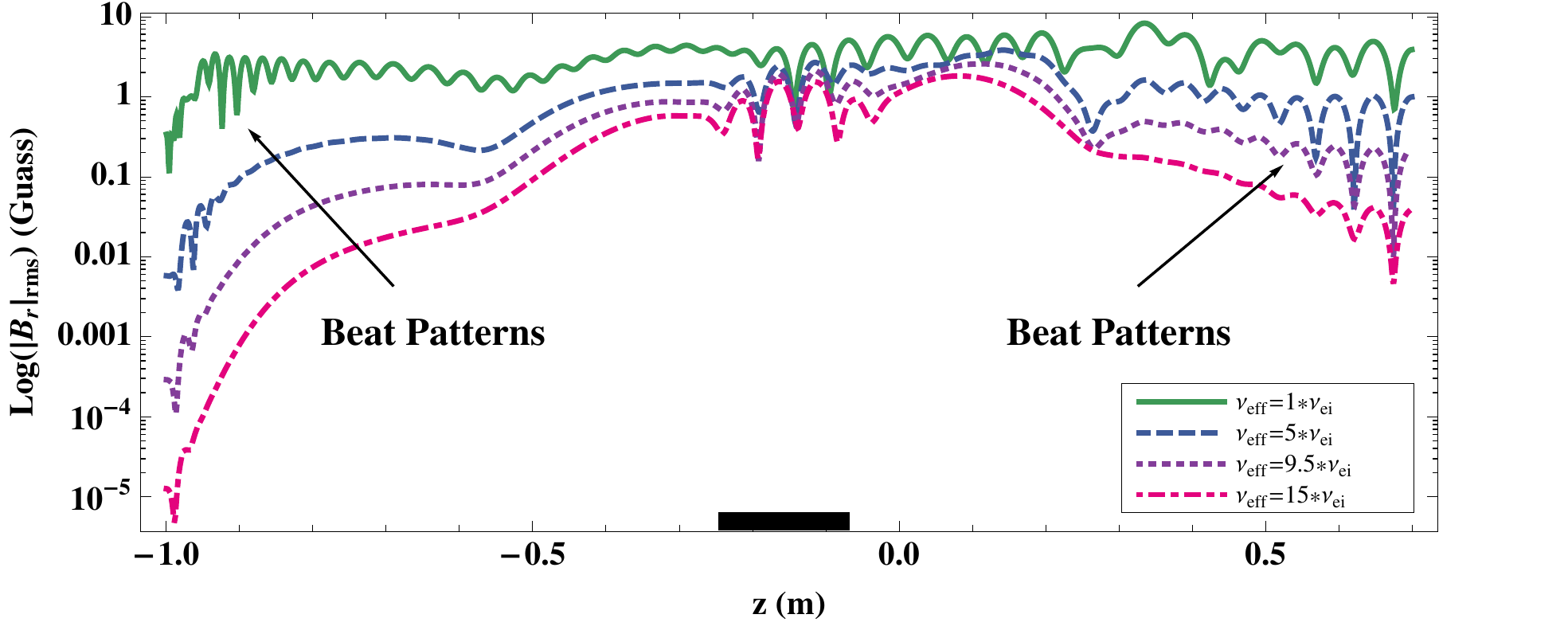}\\
\end{array}$
\end{center}
\caption{Variations of on-axis wave amplitudes in axial direction with different electron-ion collision frequencies: log scale of $|B_r|_{\mathrm{rms}}$.}
\label{fg3_8}
\end{figure*}

In a similar manner to other work,\cite{Zhang:2008aa, Lee:2011aa} we have used an enhancement to $\nu_{ei}$ (here $\nu_{\mathrm{eff}}\approx 9.5\nu_{ei}$), in order to find a qualitative match of simulated wave field to the data. In this section, we explore the physical consequences of scaling $\nu_{\mathrm{eff}}/\nu_{ei}$ in simulations while keeping the adjustment $\iota_\mathrm{ad}=0.88$ in the antenna radius. Variations of wave amplitude on axis in the axial direction for different collision frequencies are shown in Fig.~\ref{fg3_8}. 
As the collision frequency is increased from $\nu_{ei}$ to $15\nu_{ei}$, the wave amplitude decreases nearly everywhere, and the wave decay length is shortened. This indicates that the wave energy or power coupled from the antenna to the core plasma drops as the collision frequency becomes higher, and the power is more absorbed under the antenna. This is consistent with a previous conclusion that the RF energy is almost all absorbed in the near region of the antenna rather than in the far region.\cite{Chen:1996aa} The oscillations near the downstream and upstream ends at low $\nu_{\mathrm{eff}}$ are caused by reflections from the ideally conducting endplates, which disappear if the endplates are moved further away.

As suggested by Lee et al.,\cite{Lee:2011aa} an enhanced electron-ion collision frequency may be due to ion-acoustic turbulence which can happen if the electron drift velocity exceeds the speed of sound in magnetised plasmas. Based on the experimental conditions in MAGPIE, we have calculated the threshold field strength $B_{\mathrm{T}}$, below which ion-acoustic turbulence can happen. This threshold is given by $v_D \geq C_s$, where $v_D \approx k_B T_e/|e| B_0 R_p$ is the electron drift velocity with $k_B$ Boltzmann's constant and $C_s=\sqrt{k_B T_e/m_i} $ the speed of sound in magnetised plasmas. Application to MAGPIE conditions yields $B_{\mathrm{T}} \leq 0.0224~\mathrm{T}$. Thus, the whole source region which produces helicon plasmas and waves, has a magnetic field below this threshold. The ion-acoustic turbulence has the effect of providing additional electron-ion collisions within a dielectric tensor model, and thereby improves the agreement with observations. Other possible reason for this enhanced electron-ion collisionality includes kinetic effects which are beyond the reach of the employed cold-plasma model, e.g. Landau damping. The Trivelpiece-Gould mode, which could be present in the low field region ($B_0(z)<0.01$~T) and is consistent with the strong edge heating observed at enhanced collision frequencies,\cite{Arnush:2000aa, Shinohara:2002aa} is included in the present full wave simulations. For the present conditions of MAGPIE, the electron drift velocity is well blow the electron thermal speed, therefore, the Buneman instability does not occur.

\subsection{Direction of static magnetic field}\label{dir3}
\begin{figure*}[h]
\begin{center}$
\begin{array}{l}
(a)\\
\hspace{0.15cm}\includegraphics[width=0.65\textwidth,angle=0]{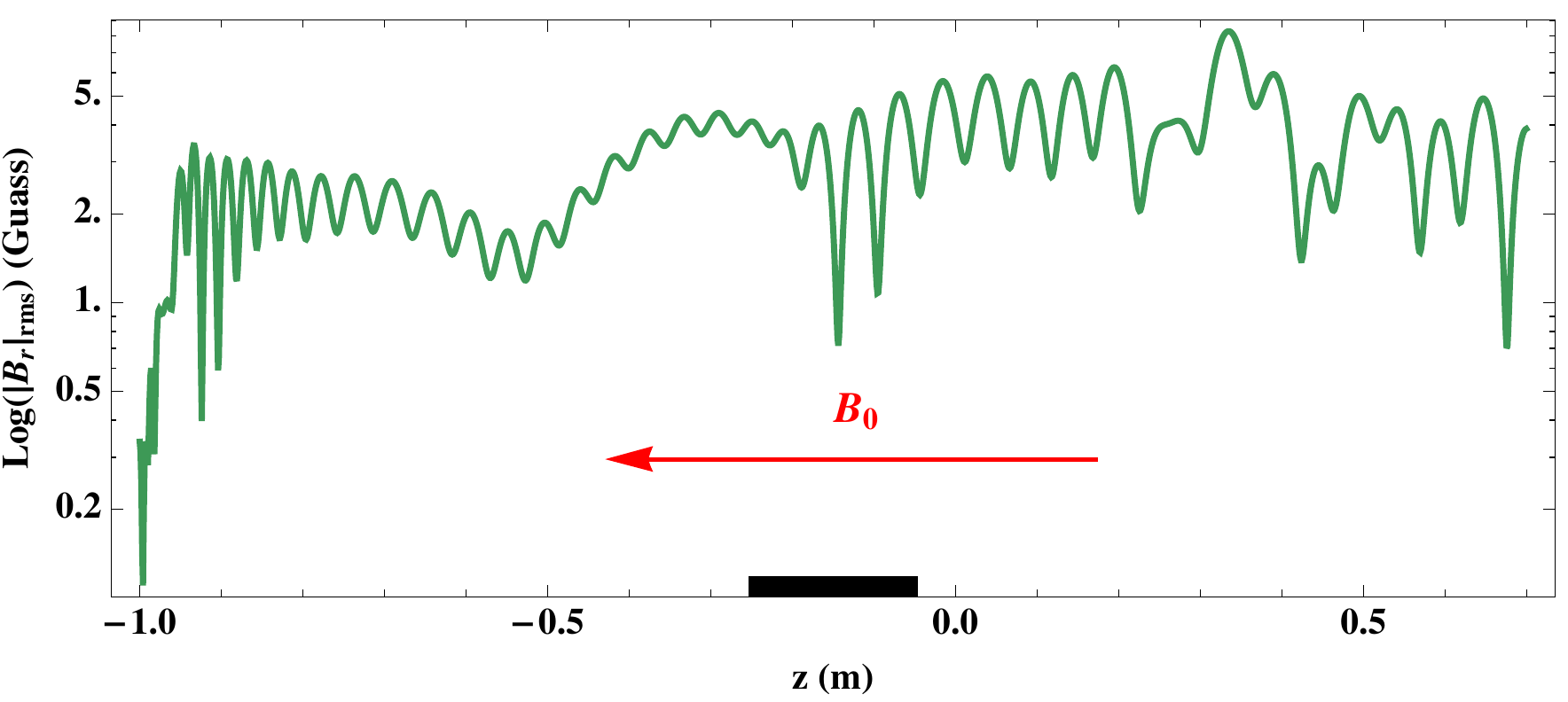}\\
(b)\\
\hspace{-0.15 cm}\includegraphics[width=0.76\textwidth,angle=0]{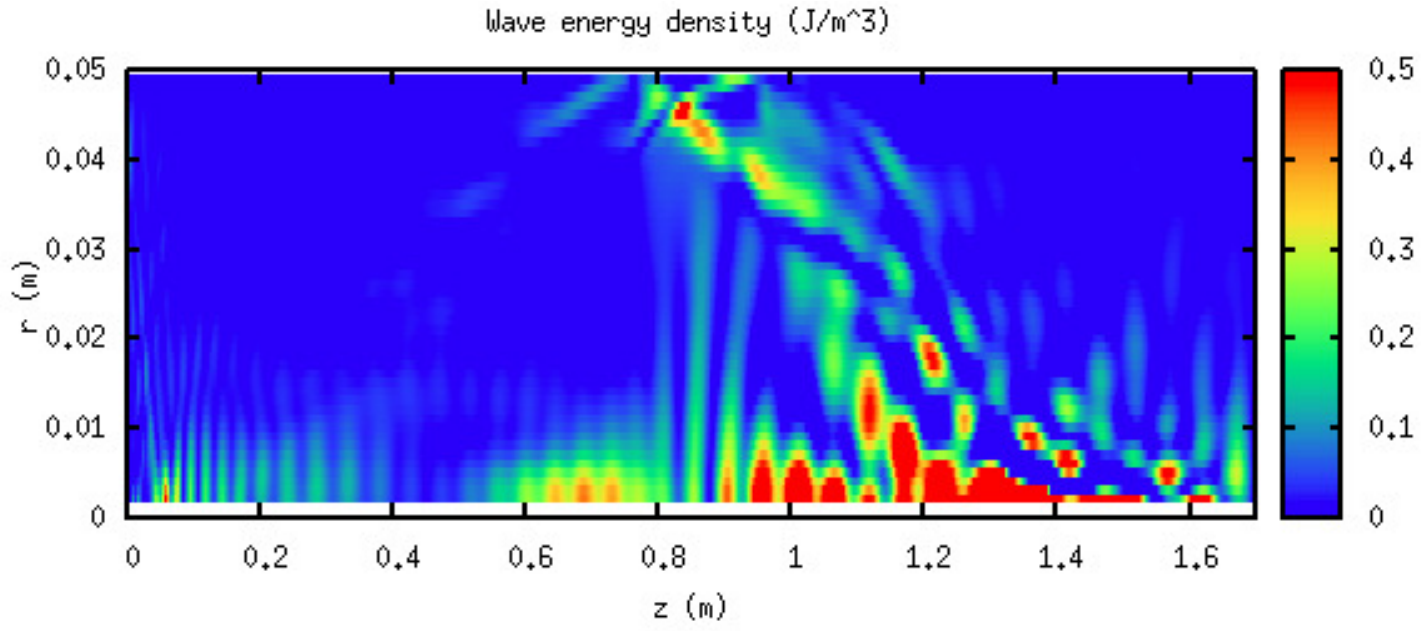}\\
(c)\\
\hspace{0.15cm}\includegraphics[width=0.655\textwidth,angle=0]{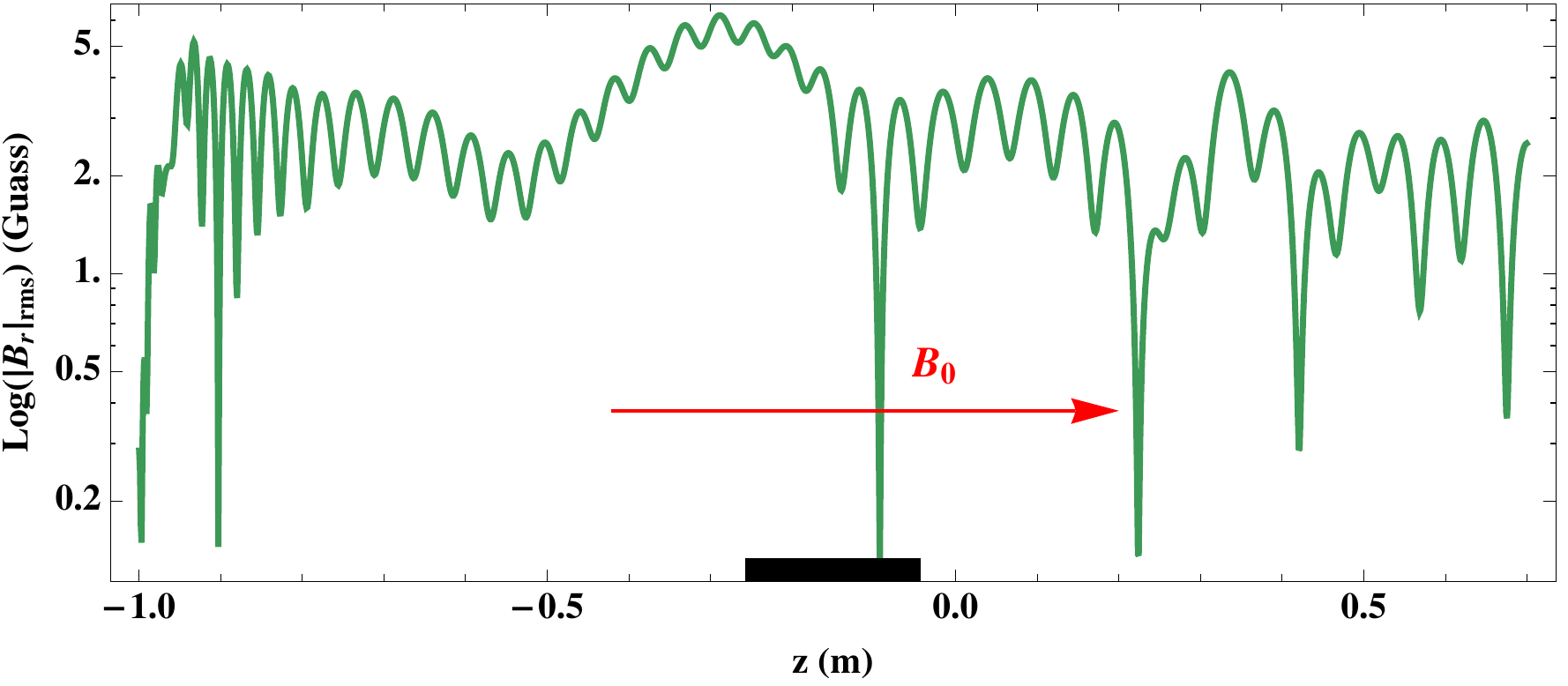}\\
(d)\\
\hspace{-0.15 cm}\includegraphics[width=0.76\textwidth,angle=0]{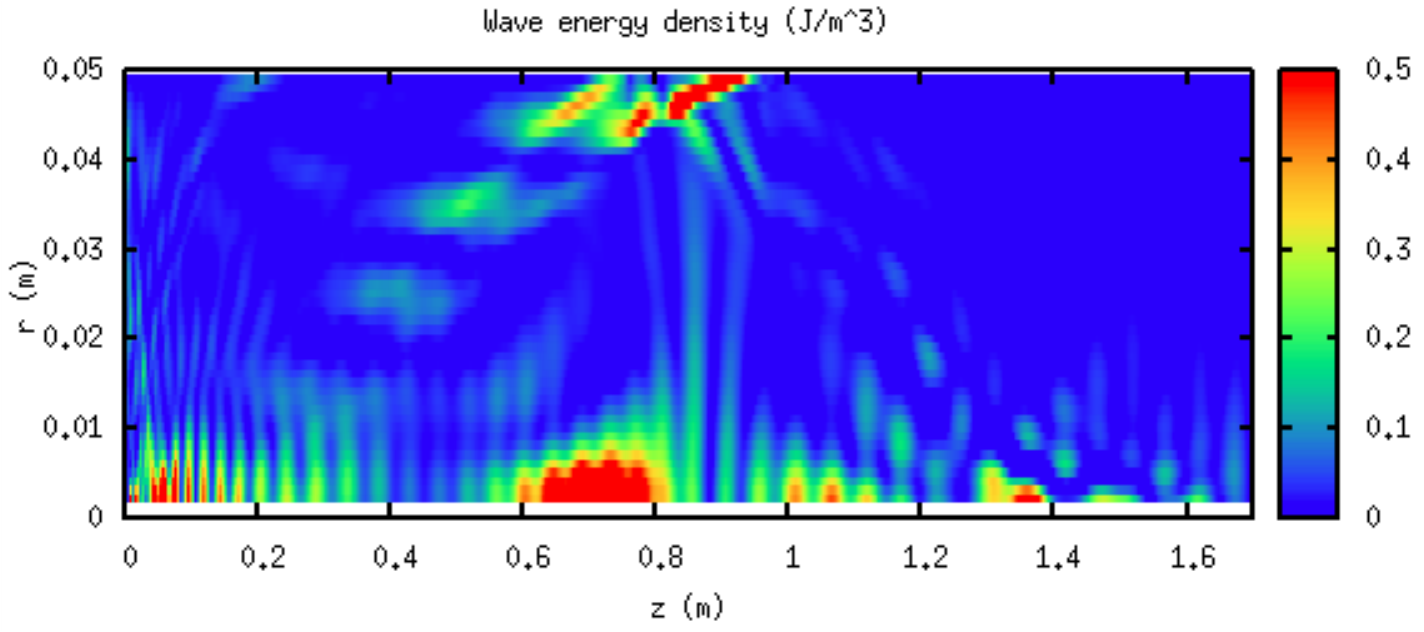}
\end{array}$
\end{center}
\caption{Axial profiles of magnetic wave field (on-axis) and contour plots of wave energy density in ($z$, $r$) space for a non-uniform plasma density: (a), log scale of $|B_r|_{\mathrm{rms}}$ for upstream $B_0(z)$; (b), wave energy density for upstream $B_0(z)$; (c), log scale of $|B_r|_{\mathrm{rms}}$ for downstream $B_0(z)$; (d) wave energy density for downstream $B_0(z)$.}
\label{fg3_9}
\end{figure*}

Observations have been made previously that the directionality of helicon wave propagations is dependent on the direction of static magnetic field in helicon discharges using helical antennas, but all in uniform field configurations.\cite{Chen:1996aa, Chen:1996ab, Sudit:1996aa, Lee:2011aa} In this section, we study the directionality in a non-uniform field configuration for a left-hand half turn helical antenna. Specifically, we have computed the wave amplitude and wave energy density in MAGPIE for the experimental and field reversed configurations. In MAGPIE, the field points from target to source. Figure~\ref{fg3_9} shows the computed axial profiles of wave amplitude on axis and 2D contour plots of wave energy density for both field direction pointing from target to source (Fig.~\ref{fg3_9}(a) and Fig.~\ref{fg3_9}(b)) and field direction pointing from source to target (Fig.~\ref{fg3_9} and Fig.~\ref{fg3_9}(d)). In this calculation, we have chosen $\nu_{\mathrm{eff}}=\nu_{ei}$ to see more details, and chosen the density profile to be linear with $B_0(z)$ in the axial direction and non-uniform in radius as measured in Fig.~\ref{fg3_1}(b). The field strength profile used here is shown in Fig.~\ref{fg3_1}(a). Inspection of Fig.~\ref{fg3_9} reveals that the wave energy is larger on the opposite side of the antenna, relative to the direction of the static magnetic field. This observation has been confirmed experimentally through finding that the plasma is brighter on the opposite side of the antenna relative to the direction of the applied external field. In summary, the dependence of the direction of helicon wave propagations to that of static magnetic field still exists even when the field configuration is non-uniform.

\section{Conclusions}\label{cnl3}
A RF field solver based on Maxwell's equations and a cold-plasma dielectric tensor is employed to describe the wave phenomena observed in a cylindrical non-uniform helicon discharge, MAGPIE. Here, the non-uniformity is both radial and axial: the plasma density is dependent on $r$ and $z$, the static magnetic field varies with $z$, and the electron temperature is a function of $r$. A linear fitting was conducted for radial profiles of plasma density and electron temperature, and the fitted profiles were utilised in wave field calculations. A linear relationship between the axial profile of plasma density and the static magnetic field was also assumed. Other conditions used in the simulation were taken from experiment directly, including filling gas (argon), antenna current of $38.8$~A, driving frequency of $13.56$~MHz, and a left hand half-turn helical antenna. 

With an enhancement factor of $9.5$ to the electron-ion Coulomb collision frequency $\nu_{ei}$ to approximate the observed attenuation, and a $12\%$ reduction in the antenna radius to match the amplitude of the wave field, the wave solver produced consistent wave fields compared to experimental data, including the axial and radial profiles of wave amplitude and phase. A local minimum in the axial profiles of wave amplitude was observed both experimentally and numerically, agreeing with previous studies.\cite{Guo:1999aa, Degeling:2004aa, Light:1995aa, Mori:2004aa} Mode structure of $m=1$ is consistent with the left hand half-turn helical antenna being used. A possible explanation for the enhanced electron-ion collision frequency has been offered through ion-acoustic turbulence, which can happen if the electron drift velocity exceeds the speed of sound in magnetised plasmas.\cite{Lee:2011aa} By calculating these two speeds based on MAGPIE conditions, we found that it is indeed satisfied in the source region of MAGPIE where the helicon plasmas and waves are produced. For the field strength under the driving antenna, the ratio of electron drift velocity and the speed of sound is around $2$. Other possible candidate explanation may also include kinetic effects which are neglected in the cold-plasma model employed here.

A numerical study on the effects of axial gradients in plasma density and static magnetic field on wave propagations was carried out. This showed that density increasing in proportion to the static magnetic field has little effect on RF absorption, while the density level near the antenna affects the wave amplitude significantly at all axial locations. The axial gradient in magnetic field increases the decay length of helicon waves in the target region. The relationship between plasma density, static magnetic field and axial wavelength is consistent with a simple theory developed previously.\cite{Chen:1996ab} 

A numerical scan of the enhancement factor to $\nu_{ei}$ reveals that with increased electron-ion collision frequency the wave amplitude is lowered and more focused near the antenna. This is mainly because of stronger edge heating at higher collision frequencies which prevent more energy transported from the antenna into the core plasma. The wave amplitude profile at $\nu_{\mathrm{eff}}=9.5\nu_{ei}$, which agrees with experimental data, shows consistent feature with a previous study that the RF energy is almost all absorbed in the near region of the antenna rather than in the far region.\cite{Chen:1996aa} We also studied the effect of the direction of static magnetic field on wave propagations, and found the antiparallel feature that waves propagate in the opposite direction of magnetic field for the antenna helicity existing in these experiments. This dependence of the direction of helicon wave propagations to that of static magnetic field in a non-uniform field configuration is consistent with previous observations made in uniform field configurations.\cite{Chen:1996aa, Chen:1996ab, Sudit:1996aa, Lee:2011aa} 

Physics questions raised by this work include: further explanation of exactly how axially non-uniform field might affect the radially localised helicon mode,\cite{Breizman:2000aa} inclusion of different $m$ numbers in the glass layer and any subsequent coupling to the plasma at the plasma-glass interface, and identification of independent first and second radial modes that superpose to yield a local minimum in wave field amplitude at $z=0.27$~m. Experimental measurements that might corroborate the wave field generation mechanism and associated physics include: the measurement of axial profile of density, and measurements of $n_e$, $\mathbf{B}$, and $\mathbf{E}$ with a reversed field. 

\chapter{Gap eigenmode of radially localised helicon waves in a periodic structure}\label{chp4}

The EMS (ElectroMagnetic Solver\cite{Chen:2006aa}) is employed to model a spectral gap and a gap eigenmode in a periodic structure in the whistler frequency range. A RLH (Radially Localised Helicon\cite{Breizman:2000aa}) mode is considered. We demonstrate that the computed gap frequency and gap width agree well with a theoretical analysis, and find a discrete eigenmode inside the gap by introducing a defect to the system's periodicity. The axial wavelength of the gap eigenmode is close to twice the system's periodicity, which is consistent with Bragg's law. Such an eigenmode could be excited by energetic electrons, similar to the excitation of Toroidal Alfv\'{e}n Eigenmodes (TAE) by energetic ions in tokamaks. Experimental identification of this mode is conceivable on the LAPD.\cite{Gekelman:1991aa}

\section{Introduction}\label{introduction}\label{int4}
It is a generic phenomenon that spectral gaps are formed when waves propagate in periodic media, and eigenmodes can exist with frequencies inside the spectral gaps if a defect is introduced to break the system's perfect translational symmetry, and create an effective potential well to localise these waves.\cite{Strutt:1887aa, Anderson:1958aa, Mott:1968aa, John:1987aa, Yablonovitch:1991aa, Figotin:1997aa} Fusion plasmas have a few periodicities that can produce spectral gaps: geodesic curvature of field lines,\cite{Chu:1992aa} elongation\cite{DIppolito:1980aa, Dewar:1974aa} or triangularity of flux surfaces,\cite{Betti:1992aa} helicity\cite{Nakajima:1992aa} and periodic mirroring in stellerators.\cite{Kolesnichenko:2001aa} Weakly damped eigenmodes which are readily destabilised by energetic ions often reside in these gaps, and they may degrade fast ion confinement.\cite{Duong:1993aa, White:1995aa} The most extensively studied gap eigenmode is TAE,\cite{Cheng:1985aa} however, there are also numerous other modes with similar features.\cite{Wong:1999aa, Coppi:1986aa, Gorelenkov:1995aa, Kamelander:1996aa}

Zhang et. al.\cite{Zhang:2008aa} observed a spectral gap in the shear Alfv\'{e}n wave continuum in experiments on the LAPD with a multiple magnetic mirror array, and obtained consistent results through two-dimensional numerical modelling using the finite difference code, EMS.\cite{Chen:2006aa} Although eigenmodes inside this gap were not formed, a possible experimental implementation was proposed to detect them. The idea is to use the endplate of the machine as a ``defect" that breaks the axial periodicity or to vary current in one of the independently powered magnetic coils.

In this chapter, we will use EMS to examine a gap eigenmode in a linear system with slightly broken axial periodicity. We will consider the RLH mode, whose radial structure has been described in \cite{Breizman:2000aa}. Beyond \cite{Breizman:2000aa}, which deals with uniform magnetic field, the spectral gap and gap eigenmode of the RLH mode will be theoretically and numerically studied in a rippled magnetic field in the present study, together with those for shear Alfv\'{e}n waves in Chapter~\ref{chp5}. We will show that the computed gap frequency and gap width agree well with the theoretical analysis, and that there is a discrete eigenmode inside the gap. Such an eigenmode could be excited by energetic electrons, similarly to the excitation of TAE by energetic ions in tokamaks, and possibly observed on a linear device, e. g. LAPD.

\section{Theoretical analysis}\label{thy4}
\subsection{Basic equations}\label{eqt4}
The spatial structure of RLH wave field and the corresponding dispersion relation can be found from Eq.~(\ref{eq1_52}) that applies to a whistler-type linear wave in a cold plasma cylinder.\cite{Breizman:2000aa} For a cylindrical coordinate system of ($r$, $\varphi$, $z$) with $\varphi$ the azimuthal angle and wave form of $\exp[i(m\varphi+k_z z-\omega t)]$, it has the form
\begin{equation}\label{eq4_1}
\frac{1}{r}\frac{\partial}{\partial r}r \frac{\partial E}{\partial r}-\frac{m^2}{r^2}E=\frac{m}{k_z^2 r}\frac{\omega^2}{c^2}\frac{E\partial D/\partial r}{1-(m\partial D/\partial r)/k_z^2 r P},
\end{equation}
where $E=E_z-(k_z r/m)E_\varphi$ with $E_z$ and $E_\varphi$ the axial and azimuthal components of the wave electric field, respectively, $D=-\omega_{p}^2/\omega \omega_{c}$, and $P=-\omega_{p}^2/\omega^2$. Here, $\omega_p$ and $\omega_c$ are the electron plasma frequency and electron cyclotron frequency, respectively. In this section, we will limit our consideration to the case of sufficiently dense plasma in which $\omega_{p}^2\gg c^2m/a^2$, where $a$ is the plasma radius, and nonzero values of the azimuthal mode number $m$ as the radial nonuniformity of plasma density has a surprisingly strong effect on the structure of helicon modes with $m\neq 0$.\cite{Breizman:2000aa} The resulting eigenfrequency for the mode of interest scales roughly as\cite{Breizman:2000aa}
\begin{equation}\label{eq4_2}
\omega\sim\omega_{c}\frac{k_z^2 c^2}{\omega_{p}^2},
\end{equation}
which shows that the term $-(m\partial D/\partial r)/k_z^2 r P$ in the denominator of Eq.~(\ref{eq4_1}) can be estimated as
\begin{equation}\label{eq4_3}
-(m\partial D/\partial r)/k_z^2 r P\sim m\frac{c^2}{a^2 \omega_{p}^2}\ll 1. 
\end{equation}
It was also shown in \cite{Breizman:2000aa} (see Eq. (16) there) that
\begin{equation}\label{eq4_4}
\frac{E_z}{E}=\frac{-(m\partial D/\partial r)/k_z^2 r P}{1-(m\partial D/\partial r)/k_z^2 r P}
\end{equation}
which is thus much less than unity for the mode of interest. The underlying reason is that the electron conductivity along the guiding magnetic field is much greater than the cross-field Hall conductivity. It is therefore allowable to set $E_z=0$ and ignore the term $-(m\partial D/\partial r)/k_z^2 r P$ in the denominator of Eq.~(\ref{eq4_1}), which gives
\begin{equation}\label{eq4_5}
k_z^2\left(\frac{\partial}{\partial r}r\frac{\partial}{\partial r}r E_\varphi-m^2 E_\varphi\right)=E_\varphi m \frac{\omega^2}{c^2}r\frac{\partial D}{\partial r}.
\end{equation}
We now generalise this equation to the case of slightly modulated ($z$-dependent) plasma equilibrium with a following separable form of $\omega_{p}^2/\omega_{c}$
\begin{equation}\label{eq4_6}
\frac{\omega_{p}^2}{\omega_{c}}=\frac{\omega_{p0}^2(r)}{\omega_{c0}}[1-\epsilon(z)\cos( q z)]. 
\end{equation}
Here, $\epsilon\ll 1$ and $q$ are the modulation envelope and wavenumber respectively. The axial scale-length of the envelope $\epsilon(z)$ is assumed to be much greater than $1/q$. We recall that $k_z^2=-\partial^2/\partial z^2$ and transform Eq.~(\ref{eq4_5}) to 
\begin{equation}\label{eq4_7}
\frac{\partial^2}{\partial z^2}\left(\frac{\partial}{\partial r}r\frac{\partial}{\partial r}r E_\varphi-m^2 E_\varphi\right)=E_\varphi m \frac{\omega}{c^2}[1-\epsilon(z)\cos( q z)]r\frac{\partial}{\partial r}\left(\frac{\omega_{p0}^2}{\omega_{c0}}\right).
\end{equation}
The modulated equilibrium introduces resonant coupling between the modes with $k_z=q/2$ and $k_z=-q/2$, which suggests the following form for $E_\varphi$
\begin{equation}\label{eq4_8}
E_\varphi=A_+e^{iqz/2}+A_-e^{-iqz/2}, 
\end{equation}
where $A_+$ and $A_-$ are slow functions of $z$ compared to $\cos(qz)$. Let $\omega_0$ be an eigenfrequency of Eq.~(\ref{eq4_7}) for $\epsilon=0$ and $|k_z|=q/2$. It is then straightforward to separate spatial scales in Eq.~(\ref{eq4_7}) and obtain a set of coupled equations for $A_+$ and $A_-$:
\begin{equation}\label{eq4_9}
\begin{array}{c}
\vspace{0.3cm}-\frac{q^2}{4}\left(\frac{\partial}{\partial r}r\frac{\partial}{\partial r}r A_+-m^2 A_+\right)-A_+m\frac{\omega_0}{c^2}r\frac{\partial}{\partial r}\left(\frac{\omega_{p0}^2}{\omega_{c0}}\right)=\\
-i q\frac{\partial}{\partial z}(\frac{\partial}{\partial r}r \frac{\partial}{\partial r}r A_+-m^2 A_+)+A_+ m \frac{\omega-\omega_0}{c^2}r\frac{\partial}{\partial r}\left(\frac{\omega_{p0}^2}{\omega_{c0}}\right)-\frac{1}{2}A_-m \frac{\omega_0}{c^2}\epsilon(z)r\frac{\partial}{\partial r}\left(\frac{\omega_{p0}^2}{\omega_{c0}}\right),
\end{array}
\end{equation}
\begin{equation}\label{eq4_10}
\begin{array}{c}
\hspace{-0.16cm}\vspace{0.3cm}-\frac{q^2}{4}\left(\frac{\partial}{\partial r}r\frac{\partial}{\partial r}r A_--m^2 A_-\right)-A_- m\frac{\omega_0}{c^2}r\frac{\partial}{\partial r}\left(\frac{\omega_{p0}^2}{\omega_{c0}}\right)=\\
\hspace{-0.16cm}+i q\frac{\partial}{\partial z}\left(\frac{\partial}{\partial r}r \frac{\partial}{\partial r}r A_--m^2 A_-\right)+A_- m \frac{\omega-\omega_0}{c^2}r\frac{\partial}{\partial r}\left(\frac{\omega_{p0}^2}{\omega_{c0}}\right)-\frac{1}{2}A_+ m \frac{\omega_0}{c^2}\epsilon(z)r\frac{\partial}{\partial r}\left(\frac{\omega_{p0}^2}{\omega_{c0}}\right).
\end{array}
\end{equation}
We have neglected second axial derivatives of $A_+$ and $A_-$ and arranged Eq.~(\ref{eq4_9}) and Eq.~(\ref{eq4_10}) so that their right hand sides represent relatively small terms compared to the dominant terms on the left hand sides. With this ordering, we conclude that the radial dependencies of $A_+$ and $A_-$ need to be close to the eigenfunction $\Psi(r)$ of the lowest order ODE
\begin{equation}\label{eq4_11}
\frac{q^2}{4}\left(\frac{\partial}{\partial r}r\frac{\partial}{\partial r}r \Psi-m^2 \Psi\right)+\Psi m\frac{\omega_0}{c^2}r\frac{\partial}{\partial r}\left(\frac{\omega_{p0}^2}{\omega_{c0}}\right)=0.
\end{equation}
The differential operator in this equation is self-adjoint. As a result, multiplication of Eq.~(\ref{eq4_9}) and Eq.~(\ref{eq4_10}) by $r\Psi$ and integration over radius lead to:
\begin{equation}\label{eq4_12}
\begin{array}{c}
\vspace{0.3cm}0=-i q\frac{\partial}{\partial z}\int r\Psi\left(\frac{\partial}{\partial r}r \frac{\partial}{\partial r}r A_+-m^2 A_+\right)dr+\int r\Psi A_+ m \frac{\omega-\omega_0}{c^2}r\frac{\partial}{\partial r}\left(\frac{\omega_{p0}^2}{\omega_{c0}}\right)dr\\
-\frac{1}{2}\int r\Psi A_-m \frac{\omega_0}{c^2}\epsilon(z)r\frac{\partial}{\partial r}\left(\frac{\omega_{p0}^2}{\omega_{c0}}\right)dr,
\end{array}
\end{equation}
\begin{equation}\label{eq4_13}
\begin{array}{c}
\vspace{0.3cm}0=i q\frac{\partial}{\partial z}\int r\Psi\left(\frac{\partial}{\partial r}r \frac{\partial}{\partial r}r A_--m^2 A_-\right)dr+\int r\Psi A_- m \frac{\omega-\omega_0}{c^2}r\frac{\partial}{\partial r}\left(\frac{\omega_{p0}^2}{\omega_{c0}}\right)dr\\
-\frac{1}{2}\int r\Psi A_+ m \frac{\omega_0}{c^2}\epsilon(z)r\frac{\partial}{\partial r}\left(\frac{\omega_{p0}^2}{\omega_{c0}}\right)dr.
\end{array}
\end{equation}
Having eliminated the lowest order terms, we now set:
\begin{equation}\label{eq4_14}
A_+=F(z)\Psi(r),~A_-=G(z)\Psi(r)
\end{equation}
in Eq.~(\ref{eq4_12}) and Eq.~(\ref{eq4_13}) to obtain the following set of coupled equations for $F(z)$ and $G(z)$: 
\begin{equation}\label{eq4_15}
0=\frac{4 i}{q}\frac{\partial F}{\partial z}+F\frac{\omega-\omega_0}{\omega_0}-\frac{G}{2}\epsilon(z), 
\end{equation}
\begin{equation}\label{eq4_16}
0=-\frac{4 i}{q}\frac{\partial G}{\partial z}+G\frac{\omega-\omega_0}{\omega_0}-\frac{F}{2}\epsilon(z). 
\end{equation}

\subsection{Spectral gap and continuum}\label{spc4}
In the case of $z$-independent $\epsilon$, Eq.~(\ref{eq4_15}) and Eq.~(\ref{eq4_16}) have exponential solutions
\begin{equation}\label{eq4_17}
F\propto G\propto e^{i\kappa z}
\end{equation}
with $\kappa$ the wave number. The corresponding dispersion relation has two roots: 
\begin{equation}\label{eq4_18}
\begin{array}{l}
\hspace{-0.1cm}\vspace{0.3cm}\omega_+(\kappa)=\omega_0[1+\sqrt{(4\kappa/q)^2+(\epsilon/2)^2}],\\
\hspace{-0.1cm}\omega_-(\kappa)=\omega_0[1-\sqrt{(4\kappa/q)^2+(\epsilon/2)^2}].
\end{array}
\end{equation}
Here, $\omega_+(\kappa)$ and $\omega_-(\kappa)$ are the continuum frequencies above and below the spectral gap, respectively. The upper and lower tips of the spectral gap correspond to $\kappa=0$:
\begin{equation}\label{eq4_19}
\begin{array}{l}
\hspace{-0.1cm}\vspace{0.3cm}\omega_+(0)=\omega_0\left(1+\frac{\epsilon}{2}\right),\\
\hspace{-0.1cm}\omega_-(0)=\omega_0\left(1-\frac{\epsilon}{2}\right).
\end{array}
\end{equation}
The normalised width of the gap, $\Delta\omega\equiv[\omega_+(0)-\omega_-(0)]/\omega_0$, is therefore equal to the modulation amplitude $\epsilon$. The central frequency of the gap $\omega_0$ needs to be found from Eq.~(\ref{eq4_11}) as the RLH eigenfrequency for an axially uniform plasma cylinder and it can be written as
\begin{equation}\label{eq4_20}
\omega_0=\Gamma \frac{\omega_{c0}(0)c^2 q^2}{4\omega_{p0}(0)^2},
\end{equation} 
where $\omega_{c0}(0)$ and $\omega_{p0}(0)$ are the on-axis values and the numerical form-factor $\Gamma$ is determined by the plasma radial profile. 

\subsection{Wall-localised eigenmodes}\label{wll4}
\begin{figure*}[ht]
\begin{center}$
\begin{array}{ll}
(a)&(b)\\
\includegraphics[width=0.47\textwidth,angle=0]{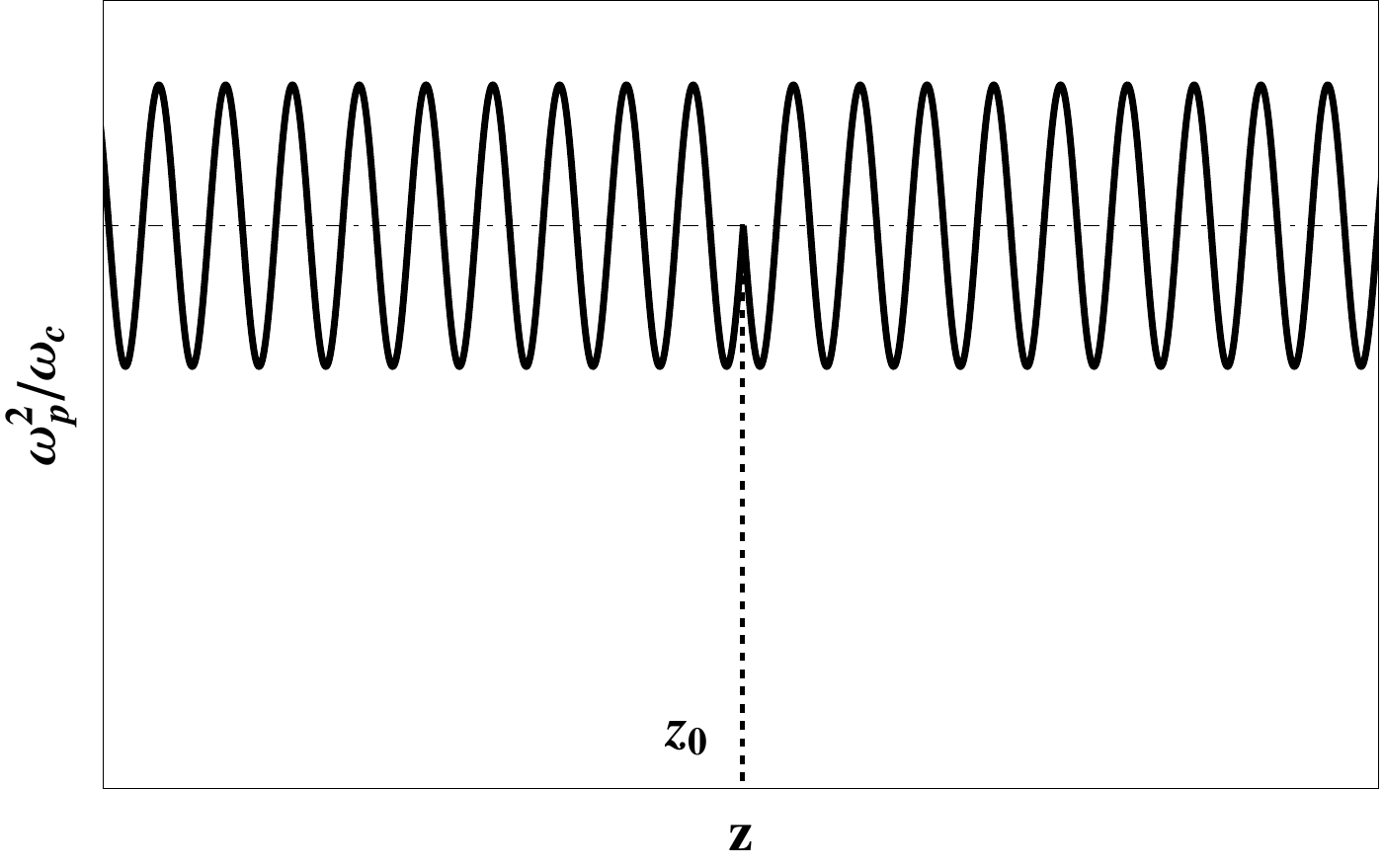}&\includegraphics[width=0.47\textwidth,angle=0]{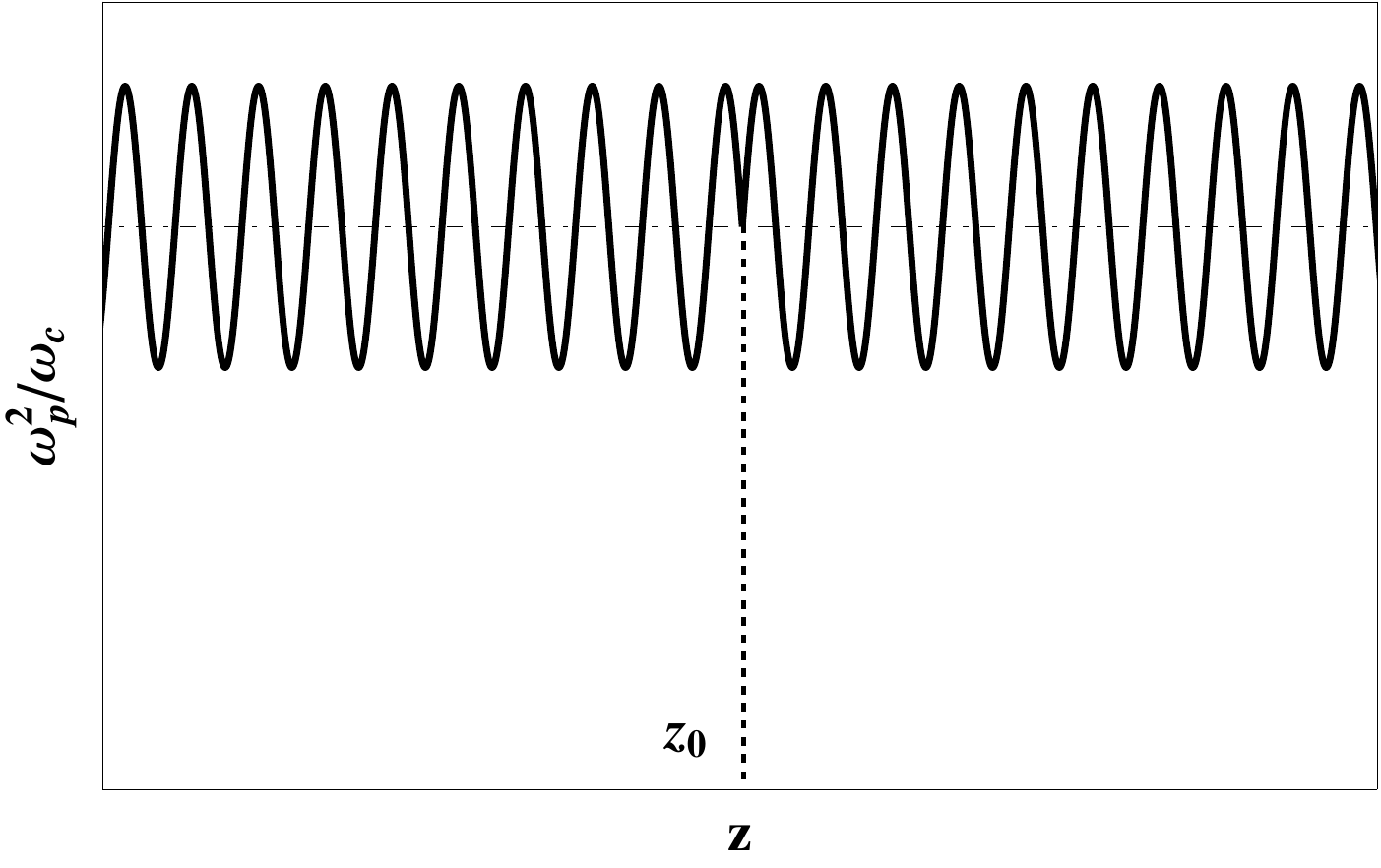}
\end{array}$
\end{center}
\caption{Axial profiles of $\omega_{p}^2/\omega_{c}$ with defects: (a) defect location at $\sin(q z_0)=1$, (b) defect location at $\sin(q z_0)=-1$.}
\label{fg4_1}
\end{figure*}

A discrete-spectrum eigenmode can be created inside the spectral gap if the system's periodicity is broken. We illustrate that by considering an ideally conducting endplate located at $z=z_0$, so that the plasma now occupies only a half cylinder to the right of the endplate ($z>z_0$). The boundary condition at the endplate is
\begin{equation}\label{eq4_21}
E_\varphi(r, z_0)=0
\end{equation}
or, equivalently, 
\begin{equation}\label{eq4_22}
F(z_0)e^{i q z_0/2}+G(z_0)e^{-i q z_0/2}=0
\end{equation}
(see Eq.~(\ref{eq4_8}) and Eq.~(\ref{eq4_14})). The electric field of the discrete-spectrum mode must also vanish at $z\rightarrow\infty$, i. e., 
\begin{equation}\label{eq4_23}
F(\infty)=G(\infty)=0. 
\end{equation}
Equation~(\ref{eq4_15}) and Eq.~(\ref{eq4_16}) (with $\epsilon=const$) admit an exponential solution
\begin{equation}\label{eq4_24}
F\propto G\propto e^{-\lambda z}
\end{equation}
with a positive value of $\lambda$ that satisfies boundary conditions in Eq.~(\ref{eq4_22}) and Eq.~(\ref{eq4_23}). Indeed, the exponential ansatz, Eq.~(\ref{eq4_24}), reduces Eq.~(\ref{eq4_15}), Eq.~(\ref{eq4_16}) and Eq.~(\ref{eq4_22}) to an algebraic set:
\begin{equation}\label{eq4_25}
0=-\frac{4 i\lambda}{q}F+F\frac{\omega-\omega_0}{\omega_0}-\frac{G}{2}\epsilon, 
\end{equation}
\begin{equation}\label{eq4_26}
0=\frac{4 i\lambda}{q}G+G\frac{\omega-\omega_0}{\omega_0}-\frac{F}{2}\epsilon,
\end{equation}
\begin{equation}\label{eq4_27}
F+G e^{-i q z_0}=0.
\end{equation}
The solvability conditions for this set determine $\omega$ and $\lambda$ as functions of $z_0$ as follows: 
\begin{equation}\label{eq4_28}
\frac{\omega-\omega_0}{\omega_0}=-\frac{\epsilon}{2}\cos(q z_0),
\end{equation}
\begin{equation}\label{eq4_29}
\lambda=\frac{\epsilon}{8}q\sin(q z_0). 
\end{equation}
We observe that $\lambda$ is positive for $0<q z_0<\pi$, and the wave field therefore vanishes at infinity in this case to represent a discrete eigenmode whose frequency is automatically inside the gap. In particular, the eigenfrequency is exactly at the gap centre when $q z_0=\pi/2$. We further note that this wall-localised solution can also be viewed as an odd-parity eigenmode ($E_\varphi(z)=-E_\varphi(2 z_0-z)$) in the entire periodic cylinder with a defect at $z=z_0$. An axial profile of $\omega_{p}^2/\omega_{c}$ with such defect is shown in Fig.~\ref{fg4_1}(a). 

\subsection{Even-parity eigenmode}\label{evn4}
In contrast with Fig.~\ref{fg4_1}(a), Fig.~\ref{fg4_1}(b) illustrates a defect with $\pi<q z_0<2\pi$ that does not produce an odd-parity mode. Instead, an even-parity mode exists in this case. This mode is still described by Eq.~(\ref{eq4_15}) and Eq.~(\ref{eq4_16}), but the boundary condition at $z=z_0$ now changes from Eq.~(\ref{eq4_21}) to
\begin{equation}\label{eq4_30}
\left[\frac{\partial E_\varphi (r, z)}{\partial z}\right]_{z=z_0}=0,
\end{equation}
or, equivalently,
\begin{equation}\label{eq4_31}
F-Ge^{-i q z_0}=0.
\end{equation}
The solvability conditions for Eq.~(\ref{eq4_25}), Eq.~(\ref{eq4_26}) and Eq.~(\ref{eq4_31}) now give:
\begin{equation}\label{eq4_32}
\frac{\omega-\omega_0}{\omega_0}=\frac{\epsilon}{2}\cos(q z_0),
\end{equation}
\begin{equation}\label{eq4_33}
\lambda=-\frac{\epsilon}{8}q\sin(q z_0),
\end{equation}
and we observe that $\lambda$ is positive (the wave field vanishes at infinity) when $\sin(q z_0)<0$, i. e. this even-parity eigenmode indeed requires $\pi<q z_0<2\pi$ for its existence. 

\subsection{Schr\"{o}dinger equation for gap eigenmodes}\label{sch4}
Equation~(\ref{eq4_15}) and Eq.~(\ref{eq4_16}) can be straightforwardly transformed into two independent second-order equations for $F-G$ and $F+G$: 
\begin{equation}\label{eq4_34}
\frac{16}{q^2}\frac{\partial}{\partial z}\frac{1}{\left(\frac{\omega-\omega_0}{\omega_0}-\frac{\epsilon}{2}\right)}\frac{\partial (F-G)}{\partial z}+(F-G)\left(\frac{\omega-\omega_0}{\omega_0}+\frac{\epsilon}{2}\right)=0,
\end{equation}
\begin{equation}\label{eq4_35}
\frac{16}{q^2}\frac{\partial}{\partial z}\frac{1}{\left(\frac{\omega-\omega_0}{\omega_0}+\frac{\epsilon}{2}\right)}\frac{\partial (F+G)}{\partial z}+(F+G)\left(\frac{\omega-\omega_0}{\omega_0}-\frac{\epsilon}{2}\right)=0.
\end{equation}
Both of them can be further reduced to a time-independent Schr\"{o}dinger equation when $\epsilon$ is nearly constant. We assume $\epsilon=\langle\epsilon\rangle+u(z)$ with $u(z)\ll\langle\epsilon\rangle$ and $u(\pm\infty)=0$. The discrete-spectrum modes in this case are very close to the tips of the gap. For the lower tip, we have
\begin{equation}\label{eq4_36}
\frac{\omega-\omega_0}{\omega_0}=-\frac{\langle\epsilon\rangle}{2}+\delta_-
\end{equation}
with $|\delta_-|\ll\langle\epsilon\rangle$. We can then neglect $u(z)$ and $\delta_-$ in the derivative term of Eq.~(\ref{eq4_34}) and get
\begin{equation}\label{eq4_37}
-\frac{\partial^2 (F-G)}{\partial z^2}+\frac{q^2\langle\epsilon\rangle}{16}(F-G)\left[\delta_-+\frac{u(z)}{2}\right]=0. 
\end{equation}
Similar procedure for the upper tip gives
\begin{equation}\label{eq4_38}
-\frac{\partial^2 (F+G)}{\partial z^2}+\frac{q^2\langle\epsilon\rangle}{16}(F+G)\left[-\delta_++\frac{u(z)}{2}\right]=0. 
\end{equation}
We note that any negative function $u(z)$ acts as a potential well that supports spatially localised eigenmodes of Eq.~(\ref{eq4_37}) and Eq.~(\ref{eq4_38}) with $\delta_->0$ and $\delta_+<0$, respectively. The eigenfrequencies of these modes belong to the spectral gap in the continuum. 

\section{Numerical results and discussion}\label{rst4}
\begin{figure*}[h]
\begin{center}
\includegraphics[width=0.8\textwidth,angle=0]{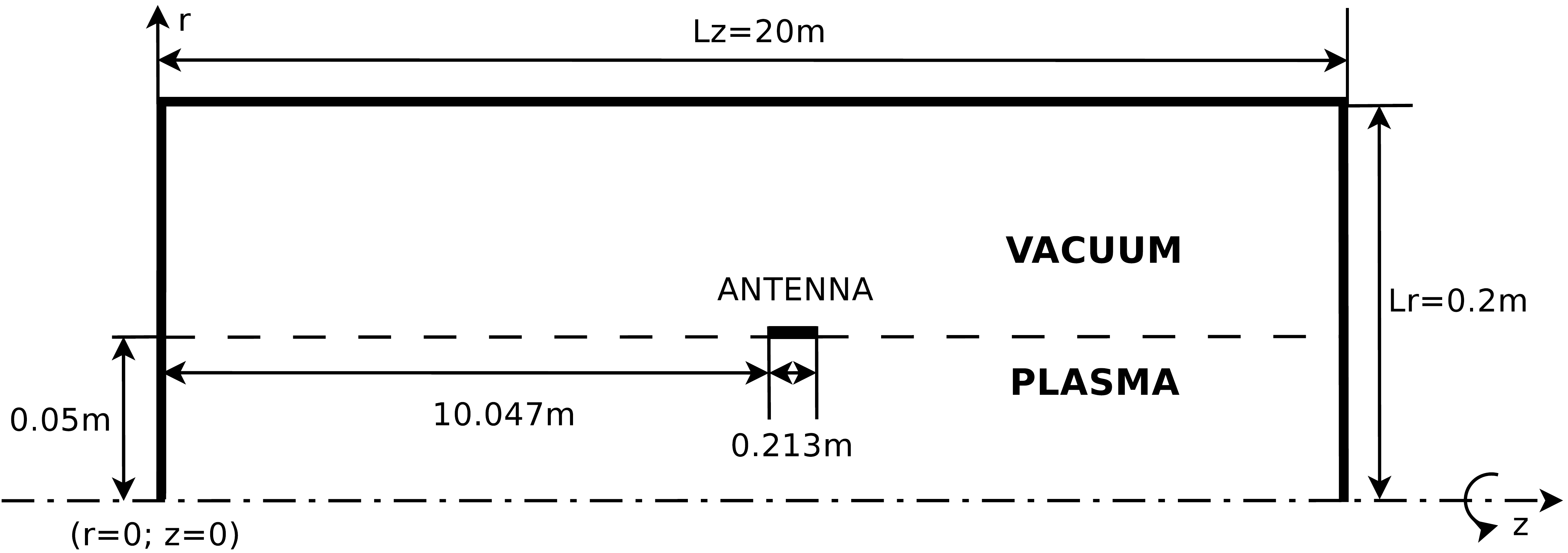}
\end{center}
\caption{Computational domain. The solid bar denotes a half-turn helical antenna. The dot-dashed line is the machine and coordinate system axis ($r=0$). The coordinate system ($r$, $\varphi$, $z$) is right-handed with an azimuthal angle $\varphi$.}
\label{fg4_2}
\end{figure*}

The EMS,\cite{Chen:2006aa} as introduced in Sec.~\ref{ems3}, is employed to study the RLH spectral gap and gap eigenmode inside. The computational domain is shown in Fig.~\ref{fg4_2}. A helical antenna is employed to excite an $m=1$ mode in the plasma. The enclosing chamber is assumed to be ideally conducting so that the boundary conditions are the same to those in Eq.~(\ref{eq3_7}). 

\subsection{RLH mode in a straight cylinder}\label{unf4}
\begin{figure}[ht]
\begin{center}$
\begin{array}{ll}
(a)&(b)\\
\includegraphics[width=0.45\textwidth,angle=0]{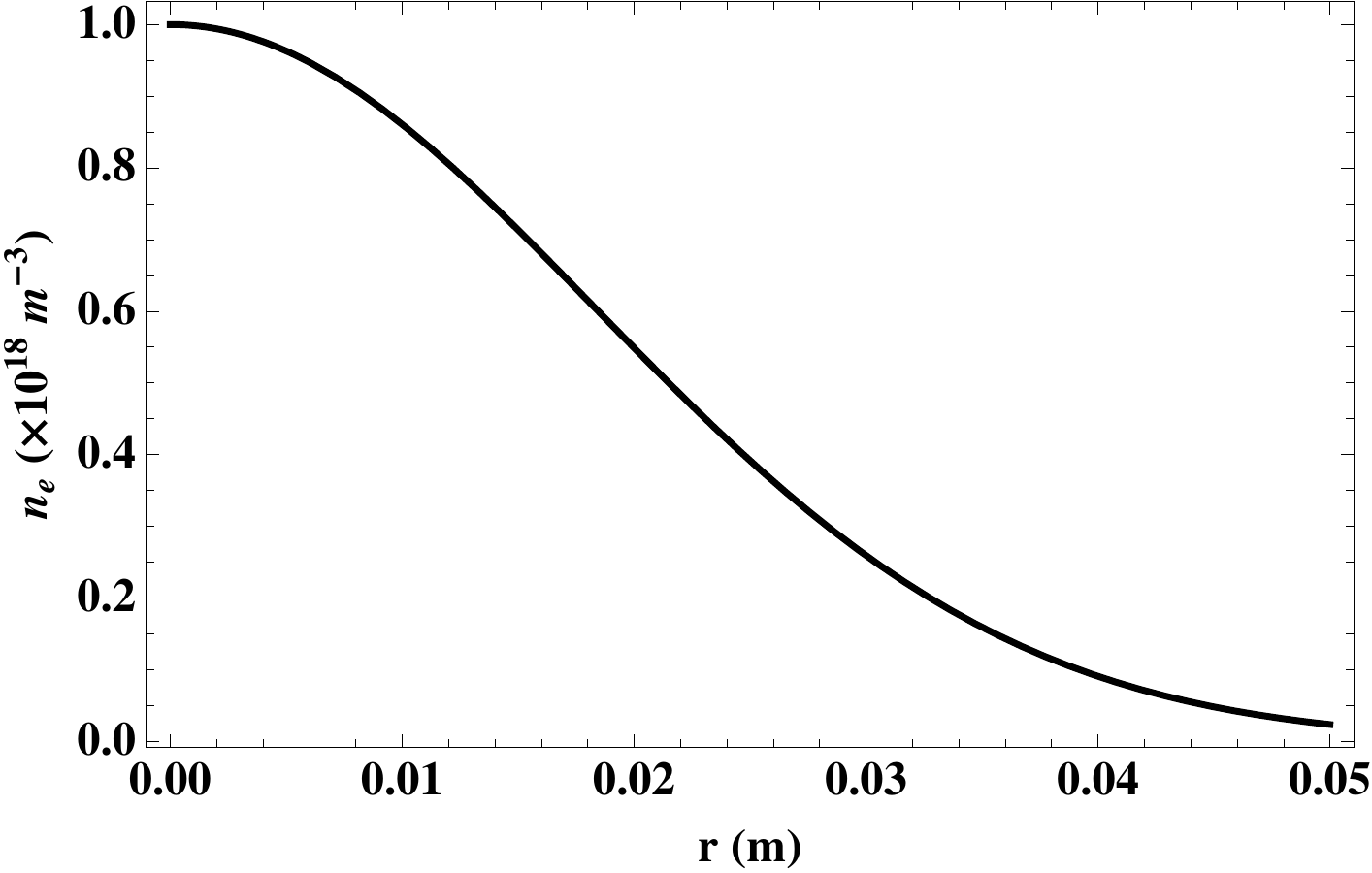}&\hspace{0 cm}\includegraphics[width=0.46\textwidth,angle=0]{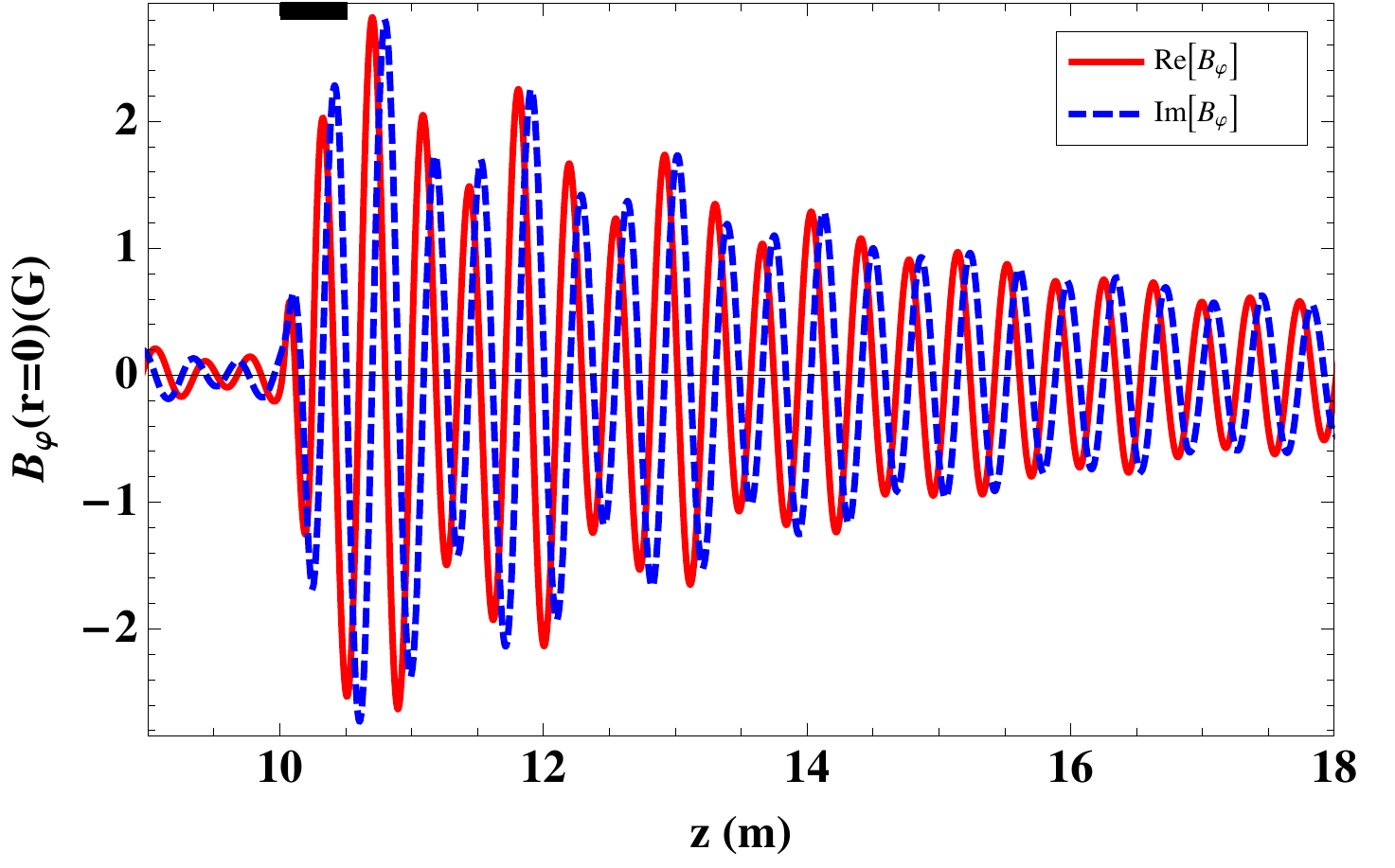}\\
(c)&(d)\\
\includegraphics[width=0.45\textwidth,angle=0]{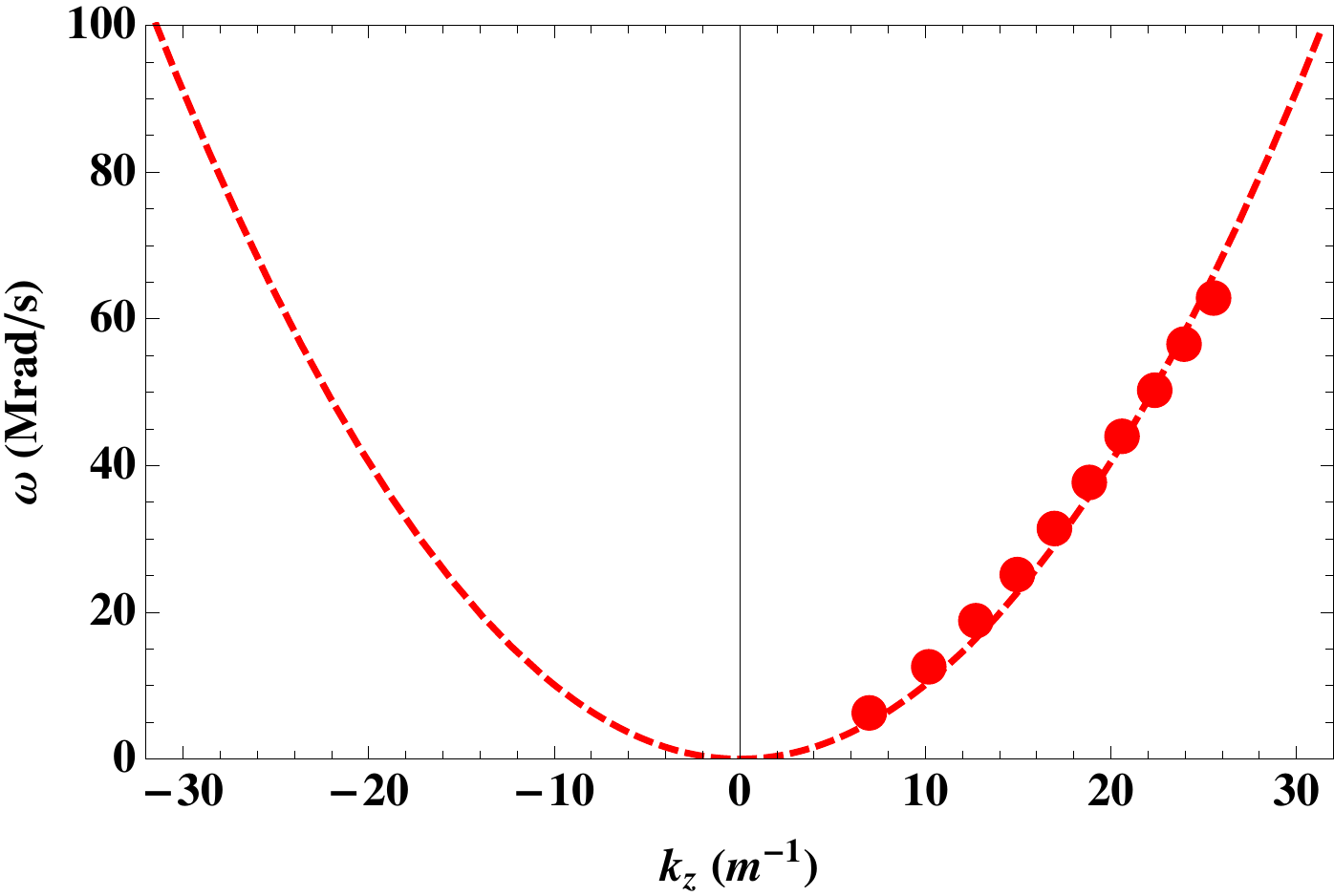}&\hspace{-0.14cm}\includegraphics[width=0.46\textwidth,height=0.295\textwidth,angle=0]{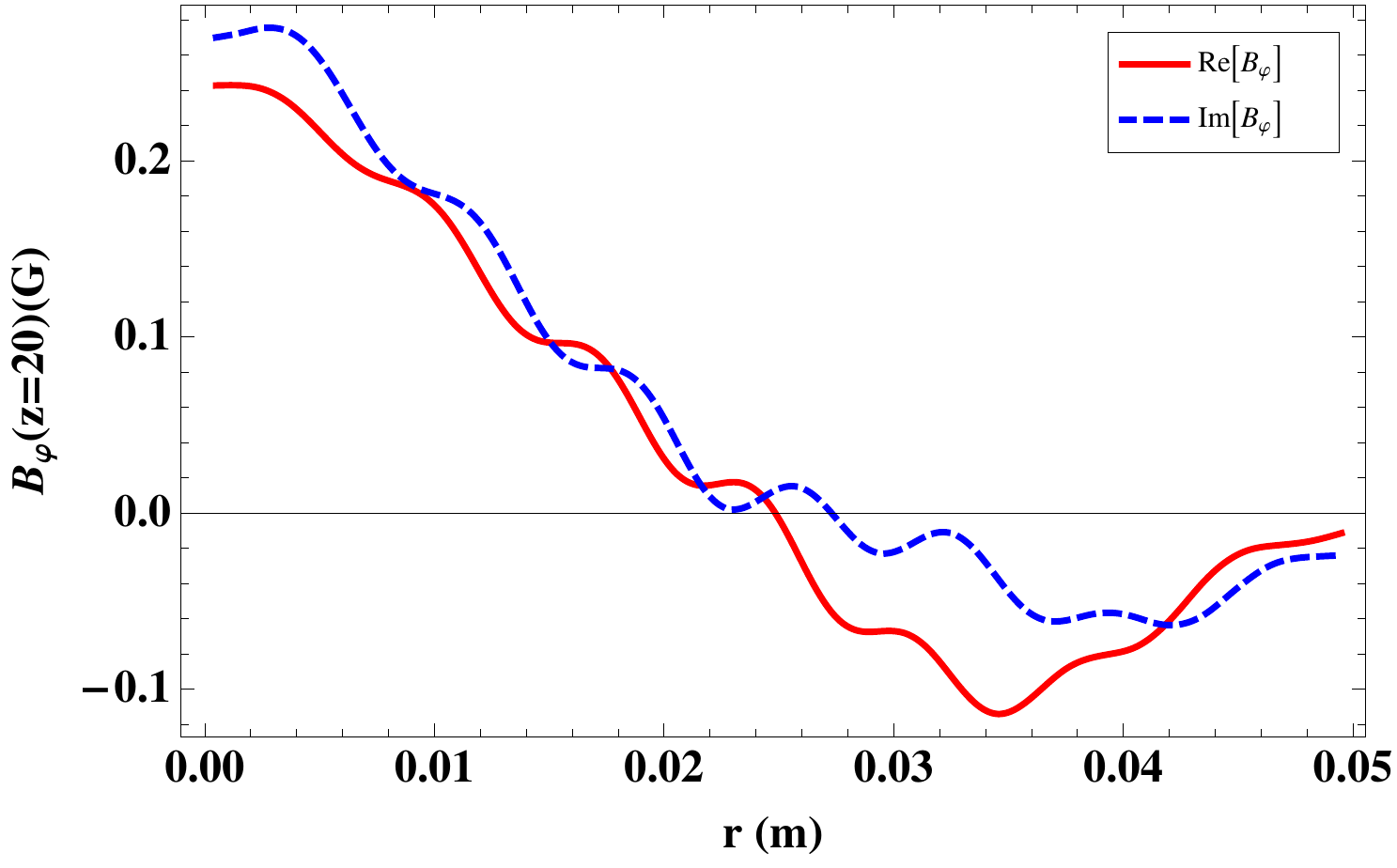}
\end{array}$
\end{center}
\caption{Plasma density profile and RLH wave field in a straight cylinder: (a) radial profile of unperturbed plasma density, (b) azimuthal magnetic field of the $m=1$ mode for an illustrative frequency $\omega=31.4$~Mrad/s (solid bar shows antenna location, and the endplates are located at $z=0$~m and $z=30$~m respectively), (c) computed dispersion relation for the $m=1$ mode (dots) and analytical scaling from Eq.~(\ref{eq4_20}) with $q/2=k_z$ and $\Gamma=-2.03$ (dashed curve), (d) radial profile of the $m=1$ mode at  $z = 20$~m for $\omega=31.4$~Mrad/s (the profile shows some modulations due to coupling to other weaker modes produced by the antenna). }
\label{fg4_3}
\end{figure}

We first use EMS to calculate the radial structure of the RLH mode in a straight cylinder with a uniform static magnetic field and recover the mode dispersion relation. We consider a single-ionised argon plasma with a radial density profile shown in Fig.~\ref{fg4_3}(a).
The surrounding vacuum chamber and the position of the rf-antenna are shown in Fig.~\ref{fg4_2}. The static magnetic field in this simulation is $B_0=0.01$~T and the electron-ion collision frequency on axis is $\nu_{ei}|_{r=0}=4\times10^6~\rm{s}^{-1}$. We run EMS for various frequencies of the antenna current, select the $m=1$ azimuthal component of the plasma response and calculate the dominant axial wavenumber, $k_z$, of this component via Fourier decomposition. We observe an apparently preferred axial coupling direction for the $m=1$ wave generated by the helical antenna. The collision frequency is sufficient to damp the radiated waves significantly towards the endplates but the dominant wavelength is still seen clearly in the plot of the excited field (Fig.~\ref{fg4_3}(b)). The resulting relation between $k_z$ and $\omega$ is shown in Fig.~\ref{fg4_3}(c). We note that $k_z$ scales as $\sqrt{\omega}$ with frequency, in agreement with analytical scaling obtained in \cite{Breizman:2000aa}, namely $\omega\sim\omega_{c} k_z^2 c^2/\omega_{p}^2$. A least squares fit to the dispersion relation given by Eq.~(\ref{eq4_20}) with $q/2=k_z$ shows that the form factor $\Gamma$ equals $-2.03$ for the selected density profile. The radial profile of this $m=1$ mode is presented in Fig.~\ref{fg4_3}(d). 

\subsection{Spectral gap in a periodic structure}\label{gap4}
\begin{figure*}[ht]
\begin{center}$
\vspace{-0.3cm}
\begin{array}{l}
(a)\\
\vspace{-0.3cm}\includegraphics[width=0.655\textwidth,angle=0]{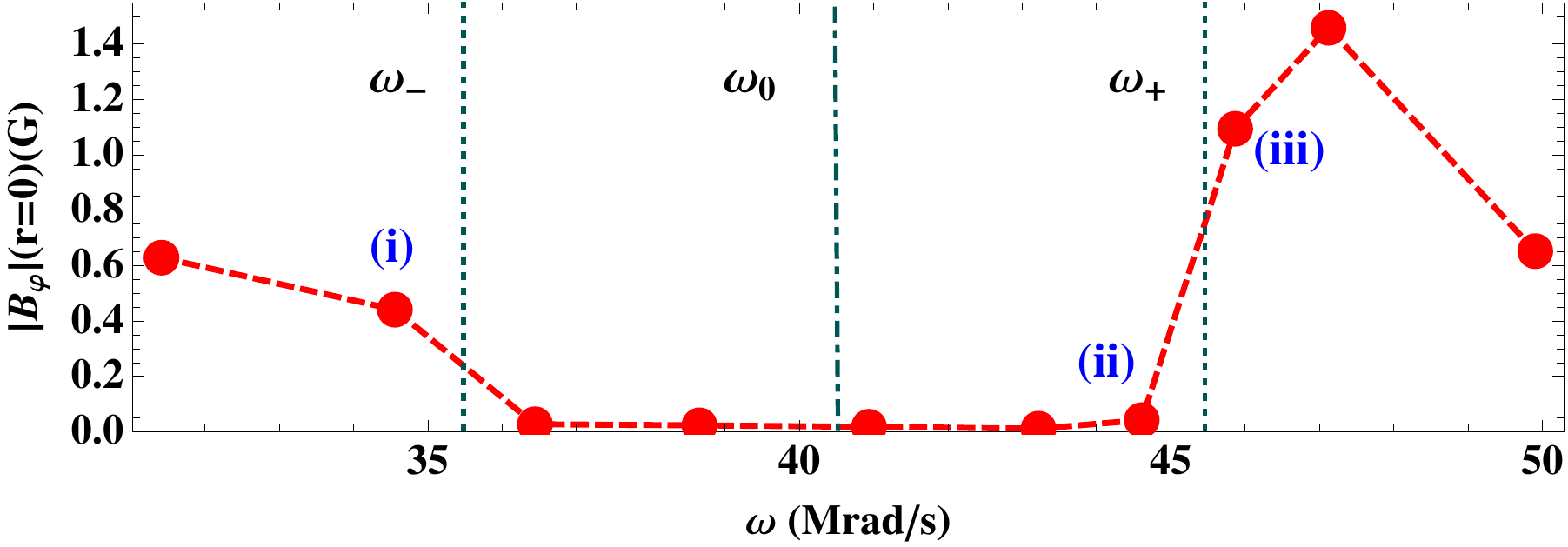}\\
(b)\\
\includegraphics[width=0.7\textwidth,angle=0]{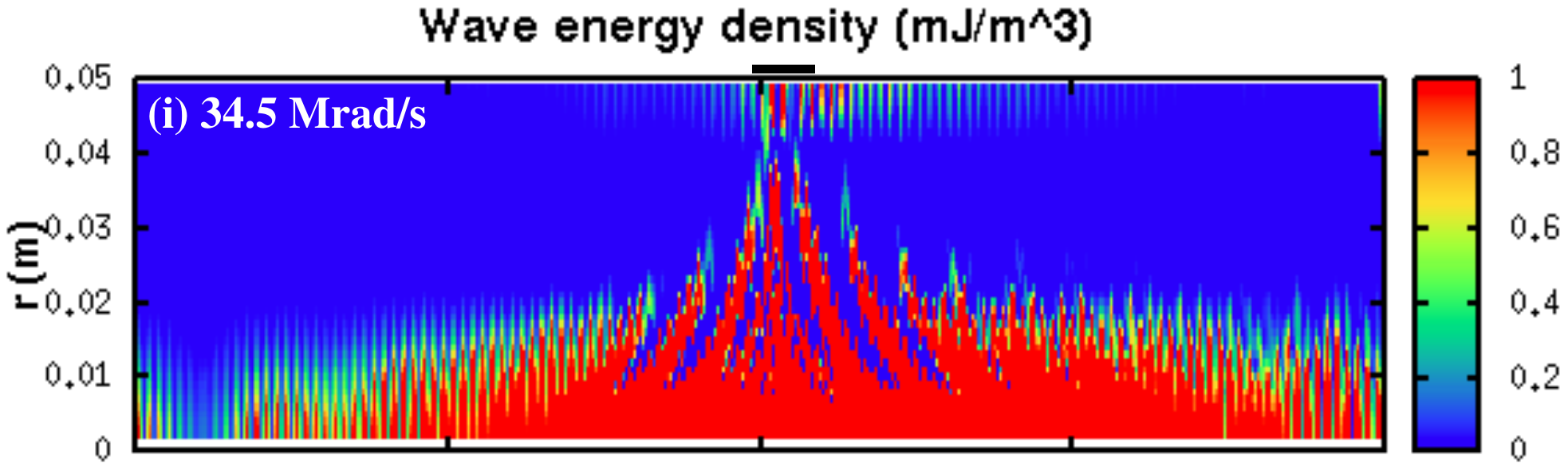}\\
\vspace{-0.2cm}\includegraphics[width=0.7\textwidth,angle=0]{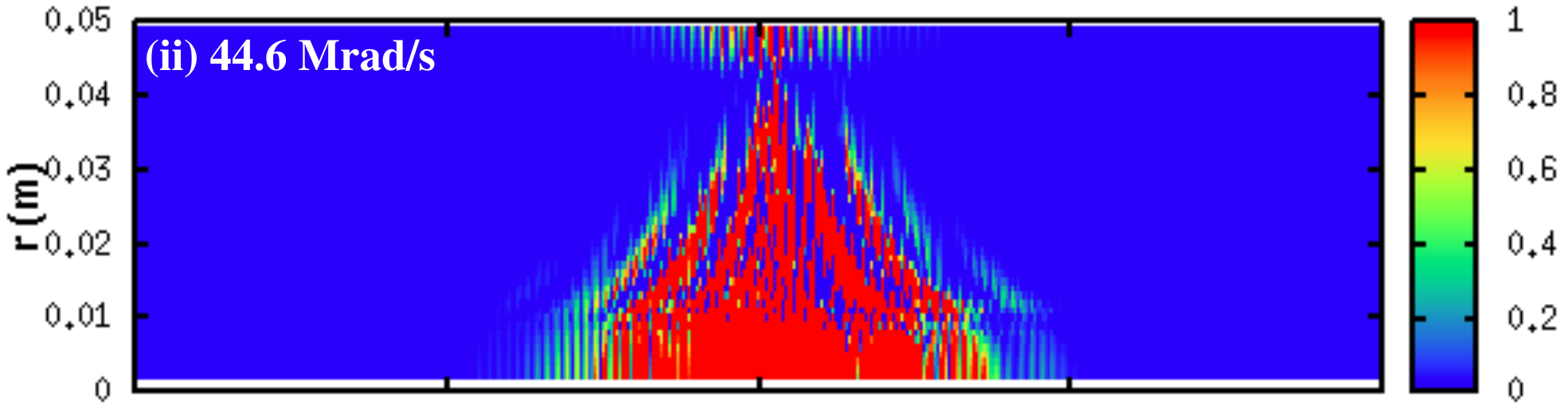}\\
\includegraphics[width=0.7\textwidth,angle=0]{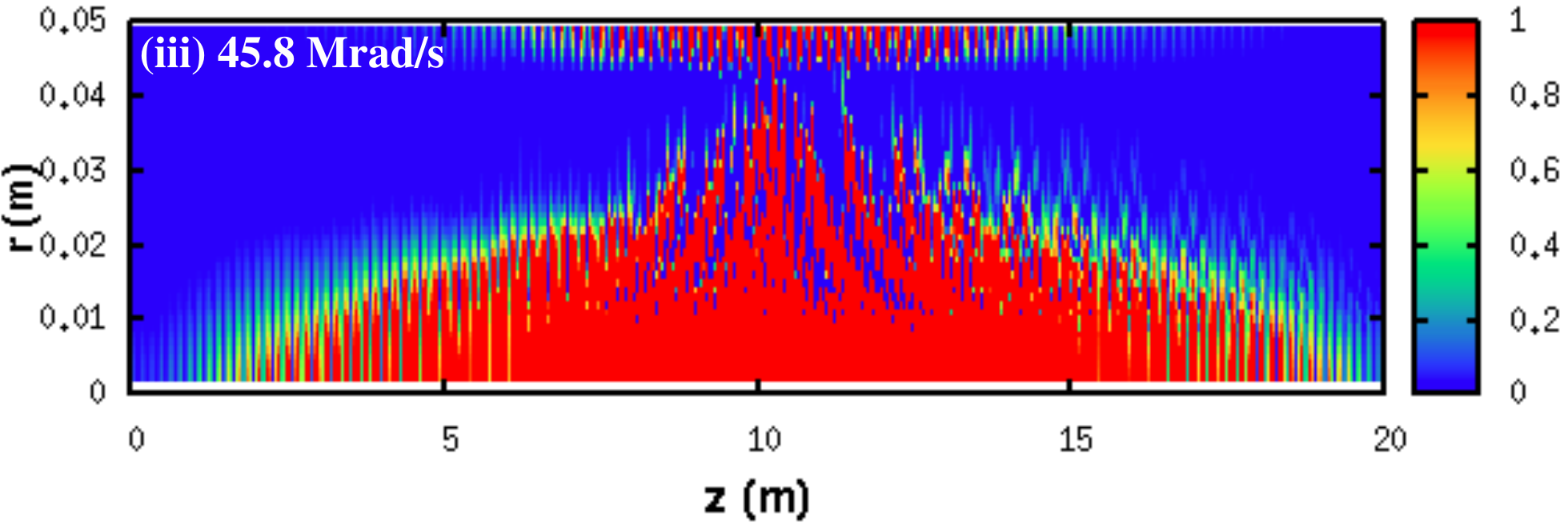}
\end{array}$
\end{center}
\caption{Identification of the spectral gap in simulations: (a) on-axis wave field at $z =15$~m as a function of driving frequency; vertical lines show analytical predictions for the gap centre $\omega_0$ and the upper and lower tips of the continuum $\omega_+$ and $\omega_-$, (b) spatial distributions of wave energy for antenna frequencies inside and outside the spectral gap (solid bar denotes the antenna location).}
\label{fg4_4}
\end{figure*}

In order to demonstrate the spectral gap, we run EMS for the same conditions as in Sec.~\ref{unf4} except that the static magnetic field is weakly modulated along $z$ and has the form $B_{0z} = B_0[1+0.25\cos(qz)]$ with $B_0=0.01$~T and $q=40~\rm{m}^{-1}$. The radial component of the static field is determined by Eq.~(\ref{eq3_5}). We scan the antenna frequency (for a fixed amplitude of the antenna current) and analyse the wave propagation from the antenna. Figure~\ref{fg4_4} illustrates the results. Figure~\ref{fg4_4}(a) is a plot of the on-axis signal at a $4.8$~m distance from the antenna, and shows an interval of strong suppression (from $36$~Mrad/s to $44$~Mrad/s), which represents the anticipated spectral gap. The location and width of this gap agree well with analytic estimates based on Eq.~(\ref{eq4_19}) and Eq.~(\ref{eq4_20}). Figure~\ref{fg4_4}(b) presents spatial distributions of wave energy for frequencies that are inside and outside of the spectral gap, from which we also see that wave propagation is evanescent inside the gap.

\subsection{Gap eigenmodes}\label{gmd4}
\begin{figure}[ht]
\begin{center}$
\begin{array}{ll}
(a)&(b)\\
\includegraphics[width=0.451\textwidth,angle=0]{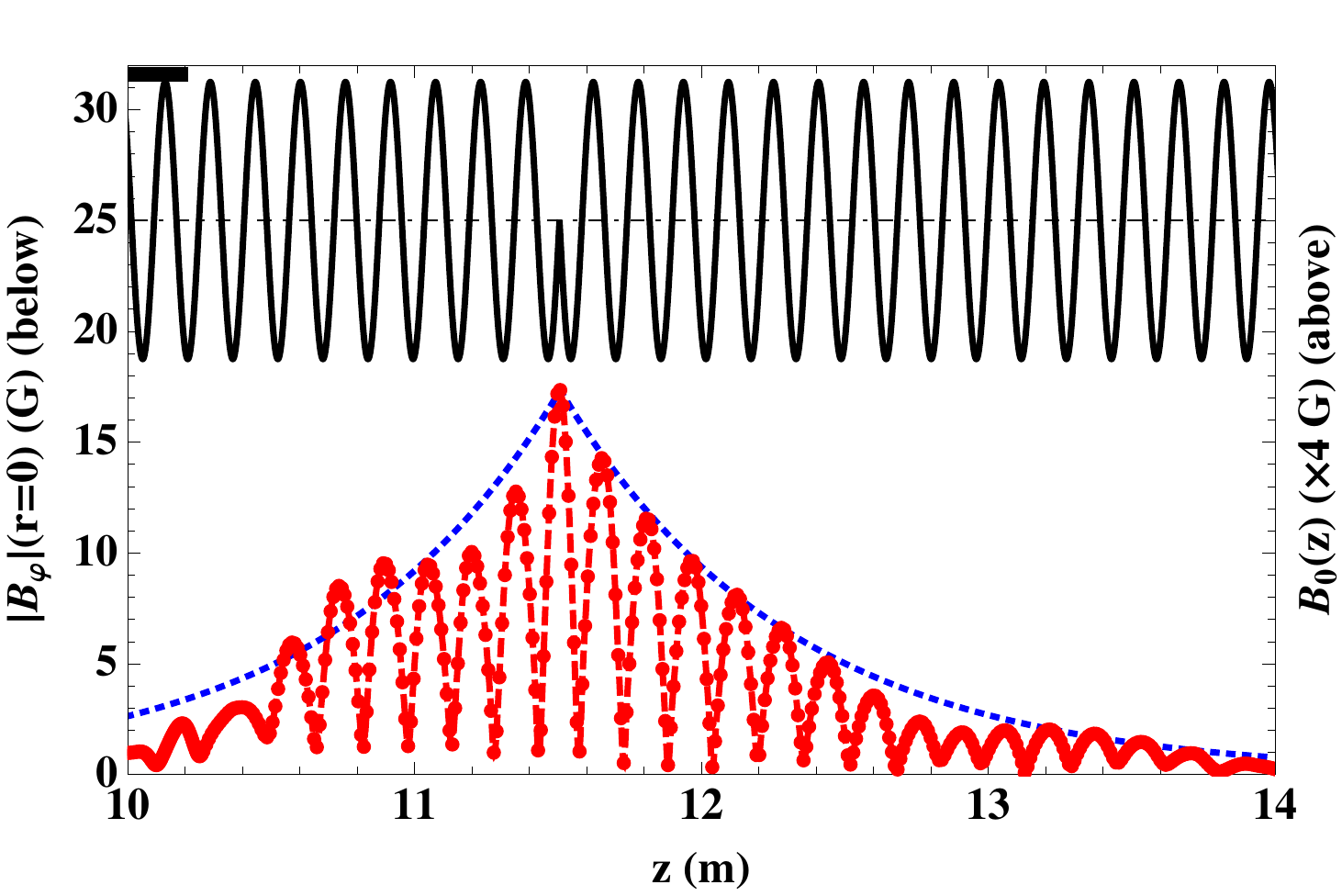}&\hspace{-0.1cm}\includegraphics[width=0.45\textwidth,angle=0]{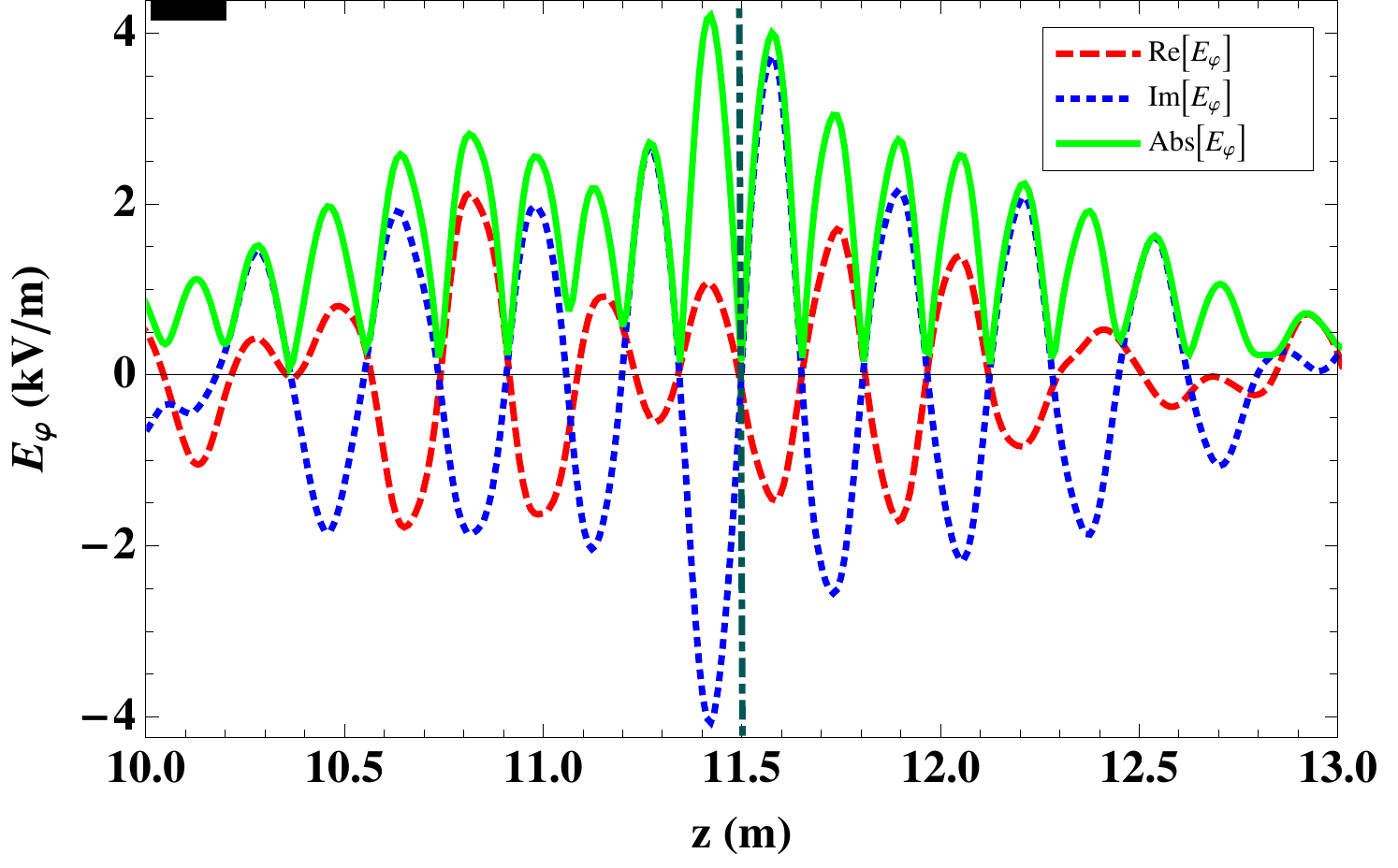}\\
(c)&(d)\\
\includegraphics[width=0.435\textwidth,angle=0]{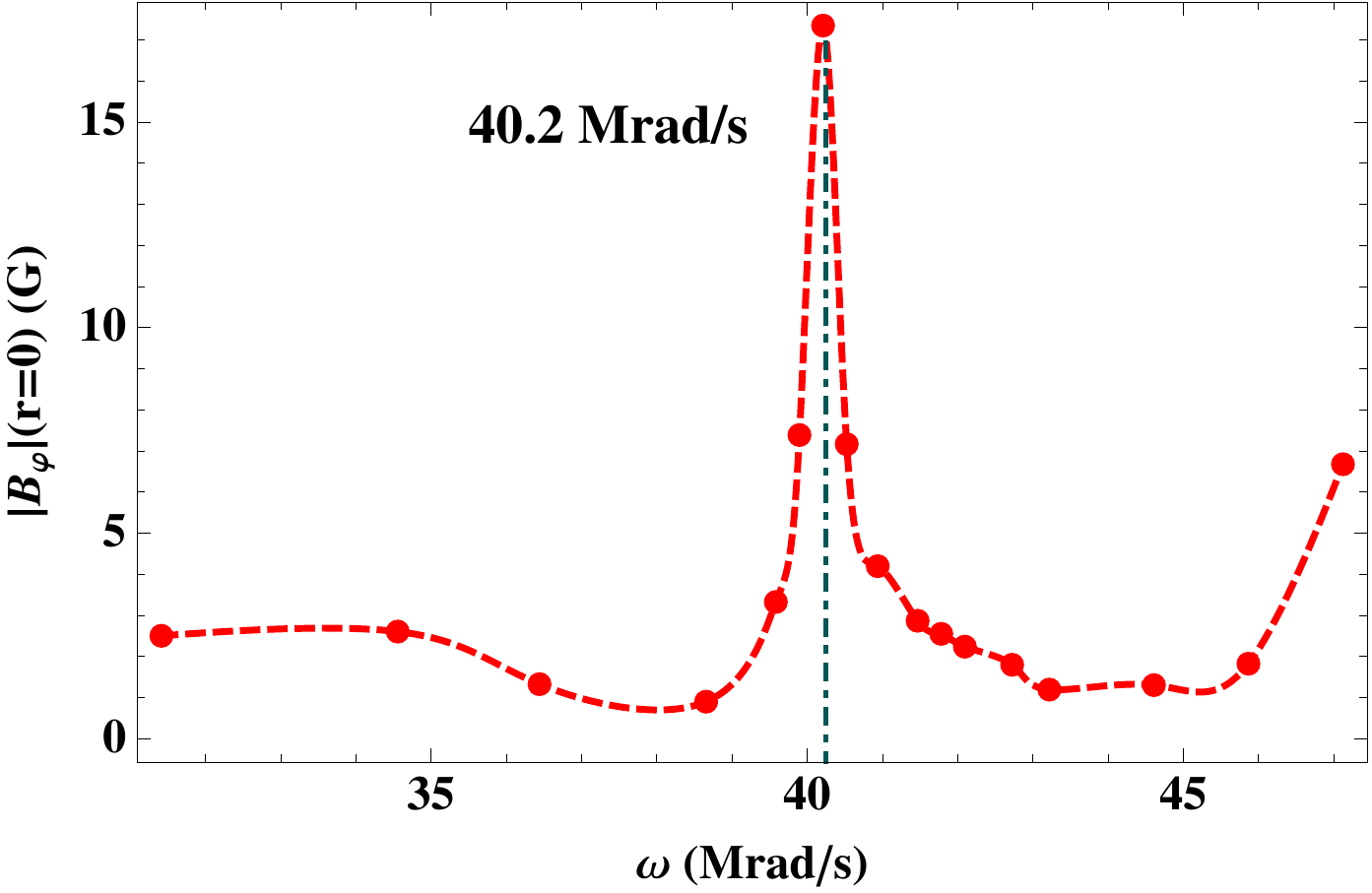}&\hspace{-0.3cm}\includegraphics[width=0.455\textwidth,angle=0]{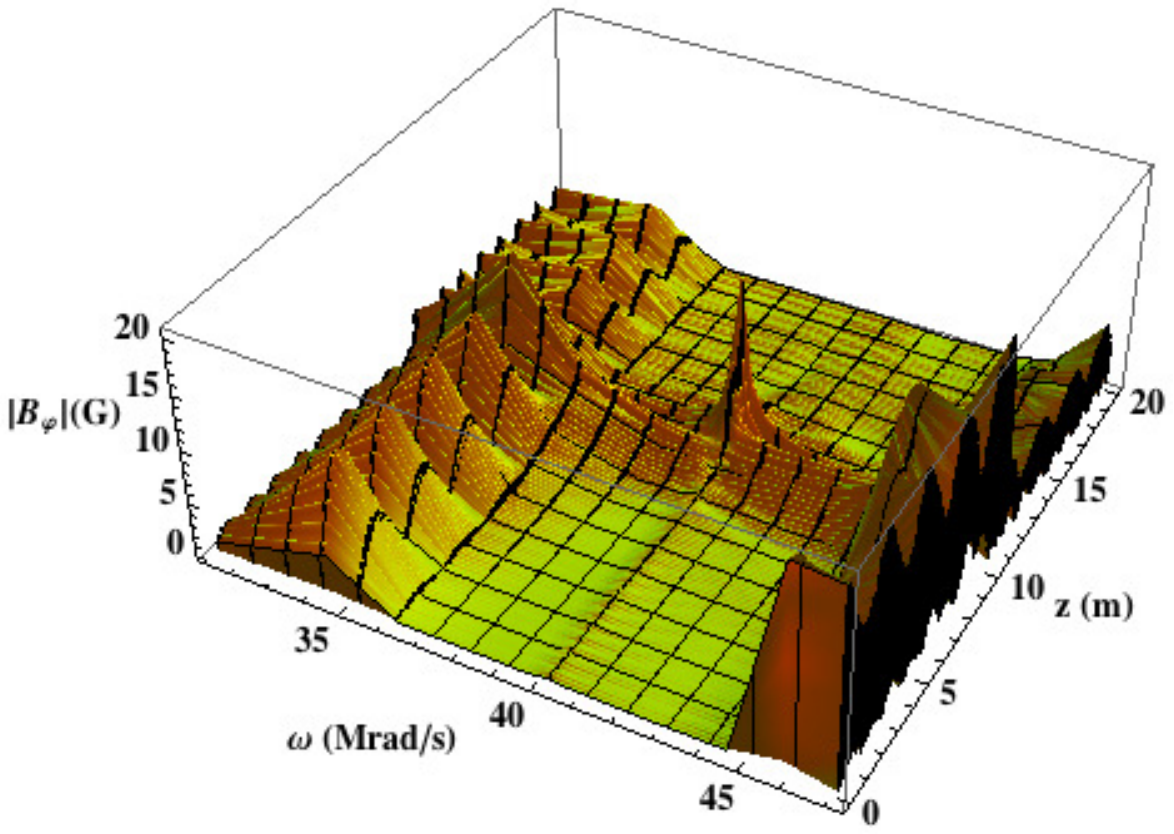}\\
\end{array}$
\end{center}
\caption{Odd-parity gap eigenmode: (a) longitudinal profiles of the static magnetic field (solid line) and the RF magnetic field on axis for $\omega=40.2$~Mrad/s (dots), together with the theoretically calculated envelope (dotted line), (b) longitudinal profile of $E_\varphi$ (on-axis) at $\omega=40.2$~Mrad/s (vertical dot-dashed line marks the defect location, and solid horizontal bar marks the antenna), (c) resonance in the dependence of the on-axis amplitude of the RF magnetic field on driving frequency at the location of the defect, (d) 3D plot of the on-axis wave field strength as a function of $z$ and $\omega$. }
\label{fg4_5}
\end{figure}

\begin{figure}[ht]
\begin{center}$
\begin{array}{ll}
(a)&(b)\\
\includegraphics[width=0.453\textwidth,angle=0]{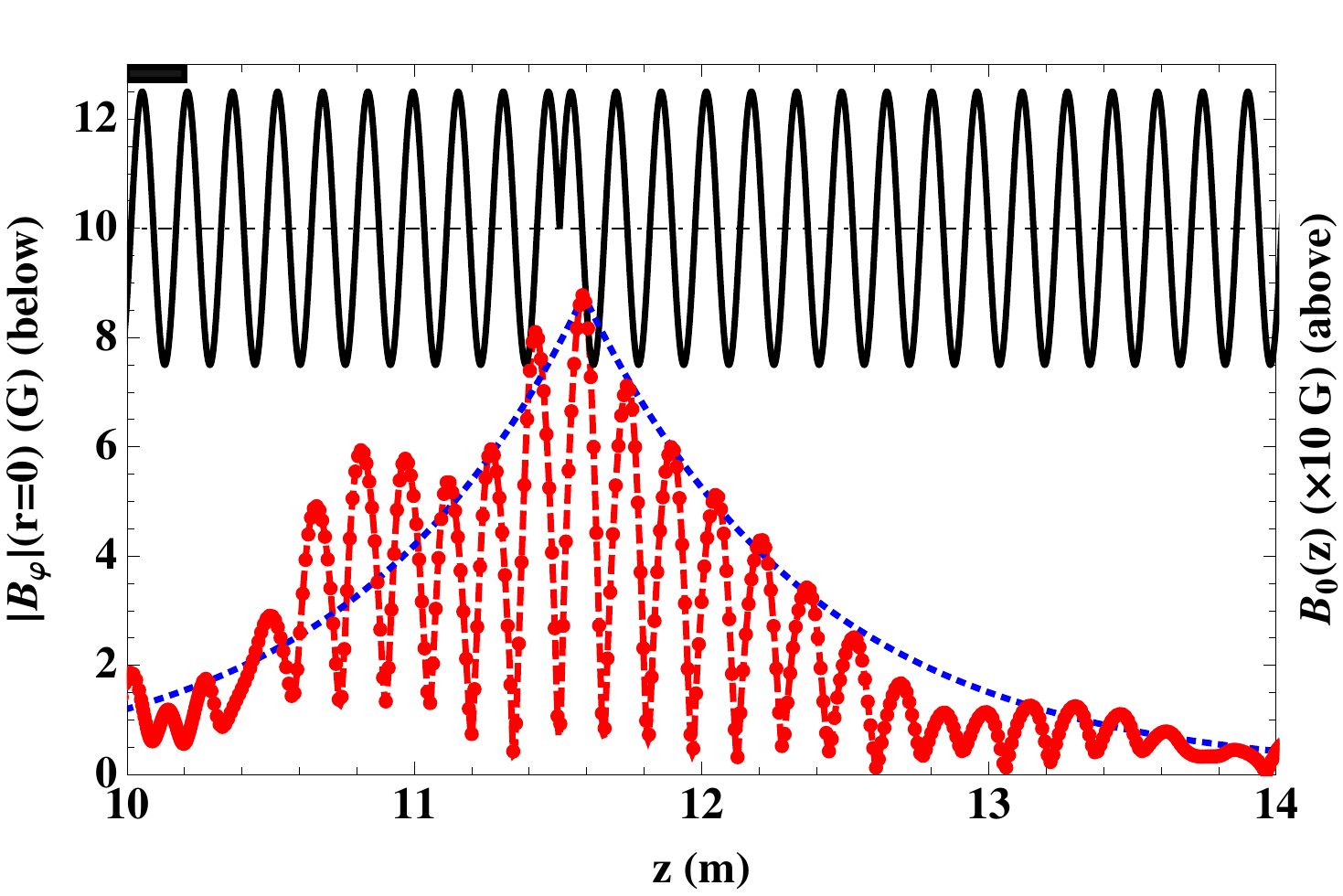}&\hspace{-0.1cm}\includegraphics[width=0.456\textwidth,angle=0]{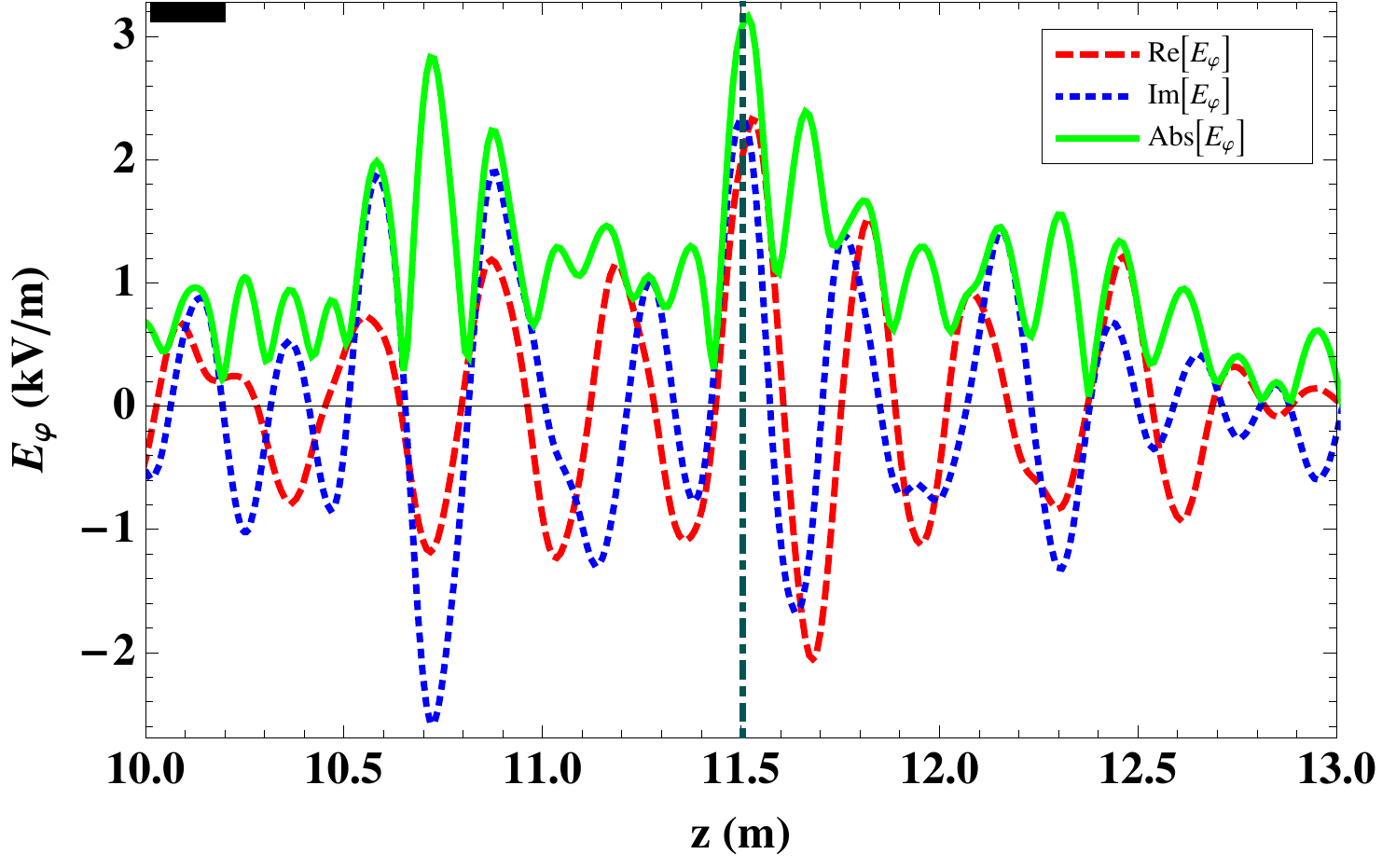}\\
(c)&(d)\\
\includegraphics[width=0.435\textwidth,angle=0]{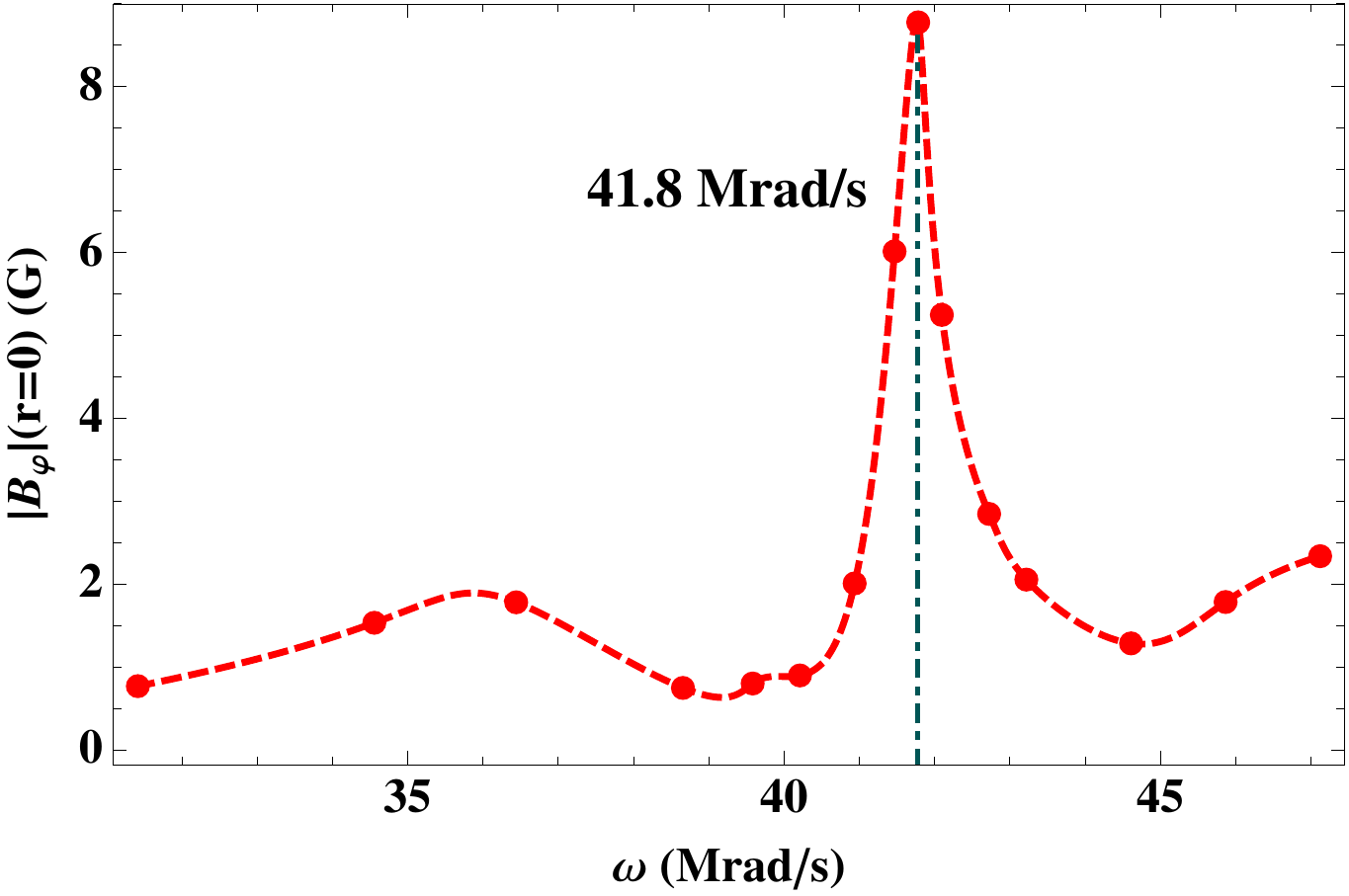}&\hspace{-0.3cm}\includegraphics[width=0.455\textwidth,angle=0]{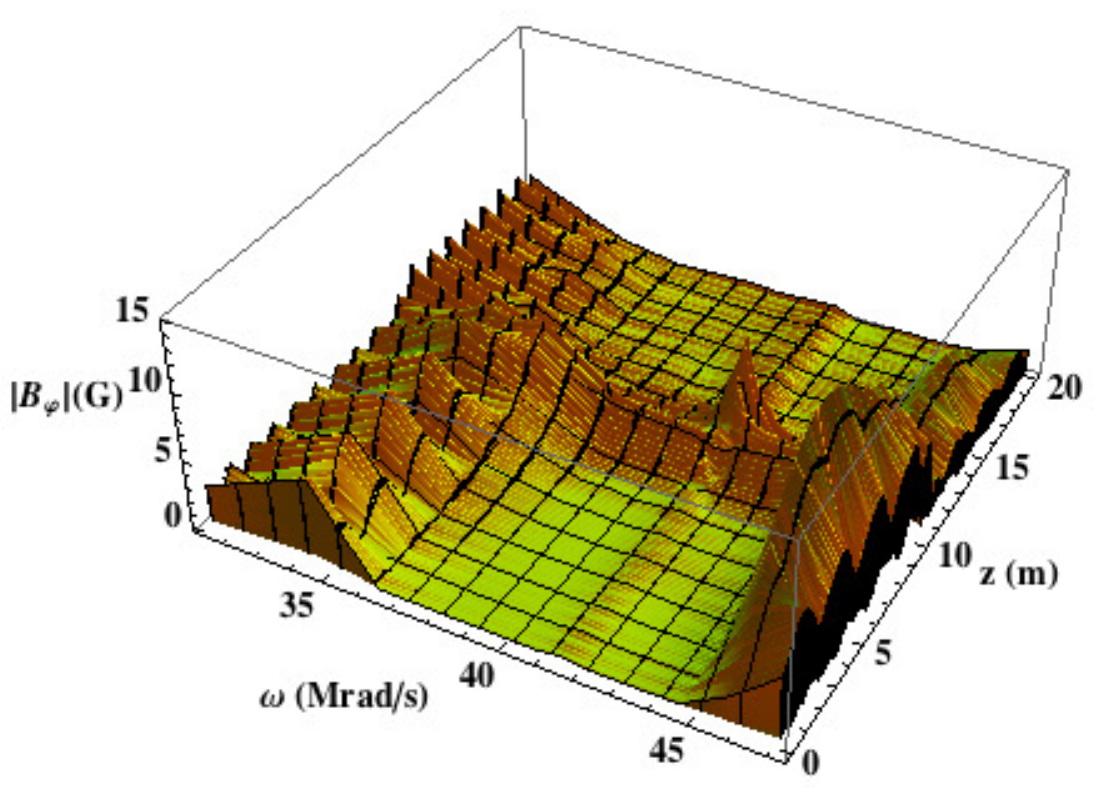}\\
\end{array}$
\end{center}
\caption{Even-parity gap eigenmode: (a) longitudinal profiles of the static magnetic field (solid line) and the RF  magnetic field on axis for $\omega=41.8$~Mrad/s (dots), together with the theoretically calculated envelope (dotted line), (b) longitudinal profile of $E_\varphi$ (on-axis) at $\omega=41.8$~Mrad/s (vertical dot-dashed line marks the defect location, and solid horizontal bar marks the antenna), (c) resonance in the dependence of the on-axis amplitude of the RF magnetic field on driving frequency near the location of the defect, (d) 3D plot of the on-axis wave field strength as a function of $z$ and $\omega$. }
\label{fg4_6}
\end{figure}

To form an eigenmode inside the spectral gap shown in Fig.~\ref{fg4_4}, we introduce a defect in the otherwise periodic static magnetic field. Equation~(\ref{eq4_28}) and Eq.~(\ref{eq4_32}) suggest that this defect should be located at $\cos(q z_0)= 0$, in order for the eigenmode frequency to be at the gap centre. The mode is then expected to have the shortest possible width (see Eq.~(\ref{eq4_29}) and Eq.~(\ref{eq4_33})). Moreover, Sec.~\ref{wll4} and Sec.~\ref{evn4} indicate that the gap eigenmode can have either odd parity or even parity, depending on the defect profile. We will confirm this analysis by simulation results shown in Fig.~\ref{fg4_5} and Fig.~\ref{fg4_6}.

Figure~\ref{fg4_5}(a)-(d) illustrates the odd-parity gap eigenmode associated with the defect shown in Fig.~\ref{fg4_1}(a). In order to minimise the role of collisional dissipation and thereby obtain a sharper resonant peak in Fig.~\ref{fg4_5}(c), we have reduced the collision frequency to $0.1\nu_{ei}$ in Fig.~\ref{fg4_5} simulation results. Figure~\ref{fg4_5}(a) shows that the eigenmode is a standing wave localised around the defect. Its exponential envelope with a decay length of $1.25$~m is consistent with analytical expectation from Eq.~(\ref{eq4_29}). The evanescent field of the mode is only weakly coupled to the distant antenna, but the resonance with the antenna frequency still allows the mode to be excited easily. The internal scale of this mode is close to twice the system's periodicity, consistent with Bragg's law. Figure~\ref{fg4_5}(b) shows the on-axis profile of the mode electric field, implying an odd function of $E_\varphi$ at $z_0$. The corresponding magnetic field $B_\varphi$ is, instead, an even function. Figure~\ref{fg4_5}(d) presents a full view of the plasma response in ($z$, $\omega$) space with a clear eigenmode peak inside the spectral gap.

We have also performed similar calculations for the defect shown in Fig.~\ref{fg4_1}(b). This defect produces an even-parity mode seen in Fig.~\ref{fg4_6}(a)-(d). Except for different symmetries, the features of the even and odd modes are  apparently similar.

\begin{figure}[ht]
\begin{center}$
\begin{array}{ll}
(a)&(b)\\
\includegraphics[width=0.495\textwidth,angle=0]{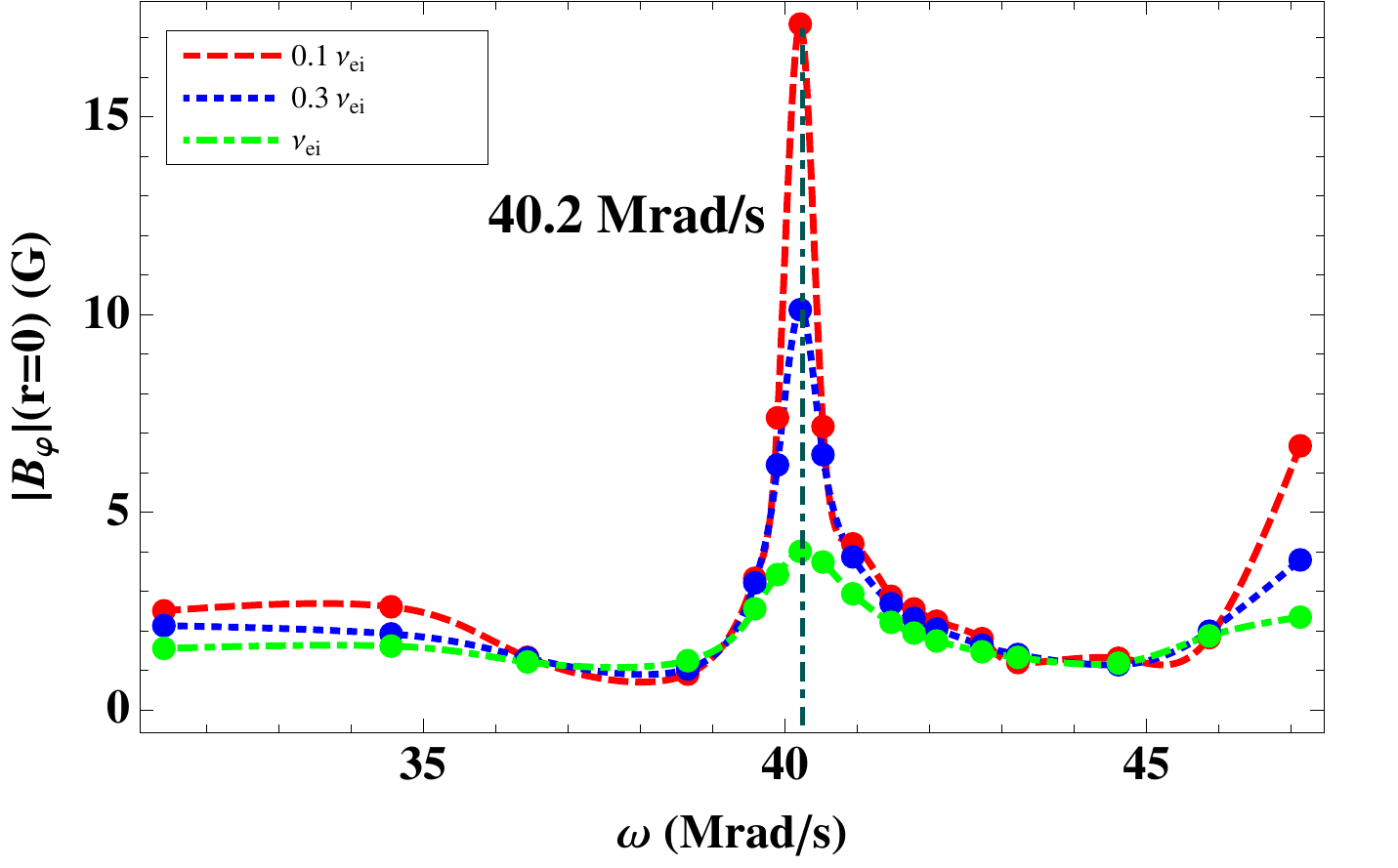}&\includegraphics[width=0.495\textwidth,angle=0]{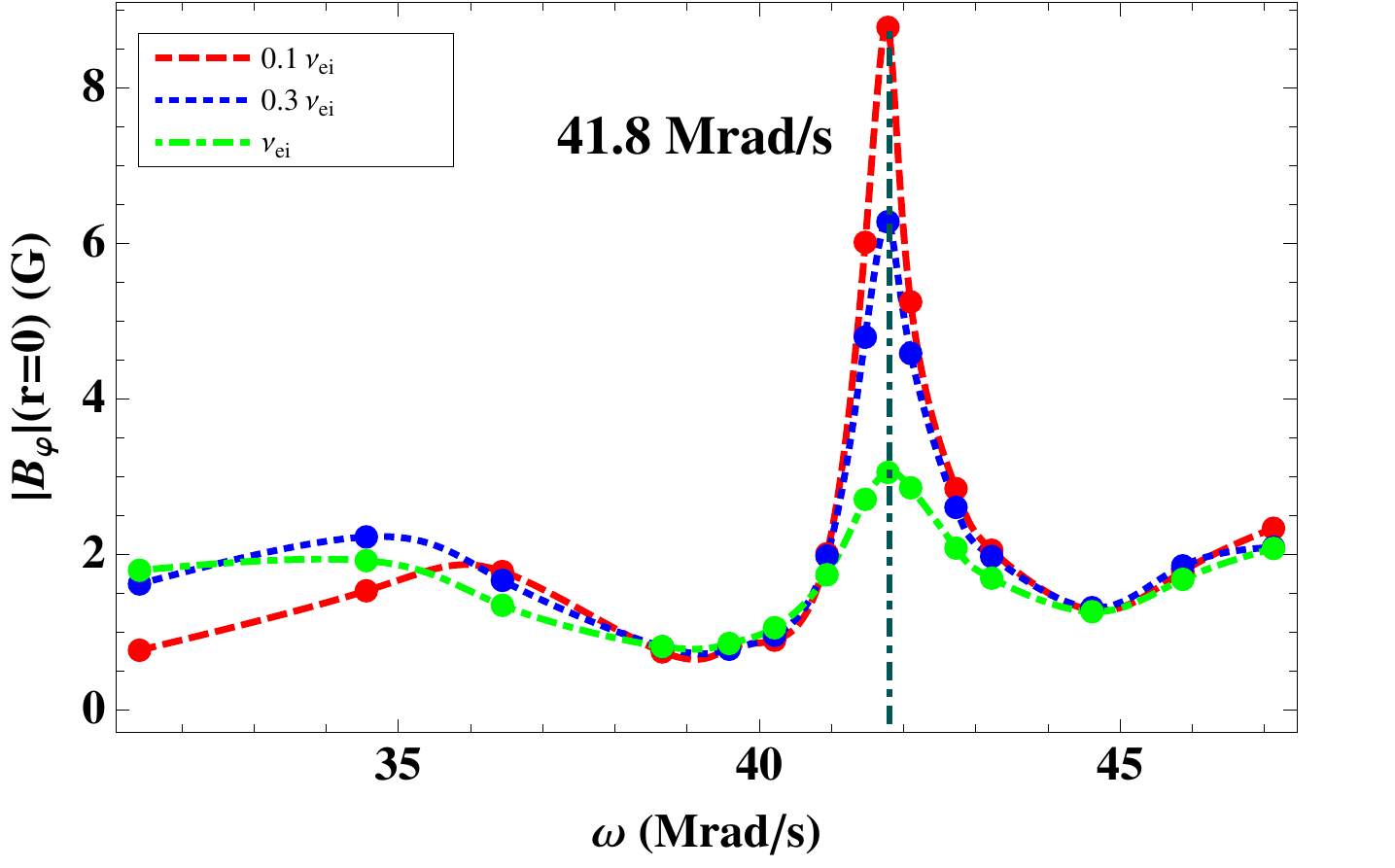}\\
\end{array}$
\end{center}
\caption{Effects of collisionality on gap eigenmodes: (a) results for odd-parity mode (Fig.~\ref{fg4_5}(c)), (b) results for even-parity mode (Fig.~\ref{fg4_6}(c)).}
\label{fg4_7}
\end{figure}

The effects of collisionality on gap eigenmodes are shown in Fig.~\ref{fg4_7}, from which we could see clearly a strength drop as the collisionality is increased. Although the resonant peak decreases and broadens for larger values of $\nu_{ei}$, it is still clearly visible even at the highest collision frequency.

\section{Summary and thoughts about possible experiments}\label{smy4}
We have shown that longitudinal modulation of the guiding magnetic field in a plasma column creates a spectral gap for radially localised helicon waves. Calculations performed with an EMS code reveal that this gap prohibits wave propagation along the column when the driving frequency of the RF antenna is in the forbidden range. The calculated width of the gap is consistent with analytical estimates. We have also shown that a discrete eigenmode can been formed inside the spectral gap by introducing a local defect to the periodic structure. Both the theoretical analysis and simulations demonstrate two types of gap eigenmode in the ``imperfect" system: odd-parity mode and even-parity mode, depending the type of the defect employed. The gap mode is localised around the defect and represents a standing wave rather than travelling wave. Its distinctive feature is a resonant peak in the plasma response to the antenna current. The gap eigenmode has two characteristic spatial scales: a short inner scale and a smooth envelope. The inner scale is nearly twice the system's periodicity, which is characteristic for Bragg's reflection; the envelope depends on the modulation amplitude and it scales roughly as the inverse width of the spectral gap. 

A plausible way to identify the gap mode in a linear device with multi-mirror configuration would be to use the end-plate of the machine as a controllable defect in the periodic system. This could make the mode observable with a modest number of mirrors in the machine. The LAPD is an apparent candidate for such experiments, provided that dissipative processes in the plasma do not destroy the gap-mode resonance. We have estimated these dissipative processes for LAPD plasma conditions. Collisional damping rate of the $m=1$ RLH mode can be roughly estimated as $\nu_e c^2/(a^2 \omega_{p}^2)$,\cite{Chen:1991aa} where $\nu_e$ is the collision frequency for electrons (a sum of electron-ion and electron-neutral collisions). Collisional damping can therefore be controlled by varying the plasma radius. Our illustrative simulation is for $a=0.05$~m, but the LAPD vacuum chamber allows to produce plasmas with considerably larger value of $a$ (up to $0.5$~m\cite{Zhang:2008aa, Gekelman:1991aa}) and thereby reduce the damping significantly. It may be preferable to choose plasma parameters in such a way that the gap mode frequency is in the range of operational frequencies for standard RF generators ($6-28$~MHz\cite{Boswell:1997aa}) through Eq.~(\ref{eq4_20}). The LAPD magnetic coils are about $0.61$~m in radius (inner),\cite{Gekelman:1991aa} which limits the number of magnetic field periods within the machine length to about $20$. This number seems to be large enough for the mode identification, especially because the axial extension of the mode is expected to decrease when the modulation amplitude increases. Stronger modulation of the magnetic field broadens the gap and makes it easier to identify the mode in the presence of finite damping. Thus, the amplitude of modulation is an apparent control knob for the conceivable experiment. The LAPD experiment can use larger plasma radius and stronger magnetic field (with the same plasma density $n_e\sim 10^{18}~\rm{m}^{-3}$) to reduce the damping but still keep the mode frequency in the convenient operational range for the RF-antenna.

\chapter{Gap eigenmode of shear Alfv\'{e}n waves in a periodic structure}\label{chp5}

In this chapter, a gap in the shear Alfv\'{e}n wave spectrum is predicted for a cold cylindrical plasma by modulating the magnetic field periodically. A discrete eigenmode is generated inside the gap by breaking the field periodicity through an introduced defect. We find that the axial wavelength of the gap eigenmode is nearly twice the system's periodicity, which is characteristic of Bragg's reflection, and the decay length of the eigenmode agrees well with the analytical estimate. Our treatment is based on numerical and analytical solutions to the cold plasma dielectric tensor, and complements Chapter~\ref{chp4} on the gap eigenmode generation in the whistler frequency range.

\section{Introduction}\label{int5}
In Chapter~\ref{chp4}, a discrete eigenmode of RLH waves was generated inside a spectral gap by introducing a defect to the system's periodicity.\cite{Chang:2013aa} This gap eigenmode is in the whistler frequency range and could be excited by energetic electrons, similar to the excitation of toroidal Alfv\'{e}n eigenmodes by energetic ions in tokamaks. We concluded that this eigenmode is observable on the LAPD, given that dissipative processes in the plasma do not destroy the gap-mode resonance. The LAPD, which has $10$ sets of pancake electromagnets ($90$ in total) surrounding a $17.56$-m-long, $1$-m-diam plasma column, is capable of forming multiple magnetic mirrors and identifying a spectral gap of plasma waves.\cite{Zhang:2008aa} This makes it a promising candidate to identify a gap eigenmode. However, to observe the RLH gap eigenmode, plasma parameters on the LAPD have to be chosen carefully to keep the mode frequency within the convenient operational range of standard RF generators. In 2008, Zhang et al.\cite{Zhang:2008aa} observed spectral gaps and continua of SAW on the LAPD, with the gap frequency about $0.3$ of the ion cyclotron frequency. Although eigenmodes were not formed inside this gap, they proposed a possible experimental implementation to detect them:  varying current in one of the independently powered magnetic coils to introduce a strong defect that breaks the system's periodicity. Therefore, we are interested in generalising the gap-mode analysis from the whistler frequency range to ion cyclotron range of frequencies. The SAW, which have frequencies below the ion cyclotron frequency and are fundamentally different from those with frequencies above, will be focused on with reference to Zhang et al.'s experiment.\cite{Zhang:2008aa} We shall numerically form a gap eigenmode of SAW by introducing a defect to the system's periodicity. 

\section{Theoretical analysis}\label{thy5}
Although the focus of the study is on SAW, the theoretical analysis presented below is also valid for waves with frequencies slightly above the ion cyclotron frequency. 
\subsection{Basic equations}\label{eqn5}
We consider an electromagnetic wave in a cold plasma cylinder with uniform static magnetic field and radially non-uniform plasma density. The cold plasma approximation implies a much greater longitudinal wave phase velocity than particle thermal velocity, and assures low Landau damping. We start from the linear wave equation (see Eq.~(\ref{eq1_44}))
\begin{equation}\label{eq5_3}
\nabla\times\nabla\times\mathbf{E}=\frac{\omega^2}{c^2}\mathbf{\tilde{\epsilon}}\cdot\mathbf{E}.
\end{equation}
For $\omega\approx\omega_{ci}$, the components of $\mathbf{\tilde{\epsilon}}$ become
\begin{equation}\label{eq5_5}
\begin{array}{l}
\vspace{0.3cm}S=1-\frac{\omega_{pe}^2}{\omega^2-\omega_{ce}^2}-\frac{\omega_{pi}^2}{\omega^2-\omega_{ci}^2}\approx-\frac{\omega_{pi}^2}{\omega^2-\omega_{ci}^2},\\
\vspace{0.3cm}D=\frac{\omega_{ce}\omega_{pe}^2}{\omega(\omega^2-\omega_{ce}^2)}+\frac{\omega_{ci}\omega_{pi}^2}{\omega(\omega^2-\omega_{ci}^2)}\approx\frac{\omega\omega_{pi}^2}{\omega_{ci}(\omega^2-\omega_{ci}^2)},\\
P=1-\frac{\omega_{pe}^2}{\omega^2}-\frac{\omega_{pi}^2}{\omega^2}\approx-\frac{\omega_{pe}^2}{\omega^2}. 
\end{array}
\end{equation}
Here, the relation $\omega_{pe}^2/\omega_{ce}=-\omega_{pi}^2/\omega_{ci}$ has been used. As in Chapter~\ref{chp4}, we will limit our consideration to the case of sufficiently dense plasma and nonzero values of the azimuthal mode number $m$, because the radial nonuniformity of plasma density has a surprisingly strong effect on the structure of $m\neq 0$ modes.\cite{Breizman:2000aa} Due to $\omega\ll\omega_{pe}$, the parallel conductivity $P$ is very large so that the axial component of the electric field vanishes, namely, $E_z=0$.  This simplifies the radial and azimuthal components of Eq.~(\ref{eq5_3}) as:
\begin{equation}\label{eq5_6}
\frac{i m}{r}\left[\frac{1}{r}\frac{\partial}{\partial r}(r E_\varphi)-\frac{i m}{r}E_r\right]+k_z^2 E_r=\frac{\omega^2}{c^2}\left(S E_r-i D E_\varphi\right),
\end{equation}
\begin{equation}\label{eq5_7}
-\frac{\partial}{\partial r}\left[\frac{1}{r}\frac{\partial}{\partial r}(r E_\varphi)-\frac{i m}{r}E_r\right]+k_z^2 E_\varphi=\frac{\omega^2}{c^2}\left(i D E_r+S E_\varphi\right).
\end{equation}
It is convenient to combine Eq.~(\ref{eq5_6}) and Eq.~(\ref{eq5_7}) into 
\begin{equation}\label{eq5_8}
0=\frac{\omega^2}{c^2}\left(S E_\varphi+i D E_r\right)-k_z^2 E_\varphi+\frac{\partial}{\partial r}\left\{\frac{r}{i m}\left[\frac{\omega^2}{c^2}\left(S E_r-i D E_\varphi\right)-k_z^2 E_r\right]\right\},
\end{equation}
and use Eq.~(\ref{eq5_6}) and Eq.~(\ref{eq5_8}) instead of Eq.~(\ref{eq5_6}) and Eq.~(\ref{eq5_7}). It follows from Eq.~(\ref{eq5_6}) that
\begin{equation}\label{eq5_9}
\frac{1}{r}\frac{\partial}{\partial r}(r E_\varphi)-\frac{i m}{r}E_r\approx 0
\end{equation}
for modes with sufficiently small values of $k_z$ and $\omega$. We choose small values of $k_z$ to see clear effects of the radial non-uniformity in plasma density on the wave mode structure for non-zero azimuthal mode numbers.\cite{Breizman:2000aa} We can then use Eq.~(\ref{eq5_9}) to eliminate $E_r$ from Eq.~(\ref{eq5_8}), and obtain the wave equation of SAW
\begin{equation}\label{eq5_10}
0=\left(\frac{\omega^2}{c^2}S-k_z^2\right)r E_\varphi-\frac{r E_\varphi}{m}\frac{\omega^2}{c^2}r\frac{\partial D}{\partial r}+r\frac{\partial}{\partial r}\left[-\frac{r}{m^2}\left(\frac{\omega^2}{c^2}S-k_z^2\right)\frac{\partial}{\partial r}(r E_\varphi)\right].
\end{equation}
Equation~(\ref{eq5_10}) is identical to the wave equation of RLH waves when $S=0$ (see Eq.~(\ref{eq1_52})) and can be solved in a similar way. 

\subsection{Uniform magnetic field: dispersion relation}\label{uni5}
We first solve Eq.~(\ref{eq5_10}) for a uniform magnetic field to derive the dispersion relation of SAW. We shall construct a surface-wave solution through considering a step-like radial profile of plasma density (see Eq.~(\ref{eq1_54a}))
\begin{equation}\label{eq5_11}
n_i=\left\{
\begin{array}{cc}
n_-,& r<r_0\\
n_+,& r>r_0
\end{array}
\right.
\end{equation}
with $r_0$ the radius of the density discontinuity. The width of the discontinuity layer at $r=r_0$ is assumed to be larger than the skin depth, namely $2\delta_{\mathrm{w}}>c/\omega_{pe}$ with $\delta_{\mathrm{w}}$ half width of the layer. This density profile leads to:
\begin{equation}\label{eq5_12}
\begin{array}{l}
\vspace{0.3cm}\frac{\partial S}{\partial r}=[S(n_+)-S(n_-)]\delta(r-r_0),\\
\frac{\partial D}{\partial r}=[D(n_+)-D(n_-)]\delta(r-r_0).
\end{array}
\end{equation}
We then assume a specific form of $E_\varphi$ (see Eq.~(\ref{eq1_54b}))
\begin{equation}\label{eq5_13}
E_\varphi=\frac{E_0}{r_0}\left\{
\begin{array}{cc}
\vspace{0.3cm}(r/r_0)^{|m|-1},& r<r_0\\
(r/r_0)^{-|m|-1},& r>r_0
\end{array}
\right.
\end{equation}
with $E_0$ a constant electric field. Equation~(\ref{eq5_10}) is satisfied at $r\neq r_0$ for any $\omega$ and $k_z$ for the density profile constructed. For modes caused by the density jump, we integrate Eq.~(\ref{eq5_10}) across the discontinuity layer $r_0-\delta_{\mathrm{w}}\rightarrow r_0+\delta_{\mathrm{w}}$
\begin{equation}\label{eq5_14}
\begin{array}{c}
\vspace{0.3cm}\left[\left(\frac{\omega^2}{k_z^2 c^2}S-1\right)r\frac{\partial}{\partial r}(r E_\varphi)\right]_{r_0-\delta_{\mathrm{w}}}^{r_0+\delta_{\mathrm{w}}}-\int_{r_0-\delta_{\mathrm{w}}}^{r_0+\delta_{\mathrm{w}}}{\left(\frac{\omega^2}{k_z^2 c^2}S-1\right)m^2 E_\varphi}dr\\
=-\int_{r_0-\delta_{\mathrm{w}}}^{r_0+\delta_{\mathrm{w}}}{\frac{\omega^2}{k_z^2 c^2}m r E_\varphi \frac{\partial D}{\partial r}}dr.
\end{array}
\end{equation}
The right hand side of Eq.~(\ref{eq5_14}) becomes
\begin{equation}\label{eq5_15}
-\frac{\omega^2}{k_z^2 c^2}m E_0 [D(n_+)-D(n_-)]
\end{equation}
due to Eq.~(\ref{eq5_12}), and the second term on the left hand side vanishes because of the small value of $|\delta_{\mathrm{w}}|$ and finite $E_0$,
\begin{equation}\label{eq5_16}
\int_{r_0-\delta_{\mathrm{w}}}^{r_0+\delta_{\mathrm{w}}}{\left(\frac{\omega^2}{k_z^2 c^2}S-1\right)m^2 E_\varphi}dr=0.
\end{equation}
Substituting Eq.~(\ref{eq5_5}), Eq.~(\ref{eq5_12}) and Eq.~(\ref{eq5_13}), Eq.~(\ref{eq5_14}) can then be rewritten as
\begin{equation}\label{eq5_18}
\frac{2k_z^2c^2}{\omega_{pi-}^2+\omega_{pi+}^2}=\Psi\frac{|m|}{m}\frac{x^3}{x^2-1}-\frac{x^2}{x^2-1}.
\end{equation}
Here, $\Psi=(\omega_{pi-}^2-\omega_{pi+}^2)/(\omega_{pi-}^2+\omega_{pi+}^2)$ with $\omega_{pi-}$ and $\omega_{pi+}$ the ion plasma frequencies inside and outside of the discontinuity layer, respectively, and $x=\omega/\omega_{ci}$ is the normalised wave frequency. Equation~(\ref{eq5_18}) describes the dispersion relation of SAW propagating in a plasma cylinder with uniform magnetic field and step-like radial profile of plasma density. For SAW with $\omega\ll\omega_{ci}$ and propagating in a uniform plasma, Eq.~(\ref{eq5_18}) becomes $\omega_{\mathrm{A}}=k_z \nu_\mathrm{A}$ due to $x\ll1$ and $\Psi=0$. This is the well-known dispersion relation for SAW in a slab geometry.

\subsection{Periodic magnetic field: spectral gap and continuum}\label{per5}
We now introduce a slight periodic modulation to the static magnetic field, and solve Eq.~(\ref{eq5_10}) for the corresponding SAW spectrum. The modulated field is of the form $B_z=B_0[1+\epsilon\cos(qz)]$ with $B_0$ the equilibrium field strength, $\epsilon\ll 1$ and $q$ the modulation factor and wavenumber, respectively. This modulated equilibrium introduces resonant coupling between the modes with $k_z=\pm q/2$, whose frequencies are the same and can be solved from Eq.~(\ref{eq5_18}). Note from Eq.~(\ref{eq5_10}) that this modulation takes effect through $(\omega^2/c^2)S$ and $-(\omega^2/c^2)r\partial D/\partial r$, which are functions of $x$ according to Eq.~(\ref{eq5_5}), we thereby have their linear Taylor expansions about $x$:
\begin{equation}\label{eq5_19}
\begin{array}{cc}
\vspace{0.3cm}\frac{\omega^2}{c^2}S=\frac{\omega_0^2}{c^2}S_0+\left[x\frac{d}{dx}\left(\frac{\omega^2}{c^2}S\right)\right]_{x\rightarrow x_0}\left(\frac{\omega-\omega_0}{\omega_0}-\frac{\omega_{ci}-\omega_{ci0}}{\omega_{ci0}}\right),\\
-\frac{\omega^2}{c^2}r\frac{\partial D}{\partial r}=-\frac{\omega_0^2}{c^2}r\frac{\partial D_0}{\partial r}-\left[x\frac{d}{dx}\left(\frac{\omega^2}{c^2}r\frac{\partial D}{\partial r}\right)\right]_{x\rightarrow x_0}\left(\frac{\omega-\omega_0}{\omega_0}-\frac{\omega_{ci}-\omega_{ci0}}{\omega_{ci0}}\right).
\end{array}
\end{equation}
Here, $\omega_0$, $\omega_{ci0}$, $S_0$ and $D_0$ are the corresponding values of $\omega$, $\omega_{ci}$, $S$ and $D$ for a uniform static magnetic field. The dependences of $S$ and $-D$ on the frequency difference ($\omega-\omega_0$) and modulation ($\omega_{ci}-\omega_{ci0}$) have been separated. By introducing variables: 
\begin{equation}\label{eq5_20}
\begin{array}{cc}
\vspace{0.3cm}P_S=\left[x\frac{d}{dx}\rm{ln}\left(\omega^2S\right)\right]\frac{\omega-\omega_0}{\omega_0},\\
\vspace{0.3cm}P_D=-\left[x\frac{d}{dx}\rm{ln}\left(\omega^2 D\right)\right]\frac{\omega-\omega_0}{\omega_0},\\
\vspace{0.3cm}\mu_S\cos(q z)=-\left[x\frac{d}{dx}\rm{ln}\left(\omega^2S\right)\right]\frac{\omega_{ci}-\omega_{ci0}}{\omega_{ci0}},\\
\mu_D\cos(q z)=\left[x\frac{d}{dx}\rm{ln}\left(\omega^2 D\right)\right]\frac{\omega_{ci}-\omega_{ci0}}{\omega_{ci0}},\\
\end{array}
\end{equation}
we rewrite Eq.~(\ref{eq5_19}) as:
\begin{equation}\label{eq5_21}
\begin{array}{cc}
\vspace{0.3cm}\frac{\omega^2}{c^2}S=\frac{\omega_0^2}{c^2}S_0\left[1+\mu_S\cos(q z)+P_S\right],\\
-\frac{\omega^2}{c^2}r\frac{\partial D}{\partial r}=-\frac{\omega_0^2}{c^2}r\frac{\partial D_0}{\partial r}\left[1+\mu_D\cos(q z)+P_D\right].
\end{array}
\end{equation}
Assuming $[(\omega_{ci}-\omega_{ci0})/\omega_{ci0}]=\epsilon\cos(q z)$ and substituting Eq.~(\ref{eq5_21}) into Eq.~(\ref{eq5_10}) result in
\begin{equation}\label{eq5_22}
\begin{array}{c}
\vspace{0.3cm}0=\left(\frac{\omega_0^2}{c^2}S_0-\frac{q^2}{4}\right)r E_\varphi-\frac{r E_\varphi}{m}\frac{\omega_0^2}{c^2}r\frac{\partial D_0}{\partial r}+r\frac{\partial}{\partial r}\left[-\frac{r}{m^2}\left(\frac{\omega_0^2}{c^2}S_0-\frac{q^2}{4}\right)\frac{\partial}{\partial r}r E_\varphi\right]\\
\vspace{0.3cm}+\left\{\frac{\omega_0^2}{c^2}S_0\left[\mu_S\cos(q z)+P_S\right]+\frac{\partial^2}{\partial z^2}+\frac{q^2}{4}\right\}r E_\varphi-\left[\mu_D\cos(q z)+P_D\right]\frac{r E_\varphi}{m}\frac{\omega_0^2}{c^2}r\frac{\partial D_0}{\partial r}\\
\vspace{0.3cm}+r\frac{\partial}{\partial r}\left\{-\frac{r}{m^2}\left[\frac{\omega_0^2}{c^2}S_0\left(\mu_S+P_S\right)+\frac{\partial^2}{\partial z^2}+\frac{q^2}{4}\right]\frac{\partial}{\partial r}r E_\varphi\right\}.
\end{array}
\end{equation}
Letting $\Psi(r)$ be the eigenfunction of the wave equation for the equilibrium magnetic field $B_0$
\begin{equation}\label{eq5_23}
0=\left(\frac{\omega_0^2}{c^2}S_0-\frac{q^2}{4}\right)r \Psi-\frac{r\Psi}{m}\frac{\omega_0^2}{c^2}r\frac{\partial D_0}{\partial r}+r\frac{\partial}{\partial r}\left[-\frac{r}{m^2}\left(\frac{\omega_0^2}{c^2}S_0-\frac{q^2}{4}\right)\frac{\partial}{\partial r}r \Psi\right],
\end{equation}
we multiply Eq.~(\ref{eq5_22}) by $\Psi(r)$ and integrate over radius to eliminate the equilibrium terms via integration by parts,
\begin{equation}\label{eq5_24}
\begin{array}{c}
\vspace{0.3cm}0=\int\left\{\frac{\omega_0^2}{c^2}S_0\left[\mu_S\cos(q z)+P_S\right]+\frac{\partial^2}{\partial z^2}+\frac{q^2}{4}\right\}r E_\varphi\Psi dr\\
\vspace{0.3cm}-\int\left[\mu_D\cos(q z)+P_D\right]\frac{r E_\varphi}{m}\frac{\omega_0^2}{c^2}r\frac{\partial D_0}{\partial r}\Psi dr\\
+\int r\frac{\partial}{\partial r}\left\{-\frac{r}{m^2}\left(\frac{\omega_0^2}{c^2}S_0\left[\mu_S\cos(q z)+P_S\right]+\frac{\partial^2}{\partial z^2}+\frac{q^2}{4}\right)\frac{\partial}{\partial r}r E_\varphi\right\}\Psi dr.
\end{array}
\end{equation}
We try the following form of $E_\varphi$
\begin{equation}\label{eq5_25}
E_\varphi=F(z)\Psi(r) \rm{exp}(i q z/2+i\kappa z)+G(z)\Psi(r) \rm{exp}(-i q z/2+i\kappa z)
\end{equation}
with $F(z)$ and $G(z)$ slow functions of $z$ compared with $\cos(q z)$, and $\kappa$ the wave number in the modulated field. Separating the longitudinal Fourier harmonics in Eq.~(\ref{eq5_24}), we obtain two coupled equations for $F$ and $G$:
\begin{equation}\label{eq5_26}
\begin{array}{c}
\vspace{0.3cm}0=\int\left(\frac{\omega_0^2}{c^2}S_0P_S-q\kappa\right)r \Psi F\Psi dr+\int\frac{\omega_0^2}{2c^2}S_0\mu_S r \Psi G\Psi dr\\
\vspace{0.3cm}-\int P_D\frac{r \Psi F}{m}\frac{\omega_0^2}{c^2}r\frac{\partial D_0}{\partial r}\Psi dr-\int\mu_D\frac{r \Psi G}{m}\frac{\omega_0^2}{2c^2}r\frac{\partial D_0}{\partial r}\Psi dr\\
\vspace{0.3cm}+\int r\frac{\partial}{\partial r}\left[-\frac{r}{m^2}\left(\frac{\omega_0^2}{c^2}S_0P_S-q\kappa\right)\frac{\partial}{\partial r}r \Psi F\right]\Psi dr\\
+\int r\frac{\partial}{\partial r}\left(-\frac{r}{m^2}\frac{\omega_0^2}{2c^2}S_0\mu_S\frac{\partial}{\partial r}r \Psi G\right)\Psi dr,
\end{array}
\end{equation}
\begin{equation}\label{eq5_27}
\begin{array}{c}
\vspace{0.3cm}0=\int\left(\frac{\omega_0^2}{c^2}S_0P_S+q\kappa\right)r \Psi G\Psi dr+\int\frac{\omega_0^2}{2c^2}S_0\mu_S r \Psi F\Psi dr\\
\vspace{0.3cm}-\int P_D\frac{r \Psi G}{m}\frac{\omega_0^2}{c^2}r\frac{\partial D_0}{\partial r}\Psi dr-\int\mu_D\frac{r \Psi F}{m}\frac{\omega_0^2}{2c^2}r\frac{\partial D_0}{\partial r}\Psi dr\\
\vspace{0.3cm}+\int r\frac{\partial}{\partial r}\left[-\frac{r}{m^2}\left(\frac{\omega_0^2}{c^2}S_0P_S+q\kappa\right)\frac{\partial}{\partial r}r \Psi G\right]\Psi dr\\
+\int r\frac{\partial}{\partial r}\left(-\frac{r}{m^2}\frac{\omega_0^2}{2c^2}S_0\mu_S\frac{\partial}{\partial r}r \Psi F\right)\Psi dr.
\end{array}
\end{equation}
Introducing the following integrals of the eigenfunction:
\begin{equation}\label{eq5_28}
\begin{array}{l}
\vspace{0.3cm}I_1=\int\frac{\omega_0^2}{c^2}S_0 r\Psi\Psi dr,\\
\vspace{0.3cm}I_2=\int q r \Psi\Psi dr,\\
\vspace{0.3cm}I_3=-\int\frac{r\Psi}{m}\frac{\omega_0^2}{c^2}r\frac{\partial D_0}{\partial r}\Psi dr,\\
\vspace{0.3cm}I_4=\int \left[\frac{r}{m^2}\frac{\omega_0^2}{c^2}S_0\left(\frac{\partial}{\partial r}r\Psi\right)^2\right] dr,\\
I_5=\int \left[\frac{r}{m^2}q\left(\frac{\partial}{\partial r}r\Psi\right)^2\right] dr,
\end{array}
\end{equation}
we rewrite Eq.~(\ref{eq5_26}) and Eq.~(\ref{eq5_27}) as:
\begin{equation}\label{eq5_29}
F\left(I_1P_S-I_2\kappa+I_3P_D+I_4P_S-I_5\kappa\right)=-\frac{G}{2}\left(\mu_S I_1+\mu_D I_3+\mu_S I_4\right),
\end{equation}
\begin{equation}\label{eq5_30}
G\left(I_1P_S+I_2\kappa+I_3P_D+I_4P_S+I_5\kappa\right)=-\frac{F}{2}\left(\mu_S I_1+\mu_D I_3+\mu_S I_4\right).
\end{equation}
Recalling Eq.~(\ref{eq5_20}), we take the product of Eq.~(\ref{eq5_29}) and Eq.~(\ref{eq5_30}), and obtain
\begin{equation}\label{eq5_31}
\left(\frac{\omega-\omega_0}{\omega_0}\right)^2=\frac{\left(I_2+I_5\right)^2\kappa^2}{\left[\left(I_1+I_4+I_3\right)x \frac{d}{dx}\rm{ln}\left(\omega^2S\right)+I_3\right]^2}+\frac{\epsilon^2}{4}.
\end{equation}
Multiplying Eq.~(\ref{eq5_23}) by $\Psi(r)$ and integrating over radius lead to 
\begin{equation}\label{eq5_32}
I_1-\frac{q}{4}I_2+I_3+I_4-\frac{q}{4}I_5=0.
\end{equation}
We then rewrite Eq.~(\ref{eq5_31}) as
\begin{equation}\label{eq5_33}
\left(\frac{\omega-\omega_0}{\omega_0}\right)^2=\frac{16\kappa^2/q^2}{\left[x \frac{d}{dx}\rm{ln}\left(\omega^2S\right)+\frac{I_3}{I_1+I_4+I_3}\right]^2}+\frac{\epsilon^2}{4}.
\end{equation}
Equation~(\ref{eq5_33}) is the dispersion relation of SAW propagating in a plasma cylinder with periodic static magnetic field and arbitrary density profiles. 

This dispersion relation can be evaluated with a specific choice of density profile. We investigate a step-like radial profile of plasma density shown in Eq.~(\ref{eq5_11}) to avoid the continuum damping of SAW, which may occur if plasma density changes continuously in radius. Moreover, a realistic density profile gives multiple dispersion curves, the feature of coupling to other modes, which makes it difficult to study the propagating feature of SAW. The aim is to acquire an insight into the problem first. Recalling $x=\omega/\omega_{ci}$, we have:
\begin{equation}\label{eq5_34}
\begin{array}{l}
\vspace{0.3cm}\omega^2S\approx-\frac{\omega^2\omega_{pi}^2}{\omega^2-\omega_{ci}^2}=-\omega_{pi}^2\frac{x^2}{x^2-1},\\
\omega^2D\approx\frac{\omega^3\omega_{pi}^2}{\omega_{ci}(\omega^2-\omega_{ci}^2)}=\omega_{pi}^2\frac{x^3}{x^2-1},
\end{array}
\end{equation}
which leads to
\begin{equation}\label{eq5_35}
x\frac{d}{dx}\rm{ln}\left(\omega^2S\right)=-\frac{2}{\mathit{x}^2-1}.
\end{equation}
Through Eq.~(\ref{eq5_25}), the eigenfunction $\Psi$ corresponding to the step-like density profile is
\begin{equation}\label{eq5_36}
\Psi=\frac{\Psi_0}{r_0}\left\{
\begin{array}{cc}
\left(r/r_0\right)^{|m|-1},&r<r_0\\
\left(r/r_0\right)^{-|m|-1},&r>r_0,
\end{array}
\right.
\end{equation}
and so we evaluate
\begin{equation}\label{eq5_37}
\hspace{-1cm}I_4=\int\left[\frac{1}{m^2}\frac{\omega_0^2}{c^2}S_0 r\left(\frac{\partial}{\partial r}r\Psi\right)\left(\frac{\partial}{\partial r}r\Psi\right)\right]dr=\int\frac{\omega_0^2}{c^2}S_0 r\Psi\Psi dr=I_1.
\end{equation}
Equation~(\ref{eq5_34}) further gives: 
\begin{equation}\label{eq5_38}
\begin{array}{l}
\vspace{0.3cm}\frac{\partial}{\partial r}\left(\omega^2S\right)=-\frac{x^2}{x^2-1}\left(\omega_{pi+}^2-\omega_{pi-}^2\right)\delta\left(r-r_0\right),\\
\frac{\partial}{\partial r}\left(\omega^2D\right)=\frac{x^3}{x^2-1}\left(\omega_{pi+}^2-\omega_{pi-}^2\right)\delta\left(r-r_0\right).
\end{array}
\end{equation}
We then calculate $I_1$ and $I_3$:
\begin{equation}\label{eq5_39}
\begin{array}{l}
\vspace{0.3cm}I_1=-\frac{x^2}{x^2-1}\frac{1}{c^2}\frac{\Psi_0^2}{2|m|}\left(\omega_{pi+}^2+\omega_{pi-}^2\right),\\
I_3=-\frac{x^3}{x^2-1}\frac{1}{c^2}\frac{\Psi_0^2}{m}\left(\omega_{pi+}^2-\omega_{pi-}^2\right),
\end{array}
\end{equation}
which further gives
\begin{equation}\label{eq5_40}
\frac{I_3}{I_1+I_3+I_4}=\frac{I_3}{2I_1+I_3}=\frac{x}{x+\frac{m}{|m|}\frac{\omega_{pi+}^2+\omega_{pi-}^2}{\omega_{pi+}^2-\omega_{pi-}^2}}. 
\end{equation}
Substituting Eq.~(\ref{eq5_35}) and Eq.~(\ref{eq5_40}) into Eq.~(\ref{eq5_33}), we get the dispersion relation of SAW for a step-like radial profile of plasma density
\begin{equation}\label{eq5_41}
\left(\frac{\omega-\omega_0}{\omega_0}\right)^2=\frac{16\kappa^2/q^2}{\gamma^2}+\frac{\epsilon^2}{4}
\end{equation}
with
\[\gamma=\frac{x}{x-\frac{m}{|m|}\frac{1}{\Psi}}-\frac{2}{x^2-1}.\]
When $|\omega-\omega_0|/\omega_0<\epsilon/2$, $\kappa$ is imaginary and waves are evanescent along the plasma column, forming a spectral gap. The lower ($\omega_-$) and upper ($\omega_+$) edges of the gap can be calculated by setting $\kappa=0$ in Eq.~(\ref{eq5_41}): 
\begin{equation}\label{eq5_53}
\omega_-=\omega_0(1-\frac{\epsilon}{2}),~\omega_+=\omega_0(1+\frac{\epsilon}{2}). 
\end{equation}
The gap width is thereby $\bigtriangleup\omega=\omega_+-\omega_-=\epsilon\omega_0$, which is the same to that of RLH waves (see Eq.~\ref{eq4_19}). When $|\omega-\omega_0|/\omega_0>\epsilon/2$, $\kappa$ is real and waves are propagable along the plasma column, forming the spectral continuum. 

\subsection{Defective magnetic field: gap eigenmode}\label{def5}
Now we introduce a defect to the system's periodicity, similar to that employed in Chapter~\ref{chp4}, to create the gap eigenmode of SAW inside the spectral gap indicated by Eq.~(\ref{eq5_33}) and Eq.~(\ref{eq5_41}). We assume that the gap eigenmode will decay away from the defect location as
\begin{equation}\label{eq5_42}
E_\varphi=F(z)\Psi(r) e^{iqz/2-\lambda z}+G(z)\Psi(r) e^{-iqz/2-\lambda z},
\end{equation}
with $\lambda$ the decay constant. Similarly to Eqs.~(\ref{eq5_25})-(\ref{eq5_30}) but replacing $\kappa$ with $i\lambda$, we get:
\begin{equation}\label{eq5_43}
F\left(I_1P_S-i\lambda I_2+I_3P_D+I_4P_S-i\lambda I_5\right)=-\frac{G}{2}\left(\mu_S I_1+\mu_D I_3+\mu_S I_4\right),
\end{equation}
\begin{equation}\label{eq5_44}
G\left(I_1P_S+i\lambda I_2+I_3P_D+I_4P_S+i\lambda I_5\right)=-\frac{F}{2}\left(\mu_S I_1+\mu_D I_3+\mu_S I_4\right).
\end{equation}
The system is completed with a boundary condition for $E_\varphi$. An odd-parity gap eigenmode can be formed by applying the boundary condition $E_\varphi(r, z_0)=0$, which gives $F=-G \rm{\exp}(-i q z_0)$. The sum and difference of Eq.~(\ref{eq5_43}) and Eq.~(\ref{eq5_44}) are:
\begin{equation}\label{eq5_45}
I_1P_S+I_3P_D+I_4P_S=-\frac{1}{2}\left(\mu_S I_1+\mu_D I_3+\mu_S I_4\right)\cos(qz_0),
\end{equation}
\begin{equation}\label{eq5_46}
\lambda(I_2+I_5)=-\frac{1}{2}\left(\mu_S I_1+\mu_D I_3+\mu_S I_4\right)\sin(qz_0), 
\end{equation}
respectively. By using Eq.~(\ref{eq5_20}) and Eq.~(\ref{eq5_32}), we then solve for the frequency and decay constant of the odd-parity gap eigenmode: 
\begin{equation}\label{eq5_47}
\frac{\omega-\omega_0}{\omega_0}
=-\frac{\epsilon}{2}\cos(qz_0),
\end{equation}
\begin{equation}\label{eq5_48}
\lambda=\frac{\epsilon}{8}q\left[x\frac{d}{dx}\rm{ln}\left(\omega^2S\right)+\frac{I_3}{I_1+I_4+I_3}\right]\sin(qz_0).
\end{equation}
Similarly, an even-parity gap eigenmode can be formed by applying the boundary condition $E_\varphi'(r, z_0)=0$ (with $'$ over $z$), which gives $F=G\rm{\exp}(-i q z_0)$. The corresponding frequency and decay constant of the even-parity gap eigenmode are:
\begin{equation}\label{eq5_49}
\frac{\omega-\omega_0}{\omega_0}
=\frac{\epsilon}{2}\cos(qz_0),
\end{equation}
\begin{equation}\label{eq5_50}
\lambda=-\frac{\epsilon}{8}q\left[x\frac{d}{dx}\rm{ln}\left(\omega^2S\right)+\frac{I_3}{I_1+I_4+I_3}\right]\sin(qz_0).
\end{equation}
By using Eq.~(\ref{eq5_35}) and Eq.~(\ref{eq5_40}), Eq.~(\ref{eq5_48}) and Eq.~(\ref{eq5_50}) can be evaluated for the step-like radial profile of plasma density shown in Eq.~(\ref{eq5_11}):
\begin{equation}\label{eq5_51}
\lambda=\frac{\epsilon}{8}q\gamma\sin(qz_0),
\end{equation}
\begin{equation}\label{eq5_52}
\lambda=-\frac{\epsilon}{8}q\gamma\sin(qz_0),
\end{equation}
for the odd-parity and even-parity gap eigenmodes, respectively. 

\section{Numerical results}\label{num5}
\begin{figure}[ht]
\begin{center}
\hspace{0.6cm}\includegraphics[width=0.8\textwidth,angle=0]{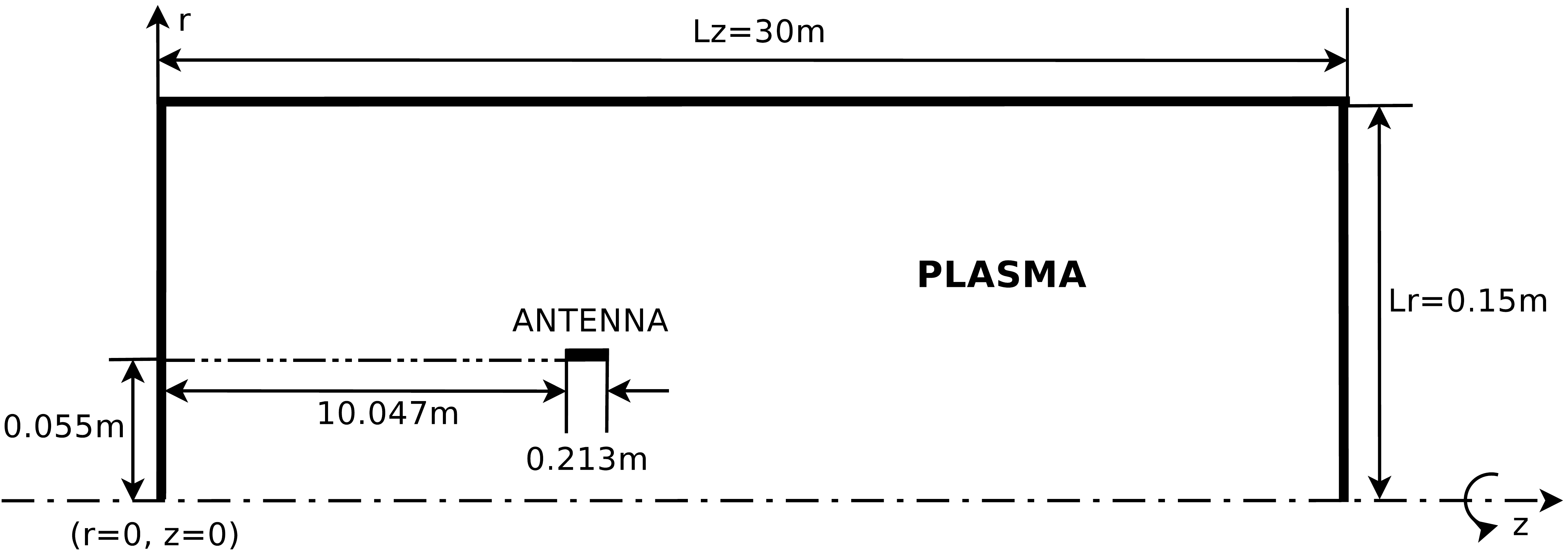}
\end{center}
\caption{Computational domain. The solid bar denotes a left-hand half-turn helical antenna. The dot-dashed line is the machine and coordinate system axis ($r=0$). The coordinate system ($r$, $\varphi$, $z$) is right-handed with $\varphi$. The double-dot-dashed line labels the inner radius of the antenna. }
\label{fg5_1}
\end{figure}

To validate the theoretical analysis in Sec.~\ref{thy5}, numerical computations are conducted in this section. We use the EMS code\cite{Chen:2006aa} to model the spectral gap and gap eigenmode of SAW. Figure~\ref{fg5_1} shows the computational domain, which is similar to that in Chapter~\ref{chp4} but with glass tube removed.
This will, for the step-like density profile employed below, ensure that there is only one density discontinuity in the whole chamber, and it is the density jump constructed. The enclosing chamber is assumed to be ideally conducting, and the RF antenna is placed inside the plasma. To well couple SAW, which are left-hand circularly polarised, the RF antenna is twisted in the left-hand direction to the static magnetic field, and it is half-turn helical to mainly excite non-axisymmetric modes, as considered in the analysis. We choose step-like radial profiles of plasma density to precisely compare with the analytical results in Sec.~\ref{thy5}, and eliminate potential continuum damping of SAW. Illustrative results for broadened density profiles will be given in Sec.~\ref{cnt5} 

\begin{figure}[ht]
\begin{center}$
\begin{array}{ll}
(a)&(b)\\
\includegraphics[width=0.45\textwidth,angle=0]{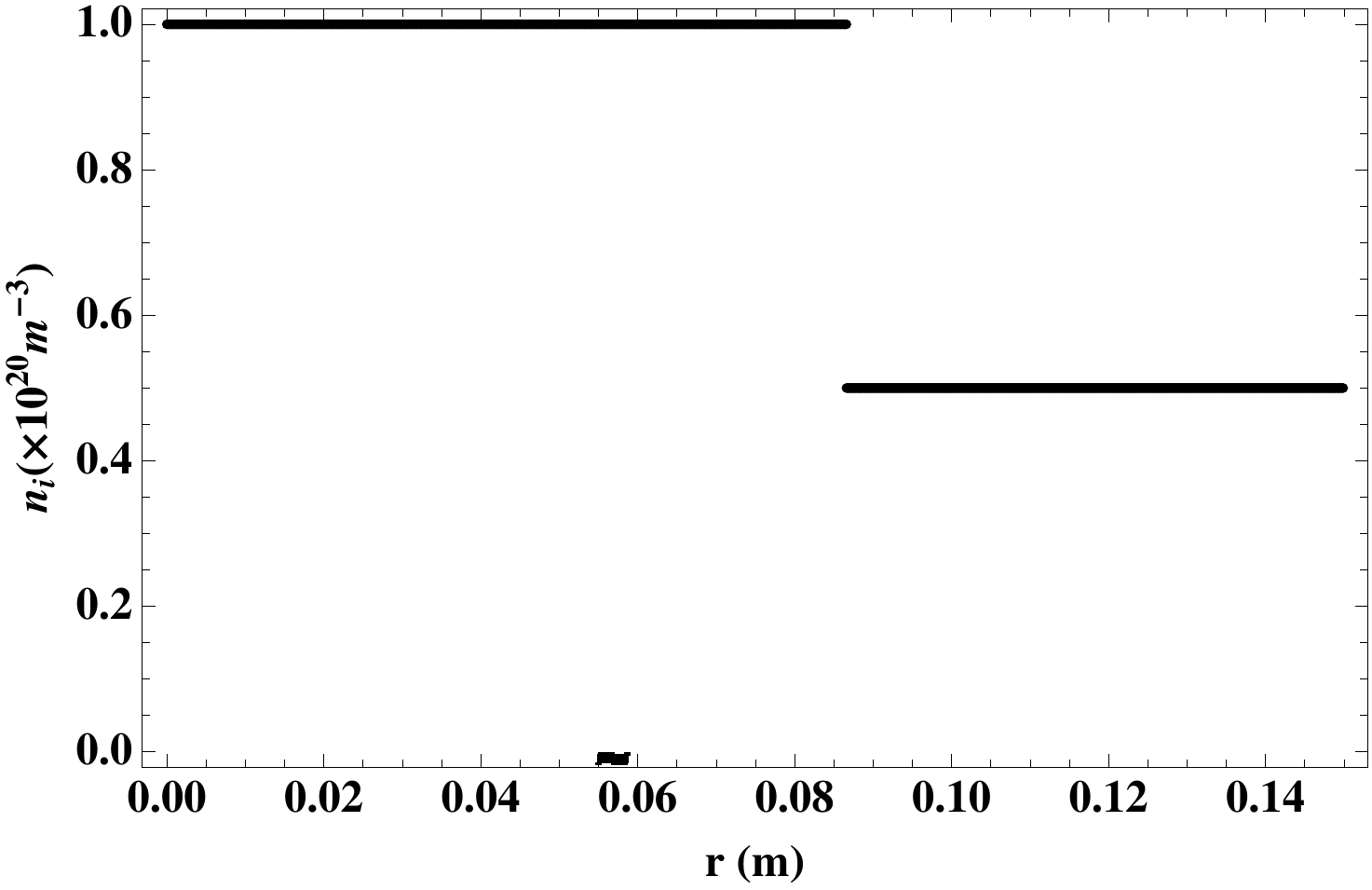}&\includegraphics[width=0.47\textwidth,angle=0]{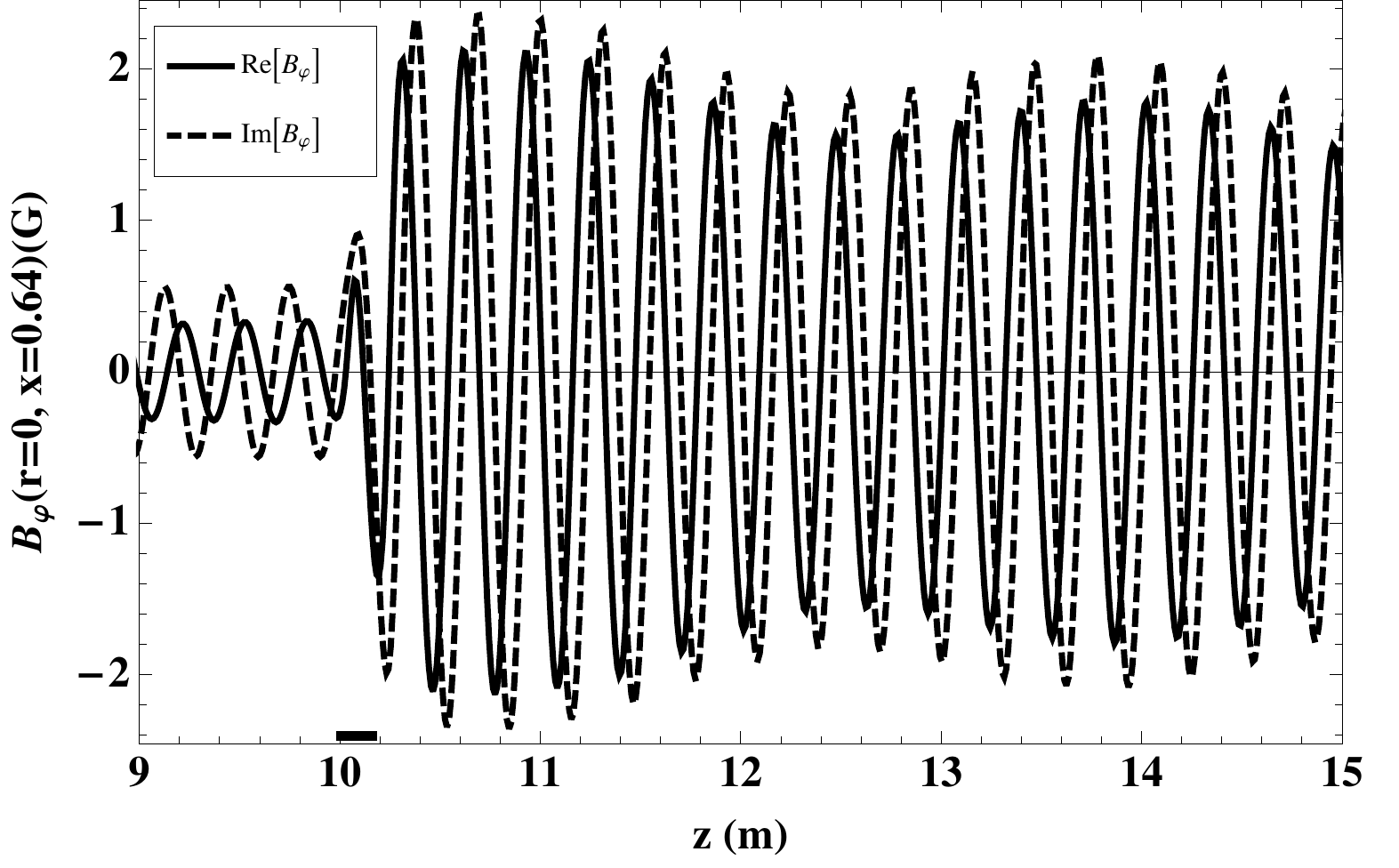}\\
(c)&(d)\\
\hspace{0.18cm}\includegraphics[width=0.44\textwidth,angle=0]{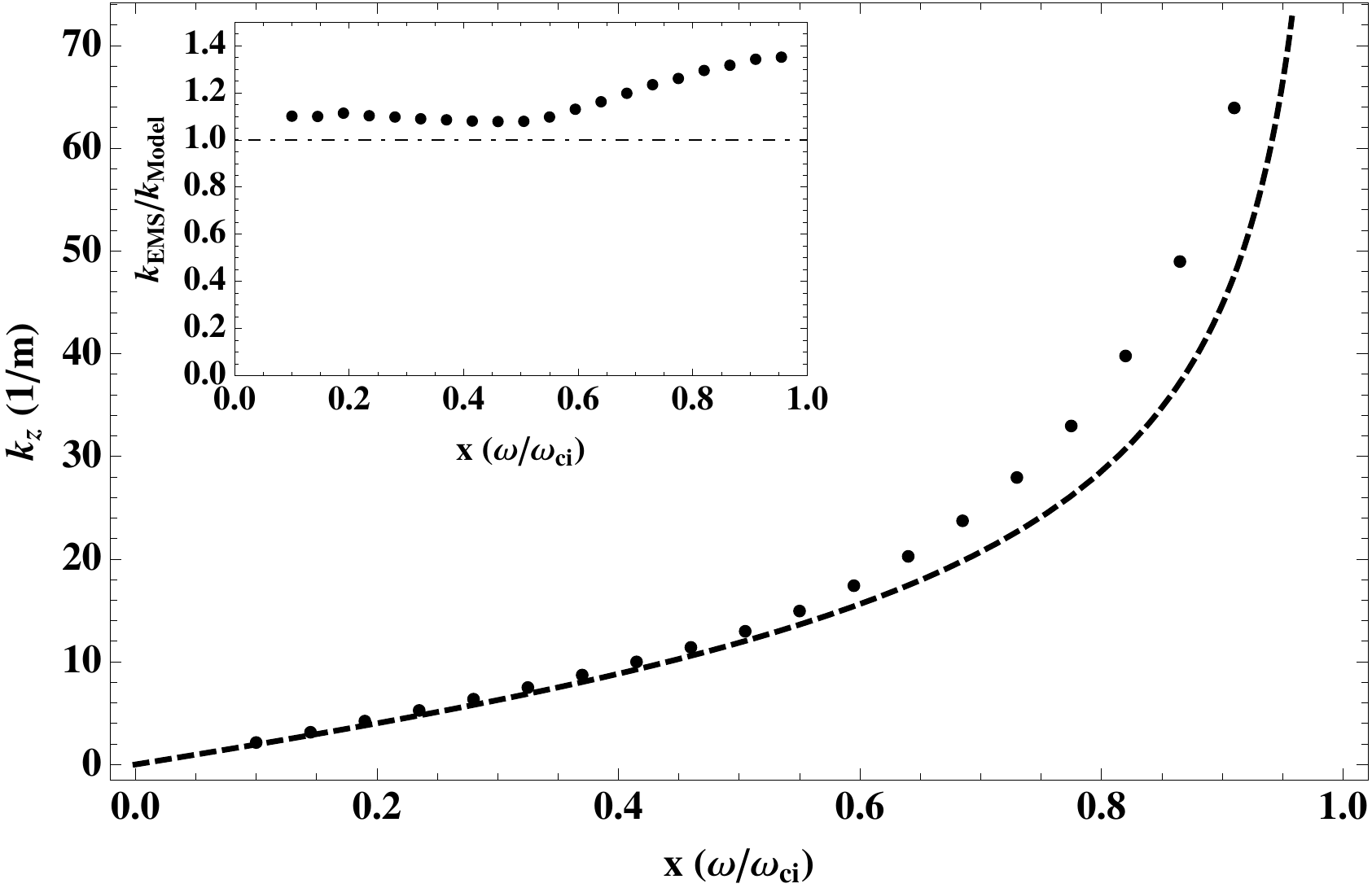}&\includegraphics[width=0.47\textwidth,angle=0]{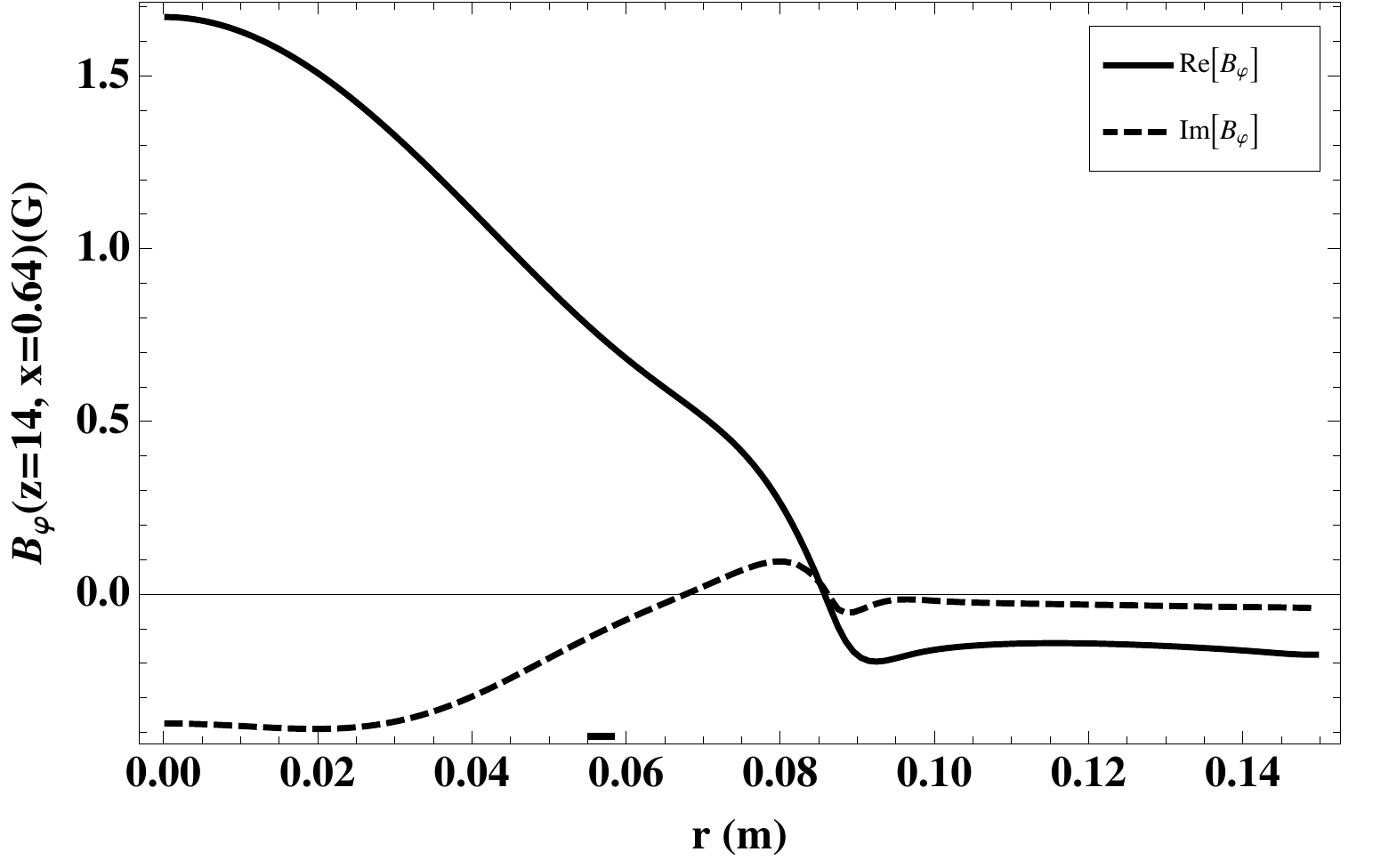}
\end{array}$
\end{center}
\caption{Plasma density profile and computed wave field in a straight plasma cylinder: (a) radial profile of unperturbed plasma density, (b) azimuthal magnetic field of the $m=-1$ mode for an illustrative frequency at $x=0.64$, (c) computed dispersion relation for the $m=-1$ mode (dots) and analytical result from Eq.~(\ref{eq5_18}) (dashed), together with the ratio between them (inset figure), (d) radial profile of the $m=-1$ mode at $z = 14$~m for $x=0.64$. Solid bar shows the antenna location, and the endplates are located at $z=0$~m and $z=30$~m.}
\label{fg5_2}
\end{figure}

For a step-like-peak density profile, as shown in Fig.~\ref{fg5_2}(a), the discontinuity is chosen at $r_0=0.087$~m so as to ensure the total number of particles in each volume is the same. 
The on-axis plasma density is chosen to be $10^{20}~\rm{m}^{-3}$, which is sufficiently dense as assumed for Eq.~(\ref{eq5_3}). We consider a single-ionised helium plasma, as did by Zhang et al.,\cite{Zhang:2008aa} so that the on-axis electron-ion collision frequency is $\nu_{ei}(0)=4\times 10^8~\rm{s}^{-1}$. The wave field structure and dispersion relation of SAW are studied in a plasma cylinder with uniform static magnetic field.  Figure~\ref{fg5_2}(b) and Fig.~\ref{fg5_2}(d) show the axial and radial profiles of the computed wave field, respectively, for an illustrative frequency at $x=0.64$. A preferred axial coupling direction can be seen clearly from Fig.~\ref{fg5_2}(b) due to the helical antenna being used. Figure~\ref{fg5_2}(c) shows the computed dispersion relation and its comparison with that from Eq.~(\ref{eq5_18}). The inset figure shows the ratio between the numerical and analytical results. The computed data were obtained by running the EMS for various frequencies, for the same conditions, and calculating the dominant axial mode number through a Fourier decomposition.  We can see that they qualitatively agree but diverge more when the wave frequency approaches the ion cyclotron frequency. This may be due to that the computed values of $k$ and $\omega$, which become bigger when $\omega\rightarrow\omega_{ci}$, are not sufficiently small as assumed in the theoretical analysis (see Eq.~(\ref{eq5_9})). 

\begin{figure}[ht]
\begin{center}$
\vspace{-0.3cm}
\begin{array}{ll}
(a)&(b)\\
\includegraphics[width=0.47\textwidth,angle=0]{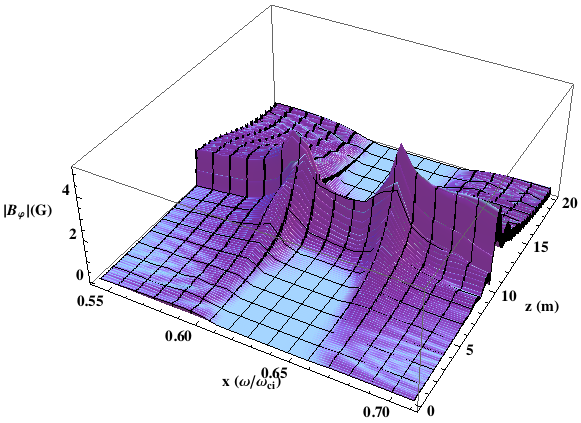}&\includegraphics[width=0.47\textwidth,angle=0]{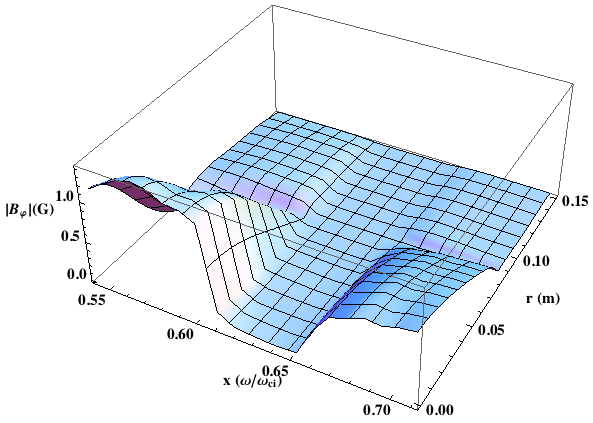}
\end{array}$
\end{center}
\caption{Surface plots of the wave field strength as a function of driving frequency: (a) in the axial direction ($r=0$~m), (b) in the radial direction ($z=14$~m). }
\label{fg5_3}
\end{figure}

Next, we introduce a small periodic modulation to the uniform magnetic field, $[B_0(z)-B_0]/B_0=0.1\cos(40z)$, to form a spectral gap of SAW. The periodicity $q=40$ was chosen to locate the gap within the frequency range of $0.5<x<0.9$ where SAW have a long decay length. Running EMS for the same conditions as above, we obtain a clear spectral gap in both axial and radial directions, as shown in Fig.~\ref{fg5_3}.
The gap width ($\bigtriangleup x\approx 0.067$) agrees well with the analytical estimate from Eq.~(\ref{eq5_53}) ($\bigtriangleup \omega/\omega_{ci0}=\epsilon x_0=0.069$), although the gap location ($x_0=0.63$) is slightly lower than that $x_0(k_z=40/2)=0.69$ from Eq.~(\ref{eq5_18}).

\begin{figure}[ht]
\begin{center}$
\begin{array}{ll}
(a)&(b)\\
\includegraphics[width=0.485\textwidth,angle=0]{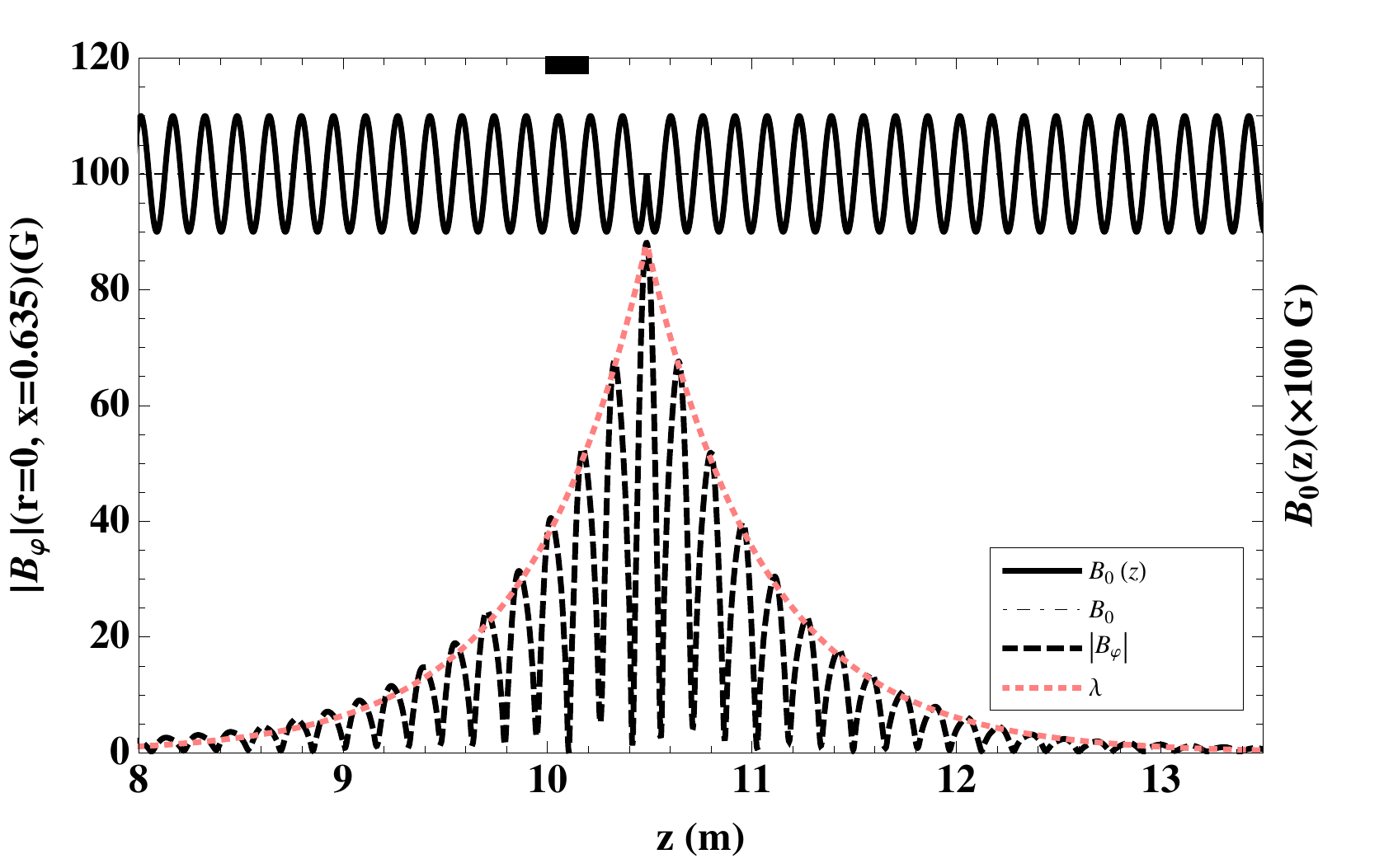}&\includegraphics[width=0.465\textwidth,angle=0]{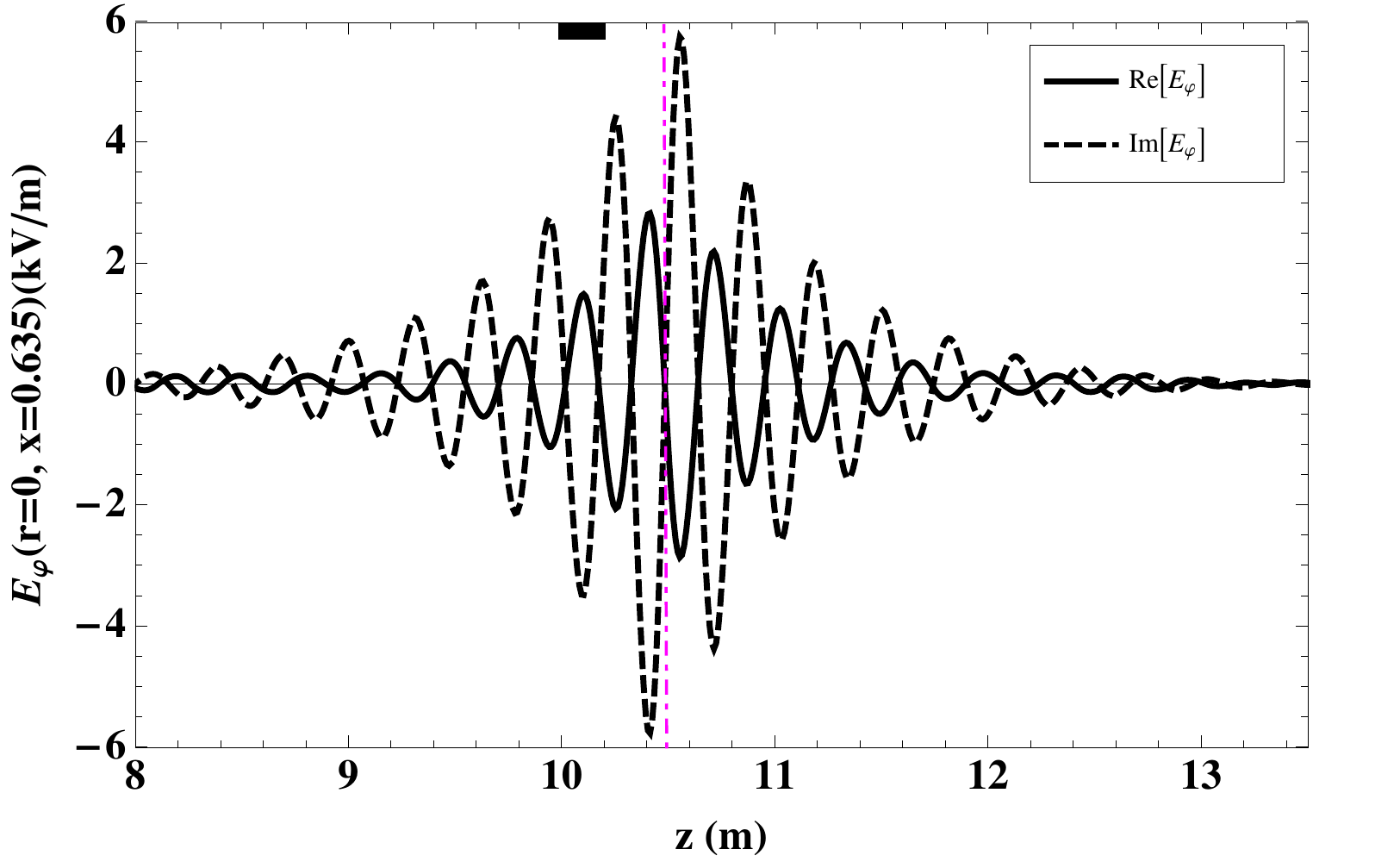}\\
(c)&(d)\\
\hspace{0.1cm}\includegraphics[width=0.437\textwidth,angle=0]{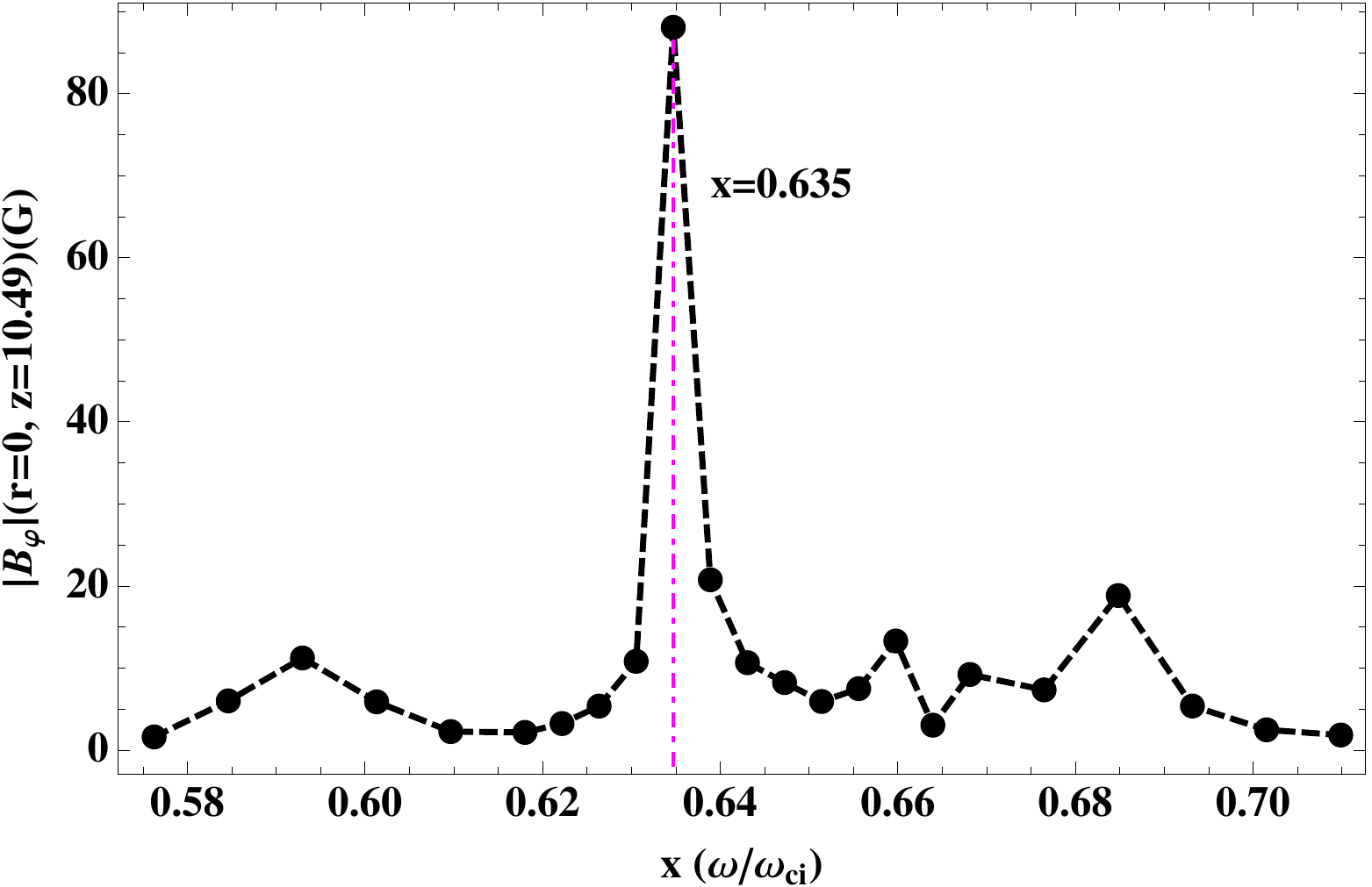}&\includegraphics[width=0.45\textwidth,angle=0]{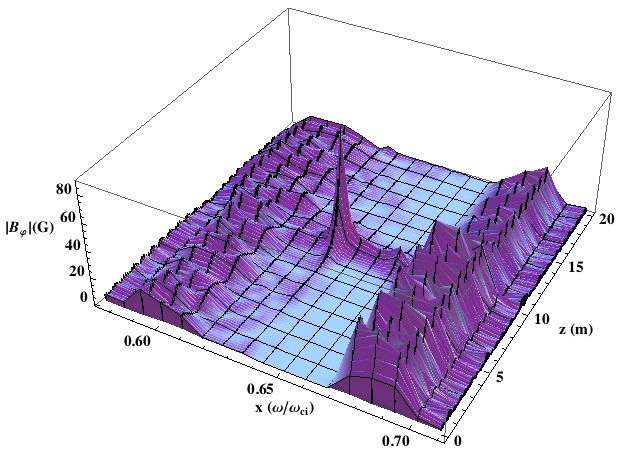}\\
\end{array}$
\end{center}
\caption{Odd-parity gap eigenmode: (a) longitudinal profiles of static magnetic field (solid line) and on-axis RF magnetic field for $x=0.635$ (dashed line), together with the decay constant from Eq.~(\ref{eq5_51}) (dotted); (b) longitudinal profile of $E_\varphi$ (on-axis) at $x=0.635$, where vertical dot-dashed line marks the defect location and solid horizontal bar marks the antenna; (c) resonance in the dependence of the on-axis amplitude of the RF magnetic field on driving frequency at the location of the defect; (d) surface plot of the on-axis wave field strength as a function of $z$ and $x$. }
\label{fg5_4}
\end{figure}

\begin{figure}[ht]
\begin{center}$
\begin{array}{ll}
(a)&(b)\\
\includegraphics[width=0.48\textwidth,angle=0]{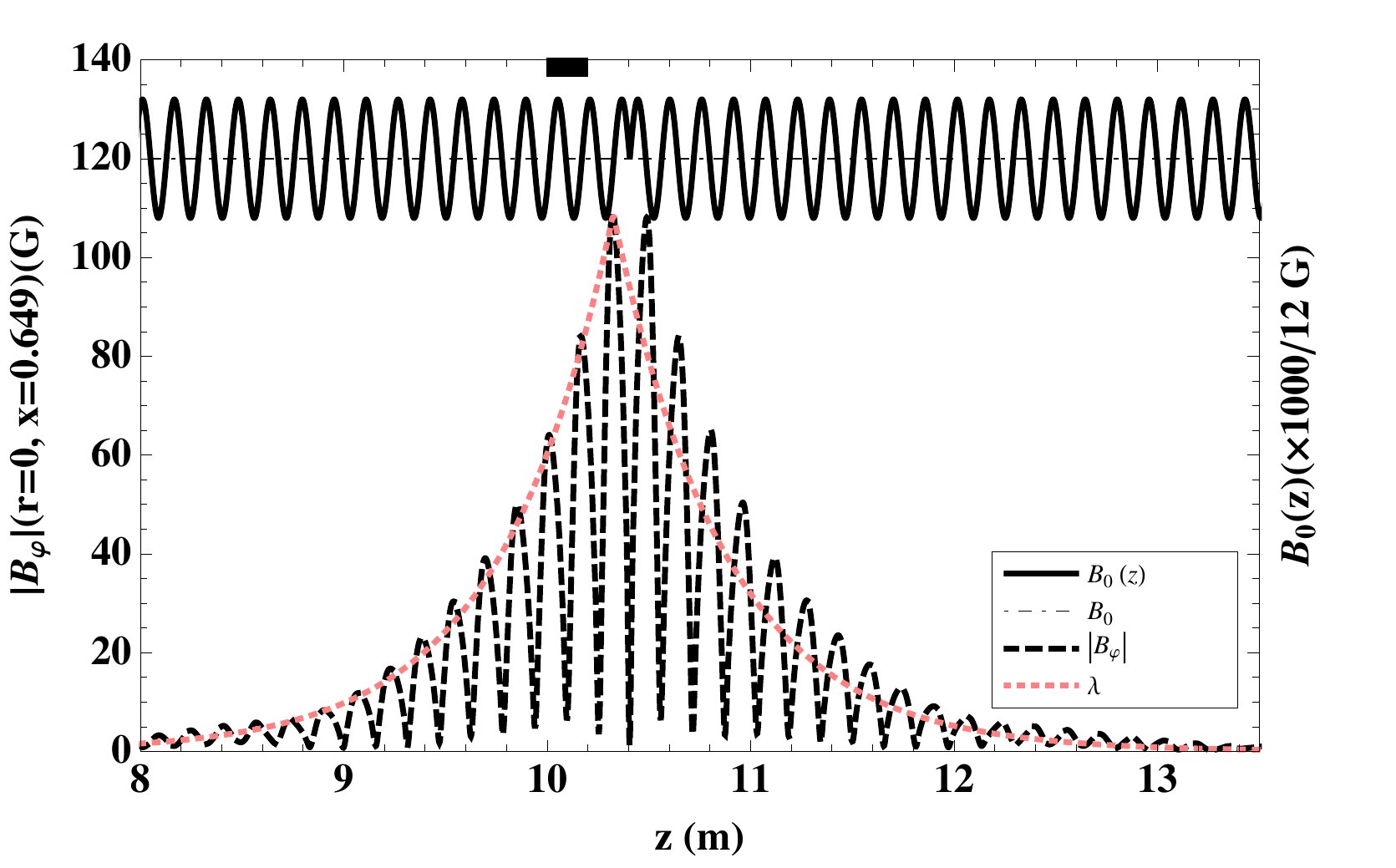}&\includegraphics[width=0.455\textwidth,angle=0]{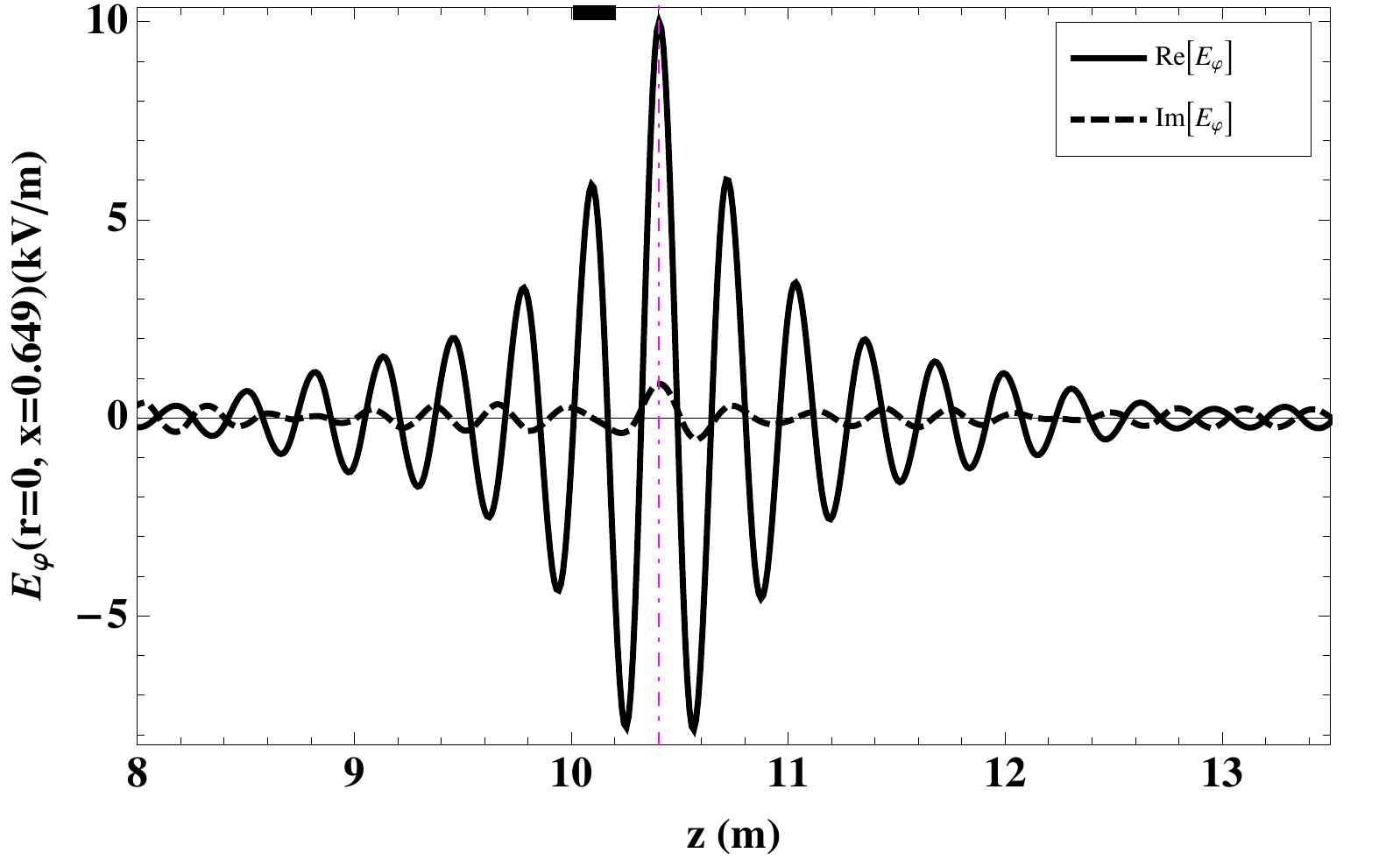}\\
(c)&(d)\\
\hspace{0.03cm}\includegraphics[width=0.434\textwidth,angle=0]{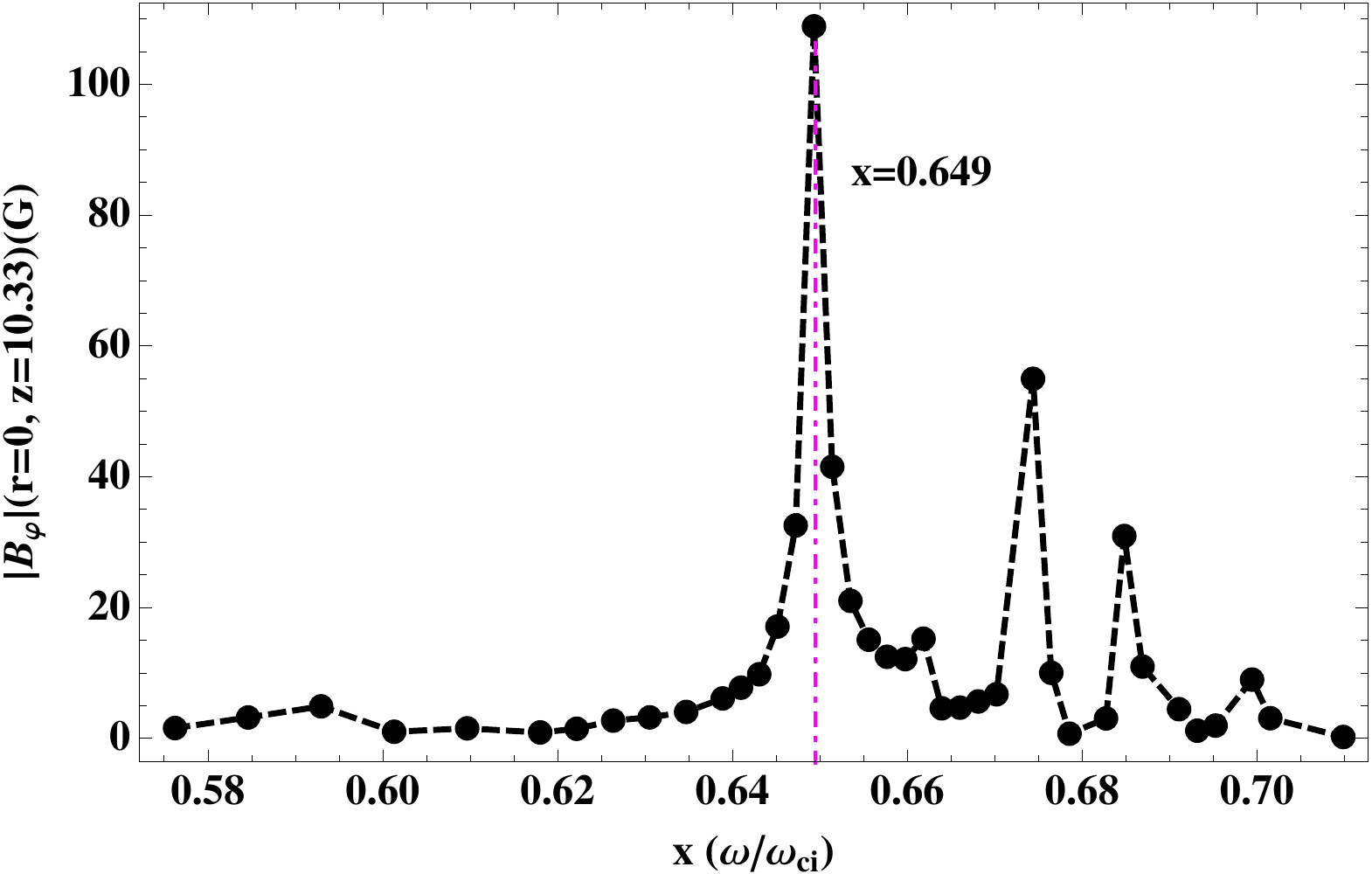}&\includegraphics[width=0.45\textwidth,angle=0]{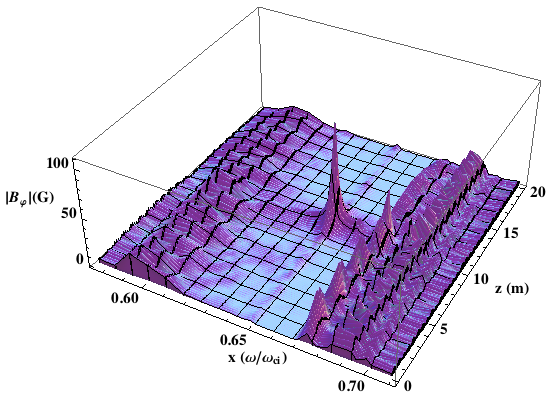}\\
\end{array}$
\end{center}
\caption{Even-parity gap eigenmode: (a) longitudinal profiles of static magnetic field (solid line) and on-axis RF magnetic field for $x=0.649$ (dashed line), together with the decay constant from Eq.~(\ref{eq5_52}) (dotted); (b) longitudinal profile of $E_\varphi$ (on-axis) at $x=0.649$, where vertical dot-dashed line marks the defect location and solid horizontal bar marks the antenna; (c) resonance in the dependence of the on-axis amplitude of the RF magnetic field on driving frequency near the location of the defect; (d) surface plot of the on-axis wave field strength as a function of $z$ and $x$. }
\label{fg5_5}
\end{figure}

Section~\ref{def5} indicates that an eigenmode could be formed inside the spectral gap if a defect is introduced to break the system's periodicity. Moreover, the defect should be located at $\cos(q z_0)=0$ in order to form the eigenmode with frequency at the gap centre (see Eq.~(\ref{eq5_47}) and Eq.~(\ref{eq5_49})) and with shortest possible width (see Eq.~(\ref{eq5_51}) and Eq.~(\ref{eq5_52})). Further, the gap eigenmode can be either odd-parity or even-parity around the defect location, depending upon the defect configurations. Figure~\ref{fg5_4} and Fig.~\ref{fg5_5} show the defect configurations employed and the corresponding odd-parity and even-parity gap eigenmodes formed.
To minimise the role of collisional dissipation and thus get a sharper resonant peak (see Fig.~\ref{fg5_4}(c) and Fig.~\ref{fg5_5}(c)), the collision frequency has been reduced to $0.0005\nu_{ei}$. Figure~\ref{fg5_4}(a) and Fig.~\ref{fg5_5}(a) show that the eigenmodes are standing waves and their wavelengths are close to twice the system's periodicity, which is characteristic of Bragg's reflection. The decay lengths indicated by the exponential envelopes agree well with the analytical estimates from Eq.~(\ref{eq5_51}) and Eq.~(\ref{eq5_52}). The odd-parity and even-parity features are indicated by the axial profiles of the azimuthal electric field around the defect location, shown in Fig.~\ref{fg5_4}(b) and Fig.~\ref{fg5_5}(b), respectively. Figure~\ref{fg5_4}(d) and Fig.~\ref{fg5_5}(d) give full views of the plasma response in ($x, z$) space with a clear eigenmode peak inside the spectral gap. 

\begin{figure}[ht]
\begin{center}$
\begin{array}{ll}
(a)&(b)\\
\includegraphics[width=0.458\textwidth,angle=0]{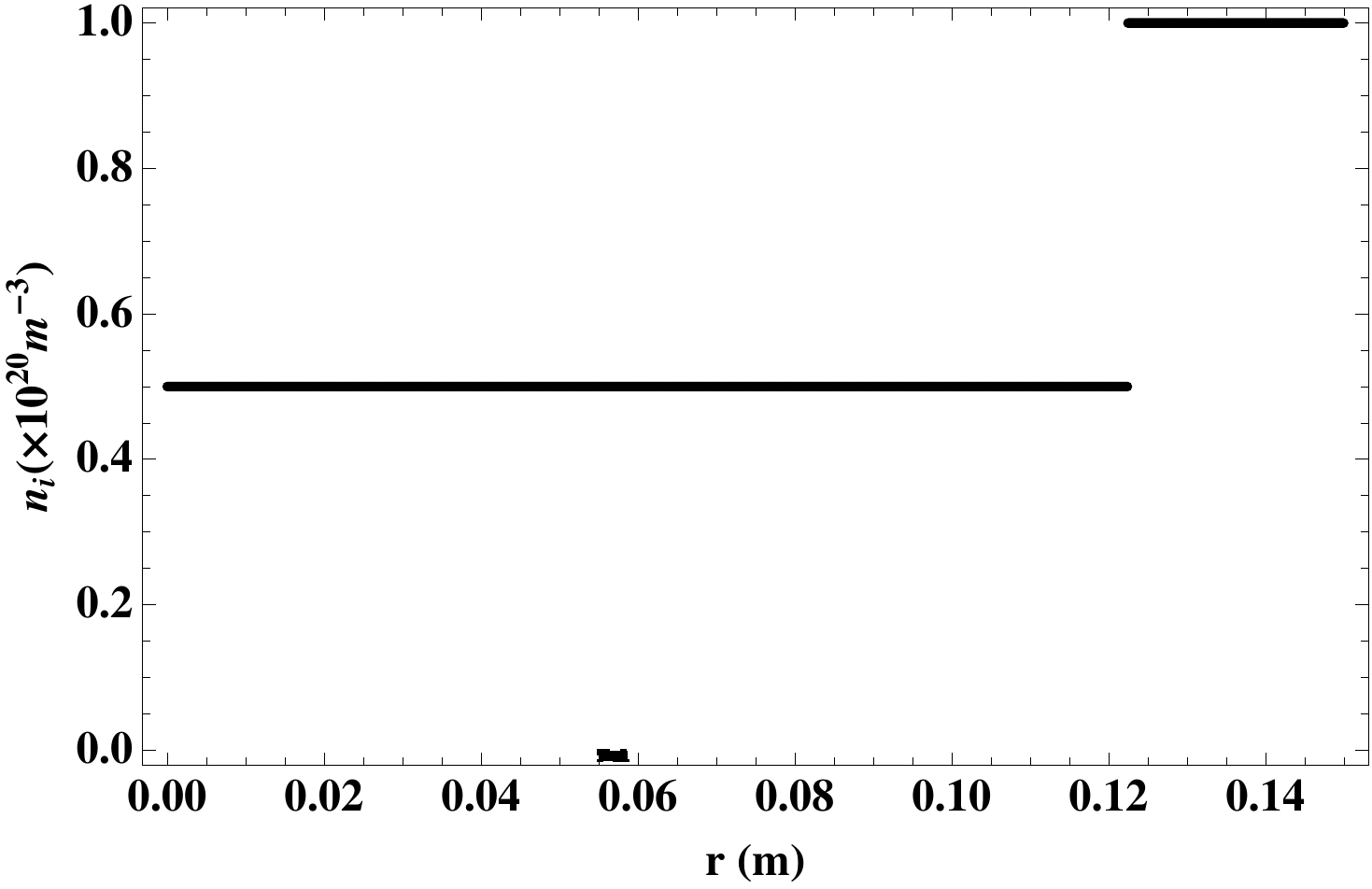}&\includegraphics[width=0.48\textwidth,angle=0]{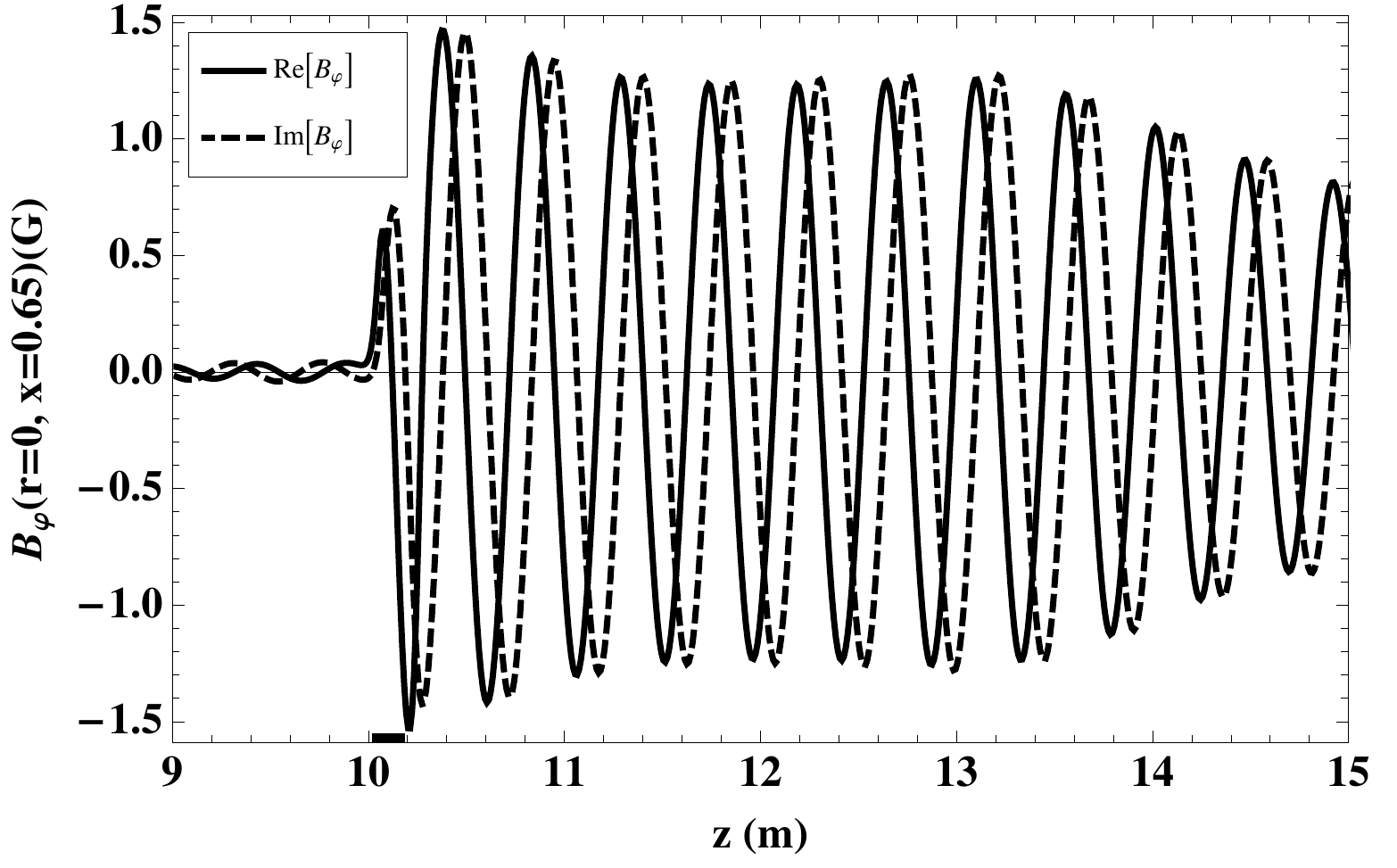}\\
(c)&(d)\\
\hspace{0.19cm}\includegraphics[width=0.447\textwidth,angle=0]{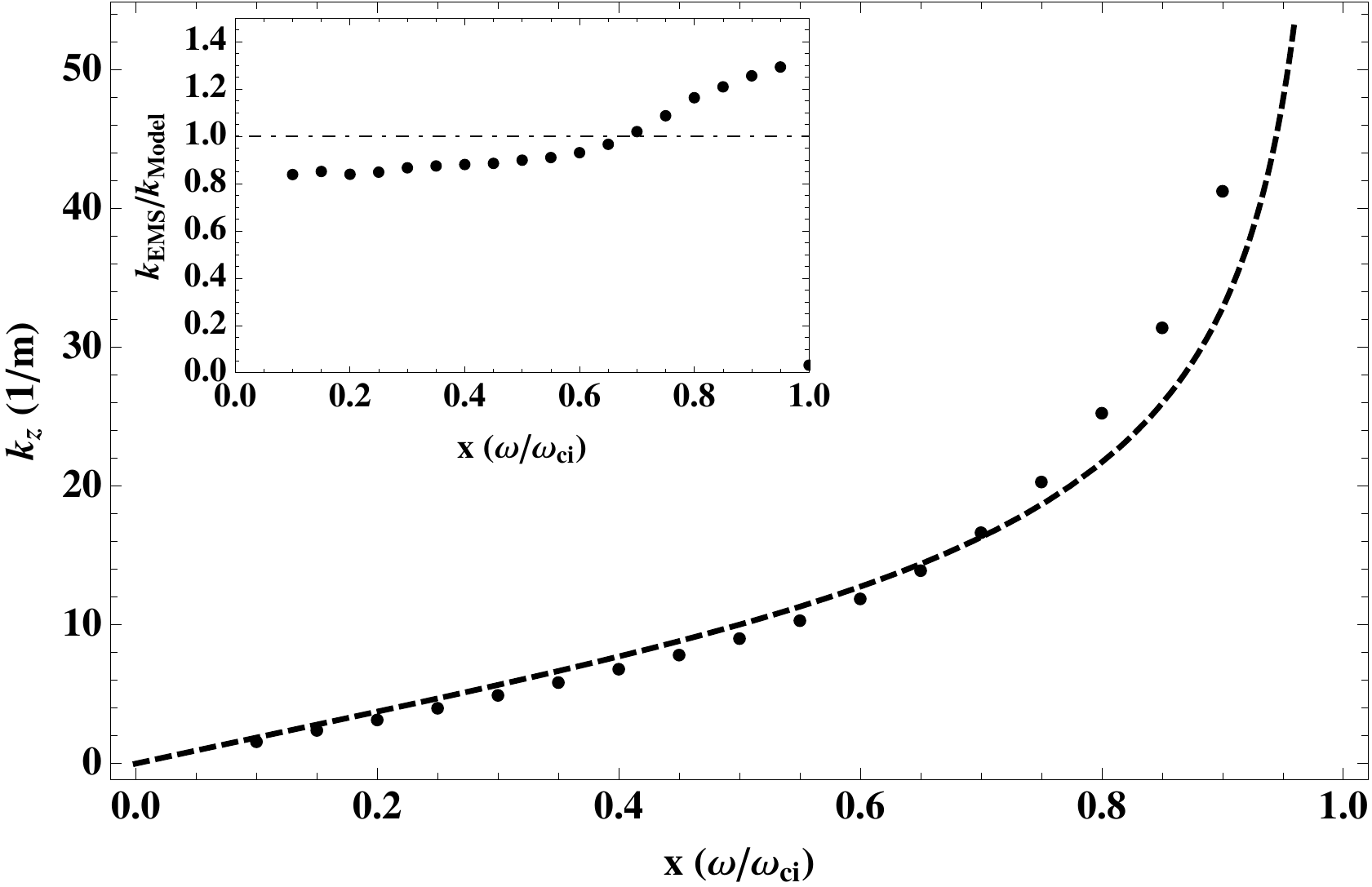}&\includegraphics[width=0.475\textwidth,angle=0]{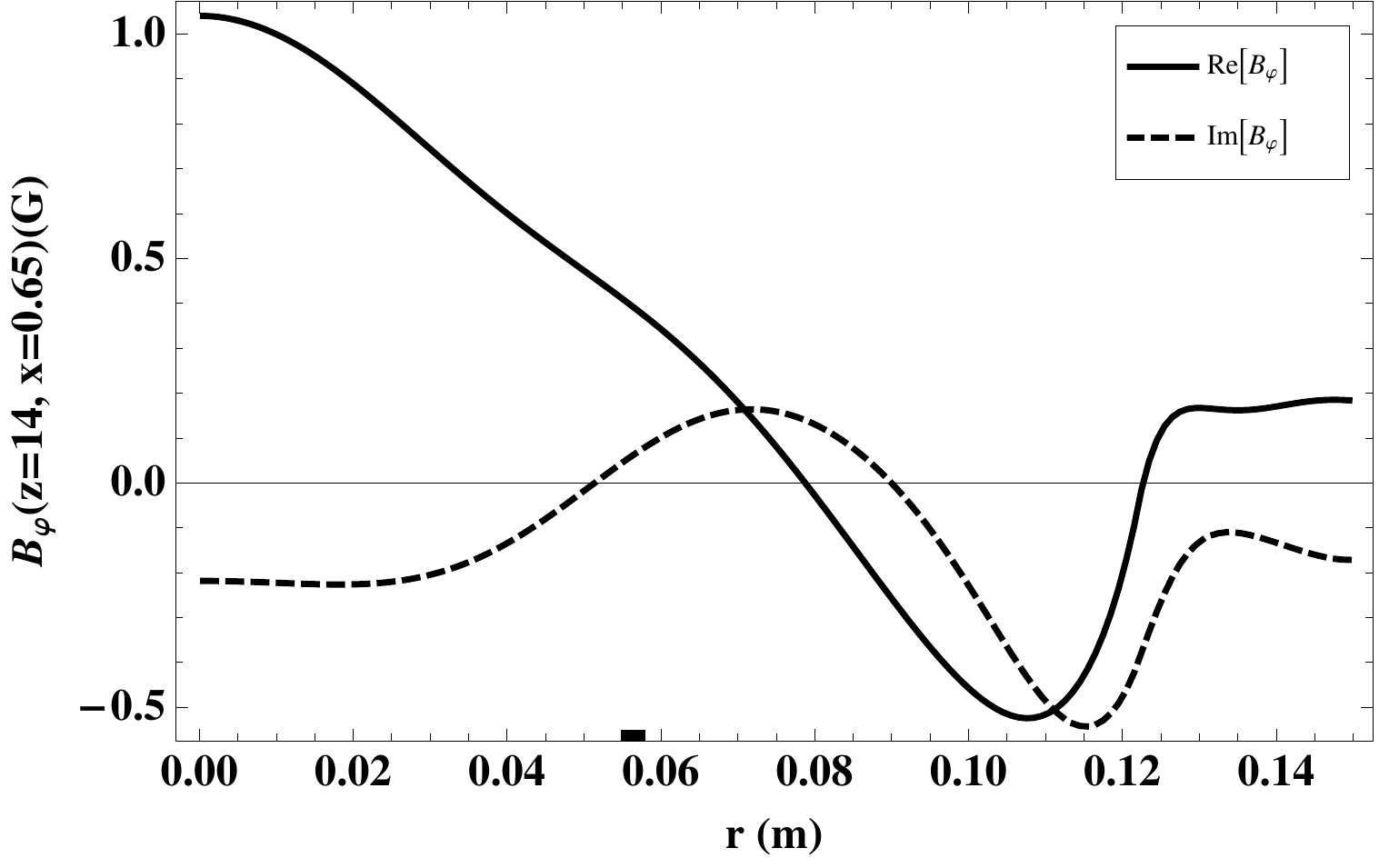}
\end{array}$
\end{center}
\caption{Plasma density profile and computed wave field in a straight plasma cylinder: (a) radial profile of unperturbed plasma density, (b) azimuthal magnetic field of the $m=-1$ mode for an illustrative frequency at $x=0.65$, (c) computed dispersion relation for the $m=-1$ mode (dots) and analytical result from Eq.~(\ref{eq5_18}) (dashed), together with the ratio between them (inset figure), (d) radial profile of the $m=-1$ mode at $z = 14$~m for $x=0.65$. Solid bar shows the antenna location, and the endplates are located at $z=0$~m and $z=30$~m.}
\label{fg5_6}
\end{figure}

\begin{figure}[ht]
\begin{center}$
\vspace{-0.3cm}
\begin{array}{l}
\hspace{2.3cm}(a)\\
\vspace{-0.3cm}\hspace{2.3cm}\includegraphics[width=0.655\textwidth,angle=0]{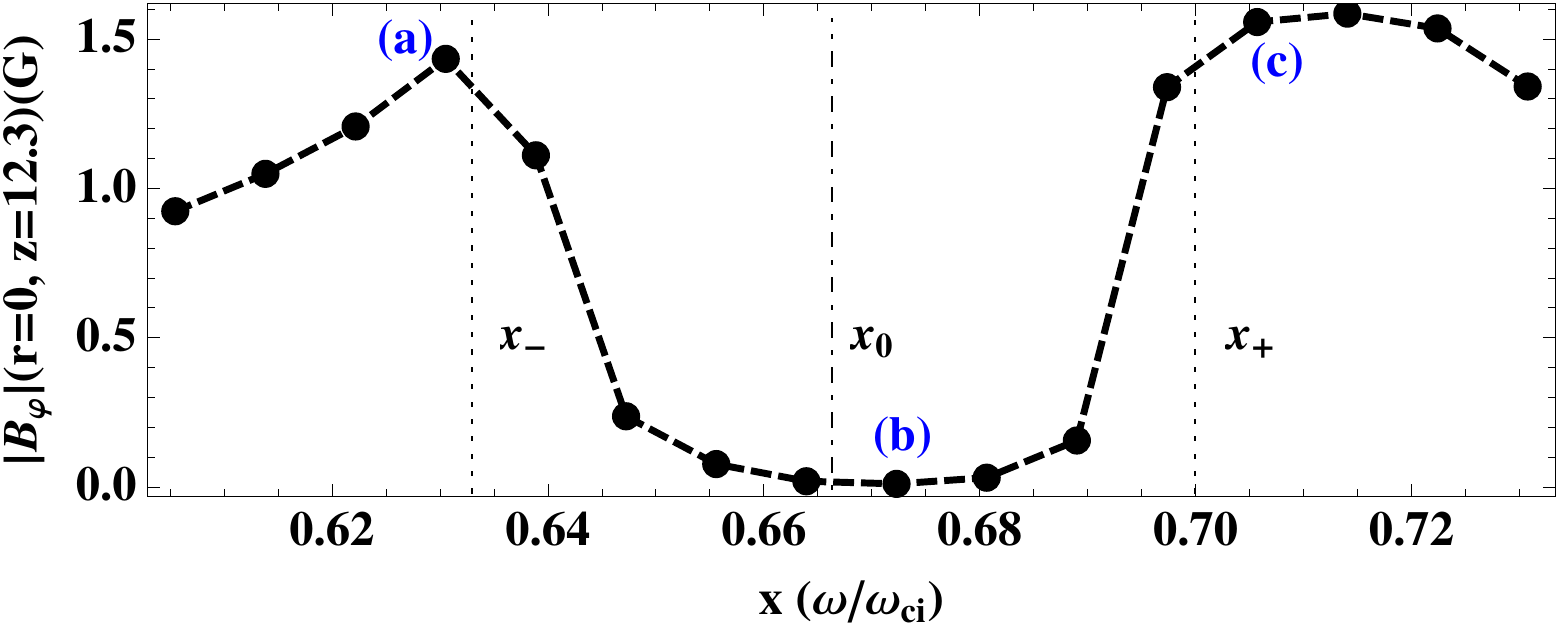}\\
\hspace{2.3cm}(b)\\
\vspace{-0.3cm}\hspace{2.2cm}\includegraphics[width=0.7\textwidth,height=0.2\textwidth,angle=0]{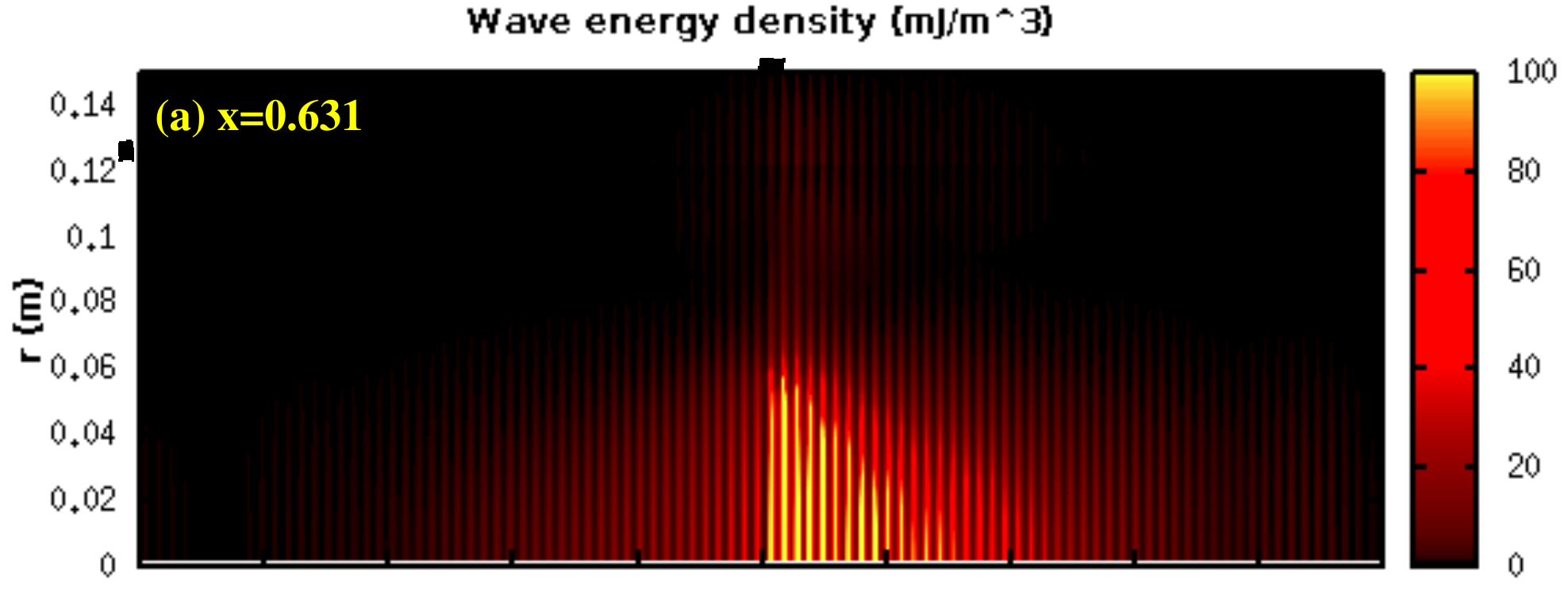}\\
\vspace{-0.3cm}\hspace{2.2cm}\includegraphics[width=0.7\textwidth,height=0.2\textwidth,angle=0]{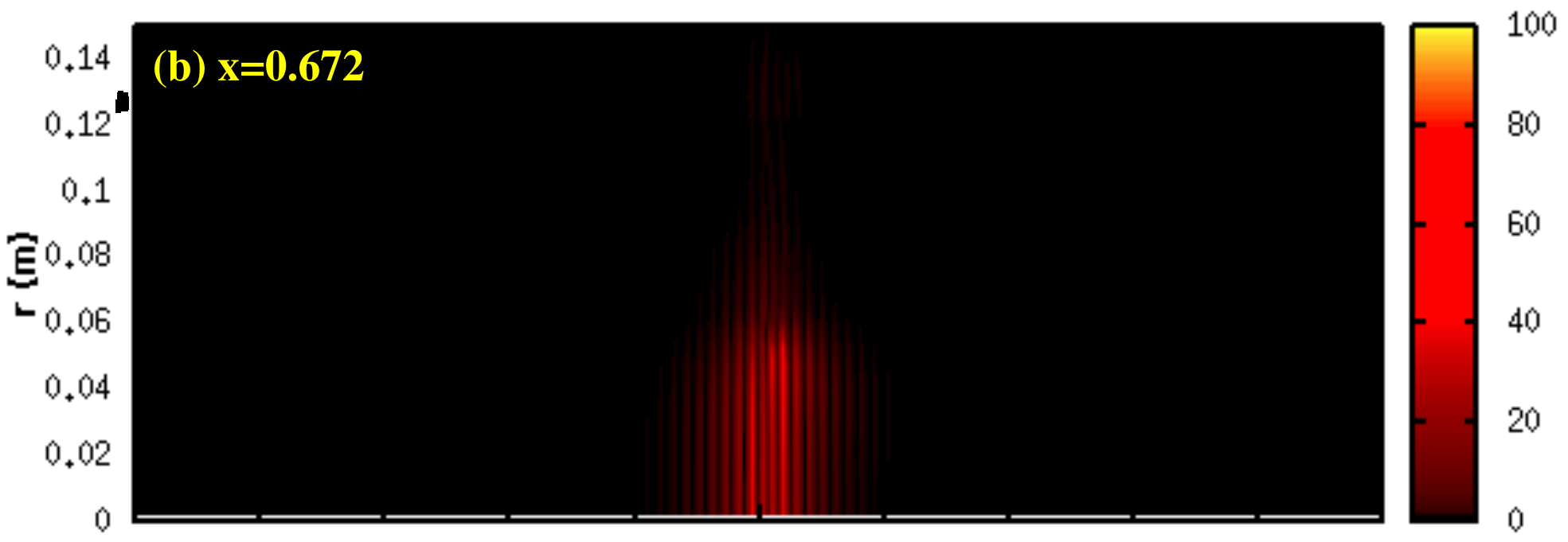}\\
\hspace{2.2cm}\includegraphics[width=0.7\textwidth,height=0.2\textwidth,angle=0]{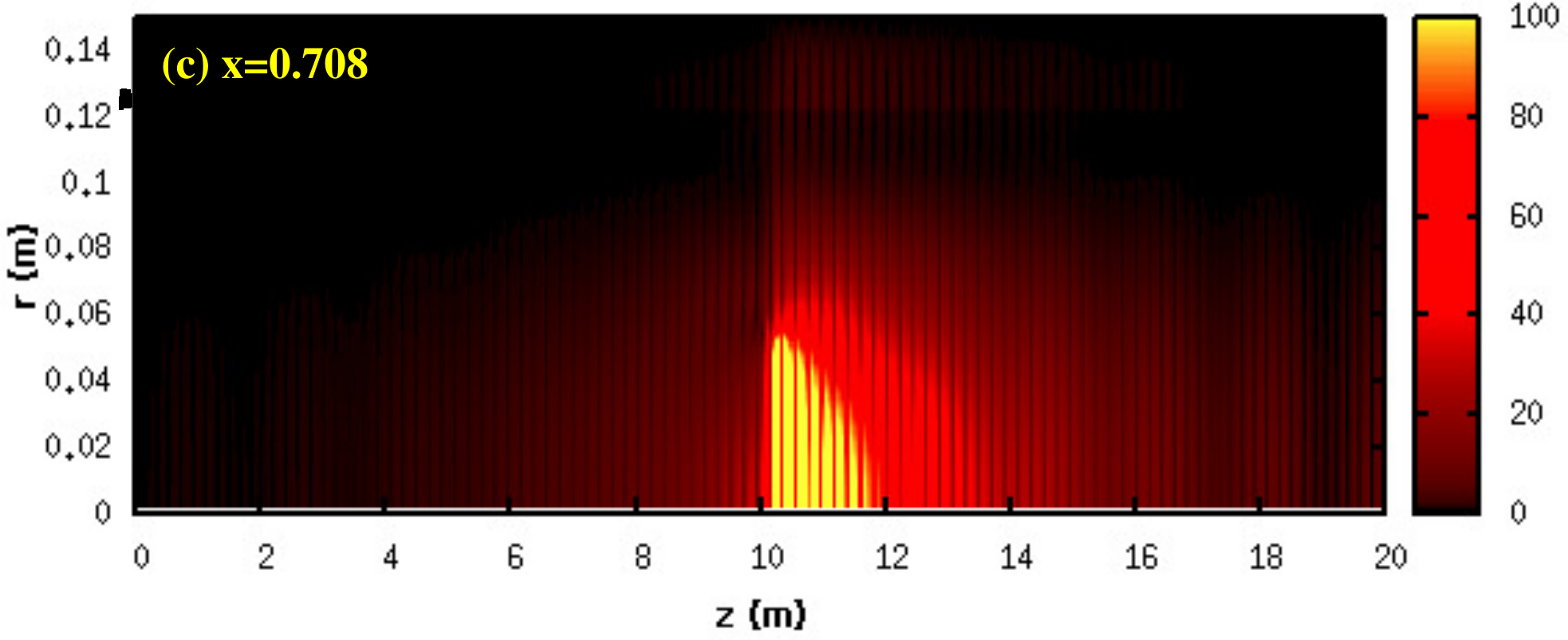}
\end{array}$
\end{center}
\caption{Identification of the spectral gap in simulations: (a) on-axis wave field at $z =12.3$~m as a function of driving frequency; vertical lines show analytical predictions for the gap centre ($x_0$) and the lower and upper edges of the gap ($x_-$ and $x_-$), (b) spatial distributions of wave energy for antenna frequencies inside and outside the spectral gap (solid bar denotes the antenna location).}
\label{fg5_7}
\end{figure}

\begin{figure}[h]
\begin{center}$
\begin{array}{ll}
(a)&(b)\\
\includegraphics[width=0.48\textwidth,angle=0]{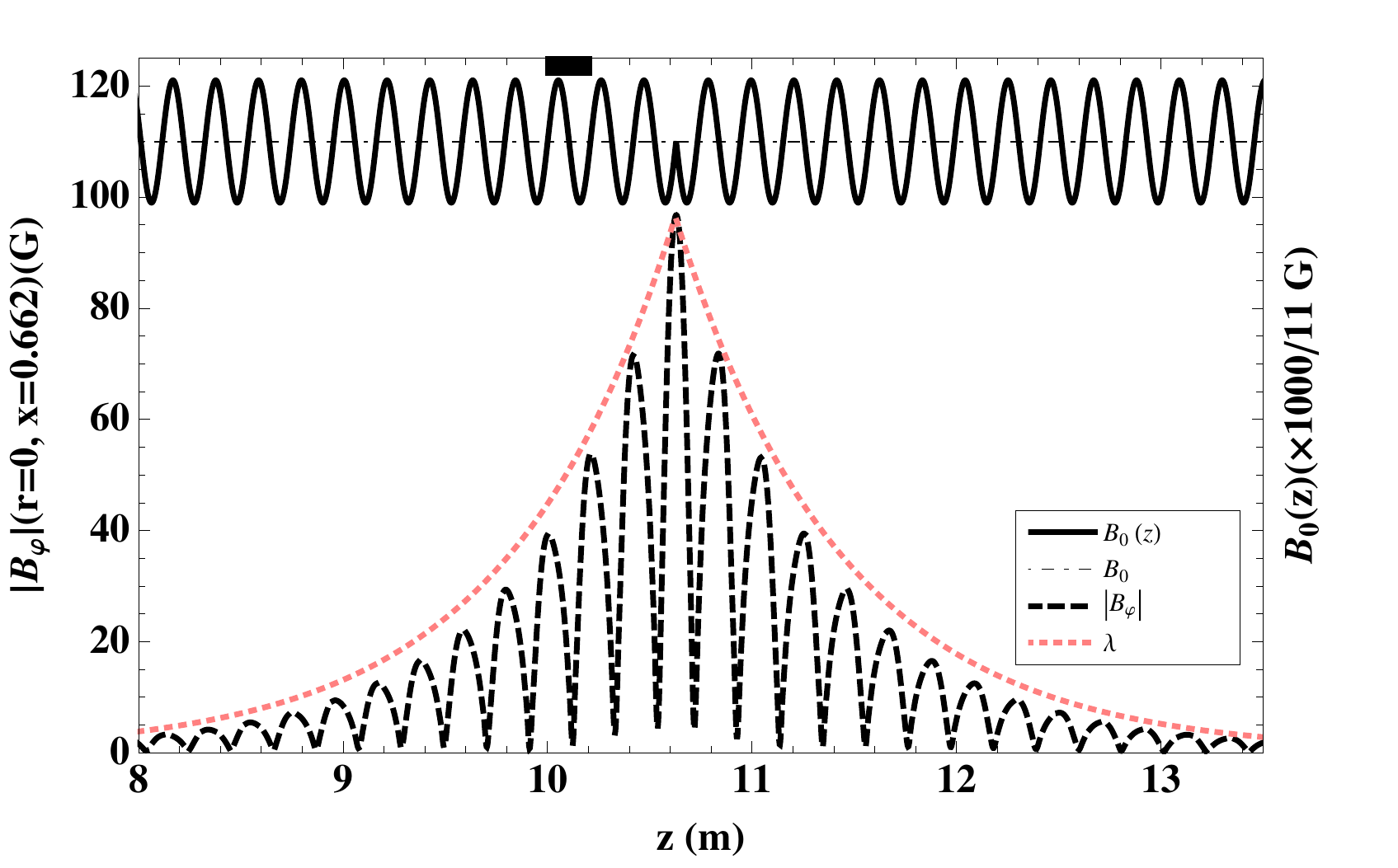}&\includegraphics[width=0.45\textwidth,angle=0]{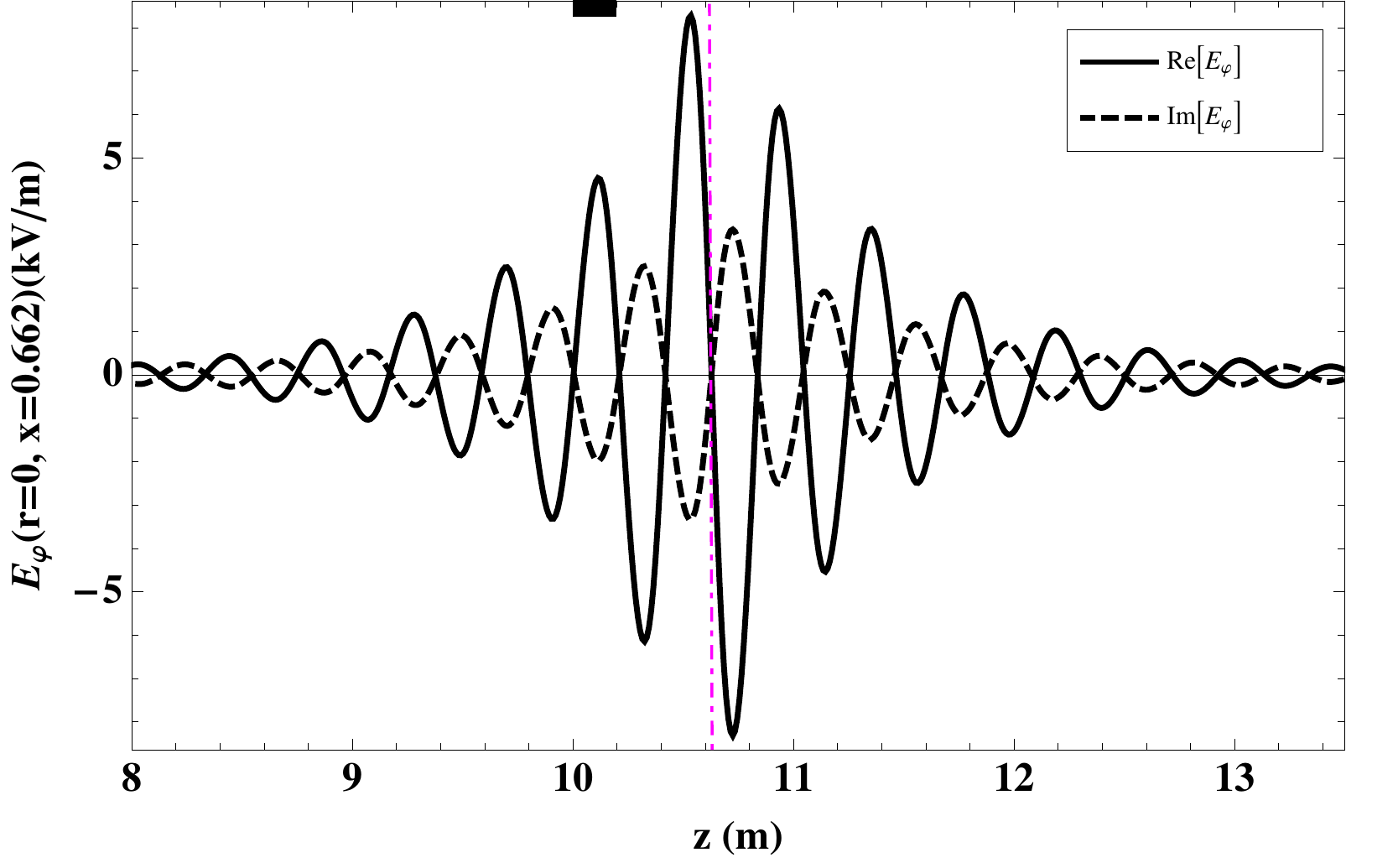}\\
(c)&(d)\\
\hspace{0.2cm}\includegraphics[width=0.434\textwidth,angle=0]{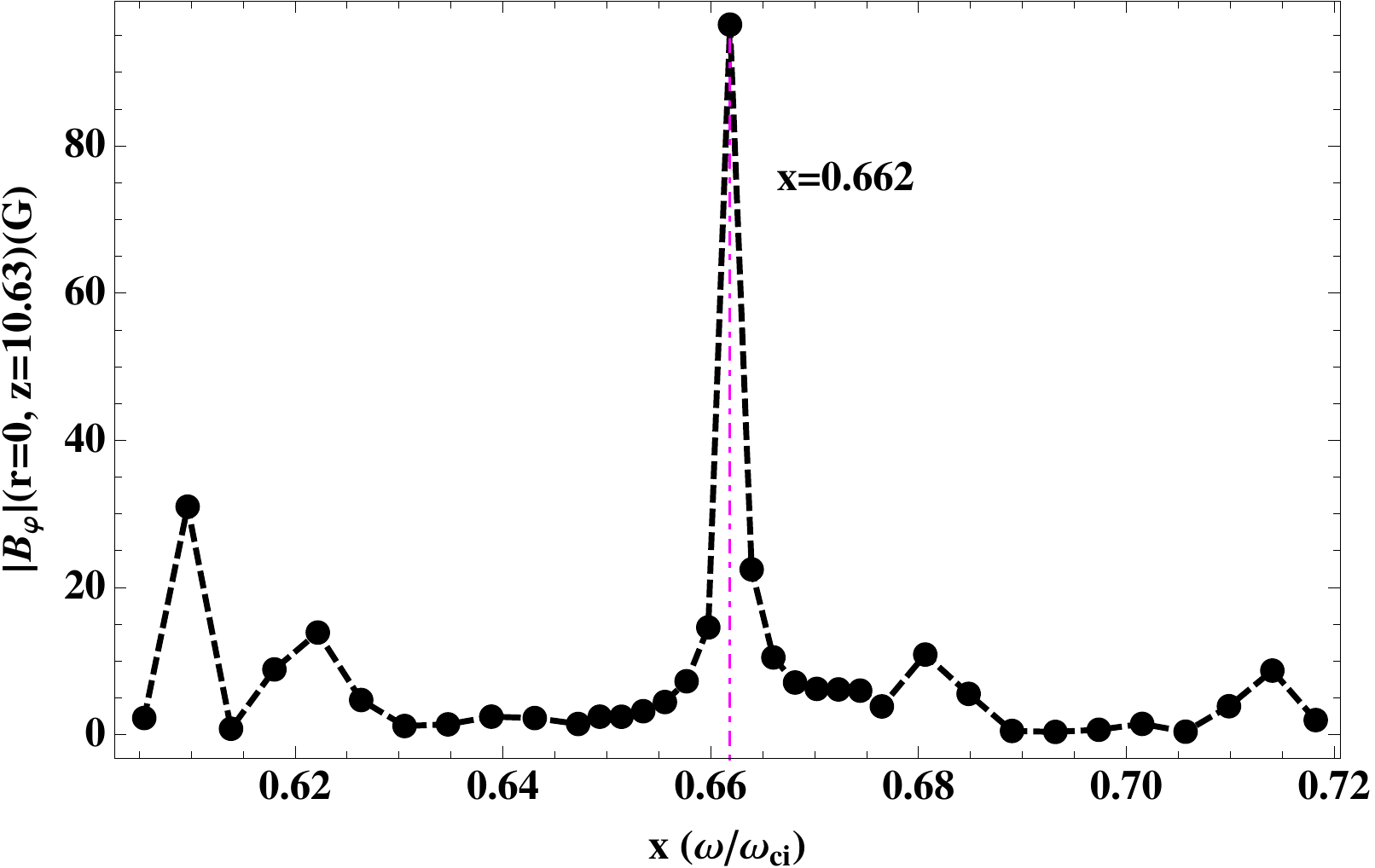}&\includegraphics[width=0.45\textwidth,angle=0]{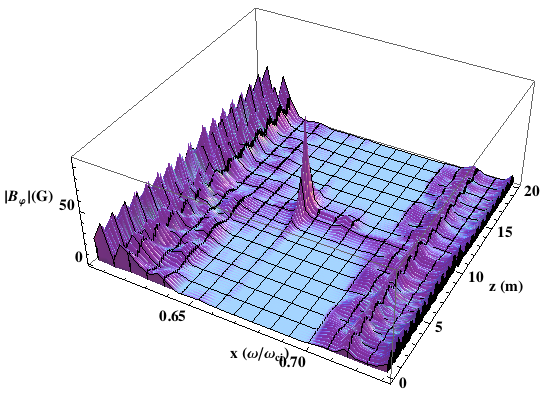}\\
\end{array}$
\end{center}
\caption{Odd-parity gap eigenmode: (a) longitudinal profiles of the static magnetic field (solid line) and the on-axis RF magnetic field for $x=0.662$ (dashed line), together with the decay constant from Eq.~\ref{eq5_51}; (b) longitudinal profile of $E_\varphi$ (on-axis) at $x=0.662$, where vertical dot-dashed line marks the defect location and solid horizontal bar marks the antenna; (c) resonance in the dependence of the on-axis amplitude of the RF magnetic field on driving frequency at the location of the defect; (d) surface plot of the on-axis wave field strength as a function of $z$ and $x$. }
\label{fg5_8}
\end{figure}

\begin{figure}[h]
\begin{center}$
\begin{array}{ll}
(a)&(b)\\
\includegraphics[width=0.48\textwidth,angle=0]{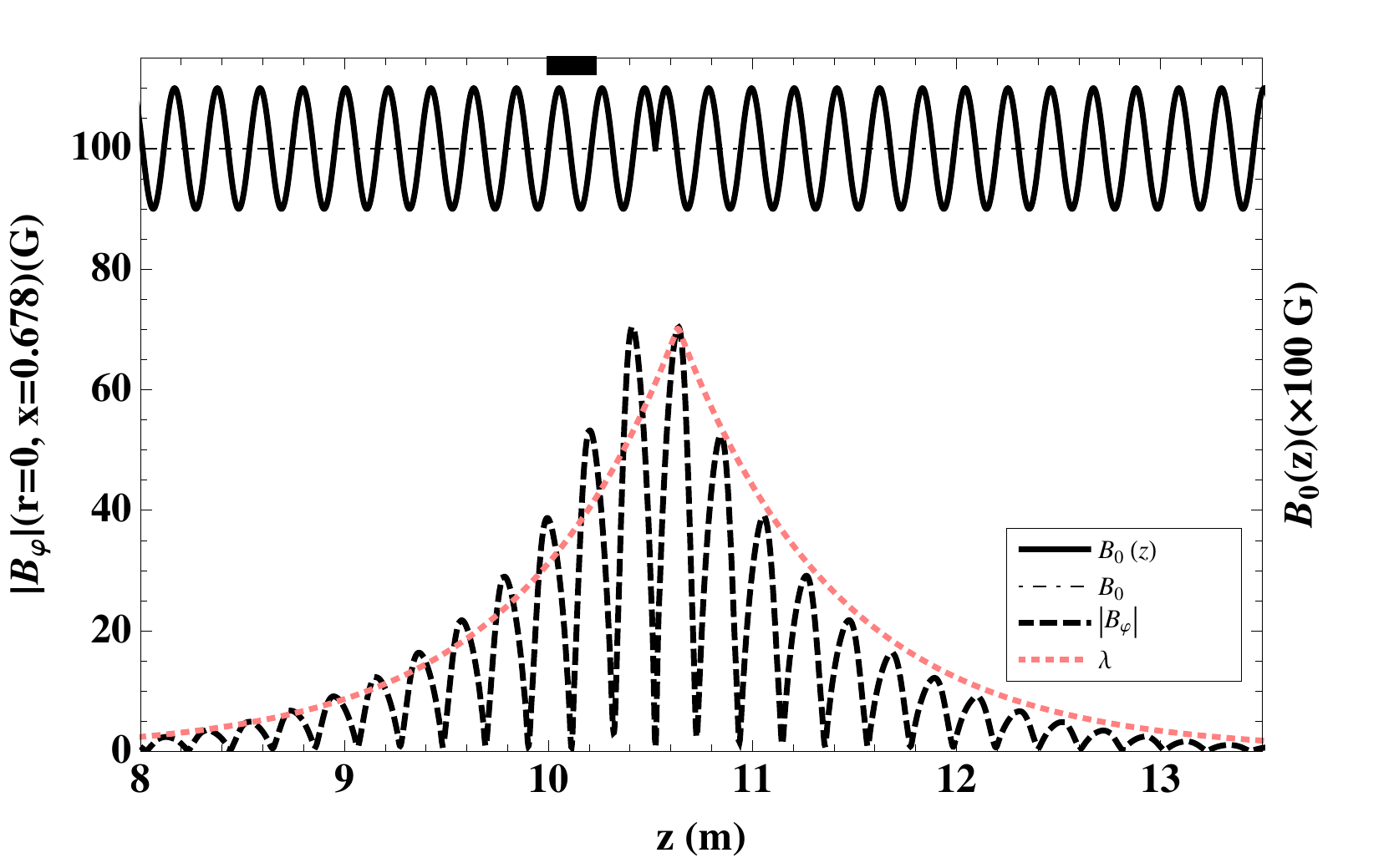}&\includegraphics[width=0.45\textwidth,angle=0]{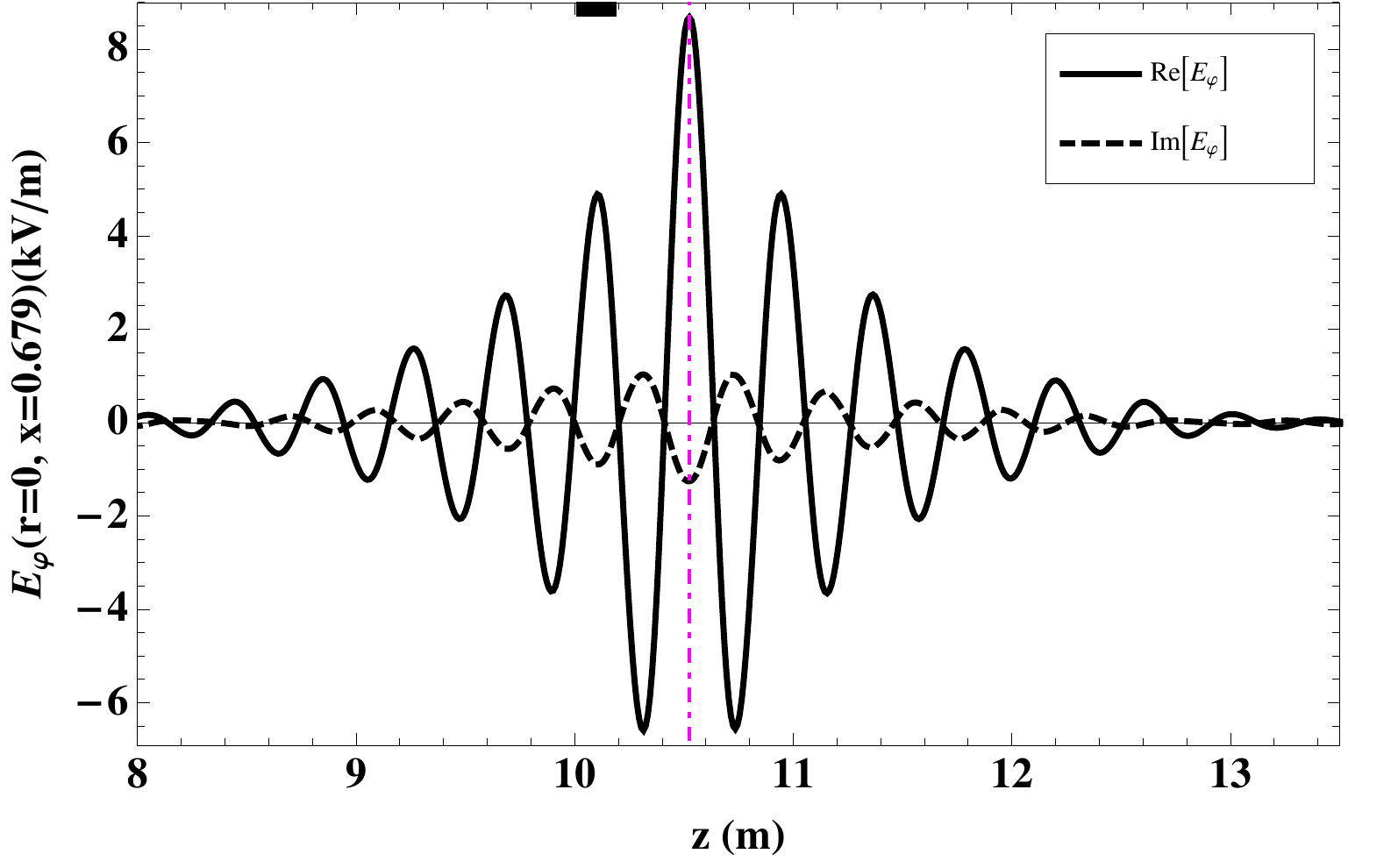}\\
(c)&(d)\\
\hspace{0.15cm}\includegraphics[width=0.44\textwidth,angle=0]{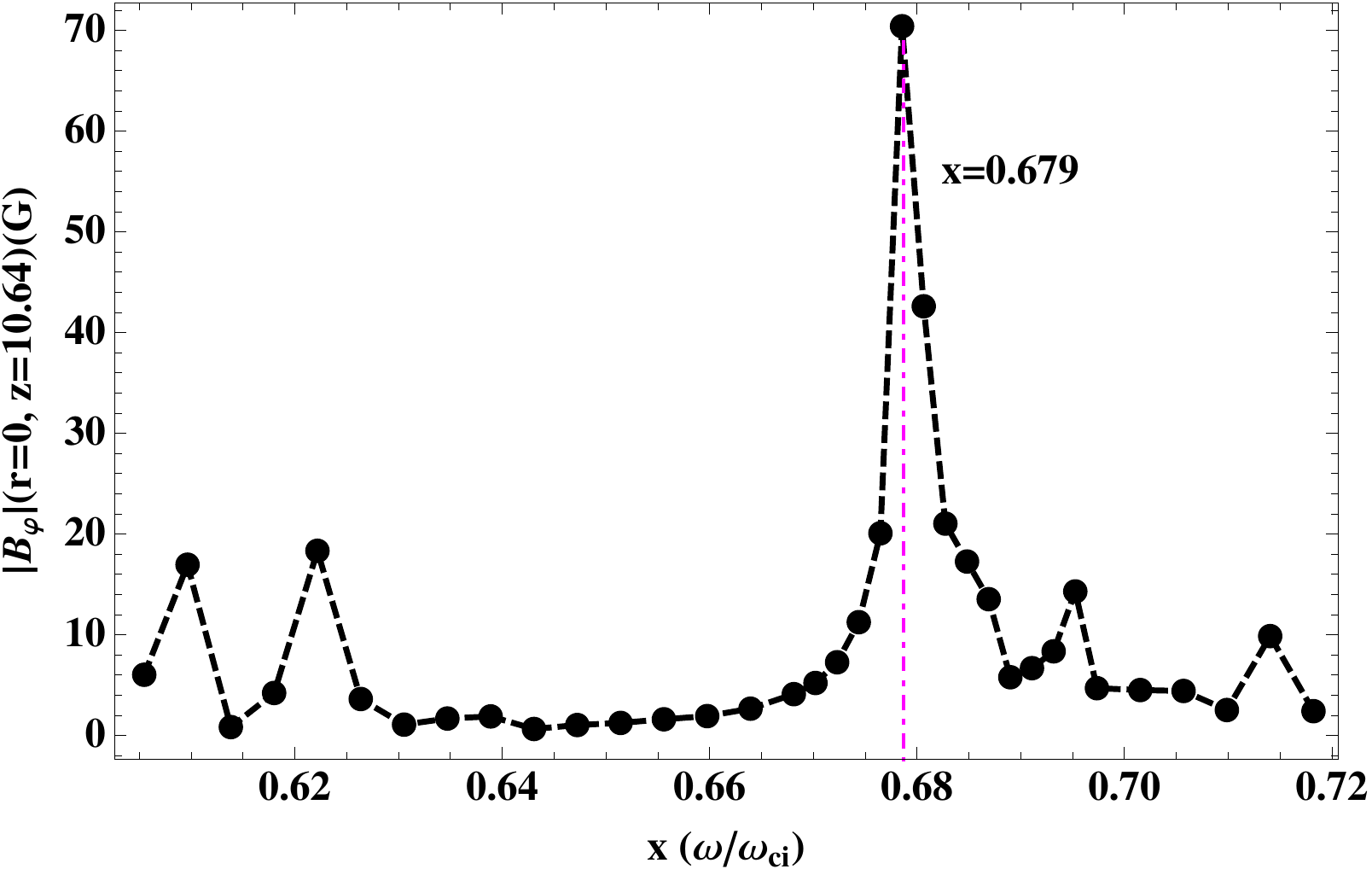}&\includegraphics[width=0.45\textwidth,angle=0]{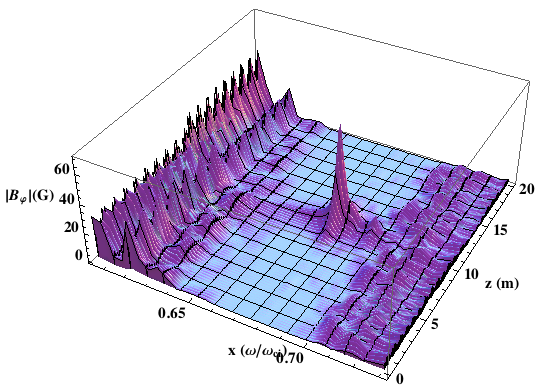}\\
\end{array}$
\end{center}
\caption{Even-parity gap eigenmode: (a) longitudinal profiles of the static magnetic field (solid line) and the on-axis RF magnetic field for $x=0.679$ (dashed line), together with the decay constant from Eq.~\ref{eq5_53}; (b) longitudinal profile of $E_\varphi$ (on-axis) at $x=0.679$, where vertical dot-dashed line marks the defect location and solid horizontal bar marks the antenna; (c) resonance in the dependence of the on-axis amplitude of the RF magnetic field on driving frequency near the location of the defect; (d) surface plot of the on-axis wave field strength as a function of $z$ and $x$. }
\label{fg5_9}
\end{figure}

We also obtained similar gap eigenmodes for a step-like-hollow density profile shown in Fig.~\ref{fg5_6}(a). 
The discontinuity location is moved radially outward ($r_0=0.122$~m) to keep the total number of particles equal on each side. For a uniform static magnetic field, illustrative axial and radial profiles of the computed wave field are shown in Fig.~\ref{fg5_6}(b) and Fig.~\ref{fg5_6}(d), respectively, for $x=0.65$. A preferred axial coupling direction is again seen clearly in Fig.~\ref{fg5_6}(b). The computed dispersion relation is compared with that from Eq.~(\ref{eq5_18}) in Fig.~\ref{fg5_6}(c). It can be seen that the two dispersion curves match at $x=0.665$. To form a spectral gap, we modulated the uniform static magnetic field in form of $[B_0(z)-B_0]/B_0=0.1\cos(30z)$. We select $q=30$ to locate the centre of the gap around the matching point by setting $k_z=q/2$ in Eq.~(\ref{eq5_18}), and get $x_0=0.667$. Figure~\ref{fg5_7} shows the computed spectral gap and its comparison with the analytical estimates at $z=12.3$~m. 
The analytical lower and upper edges of the gap are calculated from Eq.~(\ref{eq5_53}): $x_-=x_0(1-\epsilon/2)=0.634$ and $x_+=x_0(1+\epsilon/2)=0.700$. We can see that numerical and analytical spectral gaps agree quite well, in terms of the gap location and width. The spatial distribution of wave energy for frequencies inside and outside of the spectral gap is shown in Fig.~\ref{fg5_7}(b), confirming that wave propagations along the plasma column are evanescent for frequencies inside the spectral gap. The gap eigenmodes were formed by introducing two types of defect to the system's periodicity, as shown in Fig.~\ref{fg5_8} and Fig.~\ref{fg5_9} for odd-parity and even-parity respectively. We can see that they are similar to those for the step-like-peak density profile, shown in Fig.~\ref{fg5_4} and Fig.~\ref{fg5_5}. 

\section{Experimental implementation}\label{exp5}
In this section, we comment on the possible experimental realisation of the gap eigenmode of SAW in a periodic magnetic field. 

\subsection{Radial width of density jump}\label{cnt5}
\begin{figure}[ht]
\begin{center}$
\begin{array}{ll}
(a)&(b)\\
\includegraphics[width=0.49\textwidth,angle=0]{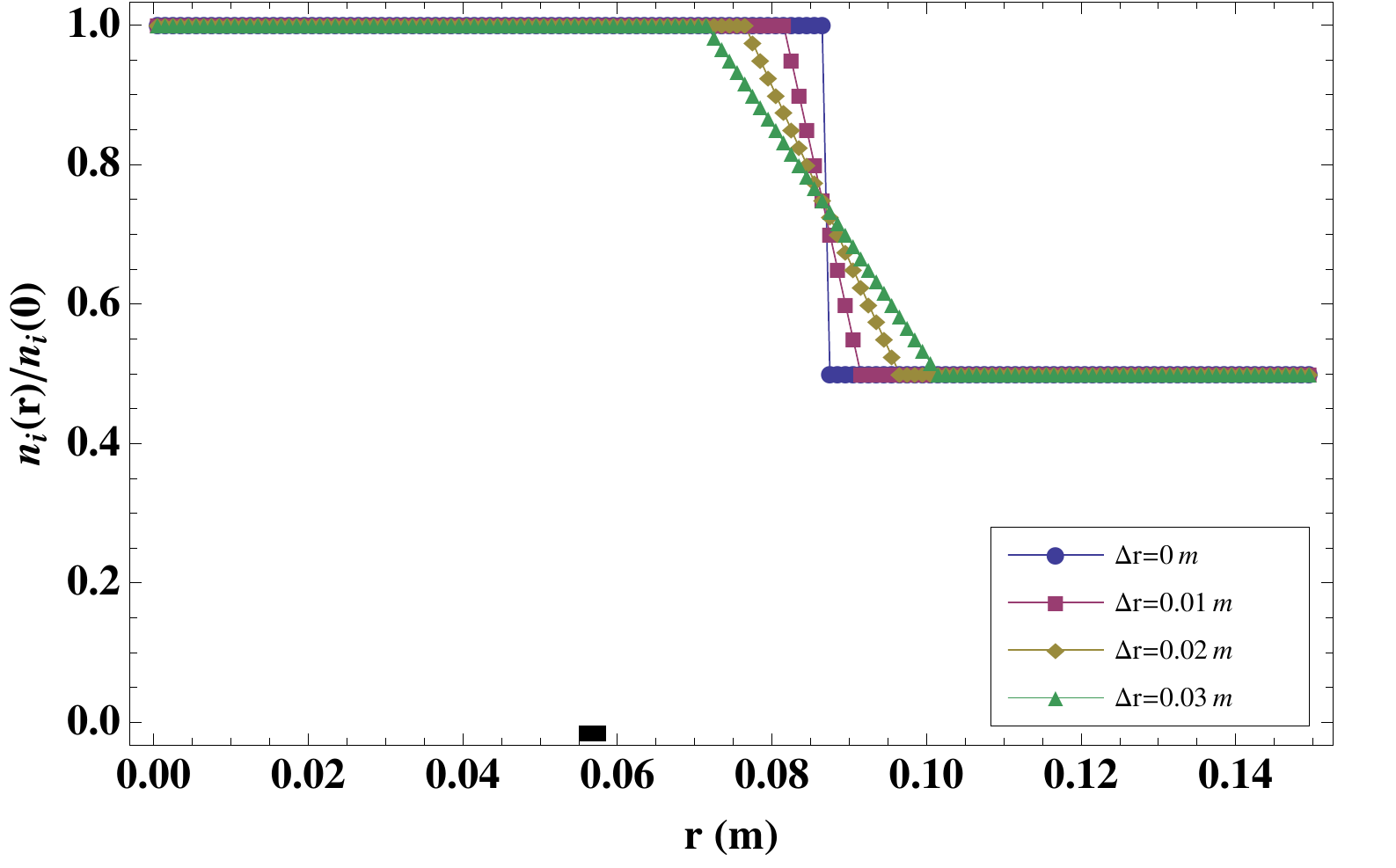}&\includegraphics[width=0.49\textwidth,angle=0]{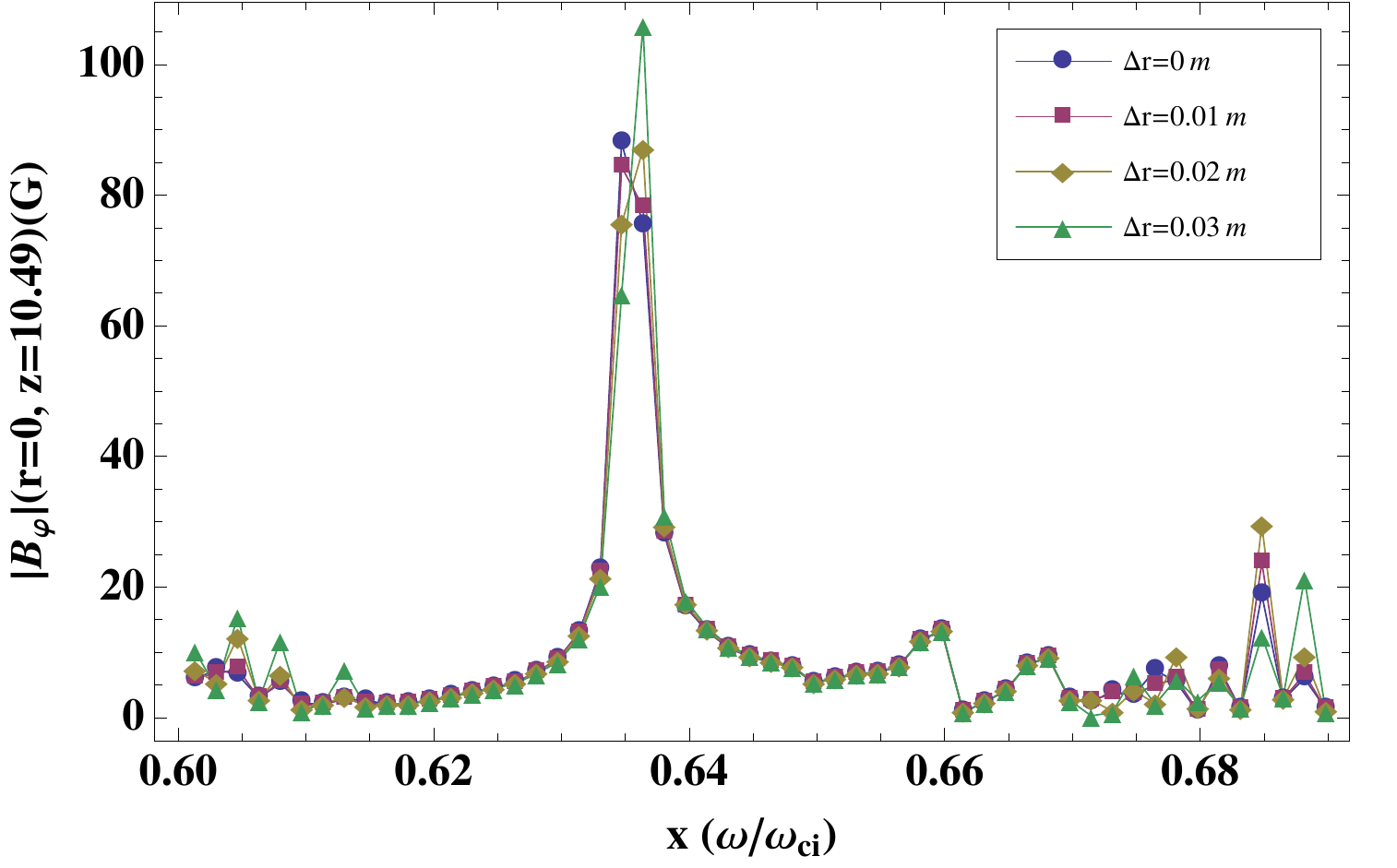}\\
\end{array}$
\end{center}
\hspace{2.2cm}(c)\\
\begin{center}
\includegraphics[width=0.75\textwidth,angle=0]{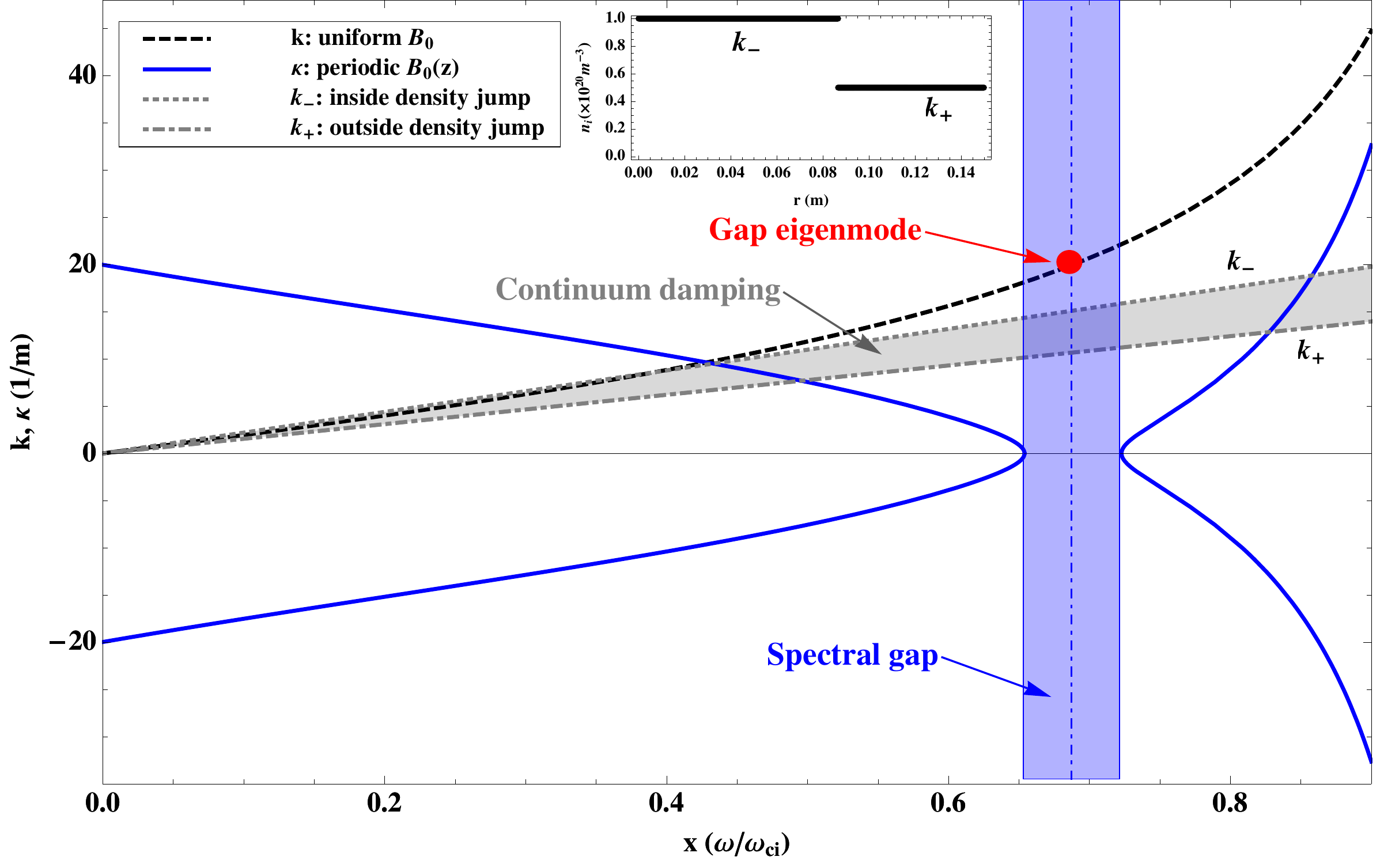}
\end{center}
\caption{Radial profile of plasma density and gap eigenmode formation: (a) density jump with different radial widths (solid bar labels the antenna location), (b) resonant peaks inside a spectral gap for the density profiles shown in (a), and (c) gap eigenmode location relative to the continuum damping region.}
\label{fg5_10}
\end{figure}

The step-like radial profile of plasma density, which has been used so far for precise comparison with the theoretical analysis in Sec.~\ref{thy5}, is not easily achievable in experiment. It clearly breaks down with additional physics such as particle diffusion and cyclotron motion, which act to broaden the radial width of density jump. We investigate the effect of this width on the gap eigenmode formation by increasing it from $\bigtriangleup r=0$~m, as used for Fig.~\ref{fg5_4}(c), to $\bigtriangleup r=0.01$~m, $\bigtriangleup r=0.02$~m and $\bigtriangleup r=0.03$~m. Figure~\ref{fg5_10}(a) shows the density profiles broadened, and Fig.~\ref{fg5_10}(b) shows the corresponding resonant peaks inside the spectral gap. 
It can be seen that the peak width and location remain the same. The peak width was found to remain the same even when the density jump width is increased to $\bigtriangleup r=0.1$~cm. Therefore, broadening the density profile shown in Fig.~\ref{fg5_2}(a) to that achievable in experiment is unlikely to affect the gap eigenmode formation. This is a sanity check that the gap eigenmode observed here is sufficiently resolved in the absence of continuum damping. Figure~\ref{fg5_10}(c) shows the location of this gap eigenmode relative to the continuum damping region. The analytical dispersion relations for uniform (same to the dashed line in Fig.~\ref{fg5_2}(c)) and periodic ($[B_0(z)-B_0]/B_0=0.1\cos(40z)$) magnetic fields are overlaid. The location of the gap eigenmode is at the joint point of the dispersion relation for uniform field and the middle line of the spectral gap, because of the chosen defect which forces $\cos(q z_0)$ to be zero in Eq.~\ref{eq5_47} and Eq.~\ref{eq5_49}. The dispersion relations for uniform plasma densities inside and outside of the density jump (for uniform field) are also shown. They label the upper and lower edges of the continuum damping region, because the continuum damping only occurs when the wave frequency experiences the local Alfv\'{e}n frequency, namely $\omega=\omega_\mathrm{A}=k_z B_0/\sqrt{\mu_0 m_i n_{i0}(r)}$.\cite{Braginskii:1965aa} We can see that the gap eigenmode is currently outside of the continuum damping region. We further found that deepening the density jump did not move the gap eigenmode location into this region. To study the impact of continuum damping on gap eigenmode formation, spectral gaps need be formed in a lower frequency region, e. g. around $x=0.2$ where the dispersion relation for uniform field is within the continuum damping region, and a very high radial resolution across the broadened density jump should be employed. This will surely increase the computation time significantly, and we leave this work for future research. 

\subsection{Implementation on the LAPD}\label{lapd5}
A modification to the experiments of Zhang et al.\cite{Zhang:2008aa} on the LAPD, in which they observed spectral gaps and continua of SAW, is a possible candidate. It is essential to ensure that dissipative processes in the plasma do not destroy the gap-mode resonance.\cite{Chang:2013aa} There exist mainly two types of dissipative mechanisms: continuum damping and collisional damping. The former needs to be avoided first of all because it will affect the mode structure globally and even prevent the propagation of SAW. Decreasing the radial gradient of plasma density to narrow the continuum damping region so that the gap eigenmode is located outside of the region will probably help. The latter needs to be small in order to form a sharp resonance peak, as pointed out in Chapter~\ref{chp4}. The collisional damping rate of Alfv\'{e}n waves (in terms of wave amplitude) is given by\cite{Braginskii:1965aa}
\begin{equation}\label{eq5_54}
\gamma_{\rm{A}}=\frac{\epsilon_0 c^2}{2}\left(\frac{k_\perp^2}{\sigma_\parallel}+\frac{k_\parallel^2}{\sigma_\perp}\right), 
\end{equation}
where $k_\parallel$ and $k_\perp$,  $\sigma_\parallel$ and $\sigma_\perp$ are components of the wave vector and electric conductivity in the directions parallel and perpendicular to the static magnetic field, respectively. Because of $k_\perp\ll k_\parallel$ and $\sigma_\perp\ll \sigma_\parallel$ for the SAW, we approximate the damping rate as
\begin{equation}\label{eq5_55}
\gamma_{\rm{A}}\approx\frac{\epsilon_0c^2}{2}\frac{k_\parallel^2}{\sigma_\perp}=\frac{3}{2}\frac{\omega^2\nu_{ei}}{\omega_{ci}|\omega_{ce}|}. 
\end{equation}
Here, the relations $k_\parallel=\omega/v_{\rm{A}}$ and $\sigma_\perp=\sigma/3=e^2 n_e/3m_e\nu_{ei}$ (in perpendicular motions, the slow electrons, which have small Larmor radii, contribute less to the conductivity than in parallel motions\cite{Chen:1984aa}) have been used. Equation~(\ref{eq5_55}) indicates that stronger magnetic field leads to larger ion cyclotron frequency and thus lower damping rate. Moreover, the damping rate can be further reduced by lowering the wave frequency. For the gap eigenmode formed here (step-like-peak density profile), the damping rate on axis is about $8.2\times 10^3~\mathrm{s}^{-1}$, which is nearly half the width of the resonance peak shown in Fig.~\ref{fg5_4}(c) (around $15\times 10^3~\mathrm{s}^{-1}$). The difference may be due to the fact that Eq.~(\ref{eq5_55}) does not consider radial boundaries but an infinite slab geometry. For the conditions employed in \cite{Zhang:2008aa} the damping rate is about $16.3~\mathrm{s}^{-1}$. Thus, the collisional dissipation on the LAPD can be controlled very low. Also, a bigger modulation factor of the periodic magnetic field broadens the spectral gap (see Eq.~(\ref{eq5_53})), and makes it easier to resolve the gap eigenmode with finite damping. Detailed comparison between the parameters employed for the present study and those in \cite{Zhang:2008aa} is given in Tab.~\ref{tb5_1}. 
\begin{table}[ht]
\caption{Parameters employed in this chapter and on the LAPD.}
\begin{center}
\begin{tabular}{lll}
Parameters&This chapter& LAPD\cite{Zhang:2008aa} \\
\hline
Length of machine&$20$~m&$17.56$~m\\
Axial periodicity&$0.16$~m&$3.63$~m\\
No. of periods&$127$&$4$\\
Radius of wall&$0.15$~m&$0.5$~m\\
$n_{i0}$& $1\times 10^{20}\ \mathrm{m^{-3}}$ & $1\times10^{18}\ \mathrm{m^{-3}}$\\
$B_0$& $1$ T& $0.12$ T\\
$T_e$& $4.22$ eV& $6$ eV\\
$m_i$& $4$ amu (He)& $4$ amu (He)\\                       
\end{tabular}
\end{center}
\label{tb5_1}
\end{table}
In this chapter, we have decreased the spatial periodicity of the magnetic field in order to form a larger number of field ripples and thereby remove conducting endplate boundary conditions, and employed a stronger magnetic field to weaken $\gamma_{\rm{A}}$ and thus the effect of collisional damping. The aim is to, for this numerical study, form a clear spectral gap and a sharp gap eigenmode, and verify the analytical model developed here before modelling the LAPD. 

To introduce a defect to the system's periodicity, besides powering one magnetic coil with extremely strong current\cite{Zhang:2008aa}, a conducting mesh could also be inserted radially into the plasma. The mesh should have a radial slot to prevent the formation of azimuthal current, and be insulated from the enclosing chamber to avoid short circuiting the plasma. Because the azimuthal electric field vanishes on the mesh ($E_\varphi(r, z_0)=0$), this will form an odd-parity gap eigenmode. To form an even-parity gap eigenmode, which requires $\partial E_\varphi(r, z_0)/\partial z=0$, a conducting ring with small radial thickness could be placed co-axially with the static magnetic field. The ring should have an axial slot and be insulated from the chamber, for the same reasons as for the mesh. 

Therefore, by controlling the collisional damping rate, varying the modulation factor, and selecting the appropriate defect, the experimental implementation of the gap eigenmode of SAW is conceivable on the LAPD.

\section{Conclusion}\label{cnl5}
The gap-mode analysis of RLH waves in the whistler frequency range\cite{Chang:2013aa} has been generalised to ion cyclotron frequencies. We applied this generalised analysis to SAW, whose spectral gaps and continua were recently observed on a linear plasma device (LAPD),\cite{Zhang:2008aa} and utilised an electromagnetic solver\cite{Chen:2006aa} to model the spectral gap and gap eigenmode. The employed conditions, such as the ratio of wave frequency and ion cyclotron frequency, ion species, and geometry, are close to those on the LAPD, except the radial profile of plasma density which is step-like and chosen for precise comparison with the analytical estimates. For a uniform magnetic field, the computed dispersion relation qualitatively agrees with that from the generalised model. However, they diverge more when the wave frequency is close to the ion cyclotron frequency. This may be because the computed values of $k$ and $\omega$ are no longer sufficiently small as assumed in the theoretical analysis. For a periodic magnetic field, a clear spectral gap is formed in both $(z, x)$ and $(r, x)$ spaces, and the gap width and location qualitatively agree with the analytical estimates. Moreover, we found a resonant peak inside the spectral gap when the field periodicity is broken with a local defect. The resonant peak stands for a gap eigenmode, which is a standing wave and with wavelength near twice the field periodicity, a characteristic of Bragg's reflection. The decay length of the gap eigenmode is consistent with the analytical prediction. We further explored the effect of the radial scale of density jump on the gap eigenmode formation, and found that the effect is ignorable. The collisional damping rate of SAW on the LAPD was also calculated, for the conditions employed in \cite{Zhang:2008aa}, and it was found much smaller than that in our simulations. Therefore, the dissipative processes on the LADP can be controlled at a very low level and will not destroy the gap-mode resonance. We conclude that experimental implementation of the gap eigenmode of SAW is conceivable on the LAPD, providing that appropriate modulation factor and defect are selected.

\chapter{Summary and recommendations for future research}\label{chp6}
The purpose of this thesis was to explore the impact of magnetic geometry on wave modes in cylindrical plasmas. This has been achieved by increasing the magnetic shaping on wave modes in helicon plasmas. A two-fluid model was applied to describing plasma-driven electrostatic drift waves in a uniform field. The model provided a consistent description of the plasma, as well as a qualitative match to the mode structure of spontaneously-driven electrostatic waves. The model is amongst the first to capture the physics of flow in a helicon discharge. The construction of a new field-pinched helicon source at the ANU afforded the opportunity to explore the impact of a ramp in the magnetic field on the waves and plasma. In the model, antenna-driven wave fields led to plasma formation through the dielectric tensor and electron-ion collisions. Wave fields in agreement with experimental data were found with a $9.5$-fold increase in the Coulomb collision frequency: this anomalous enhancement is consistent with other works. This analysis was amongst the first to describe the observed variation of wave field strength and plasma density with field strength. Finally, it was shown that the introduction of a periodic magnetic field to the system can create gaps in the continuum of whistler and shear Alfv\'{e}n modes, while the addition of a defect to the system can create gap eigenmodes. The opportunity to excite such modes in a simple magnetic confinement device offers the tantalising prospect to expand the study of gap eigenmodes of Alfv\'{e}n waves from fusion to basic plasma physics.

In the remaining part of the chapter a detailed summary of each topic is given, together with possible extensions.

\section{Uniform $B_0$: wave instabilities in a flowing plasma}\label{flw6}
\subsection{Summary}
In Chapter~\ref{chp2}, a two-fluid flowing plasma model, which was developed originally to describe wave oscillations in a vacuum arc centrifuge, \cite{Hole:2002aa, Hole:2001aa, Hole:2001ab} was employed to interpret low frequency oscillations in a RF generated linear magnetised plasma. The following conclusions were obtained: 

(1) Compared to the more rapidly rotating centrifuge plasma, the drift wave instability, unstable at larger wavelengths, has a normalised frequency much larger than the normalised frequency for the centrifuge. 

(2) The difference between dispersion curves is principally due to the low rotation speed of RF plasmas compared to those of the centrifuge. 

(3) The measured fluctuation frequency is a factor of $3.5$ lower than the predicted frequency, and the predicted trend of oscillation frequency with field strength for inferred electron temperature matches the data. 

(4) The discrepancy between the measured and predicted fluctuation frequencies may be attributable to the limitations of the model: assumed temperature uniformity across plasma column and neglected effects induced by plasma fluctuations on the externally applied field, and the lack of measurements in the experiment: simultaneously measuring plasma rotation profile and probe frequency and measuring the dependency of electron temperature with field strength.

(5) The measured and predicted perturbed density profiles have a single peak in the radial direction, indicating the perturbed mode has $n=0$, and the peak position is in the region of maximum equilibrium density gradient. 

(6) Qualitative agreement consolidates earlier claims that the mode is a drift mode, driven by the density gradient of the plasma. 

\subsection{Recommendations for future research}
Recommendations for future research regarding this topic include: 

(1) generalising the electrostatic vacuum arc centrifuge model to include magnetic field oscillations to describe electromagnetic perturbations in a cylindrical flowing plasma (this was partly done in the appendix), 

(2) including an external drive into the generalised electromagnetic model to describe driven modes in a flowing plasma, 

(3) including radial non-uniformity into the ion and electron temperatures to possibly reduce the $3.5$-factor discrepancy between computed and measured wave frequencies,

(4) and experimentally measuring the plasma rotation profile and probe frequency simultaneously, and the dependence of electron temperature with field strength. 

\section{Focused $B_0$: wave propagations in a pinched plasma}\label{pnc6}
\subsection{Summary}
In Chapter~\ref{chp3}, an electromagnetic solver based on Maxwell's equations and a cold-plasma dielectric tensor was employed to describe the wave phenomena observed in a cylindrical non-uniform helicon discharge. Here, the non-uniformity is both radial and axial: the plasma density is dependent on radial and axial positions, the static magnetic field varies with axial position, and the electron temperature is a function of radial position. The following conclusions were obtained: 

(1) With an enhancement factor of $9.5$ to the electron-ion Coulomb collision frequency to approximate the observed attenuation, and a $12\%$ reduction in the antenna radius to match the amplitude of the wave field, the wave solver produced consistent wave fields compared to experimental data, including the axial and radial profiles of wave amplitude and phase.

(2) A local minimum in the axial profiles of wave amplitude was observed both experimentally and numerically, agreeing with previous studies.\cite{Guo:1999aa, Degeling:2004aa, Light:1995aa, Mori:2004aa} 

(3) Mode structure of $m=1$ is consistent with the left hand half-turn helical antenna being used. 

(4) A possible explanation for the enhanced electron-ion collision frequency has been offered through ion-acoustic turbulence, which can happen if the electron drift velocity exceeds the speed of sound in magnetised plasmas.\cite{Lee:2011aa} 

(5) Density increasing in proportion to the static magnetic field has little effect on RF absorption, while the density level near the antenna affects the wave amplitude significantly at all axial locations. 

(6) The axial gradient in magnetic field increases the decay length of helicon waves in the target region. 

(7) The relationship between plasma density, static magnetic field, and axial wavelength is consistent with a simple theory developed previously.\cite{Chen:1996ab} 

(8) With increased electron-ion collision frequency, the wave amplitude is lowered and more focused near the antenna, which is mainly because of stronger edge heating at higher collision frequencies that prevent more energy transported from the antenna into the core plasma.

(9) The wave amplitude profile at the enhanced electron-ion collision frequency (by a factor of $9.5$), which agrees with experimental data, shows consistent feature with a previous study that the RF energy is almost all absorbed in the near region of the antenna rather than in the far region.\cite{Chen:1996aa} 

(10) The antiparallel feature that wave energy is larger on the opposite side of the antenna relative to the direction of static magnetic field has been observed both numerically and experimentally, and is consistent with previous observations made in uniform field configurations.\cite{Chen:1996aa, Chen:1996ab, Sudit:1996aa, Lee:2011aa} 

\subsection{Recommendations for future research}
Recommendations for future research regarding this topic include: 

(1) further explaining exactly how an axially non-uniform field might affect the radially localised helicon mode,\cite{Breizman:2000aa} 

(2) including different azimuthal mode numbers in the glass layer and any subsequent coupling to the plasma at the plasma-glass interface; this will possibly increase the computed wave signal which is currently weaker than measurement, 

(3) identifying independent first and second radial modes that superpose to yield a local minimum in wave field amplitude, 

(4) measuring the axial profile of plasma density, and the plasma density and electromagnetic wave field with a reversed static magnetic field, 

(5) and modelling a diverged plasma configuration which may be of interest for plasma thruster applications. 

\section{Rippled $B_0$: gap eigenmodes of RLH waves and SAW}\label{prd6}
\subsection{Summary}
In Chapter~\ref{chp4} and Chapter~\ref{chp5}, spectral gaps of RLH waves and SAW were investigated in a plasma cylinder with periodic magnetic field, and discrete eigenmodes were formed inside their gaps by introducing a defect to the field periodicity. The following conclusions were obtained:

(1) Longitudinal modulation of the guiding magnetic field in a plasma column creates a spectral gap, which prohibits wave propagation along the column when the driving frequency of the RF antenna is in the forbidden range. 

(2) The calculated width and location of the gap are consistent with analytical estimates. 

(3) A discrete eigenmode can been formed inside the spectral gap by introducing a local defect to the periodic structure. 

(4) Both the theoretical analysis and simulations demonstrate two types of gap eigenmode in the ``imperfect" system: odd-parity mode and even-parity mode, depending the type of the defect employed.  

(5) The gap mode is localised around the defect and represents a standing wave rather than travelling wave. 

(6) The distinctive feature of the gap eigenmode is a resonant peak in the plasma response to the antenna current. 

(7) The gap eigenmode has two characteristic spatial scales: a short inner scale which is nearly twice the system's periodicity and characteristic for Bragg's reflection, and a smooth envelope which depends on the modulation amplitude and scales roughly as the inverse width of the spectral gap. 

(8) Experimental identifications of the gap eigenmodes of RLH waves and SAW are conceivable on the LAPD. 

\subsection{Recommendations for future research}
Recommendations for future research regarding this topic include: 

(1) identifying the gap eigenmodes of RLH waves and SAW in a linear plasma device, e. g. LAPD, which if succeed will pave the way to further studying the interaction between gap eigenmode and energetic particles: important for energetic particle confinement in fusion plasmas, 

(2) including continuum damping into the gap-mode analysis of SAW and further studying its effect on the gap eigenmode formation.

\bibliographystyle{unsrt}

\begin{thebibliography}{100}

\bibitem{Boswell:1970aa}
R.~W. Boswell.
\newblock Plasma production using a standing helicon wave.
\newblock {\em Physics Letters A}, 33(7):457, 1970.

\bibitem{Watari:1978aa}
T.~Watari, T.~Hatori, R.~Kumazawa, S.~Hidekuma, T.~Aoki, T.~Kawamoto,
  M.~Inutake, S.~Hiroe, A.~Nishizawa, K.~Adati, T.~Sato, T.~Watanabe,
  H.~Obayashi, and K.~Takayama.
\newblock Radio-frequency plugging of a high density plasma.
\newblock {\em Physics of Fluids}, 21(11):2076, 1978.

\bibitem{Okamura:1986aa}
S.~Okamura, K.~Adati, T.~Aoki, D.R. Baker, H.~Fujita, H.R. Garner, K.~Hattori,
  S.~Hidekuma, T.~Kawamoto, R.~Kumazawa, Y.~Okubo, and T.~Sato.
\newblock Plasma production with rotating ion cyclotron waves excited by nagoya
  type-iii antennas in rfc-xx.
\newblock {\em Nuclear Fusion}, 26(11):1491, 1986.

\bibitem{Shoji:1986aa}
T.~Shoji.
\newblock {IPPJ} {A}nnu. {R}ev. (p67), {N}agoya {U}niv. ({J}apan).
\newblock 1986.

\bibitem{Shoji:1987aa}
T.~Shoji.
\newblock {IPPJ} {A}nnu. {R}ev. (p63), {N}agoya {U}niv. ({J}apan).
\newblock 1987.

\bibitem{Shoji:1988aa}
T.~Shoji.
\newblock {IPPJ} {A}nnu. {R}ev. (p83), {N}agoya {U}niv. ({J}apan).
\newblock 1988.

\bibitem{Corr:2009aa}
C.~S. Corr and R.~W. Boswell.
\newblock Nonlinear instability dynamics in a high-density, high-beta plasma.
\newblock {\em Physics of Plasmas}, 16(2):022308, 2009.

\bibitem{Blackwell:2012aa}
Boyd~D Blackwell, Juan~Francisco Caneses, Cameron~M Samuell, John Wach, John
  Howard, and Cormac Corr.
\newblock Design and characterization of the magnetized plasma interaction
  experiment (magpie): a new source for plasma--material interaction studies.
\newblock {\em Plasma Sources Science and Technology}, 21(5):055033, 2012.

\bibitem{Gekelman:1991aa}
W.~Gekelman, H.~Pfister, Z.~Lucky, J.~Bamber, D.~Leneman, and J.~Maggs.
\newblock Design, construction, and properties of the large plasma research
  device - the lapd at ucla.
\newblock {\em Review of Scientific Instruments}, 62(12):2875, 1991.

\bibitem{Zhang:2008aa}
Y.~Zhang, W.~W. Heidbrink, H.~Boehmer, R.~McWilliams, G.~Chen, B.~N. Breizman,
  S.~Vincena, T.~Carter, D.~Leneman, W.~Gekelman, P.~Pribyl, and B.~Brugman.
\newblock {{Spectral gap of shear Alfv\'{e}n waves in a periodic array of
  magnetic mirrors}}.
\newblock {\em Physics of Plasmas}, 15(1):012103, 2008.

\bibitem{Corr:2007aa}
C.~S. Corr and R.~W. Boswell.
\newblock High-beta plasma effects in a low-pressure helicon plasma.
\newblock {\em Physics of Plasmas}, 14(12):122503, 2007.

\bibitem{Chen:1984aa}
F.~F. Chen.
\newblock {\em Introduction to Plasma Physics and Controlled Fusion}, volume 1:
  Plasma Physics.
\newblock Plenum Press, New York, second edition, 1984.

\bibitem{Crookes:1879aa}
W.~Crookes.
\newblock Lecture on ``{R}adiant matter" to the {B}ritish {A}ssociation for the
  {A}dvancement of the {S}cience, 1879 (see M. N. Hirsh, H. J. Oskam (editors)
  -- Gaseous Electronics, Academic Press, 1978).

\bibitem{Langmuir:1928aa}
I.~Langmuir.
\newblock Oscillations in ionized gases.
\newblock {\em Proceedings of the National Academy of Sciences of the United
  States of America}, 14(8):627, 1928.

\bibitem{Smith:1971aa}
H.~M. MOTT-SMITH.
\newblock History of ``{P}lasmas''.
\newblock {\em Nature}, 233:219, 1971.

\bibitem{Goldston:1995aa}
R.~J. Goldston and P.~H. Rutherford.
\newblock {\em Introduction to plasma physics}.
\newblock Institute of Physics Publishing, Bristol and Philadelphia, 1995.

\bibitem{Goedbloed:2004aa}
J.~P.~(Hans) Goedbloed and Stefaan Poedts.
\newblock {\em Principles of Magnetohydrodynamics With Applications to
  Laboratory and Astrophysical Plasmas}.
\newblock Cambridge University Press, 2004.

\bibitem{Boswell:1997aa}
R.~W. Boswell and F.~F. Chen.
\newblock Helicons-the early years.
\newblock {\em Plasma Science, IEEE Transactions on}, 25(6):1229, 1997.

\bibitem{Chen:1997aa}
F.~F. Chen and R.~W. Boswell.
\newblock Helicons-the past decade.
\newblock {\em Plasma Science, IEEE Transactions on}, 25(6):1245, 1997.

\bibitem{Boswell:1987aa}
R.~W. Boswell and R.~K. Porteous.
\newblock Large volume, high density rf inductively coupled plasma.
\newblock {\em Applied Physics Letters}, 50(17):1130, 1987.

\bibitem{Heidbrink:2008aa}
W.~W. Heidbrink.
\newblock Basic physics of alfv\'{e}n instabilities driven by energetic
  particles in toroidally confined plasmas.
\newblock {\em Physics of Plasmas}, 15(5):055501, 2008.

\bibitem{Stix:1992aa}
T.~H. Stix.
\newblock {\em Waves in Plasmas}.
\newblock American Institute of Physics, New York, 1992.

\bibitem{Gurnett:2005aa}
D.~A. Gurnett and A.~Bhattacharjee.
\newblock {\em Introduction to {P}lasma {P}hysics with {S}pace and {L}aboratory
  {A}pplications}.
\newblock Cambridge University Press, 2005.

\bibitem{Ginzburg:1970aa}
V.~L. Ginzburg.
\newblock {\em The propagation of electromagnetic waves in plasmas}.
\newblock Pergamon Press, second edition, 1970.

\bibitem{Swanson:1985aa}
D.~G. Swanson.
\newblock Radio frequency heating in the ion-cyclotron range of frequencies.
\newblock {\em Physics of Fluids}, 28(9):2645, 1985.

\bibitem{Barkhausen:1919aa}
H.~Barkhausen.
\newblock Zwei mit hilfe der neuen verstarker entdeckte ersheinungen.
\newblock {\em Phys. Z}, 20:401, 1919.

\bibitem{Barkhausen:1930aa}
H.~Barkhausen.
\newblock Whistling tones from the earth.
\newblock {\em Radio Engineers, Proceedings of the Institute of}, 18(7):1155,
  1930.

\bibitem{Storey:1953aa}
L.~R.~O. Storey.
\newblock An investigation of whistling atmospherics.
\newblock {\em Phil. Trans. R. Soc. Lond. A.}, 246:113, 1953.

\bibitem{Gurnett:1965aa}
D.~A. Gurnett, S.~D. Shawhan, N.~M. Brice, and R.~L. Smith.
\newblock Ion cyclotron whistlers.
\newblock {\em Journal of Geophysical Research}, 70(7):1665, 1965.

\bibitem{Aigrain:1960aa}
P.~Aigrain.
\newblock Les 'helicons' dans les semiconducteurs.
\newblock In {\em Proceedings of the International Conference on Seminconductor
  Physics}, page 224, Prague, Czeckoslovakia, 1960.

\bibitem{Klozenberg:1965aa}
J.~P. Klozenberg, B.~McNamara, and P.~C. Thonemann.
\newblock The dispersion and attenuation of helicon waves in a uniform
  cylindrical plasma.
\newblock {\em Journal of Fluid Mechanics}, 21:545, 1965.

\bibitem{Chen:1991aa}
F.~F. Chen.
\newblock Plasma ionization by helicon waves.
\newblock {\em Plasma Physics and Controlled Fusion}, 33(4):339, 1991.

\bibitem{Chen:1997ab}
F.~F. Chen and D.~Arnush.
\newblock Generalized theory of helicon waves. i. normal modes.
\newblock {\em Physics of Plasmas}, 4(9):3411, 1997.

\bibitem{Arnush:1997aa}
D.~Arnush and F.~F. Chen.
\newblock Generalized theory of helicon waves. ii. excitation and absorption.
\newblock {\em Physics of Plasmas}, 5(5):1239, 1997.

\bibitem{Breizman:2000aa}
B.~N. Breizman and A.~V. Arefiev.
\newblock Radially localized helicon modes in nonuniform plasma.
\newblock {\em Phys. Rev. Lett.}, 84:3863, 2000.

\bibitem{Blevin:1966aa}
H.~A. Blevin and P.~J. Christiansen.
\newblock Propagation of helicon waves in a non-uniform plasma.
\newblock {\em Aust. J. Phys.}, 19:501, 1966.

\bibitem{Boswell:1972aa}
R.~W. Boswell.
\newblock Dependence of helicon wave radial structure on electron inertia.
\newblock {\em Australian Journal of Physics}, 25(4):403, 1972.

\bibitem{Blevin:1968ab}
H.~A. Blevin, P.~J. Christiansen, and B.~Davies.
\newblock Effect of electron cyclotron resonance on helicon waves.
\newblock {\em Physics Letters A}, 28(3):230, 1968.

\bibitem{Trivelpiece:1959aa}
A.~W. Trivelpiece and R.~W. Gould.
\newblock Space charge waves in cylindrical plasma columns.
\newblock {\em Journal of Applied Physics}, 30(11):1784, 1959.

\bibitem{Lehane:1965aa}
J.~A. Lehane and P.~C. Thonemann.
\newblock An experimental study of helicon wave propagation in a gaseous
  plasma.
\newblock {\em Proceedings of the Physical Society}, 85(2):301, 1965.

\bibitem{Blevin:1968aa}
H.~A. Blevin and P.~J. Christiansen.
\newblock Helicon waves in a non-uniform plasma.
\newblock {\em Plasma Physics}, 10(8):799, 1968.

\bibitem{Chen:1994aa}
F.~F. Chen, M.~J. Hsieh, and M.~Light.
\newblock Helicon waves in a non-uniform plasma.
\newblock {\em Plasma Sources Science and Technology}, 3(1):49, 1994.

\bibitem{Sudit:1994aa}
I~D Sudit and F~F Chen.
\newblock A non-singular helicon wave equation for a non-uniform plasma.
\newblock {\em Plasma Sources Science and Technology}, 3(4):602, 1994.

\bibitem{Sudan:1967aa}
R.~N. Sudan, A.~Cavaliere, and M.~N. Rosenbluth.
\newblock Nonlinear interaction of helicons (whistlers) in inhomogeneous media.
\newblock {\em Phys. Rev.}, 158:387, 1967.

\bibitem{Chang:2013aa}
L~Chang, B~N Breizman, and M~J Hole.
\newblock Gap eigenmode of radially localized helicon waves in a periodic
  structure.
\newblock {\em Plasma Physics and Controlled Fusion}, 55(2):025003, 2013.

\bibitem{Carter:2002aa}
M.~D. Carter, F.~W. Baity, G.~C. Barber, R.~H. Goulding, Y.~Mori, D.~O. Sparks,
  K.~F. White, E.~F. Jaeger, F.~R. Chang-Diaz, and J.~P. Squire.
\newblock Comparing experiments with modeling for light ion helicon plasma
  sources.
\newblock {\em Physics of Plasmas}, 9(12):5097, 2002.

\bibitem{Chen:1992aa}
F.~F. Chen.
\newblock Experiments on helicon plasma sources.
\newblock In {\em 38th National Symposium of the American Vacuum Society},
  volume~10, page 1389. AVS, 1992.

\bibitem{Chen:1996ab}
F.~F. Chen.
\newblock Physics of helicon discharges.
\newblock {\em Physics of Plasmas}, 3(5):1783, 1996.

\bibitem{Lieberman:2005aa}
M.~A. Lieberman and A.~J. Lichtenberg.
\newblock {\em Principles of Plasma Discharges and Materials Processing}.
\newblock John Wiley \& Sons, Inc., Hoboken, New Jersey, second edition
  edition, 2005.

\bibitem{Lafleur:2011aa}
T.~A. Lafleur.
\newblock {\em Helicon {W}ave {P}ropagation in {L}ow {D}iverging {M}agnetic
  {F}ields}.
\newblock PhD thesis, The Australian National University, 2011.

\bibitem{Swanson:2003aa}
D.~G. Swanson.
\newblock {\em Plasma {W}aves}.
\newblock Institute of Physics Publishing, Bristol and Philadelphia, second
  edition, 2003.

\bibitem{Scime:2008aa}
E.~E. Scime, A.~M. Keesee, and R.~W. Boswell.
\newblock Mini-conference on helicon plasma sources.
\newblock {\em Physics of Plasmas}, 15(5):058301, 2008.

\bibitem{Lee:2011aa}
Charles~A. Lee, Guangye Chen, Alexey~V. Arefiev, Roger~D. Bengtson, and
  Boris~N. Breizman.
\newblock Measurements and modeling of radio frequency field structures in a
  helicon plasma.
\newblock {\em Physics of Plasmas}, 18(1):013501, 2011.

\bibitem{Chang:2012aa}
L.~Chang, M.~J. Hole, J.~F. Caneses, G.~Chen, B.~D. Blackwell, and C.~S. Corr.
\newblock Wave modeling in a cylindrical non-uniform helicon discharge.
\newblock {\em Physics of Plasmas}, 19(8):083511, 2012.

\bibitem{Arefiev:2004ab}
A.~V. Arefiev and B.~N. Breizman.
\newblock Theoretical components of the vasimr plasma propulsion concept.
\newblock {\em Physics of Plasmas}, 11(5):2942, 2004.

\bibitem{Ziemba:2005aa}
T.~Ziemba, J.~Carscadden, J.~Slough, J.~Prager, and R.~Winglee.
\newblock High power helicon thruster.
\newblock In {\em 41st AIAA/ASME/SAE/ASEE Joint Propulsion Conference \&
  Exhibit}, Tucson, Arizona, 2005.

\bibitem{Charles:2009aa}
C.~Charles.
\newblock Plasmas for spacecraft propulsion.
\newblock {\em Journal of Physics D: Applied Physics}, 42(16):163001, 2009.

\bibitem{Batishchev:2009aa}
O.~V. Batishchev.
\newblock Minihelicon plasma thruster.
\newblock {\em Plasma Science, IEEE Transactions on}, 37(8):1563, 2009.

\bibitem{Zhu:1989aa}
P.~Zhu and R.~W. Boswell.
\newblock Ar ii laser generated by {L}andau damping of whistler waves at the
  lower hybrid frequency.
\newblock {\em Phys. Rev. Lett.}, 63:2805, 1989.

\bibitem{Boswell:1970ab}
R.~W. Boswell.
\newblock {\em A Study of Waves in Gaseous Plasmas}.
\newblock PhD thesis, Flinders University of South Australia, 1970.

\bibitem{Christopoulos:1974aa}
C.~Christopoulos, R.~W. Boswell, and P.~J. Christiansen.
\newblock Measurements of spatial cyclotron damping in a uniform magnetic
  field.
\newblock {\em Physics Letters A}, 47(3):239, 1974.

\bibitem{Loewenhardt:1991aa}
P.~K. Loewenhardt, B.~D. Blackwell, R.~W. Boswell, G.~D. Conway, and S.~M.
  Hamberger.
\newblock Plasma production in a toroidal heliac by helicon waves.
\newblock {\em Phys. Rev. Lett.}, 67:2792, 1991.

\bibitem{petrzilka:1994aa}
V~Petrzilka and J~A Tataronis.
\newblock Non-resonant currents driven by helicon waves.
\newblock {\em Plasma Physics and Controlled Fusion}, 36(6):1027, 1994.

\bibitem{Hanna:2001aa}
J.~Hanna and C.~Watts.
\newblock Alfv{\'e}n wave propagation in a helicon plasma.
\newblock {\em Physics of Plasmas}, 8(9):4251, 2001.

\bibitem{Bowers:1961aa}
R.~Bowers, C.~Legendy, and F.~Rose.
\newblock Oscillatory galvanomagnetic effect in metallic sodium.
\newblock {\em Phys. Rev. Lett.}, 7:339, 1961.

\bibitem{Rose:1962aa}
F.~E. Rose, M.~T. Taylor, and R.~Bowers.
\newblock Low-frequency magneto-plasma resonances in sodium.
\newblock {\em Phys. Rev.}, 127:1122, 1962.

\bibitem{Harding:1965aa}
G.~N. Harding and P.~C. Thonemann.
\newblock A study of helicon waves in indium.
\newblock {\em Proceedings of the Physical Society}, 85(2):317, 1965.

\bibitem{Woods:1962ab}
L.~C. Woods.
\newblock Hydromagnetic waves in a cylindrical plasma.
\newblock {\em Journal of Fluid Mechanics}, 13:570, 1962.

\bibitem{Woods:1964aa}
L.~C. Woods.
\newblock On the boundary conditions at an insulating wall for hydromagnetic
  waves in a cylindrical plasma.
\newblock {\em Journal of Fluid Mechanics}, 18:401, 1964.

\bibitem{Davies:1969ab}
B.~Davies and P.~J. Christiansen.
\newblock Helicon waves in a gaseous plasma.
\newblock {\em Plasma Physics}, 11(12):987, 1969.

\bibitem{Davies:1970aa}
B.~Davies.
\newblock Helicon wave propagation: effect of electron inertia.
\newblock {\em Journal of Plasma Physics}, 4:43, 1970.

\bibitem{Scime:2007aa}
E.~Scime, R.~Hardin, C.~Biloiu, A.~M. Keesee, and X.~Sun.
\newblock Flow, flow shear, and related profiles in helicon plasmas.
\newblock {\em Physics of Plasmas}, 14(4):043505, 2007.

\bibitem{Mori:2004aa}
Y.~Mori, H.~Nakashima, F.~W. Baity, R.~H. Goulding, M.~D. Carter, and D.~O.
  Sparks.
\newblock High density hydrogen helicon plasma in a non-uniform magnetic field.
\newblock {\em Plasma Sources Science and Technology}, 13(3):424, 2004.

\bibitem{Chen:1996aa}
F.~F Chen, I.~D. Sudit, and M.~Light.
\newblock Downstream physics of the helicon discharge.
\newblock {\em Plasma Sources Science and Technology}, 5(2):173, 1996.

\bibitem{Light:1995aa}
M.~Light, I.~D. Sudit, F.~F. Chen, and D.~Arnush.
\newblock Axial propagation of helicon waves.
\newblock {\em Physics of Plasmas}, 2(11):4094, 1995.

\bibitem{Chang:2011aa}
L.~Chang, M.~J. Hole, and C.~S. Corr.
\newblock A flowing plasma model to describe drift waves in a cylindrical
  helicon discharge.
\newblock {\em Physics of Plasmas}, 18(4):042106, 2011.

\bibitem{Hole:2002aa}
M.~J. Hole, R.~S. Dallaqua, S.~W. Simpson, and E.~Del~Bosco.
\newblock Plasma instability of a vacuum arc centrifuge.
\newblock {\em Phys. Rev. E}, 65:046409, 2002.

\bibitem{Hole:2001aa}
M.~J. Hole and S.~W. Simpson.
\newblock Analytical description of a collisional plasma column in a vacuum arc
  centrifuge.
\newblock {\em Journal of Physics D-Applied Physics}, 34(20):3028, 2001.

\bibitem{Dallaqua:1998aa}
R.~S. Dallaqua, E.~Del~Bosco, R.~P. da~Silva, and S.~W. Simpson.
\newblock Langmuir probe measurements in a vacuum arc plasma centrifuge.
\newblock {\em Ieee Transactions on Plasma Science}, 26(3):1044, 1998.

\bibitem{Gilland:1998aa}
J.~Gilland, R.~Breun, and N.~Hershkowitz.
\newblock Neutral pumping in a helicon discharge.
\newblock {\em Plasma Sources Science and Technology}, 7(3):416, 1998.

\bibitem{Guo:1999aa}
X.~M. Guo, J.~Scharer, Y.~Mouzouris, and L.~Louis.
\newblock Helicon experiments and simulations in nonuniform magnetic field
  configurations.
\newblock {\em Physics of Plasmas}, 6(8):3400, 1999.

\bibitem{Freidberg:2007aa}
J.~P. Freidberg.
\newblock {\em Plasma {P}hysics and {F}usion {E}nergy}.
\newblock Cambridge University Press, 2007.

\bibitem{Lawson:1957aa}
J.~D. Lawson.
\newblock Some criteria for a power producing thermonuclear reactor.
\newblock {\em Proceedings of the Physical Society. Section B}, 70(1):6, 1957.

\bibitem{Wesson:2011aa}
J.~Wesson.
\newblock {\em Tokamaks}.
\newblock Oxford University Press, fourth edition, 2011.

\bibitem{Artsimovich:1972aa}
L.~A. Artsimovich.
\newblock Tokamak devices.
\newblock {\em Nuclear Fusion}, 12(2):215, 1972.

\bibitem{iter:2007aa}
http://www.iter.org/.

\bibitem{Fasoli:2007aa}
A.~Fasoli, C.~Gormenzano, H.~L. Berk, B.~Breizman, S.~Briguglio, D.~S. Darrow,
  N.~Gorelenkov, W.~W. Heidbrink, A.~Jaun, S.~V. Konovalov, R.~Nazikian, J.~M.
  Noterdaeme, S.~Sharapov, K.~Shinohara, D.~Testa, K.~Tobita, Y.~Todo, G.~Vlad,
  and F.~Zonca.
\newblock Chapter 5: Physics of energetic ions.
\newblock {\em Nuclear Fusion}, 47(6):S264, 2007.

\bibitem{Rostoker:1961aa}
N.~Rostoker and A.~C. Kolb.
\newblock Fission of a hot plasma.
\newblock {\em Physical Review}, 124(4):965, 1961.

\bibitem{Freidberg:1978aa}
J.~P. Freidberg and L.~D. Pearlstein.
\newblock Rotational instabilities in a theta-pinch.
\newblock {\em Physics of Fluids}, 21(7):1207, 1978.

\bibitem{Perkins:1963aa}
W.~A. Perkins and R.~F. Post.
\newblock Observation of plasma instability with rotational effects in a mirror
  machine.
\newblock {\em Physics of Fluids}, 6(11):1537, 1963.

\bibitem{Kuo:1964aa}
L.~G. Kuo, E.~G. Murphy, M.~Petravic, and D.~R. Sweetman.
\newblock Experimental and theoretical studies of instabilities in a
  high-energy neutral injection mirror machine.
\newblock {\em Physics of Fluids}, 7(7):988, 1964.

\bibitem{Hooper:1983aa}
E.~B. Hooper, G.~A. Hallock, and J.~H. Foote.
\newblock Low-frequency oscillations in the central cell of the tmx tandem
  mirror experiment.
\newblock {\em Physics of Fluids}, 26(1):314, 1983.

\bibitem{Motley:1963aa}
R.~W. Motley and N.~Dangelo.
\newblock Excitation of electrostatic plasma oscillations near the ion
  cyclotron frequency.
\newblock {\em Physics of Fluids}, 6(2):296, 1963.

\bibitem{Krishnan:1981aa}
M.~Krishnan, M.~Geva, and J.~L. Hirshfield.
\newblock Plasma centrifuge.
\newblock {\em Phys. Rev. Lett.}, 46:36, 1981.

\bibitem{Boswell:1983aa}
R.~W. Boswell and P.~J. Kellogg.
\newblock Characteristics of 2 types of beam-plasma-discharge in a laboratory
  experiment.
\newblock {\em Geophysical Research Letters}, 10(7):565, 1983.

\bibitem{Boswell:1984aa}
R.~W. Boswell, S.~M. Hamberger, P.~J. Kellogg, I.~Morey, and R.~K. Porteous.
\newblock Direct observation of rapid impulsive electron heating during a
  beam-plasma interaction.
\newblock {\em Physics Letters A}, 101(9):501, 1984.

\bibitem{Degeling:1999aa}
A.~W. Degeling, T.~E. Sheridan, and R.~W. Boswell.
\newblock Model for relaxation oscillations in a helicon discharge.
\newblock {\em Physics of Plasmas}, 6(5):1641, 1999.

\bibitem{Greiner:1999aa}
F.~Greiner, O.~Grulke, H.~Thomsen, C.~Lechte, T.~Klinger, and A.~Piel.
\newblock Observation of coherent structures in the turbulent equilibrium of a
  toroidal helicon discharge.
\newblock In {\em Proceedings of the International Conference on Phenomena in
  Ionized Gas, Vol I}, page 175, 1999.

\bibitem{Sun:2005aa}
X.~Sun, C.~Biloiu, and E.~Scime.
\newblock Observation of resistive drift alfv[e-acute]n waves in a helicon
  plasma.
\newblock {\em Physics of Plasmas}, 12(10):102105, 2005.

\bibitem{Ellingboe:1996aa}
A.~R. Ellingboe and R.~W. Boswell.
\newblock Capacitive, inductive and helicon-wave modes of operation of a
  helicon plasma source.
\newblock {\em Physics of Plasmas}, 3(7):2797, 1996.

\bibitem{Boswell:1984ab}
R.~W. Boswell.
\newblock Very efficient plasma generation by whistler waves near the lower
  hybrid frequency.
\newblock {\em Plasma Physics and Controlled Fusion}, 26(10):1147, 1984.

\bibitem{Schroder:2004aa}
Christiane Schroder, Olaf Grulke, Thomas Klinger, and Volker Naulin.
\newblock Spatial mode structures of electrostatic drift waves in a collisional
  cylindrical helicon plasma.
\newblock {\em Physics of Plasmas}, 11(9):4249, 2004.

\bibitem{Schroder:2005aa}
Christiane Schroder, Olaf Grulke, Thomas Klinger, and Volker Naulin.
\newblock Drift waves in a high-density cylindrical helicon discharge.
\newblock {\em Physics of Plasmas}, 12(4):042103, 2005.

\bibitem{Light:2001aa}
M.~Light, F.~F. Chen, and P.~L. Colestock.
\newblock Low frequency electrostatic instability in a helicon plasma.
\newblock {\em Physics of Plasmas}, 8(10):4675, 2001.

\bibitem{Light:2002aa}
M.~Light, F.~F. Chen, and P.~L. Colestock.
\newblock Quiescent and unstable regimes of a helicon plasma.
\newblock {\em Plasma Sources Science and Technology}, 11(3):273, 2002.

\bibitem{Degeling:1999ab}
A.~W. Degeling, T.~E. Sheridan, and R.~W. Boswell.
\newblock Intense on-axis plasma production and associated relaxation
  oscillations in a large volume helicon source.
\newblock {\em Physics of Plasmas}, 6(9):3664, 1999.

\bibitem{Ellis:1980aa}
R.~F. Ellis, E.~Mardenmarshall, and R.~Majeski.
\newblock Collisional drift instability of a weakly ionized argon plasma.
\newblock {\em Plasma Physics and Controlled Fusion}, 22(2):113, 1980.

\bibitem{Grulke:2007aa}
O.~Grulke, S.~Ullrich, T.~Windisch, and T.~Klinger.
\newblock Laboratory studies of drift waves: nonlinear mode interaction and
  structure formation in turbulence.
\newblock {\em Plasma Physics and Controlled Fusion}, 49(12B):B247, 2007.

\bibitem{Horton:1984aa}
W.~Horton and J.~Liu.
\newblock Drift waves in rotating plasmas.
\newblock {\em Physics of Fluids}, 27(8):2067, 1984.

\bibitem{Sutherland:2005aa}
O.~Sutherland, M.~Giles, and R.~Boswell.
\newblock Ion cyclotron production by a four-wave interaction with a helicon
  pump.
\newblock {\em Phys. Rev. Lett.}, 94:205002, 2005.

\bibitem{Miloshevsky:2010aa}
G.~V. Miloshevsky and A.~Hassanein.
\newblock Modelling of kelvin-helmholtz instability and splashing of melt
  layers from plasma-facing components in tokamaks under plasma impact.
\newblock {\em Nuclear Fusion}, 50(11):1, 2010.

\bibitem{Hendel:1968aa}
H.~W. Hendel, T.~K. Chu, and P.~A. Politzer.
\newblock Collisional drift waves-identification stabilization and enhanced
  plasma transport.
\newblock {\em Physics of Fluids}, 11(11):2426, 1968.

\bibitem{Okabayashi:1977aa}
M.~Okabayashi and V.~Arunasalam.
\newblock Study of drift-wave turbulence by microwave-scattering in a toroidal
  plasma.
\newblock {\em Nuclear Fusion}, 17(3):497, 1977.

\bibitem{Pecseli:1983aa}
H.~L. Pecseli, T.~Mikkelsen, and S.~E. Larsen.
\newblock Drift wave turbulence in low-beta plasmas.
\newblock {\em Plasma Physics and Controlled Fusion}, 25(11):1173, 1983.

\bibitem{Liewer:1985aa}
P.~C. Liewer.
\newblock Measurements of microturbulence in tokamaks and comparisons with
  theories of turbulence and anomalous transport.
\newblock {\em Nuclear Fusion}, 25(5):543, 1985.

\bibitem{Klinger:1992aa}
T.~Klinger and A.~Piel.
\newblock Investigations of attractors arising from the interaction of drift
  waves and potential relaxation instabilities.
\newblock {\em Physics of Fluids B-Plasma Physics}, 4(12):3990, 1992.

\bibitem{Poli:2006aa}
F.~M. Poli, S.~Brunner, A.~Diallo, A.~Fasoli, I.~Furno, B.~Labit, S.~H. Muller,
  G.~Plyushchev, and M.~Podesta.
\newblock Experimental characterization of drift-interchange instabilities in a
  simple toroidal plasma.
\newblock {\em Physics of Plasmas}, 13(10):102104, 2006.

\bibitem{Hole:2001ab}
M.~J. Hole.
\newblock {\em Analysis of the Rotating Plasma Column in the Vacuum Arc
  Centrifuge}.
\newblock PhD thesis, UNIVERSITY OF SYDNEY, 2001.

\bibitem{Biloiu:2010aa}
Ioana~A. Biloiu and Earl~E. Scime.
\newblock Ion acceleration in ar--xe and ar--he plasmas. i. electron energy
  distribution functions and ion composition.
\newblock {\em Physics of Plasmas}, 17(11):113508, 2010.

\bibitem{Kline:2003aa}
J.~L. Kline, M.~M. Balkey, P.~A. Keiter, E.~E. Scime, A.~M. Keesee, X.~Sun,
  R.~Hardin, C.~Compton, R.~F. Boivin, and M.~W. Zintl.
\newblock Ion dynamics in helicon sources.
\newblock {\em Physics of Plasmas}, 10(5):2127, 2003.

\bibitem{Plihon:2005aa}
N.~Plihon, C.~S. Corr, and P.~Chabert.
\newblock Double layer formation in the expanding region of an inductively
  coupled electronegative plasma.
\newblock {\em Applied Physics Letters}, 86(9):091501, 2005.

\bibitem{Spitzer:1962aa}
L.~Spitzer.
\newblock {\em Physic of Fully Ionized Gases}.
\newblock John Wiley and Sons, 1962.

\bibitem{Shinohara:2001ab}
S.~Shinohara, N.~Matsuoka, and S.~Matsuyama.
\newblock Establishment of strong velocity shear and plasma density profile
  modification with associated low frequency fluctuations.
\newblock {\em Physics of Plasmas}, 8(4):1154, 2001.

\bibitem{Chen:2006aa}
G.~Chen, A.~V. Arefiev, R.~D. Bengtson, B.~N. Breizman, C.~A. Lee, and L.~L.
  Raja.
\newblock Resonant power absorption in helicon plasma sources.
\newblock {\em Physics of Plasmas}, 13(12):123507, 2006.

\bibitem{Takechi:1999aa}
S.~Takechi and S.~Shinohara.
\newblock Rf wave propagation in bounded plasma under divergent and convergent
  magnetic field configurations.
\newblock {\em Japanese Journal of Applied Physics}, 38(Part 2, No. 11A):L1278,
  1999.

\bibitem{Impedans:2012aa}
http://www.impedans.com/.

\bibitem{Boswell:2012aa}
R.~W. Boswell, private communication.

\bibitem{Franck:2002aa}
C.~M. Franck, O.~Grulke, and T.~Klinger.
\newblock Magnetic fluctuation probe design and capacitive pickup rejection.
\newblock {\em Review of Scientific Instruments}, 73(11):3768, 2002.

\bibitem{Light:1995ab}
M.~Light and F.~F. Chen.
\newblock Helicon wave excitation with helical antennas.
\newblock {\em Physics of Plasmas}, 2(4):1084, 1995.

\bibitem{Degeling:2004aa}
A.~W. Degeling, G.~G. Borg, and R.~W. Boswell.
\newblock Transitions from electrostatic to electromagnetic whistler wave
  excitation.
\newblock {\em Physics of Plasmas}, 11(5):2144, 2004.

\bibitem{Arnush:2000aa}
D.~Arnush.
\newblock The role of trivelpiece--gould waves in antenna coupling to helicon
  waves.
\newblock {\em Physics of Plasmas}, 7(7):3042, 2000.

\bibitem{Shinohara:2002aa}
S.~Shinohara and K.~P. Shamrai.
\newblock Effect of electrostatic waves on a rf field penetration into highly
  collisional helicon plasmas.
\newblock {\em Thin Solid Films}, page 215, 2002.

\bibitem{Sudit:1996aa}
I.~D. Sudit and F.~F Chen.
\newblock Discharge equilibrium of a helicon plasma.
\newblock {\em Plasma Sources Science and Technology}, 5(1):43, 1996.

\bibitem{Strutt:1887aa}
J.~W. (Lord~Rayleigh) Strutt.
\newblock On the maintenance of vibrations by forces of double frequency, and
  on the propagation of waves through a medium endowed with periodic structure.
\newblock {\em Philosophical Magazine}, 24:145, 1887.

\bibitem{Anderson:1958aa}
P.~W. Anderson.
\newblock Absence of diffusion in certain random lattices.
\newblock {\em Phys. Rev.}, 109:1492, 1958.

\bibitem{Mott:1968aa}
N.~F. Mott.
\newblock Conduction in non-crystalline systems .i. localized electronic states
  in disordered systems.
\newblock {\em PHILOSOPHICAL MAGAZINE}, 17(150):1259, 1968.

\bibitem{John:1987aa}
S.~John.
\newblock Strong localization of photons in certain disordered dielectric
  superlattices.
\newblock {\em Phys. Rev. Lett.}, 58:2486, 1987.

\bibitem{Yablonovitch:1991aa}
E.~Yablonovitch, T.~J. Gmitter, R.~D. Meade, A.~M. Rappe, K.~D. Brommer, and
  J.~D. Joannopoulos.
\newblock Donor and acceptor modes in photonic band structure.
\newblock {\em Phys. Rev. Lett.}, 67:3380, 1991.

\bibitem{Figotin:1997aa}
A.~Figotin and A.~Klein.
\newblock Localized classical waves created by defects.
\newblock {\em Journal of Statistical Physics}, 86:165, 1997.

\bibitem{Chu:1992aa}
M.~S. Chu, J.~M. Greene, L.~L. Lao, A.~D. Turnbull, and M.~S. Chance.
\newblock {{A numerical study of the high-n shear Alfv\'{e}n spectrum gap and
  the high-n gap mode}}.
\newblock {\em Physics of Fluids B: Plasma Physics}, 4(11):3713, 1992.

\bibitem{DIppolito:1980aa}
D.~A. D'Ippolito.
\newblock Mode coupling in a toroidal sharp-boundary plasma. i. weak-coupling
  limit.
\newblock {\em Plasma Physics}, 22(12):1091, 1980.

\bibitem{Dewar:1974aa}
R.~L. Dewar, R.~C. Grimm, J.~L. Johnson, E.~A. Frieman, J.~M. Greene, and P.~H.
  Rutherford.
\newblock Long-wavelength kink instabilities in low-pressure, uniform axial
  current, cylindrical plasmas with elliptic cross sections.
\newblock {\em Physics of Fluids}, 17(5):930, 1974.

\bibitem{Betti:1992aa}
R.~Betti and J.~P. Freidberg.
\newblock {{Stability of Alfv\'{e}n gap modes in burning plasmas}}.
\newblock {\em Physics of Fluids B: Plasma Physics}, 4(6):1465, 1992.

\bibitem{Nakajima:1992aa}
N.~Nakajima, C.~Z. Cheng, and M.~Okamoto.
\newblock {{High-n helicity-induced shear Alfv\'{e}n eigenmodes}}.
\newblock {\em Physics of Fluids B: Plasma Physics}, 4(5):1115, 1992.

\bibitem{Kolesnichenko:2001aa}
Y.~I. Kolesnichenko, V.~V. Lutsenko, H.~Wobig, Y.~V. Yakovenko, and O.~P.
  Fesenyuk.
\newblock Alfv\'{e}n continuum and high-frequency eigenmodes in optimized
  stellarators.
\newblock {\em Physics of Plasmas}, 8(2):491, 2001.

\bibitem{Duong:1993aa}
H.~H. Duong, W.~W. Heidbrink, E.~J. Strait, T.~W. Petrie, R.~Lee, R.~A. Moyer,
  and J.~G. Watkins.
\newblock Loss of energetic beam ions during tae instabilities.
\newblock {\em Nuclear Fusion}, 33(5):749, 1993.

\bibitem{White:1995aa}
R.~B. White, E.~Fredrickson, D.~Darrow, M.~Zarnstorff, R.~Wilson, S.~Zweben,
  K.~Hill, Yang Chen, and Guoyong Fu.
\newblock {{Toroidal Alfv\'{e}n eigenmode-induced ripple trapping}}.
\newblock {\em Physics of Plasmas}, 2(8):2871, 1995.

\bibitem{Cheng:1985aa}
C.~Z. Cheng~L. Chen and M.~S. Chance.
\newblock {{High-n ideal and resistive shear Alfv\'{e}n waves in tokamaks}}.
\newblock {\em Annals of Physics}, 161(1):21, 1985.

\bibitem{Wong:1999aa}
K.~L. Wong.
\newblock {{A review of Alfv\'{e}n eigenmode observations in toroidal
  plasmas}}.
\newblock {\em Plasma Physics and Controlled Fusion}, 41(1):R1, 1999.

\bibitem{Coppi:1986aa}
B.~Coppi, S.~Cowley, R.~Kulsrud, P.~Detragiache, and F.~Pegoraro.
\newblock High-energy components and collective modes in thermonuclear plasmas.
\newblock {\em Physics of Fluids}, 29(12):4060, 1986.

\bibitem{Gorelenkov:1995aa}
N.~N. Gorelenkov and C.~Z. Cheng.
\newblock Alfv\'{e}n cyclotron instability and ion cyclotron emission.
\newblock {\em Nuclear Fusion}, 35:1743, 1995.

\bibitem{Kamelander:1996aa}
G.~Kamelander and Ya.~I. Kolesnichenko.
\newblock Localized whistler eigenmodes in tokamaks.
\newblock {\em Physics of Plasmas}, 3:4102, 1996.

\bibitem{Braginskii:1965aa}
S.~I. Braginskii.
\newblock Transport processes in a plasma.
\newblock {\em Reviews of Plasma Physics}, 1:302, 1965.

\bibitem{Jones:1993aa}
D.~J. Jones.
\newblock Use of a shooting method to compute eigenvalues of fourth-order
  two-point boundary value problems.
\newblock {\em Journal of Computational and Applied Mathematics}, 47(3):395,
  1993.

\end{thebibliography}

\appendix
\chapter{A two-fluid electromagnetic flowing plasma model}\label{app}  % the * means don't put a number in the title

\section{TOEFL model}\label{tfma}
To describe the electromagnetic perturbations in a flowing plasma cylinder, a TwO-fluid Electromagnetic Flowing pLasma (TOEFL) model is developed. It is a generalisation of the electrostatic vacuum arc centrifuge (VAC) model developed by Hole et al.\cite{Hole:2002aa} to include magnetic field oscillations. The TOEFL model comprises: 
\begin{equation}\label{a_1}
\frac{\partial\mathbf{u_i}}{\partial\tau}+\left(\mathbf{u_i}\cdot\nabla\right)\mathbf{u_i}=-\psi Z\left[\left(1+\frac{\lambda_T}{Z}\right)\nabla l_i-\frac{1}{\beta_{ct} e^{l_i}}\left(\nabla\times\nabla\times\mathbf{A_n}\right)\times(\nabla\times\mathbf{A_n})\right],
\end{equation}
\begin{equation}\label{a_2}
\nabla\chi-\nabla l_i=-\frac{\partial\mathbf{A_n}}{\partial\tau}+\left[\mathbf{u_i}-\frac{1}{\beta_{ct} e^{l_i}}\left(\nabla\times\nabla\times\mathbf{A_n}\right)\right]\times(\nabla\times\mathbf{A_n})-\frac{\delta_{rs}}{\psi Z\beta_{ct}}\mathbf{\tilde{\xi}}\cdot\left(\nabla\times\nabla\times\mathbf{A_n}\right),
\end{equation}
\begin{equation}\label{a_3}
-\frac{\partial l_i}{\partial\tau}=\nabla\cdot\mathbf{u_i}+\mathbf{u_i}\cdot\nabla l_i, 
\end{equation}
\begin{equation}\label{a_4}
\nabla\cdot\mathbf{A_n}=0, 
\end{equation}
with $\mathbf{A_n}=e\mathbf{A}R\omega_{ci}/(k_B T_e)$ the normalisation form of the magnetic vector potential $\mathbf{A}$ and $\beta_{ct}=[\mu_0 e^2 Z n_i(0)/(k_B T_e)](R\omega_{ci})^2 R^2$ a convenient constant. Equation~(\ref{a_1}) is the difference between the momentum equations for ions and electrons. Equation~(\ref{a_2}) is the momentum equation for electrons in which Ampere's law
\begin{equation}\label{a_5}
\nabla\times\nabla\times\mathbf{A_n}=\beta_{ct} e^{l_i}(\mathbf{u_i}-\mathbf{u_e})
\end{equation}
has been used. Equation~(\ref{a_3}) is the continuity equation for ions, and Eq.~(\ref{a_4}) is the Coulomb gauge. The electric field $\mathbf{E}$ and magnetic field $\mathbf{B}$ are normalised as:
\begin{equation}\label{a_6}
\mathbf{E}_n=\frac{e R}{k_B T_e}\mathbf{E}=-\nabla\chi-\frac{\partial\mathbf{A_n}}{\partial\tau},
\end{equation}
\begin{equation}\label{a_7}
\mathbf{B}_n=\frac{e R^2\omega_{ci}}{(k_B T_e)}\mathbf{B}=\nabla\times\mathbf{A_n}, 
\end{equation}
respectively. Other terms are the same as defined in Sec.~\ref{eqt2}. This model can be easily benchmarked against the electrostatic VAC model by choosing a time-independent $\mathbf{A_n}$ whose curl equals zero. 

\section{Steady-state solution}\label{stda}
\begin{figure}[ht]
\begin{center}$
\begin{array}{ll}
(a)&(b)\\
\hspace{-0.07cm}\includegraphics[width=0.47\textwidth,angle=0]{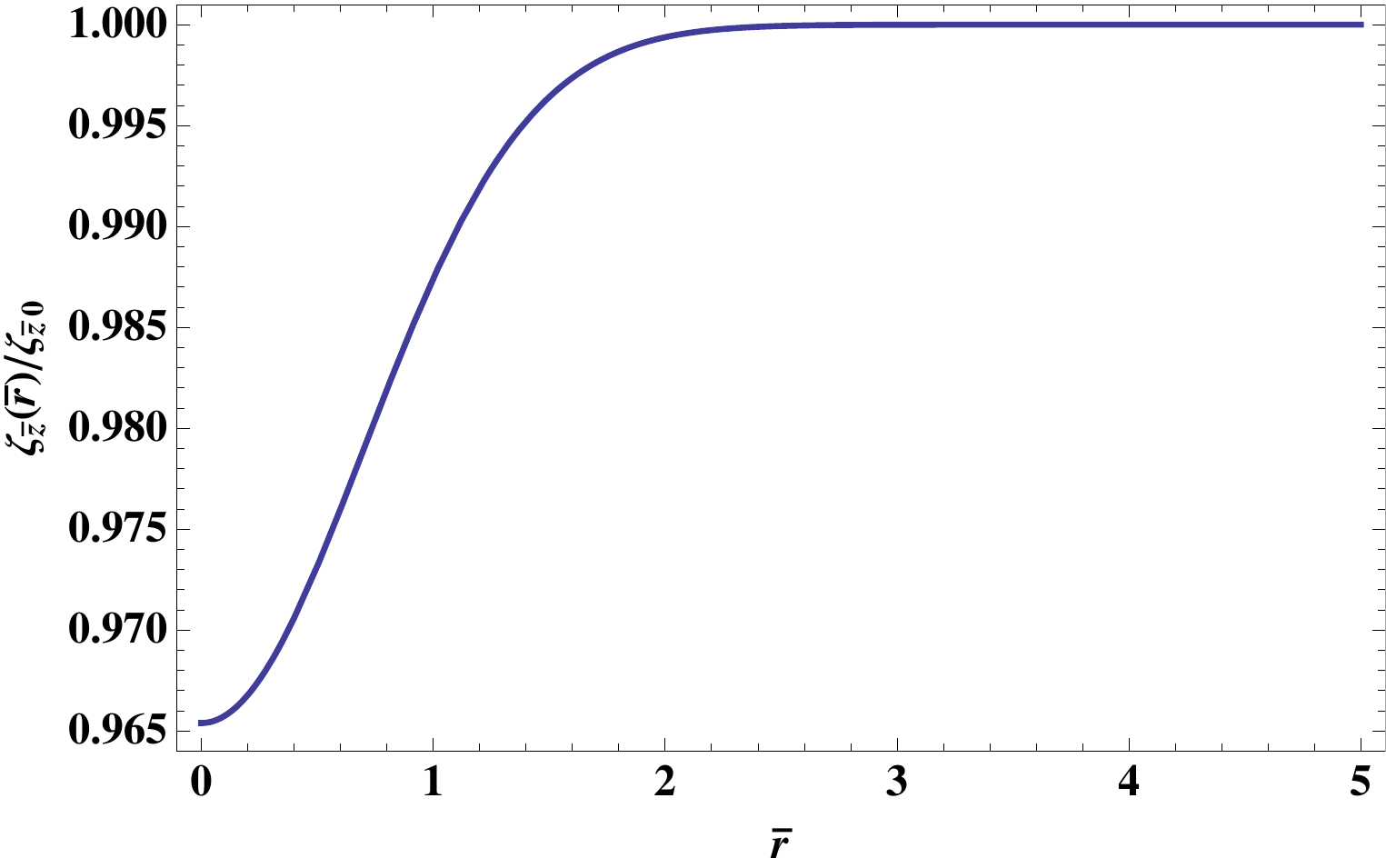}&\hspace{0.25cm}\includegraphics[width=0.47\textwidth,angle=0]{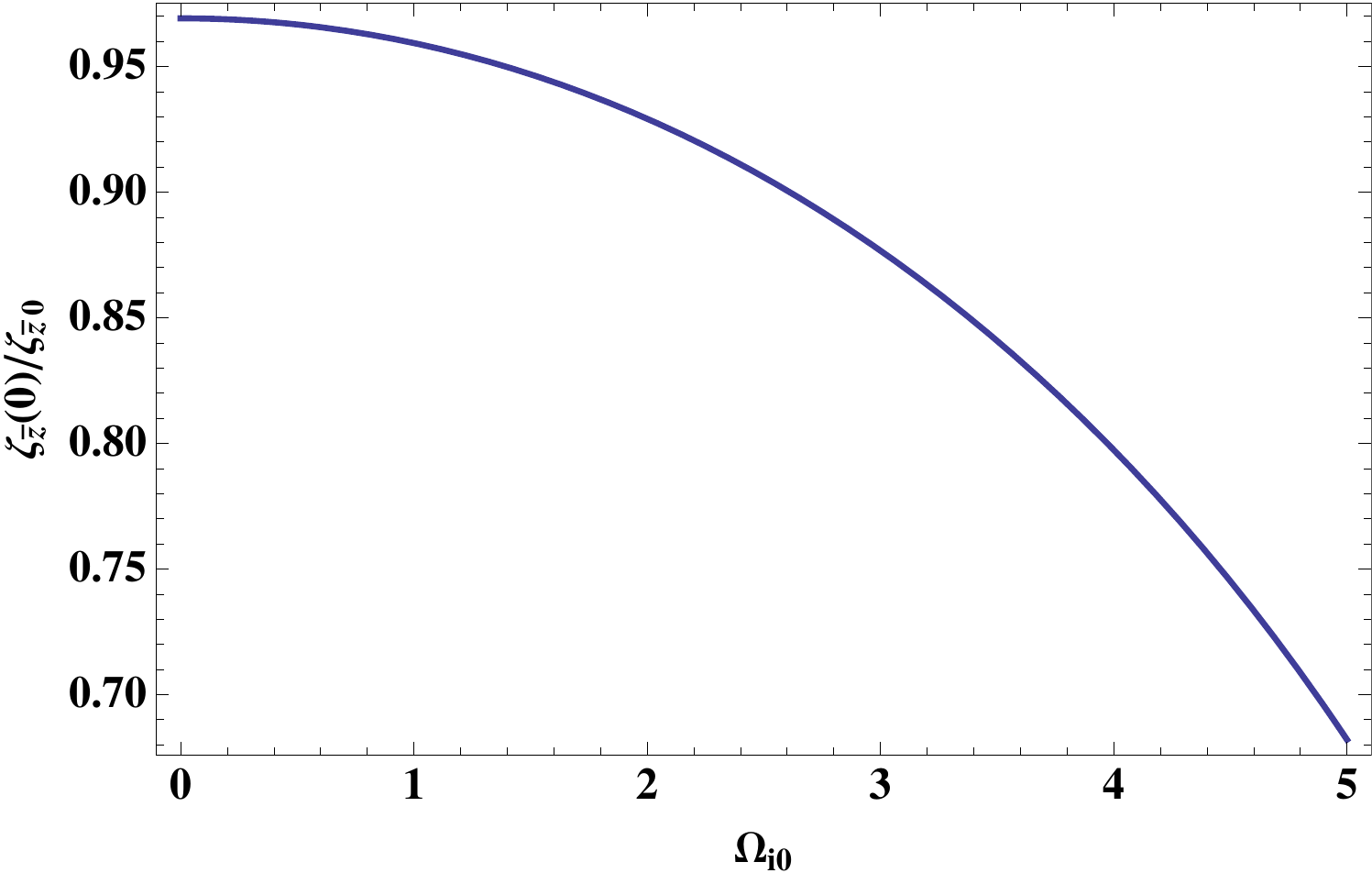}\\
\end{array}$
\end{center}
\caption{Diamagnetic effect $\zeta_{\bar{z}}(\bar{r})/\zeta_{\bar{z}0}$ as a function of: (a) $\bar{r}$, and (b) $\Omega_{i0}$.}
\label{fga_1}
\end{figure}

\begin{figure}[ht]
\begin{center}$
\begin{array}{ll}
(a)&(b)\\
\hspace{-0.07cm}\includegraphics[width=0.49\textwidth,angle=0]{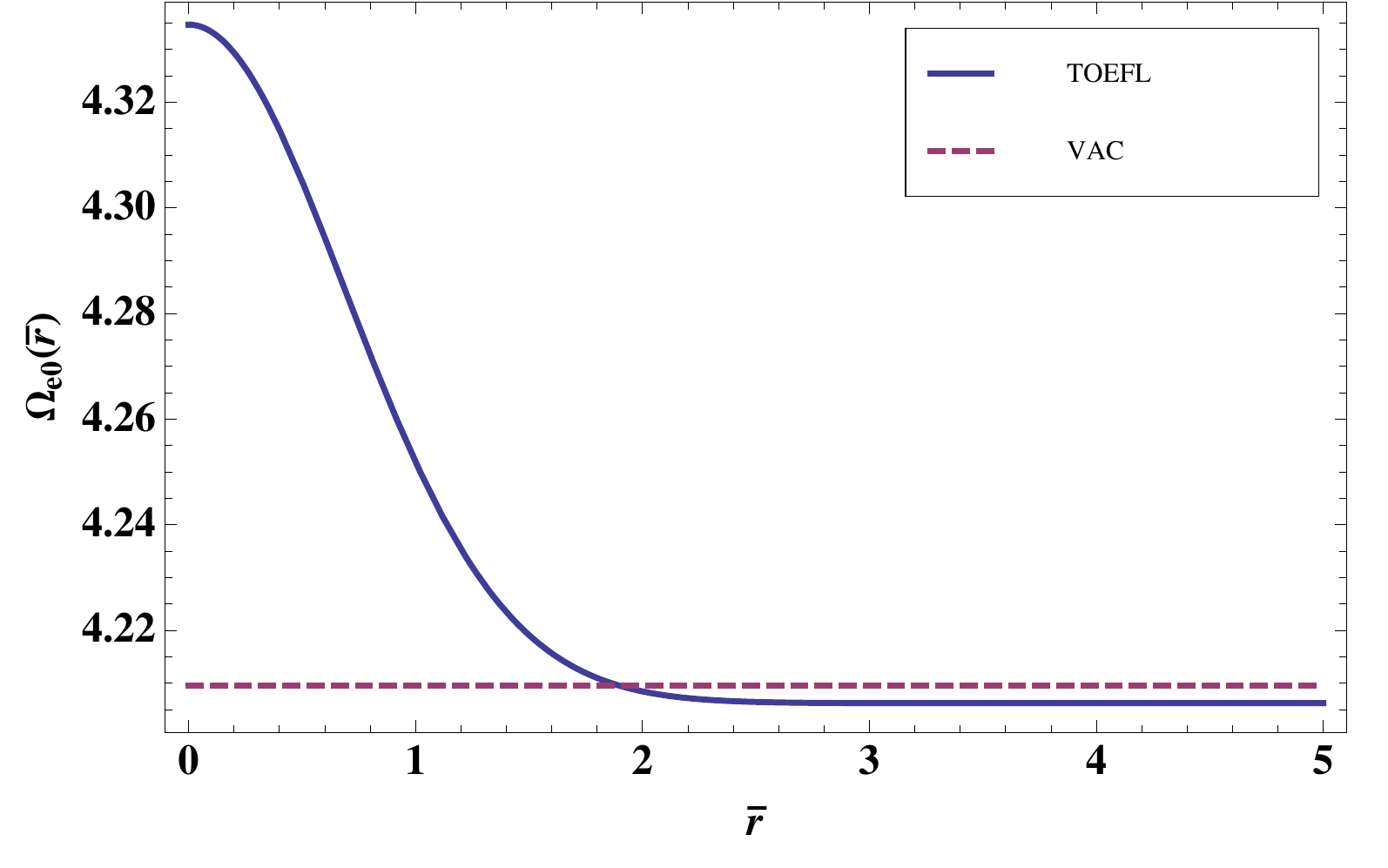}&\hspace{0.25cm}\includegraphics[width=0.49\textwidth,angle=0]{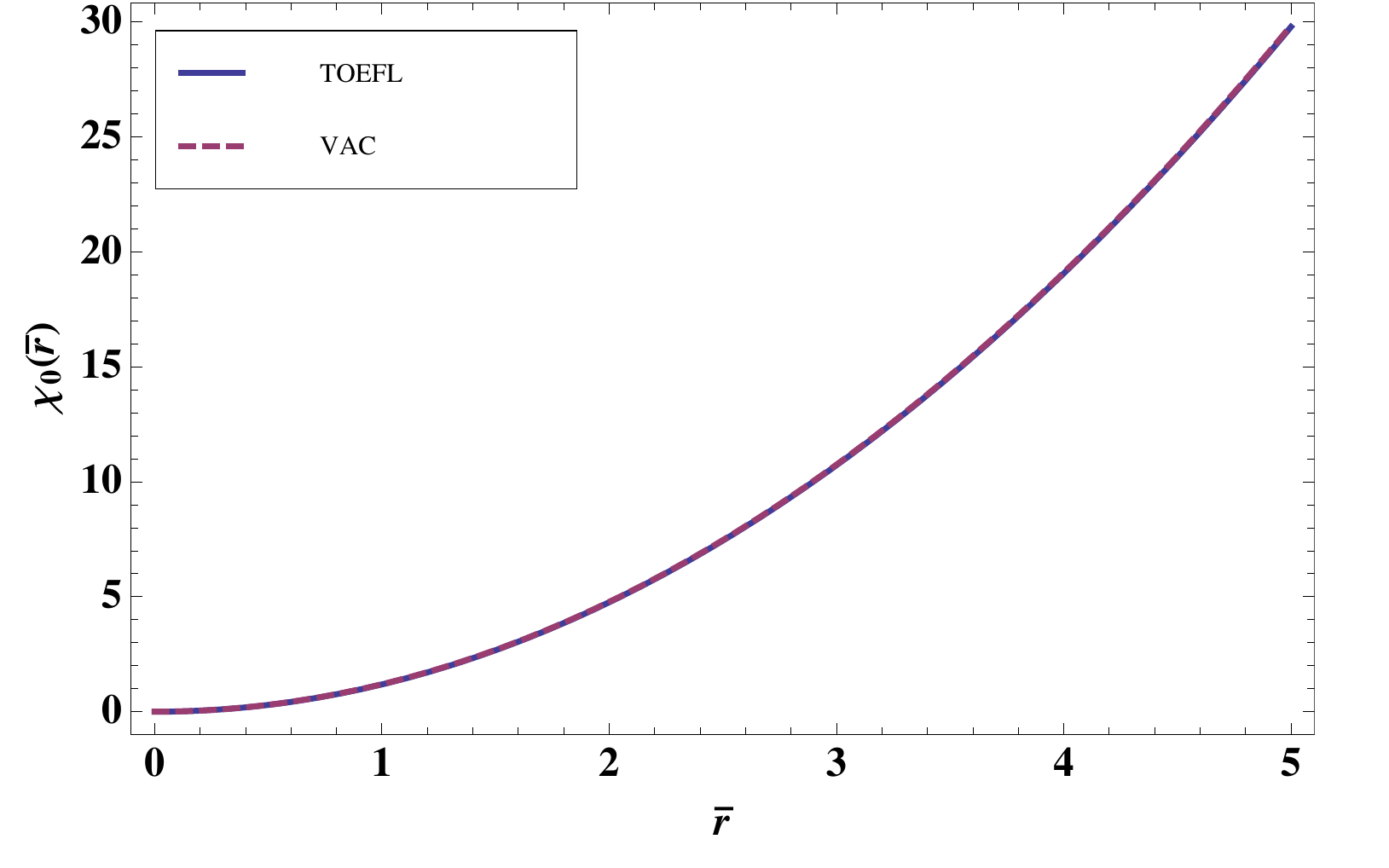}\\
\end{array}$
\end{center}
\caption{Steady-state solutions from the TOEFL model and the electrostatic VAC model: (a) $\Omega_{e0}(\bar{r})$, (b) $\chi_0(\bar{r})$.}
\label{fga_2}
\end{figure}

In the steady state, $\mathbf{A_{n0}}$ is independent on time, the ion density is $l_{i0}(\bar{r})=-\bar{r}^{2}$, and the ion and electron velocities are $\mathbf{u_{i0}}=(0, \bar{r}\Omega_{i0}, u_{i\bar{z}0})$ and $\mathbf{u_{e0}}=(0, \bar{r}\Omega_{e0}(\bar{r}), u_{e\bar{z}0}(\bar{r}))$, respectively. It can be seen from the axial component of Eq.~(\ref{a_2}) that $u_{e\bar{z}0}(\bar{r})=u_{i\bar{z}0}$ (through Eq.~(\ref{a_5})). Hence, there is no equilibrium axial current in the quasi-neutral plasma, and consequently no azimuthal magnetic field. It is thereby reasonable to assume an entire axial magnetic field $\nabla\times\mathbf{A_{n0}}=(0, 0, \zeta_{\bar{z}}(\bar{r}))$,\cite{Hole:2001aa} which gives
\begin{equation}\label{a_8}
[\Omega_{i0}-\Omega_{e0}(\bar{r})]\bar{r}=-\frac{e^{\bar{r}^2}}{\beta_{ct}}\frac{\partial\zeta_{\bar{z}}(\bar{r})}{\partial \bar{r}}
\end{equation}
through Eq.~(\ref{a_5}). Substituting Eq.~(\ref{a_8}) into Eq.~(\ref{a_1}) and integrating over $\bar{r}$ yield
\begin{equation}\label{a_9}
\zeta_{\bar{z}}(\bar{r})=\zeta_{\bar{z}0}\sqrt{1-\frac{\omega_{pi}^2(0)}{c^2}R^2[\Omega_{i0}^2+2\psi(\lambda_T+Z)]e^{-\bar{r}^2}}. 
\end{equation}
Here, $\zeta_{\bar{z}0}=e B_0 R^2 \omega_{ci}/(k_B T_e)$ is the normalised externally applied magnetic field lying along $\mathbf{\hat{\bar{z}}}$. The second term inside of the square root of Eq.~(\ref{a_9}) represents the diamagnetic effect caused by the plasma rotation and thermal motion. For typical VAC parameters described in Tab.~\ref{tb2_1}, Fig.~\ref{fga_1} shows the variation of this effect as a function of radius and ion rotation frequency. It can be seen that the diamagnetic field becomes stronger when it is closer to the plasma core and when the ion rotation frequency is higher. However, overall, it is very small compared with the externally applied magnetic field (about $3.5\%$ on axis). 
Substituting Eq.~(\ref{a_9}) into Eq.~(\ref{a_8}) gives the equilibrium electron rotation frequency
\begin{equation}\label{a_10}
\Omega_{e0}(\bar{r})=\Omega_{i0}+\frac{\zeta_{\bar{z}0}}{\beta_{ct}}\frac{\varsigma}{\sqrt{1-\varsigma e^{-\bar{r}^2}}}
\end{equation}
with
\begin{equation}\label{a_11}
\varsigma=\frac{\omega_{pi}^2(0)}{c^2}R^2[\Omega_{i0}^2+2\psi(\lambda_T+Z)].
\end{equation}
Substituting Eq.~(\ref{a_10}) and Eq.~(\ref{a_9}) into Eq.~(\ref{a_2}) and integrating over $\bar{r}$ give the equilibrium electric potential
\begin{equation}\label{a_12}
\chi_0(\bar{r})=\chi_c-\bar{r}^2+\frac{\zeta_{\bar{z}0}^2}{2\beta_{ct}}\varsigma \bar{r}^2+\Omega_{i0}\zeta_{\bar{z}0}\left\{\frac{\bar{r}^2}{2}-f(\bar{r})+\mathrm{ln}[1+f(\bar{r})]\right\}, 
\end{equation}
where $\chi_c$ is an arbitrary reference potential and $f(\bar{r})=\sqrt{1-\varsigma e^{-\bar{r}^2}}$. To compare with the electrostatic VAC model, the plasma-induced magnetic field is now negleted, namely $\nabla\times\mathbf{A_{n0}}=(0, 0, \zeta_{\bar{z}0})$, and $\Omega_{e0}(\bar{r})$ and $\chi_0(\bar{r})$ are solved via a similar method. This gives: 
\begin{equation}\label{a_13}
\Omega_{e0}(\bar{r})=\Omega_{i0}(1+\Omega_{i0})+2\psi(\lambda_T+Z),
\end{equation}
\begin{equation}\label{a_14}
\chi_0(\bar{r})=\chi_c+\frac{\Omega_{i0}}{2\psi Z}(1+\Omega_{i0})\bar{r}^2+\frac{\lambda_T}{Z}\bar{r}^{2},
\end{equation}
and so $\Omega_{e0}$ is a constant. Figure~\ref{fga_2}(a) shows the comparison between Eq.~(\ref{a_10}) and Eq.~(\ref{a_13}), and Fig.~\ref{fga_2}(b) shows the comparison between Eq.~(\ref{a_12}) and Eq.~(\ref{a_14}). 
It can be seen that the inclusion of the plasma-induced magnetic field increases the equilibrium electron rotation frequency about $3\%$ on axis, but decreases it noticeably where $\bar{r}>1.9$. The equilibrium electric potentials from the TOEFL model and the VAC model are the same.

\section{Perturbation treatment}\label{ptba}
To explore normal modes of a flowing plasma system, the TOEFL model is linearised with perturbations of the form (see Eq.~(\ref{eq2_7}))
\begin{equation}\label{a_16}
\zeta (\tau, \bar{r}, \theta, \bar{z})=\zeta_0(\bar{r})+\varepsilon \zeta_1(\bar{r}) e^{i(m \theta+k_{\bar{z}} \bar{z}-\omega \tau)}, 
\end{equation}
where $\zeta$ is a place holder for ($A_{n\bar{r}1}$, $A_{n\theta 1}$, $A_{n\bar{z}1}$), $l_{i1}$, $X_1$, and ($\varphi_{\bar{r}i1}$, $\Omega_{i1}$, $u_{i\bar{z}1}$), and the subscripts $0$ and $1$ label equilibrium and perturbed parts, respectively. A matrix form of the perturbed TOEFL model is then obtained
\begin{equation}\label{a_15}
\mathbf{M}\left(
\begin{array}{c}
A_{n\bar{r}1}'(\bar{r})\\
A_{n\theta1}'(\bar{r})\\
A_{n\bar{z}1}'(\bar{r})\\
l_{i1}'(\bar{r})\\
X_1'(\bar{r})\\
\varphi_{\bar{r}i1}'(\bar{r})\\
A_{n\theta1}''(\bar{r})\\
A_{n\bar{z}1}''(\bar{r})
\end{array}
\right)=\mathbf{N}\left(
\begin{array}{c}
A_{n\bar{r}1}(\bar{r})\\
A_{n\theta1}(\bar{r})\\
A_{n\bar{z}1}(\bar{r})\\
l_{i1}(\bar{r})\\
X_1(\bar{r})\\
\varphi_{\bar{r}i1}(\bar{r})\\
\Omega_{i1}(\bar{r})\\
u_{i\bar{z}1}(\bar{r})
\end{array}
\right),
\end{equation}
where
\tiny
\[
\mathbf{M}=\left(
\begin{array}{cccccccc}
\frac{i m f(\bar{r}) e^{\bar{r}^2}}{\beta_{ct} \bar{r}}&-\frac{[f(\bar{r})+\bar{r} f'(\bar{r})]e^{\bar{r}^2}}{\beta_{ct} \bar{r}}&0&-\Psi&0&0&-\frac{f(\bar{r})e^{\bar{r}^2}}{\beta_{ct}}&0\\
0&-\frac{i m f(\bar{r}) e^{\bar{r}^2}}{\beta_{ct} \bar{r}}&-\frac{i k_{\bar{z}} f(\bar{r}) e^{\bar{r}^2}}{\beta_{ct}}&0&0&0&0&0\\
0&0&0&0&0&0&0&0\\
-\frac{i m f(\bar{r}) e^{\bar{r}^2}}{\psi Z \beta_{ct} \bar{r}}&\frac{-i m\delta_{rs}\xi+\Omega_{i0}\psi Z\beta_{ct} \bar{r}^2+[f(\bar{r})+\bar{r} f'(\bar{r})]e^{\bar{r}^2}}{\psi Z\beta_{ct} \bar{r}}&u_{i\bar{z}0}-\frac{i k_{\bar{z}}\delta_{rs}\xi}{\psi Z\beta_{ct}}&0&\frac{\Psi}{\psi Z}&0&\frac{f(\bar{r})e^{\bar{r}^2}}{\psi Z\beta_{ct}}&0\\
-\frac{i m\delta_{rs}\xi}{\psi Z\beta_{ct} \bar{r}}&\frac{\delta_{rs}\xi+i m f(\bar{r}) e^{\bar{r}^2}}{\psi Z\beta_{ct} \bar{r}}&\frac{i k_{\bar{z}} f(\bar{r}) e^{\bar{r}^2}}{\psi Z\beta_{ct}}&0&0&0&\frac{\delta_{rs}\xi}{\psi Z\beta_{ct}}&0\\
-\frac{i k_{\bar{z}}\delta_{rs}}{\psi Z\beta_{ct}}&0&\frac{\delta_{rs}}{\psi Z\beta_{ct} \bar{r}}&0&0&0&0&\frac{\delta_{rs}}{\psi Z \beta_{ct}}\\
0&0&0&0&0&-x&0&0\\
-1&0&0&0&0&0&0&0
\end{array}
\right)
\]
\normalsize
and
\scriptsize
\[
\hspace{-1cm}\mathbf{N}=\left(
\begin{array}{cccccccc}
\frac{i m [f(\bar{r})-\bar{r} f'(\bar{r})]e^{\bar{r}^2}}{\beta_{ct} \bar{r}^2}&-\frac{\left[f(\bar{r})+k_{\bar{z}}^2 \bar{r}^2 f(\bar{r})-\bar{r} f'(\bar{r})\right]e^{\bar{r}^2}}{\beta_{ct} \bar{r}^2}\\
\frac{\left(m^2+k_{\bar{z}}^2 \bar{r}^2\right)f(\bar{r})e^{\bar{r}^2}}{\beta_{ct} \bar{r}^2}&\frac{i m f(\bar{r}) e^{\bar{r}^2}}{\beta_{ct} \bar{r}^2}\\
0&\frac{i k_{\bar{z}} f'(\bar{r}) e^{\bar{r}^2}}{\beta_{ct}}\\
\frac{\delta_{rs}\xi\left(m^2+k_{\bar{z}}^2 \bar{r}^2\right)-i \psi Z\beta_{ct}\varpi \bar{r}^2-i m [f(\bar{r})-\bar{r} f'(\bar{r})]e^{\bar{r}^2}}{\psi Z\beta_{ct} \bar{r}^2}&\frac{i m \delta_{rs}\xi-\psi Z\beta_{ct} \Omega_{i0} \bar{r}^2+\left[f(\bar{r})+k_{\bar{z}}^2 \bar{r}^2 f(\bar{r})-\bar{r} f'(\bar{r})\right]e^{\bar{r}^2}}{\psi Z\beta_{ct} \bar{r}^2}\\
-\frac{i m\delta_{rs}\xi+\left(m^2+k_{\bar{z}}^2 \bar{r}^2\right)f(\bar{r})e^{\bar{r}^2}}{\psi Z\beta_{ct} \bar{r}^2}&\frac{\delta_{rs}\xi\left(1+k_{\bar{z}}^2 \bar{r}^2\right)-i \psi Z\beta_{ct}(m\Omega_{i0}+\varpi)\bar{r}^2-i m f(\bar{r}) e^{\bar{r}^2}}{\psi Z \beta_{ct} \bar{r}^2}\\
\frac{i k_{\bar{z}} \delta_{rs}}{\psi Z\beta_{ct} \bar{r}}&-\frac{i k_{\bar{z}}\left[-i m\delta_{rs}+\psi Z\beta_{ct}\Omega_{i0}\bar{r}^2+\bar{r} f'(\bar{r}) e^{\bar{r}^2}\right]}{\psi Z\beta_{ct} \bar{r}}\\
0&0\\
\frac{1}{\bar{r}}&\frac{i m}{\bar{r}}
\end{array}
\right. ...
\]
\[
\hspace{1.5cm}...\left.
\begin{array}{cccccccc}
\frac{m k_{\bar{z}} f(\bar{r})e^{\bar{r}^2}}{\beta_{ct} \bar{r}}&-\frac{f(\bar{r})f'(\bar{r})e^{\bar{r}^2}}{\psi Z\beta_{ct}}&0&-i\varpi \bar{r}&-2\Omega_{i0}\bar{r}&0\\
0&\frac{i m\Psi}{\bar{r}}&0&2\Omega_{i0}\bar{r}&-i\varpi \bar{r}&0\\
-\frac{i m f'(\bar{r})e^{\bar{r}^2}}{\beta_{ct} \bar{r}}&i k_{\bar{z}} \Psi&0&0&0&-i \varpi\\
-\frac{m k_{\bar{z}} f(\bar{r}) e^{\bar{r}^2}}{\psi Z\beta_{ct} \bar{r}}&\frac{f(\bar{r})f'(\bar{r})e^{\bar{r}^2}}{\psi^2 Z^2\beta_{ct}}&0&0&-\frac{\bar{r} f(\bar{r})}{\psi Z}&0\\
-\frac{m(\delta_{rs}\xi k_{\bar{z}}+i\psi Z\beta_{ct} u_{i\bar{z}0})}{\psi Z\beta_{ct} \bar{r}}&0&-\frac{i m\Psi}{\psi Z \bar{r}}&\frac{\bar{r} f(\bar{r})}{\psi Z}&0&0\\
\frac{m^2\delta_{rs}-i\psi Z\beta_{ct}(k_{\bar{z}} u_{i\bar{z}0}+\varpi)\bar{r}^2+i m \bar{r} f'(\bar{r})e^{\bar{r}^2}}{\psi Z\beta_{ct} \bar{r}^2}&0&-\frac{i k_{\bar{z}}\Psi}{\psi Z}&0&0&0\\
0&-i\varpi&0&2-2 \bar{r}^2&i m &i k_{\bar{z}}\\
i k_{\bar{z}}&0&0&0&0&0
\end{array}
\right).
\]
\normalsize
This matrix includes eight equations with eight unknowns, and can be solved in a similar way as used in Chapter~\ref{chp2}: by reducing the number of equations through substituting unknowns, and shooting for the dispersion relation through matching the eigenfunction to certain boundary conditions. To shoot for the dispersion relation, the flute mode of this model, which has $k_{\bar{z}}=0$ and is the starting point of the shooting procedure, needs to be solved first. For $k_{\bar{z}}=0$, we have reduced the matrix to two coupled third-order ordinary differential equations in terms of $\varphi_{\bar{r}i1}$ and $A_{n\bar{r}1}$. Solving the flute mode through a double-shooting method is in progress. Details about the method can be found in \cite{Jones:1993aa}.

\label{app}
\end{document}